\pdfoutput=1
\documentclass[
	affil-it, 
	auth-lg, 
	british, 
	compression, 
	contents, 
	custom-packages={simpler-wick,tabularx}, 
	references={bigone}, 
]{lib/preprint}

\begin{document}

	\hypersetup{
		pdftitle = {Extremal Black Holes from Homotopy Algebras},
		pdfauthor = {Jan Gutowski,Chettha Saelim,Martin Wolf},
		pdfkeywords = {}
	}

	\date{\today}

	\email{j.gutowski@surrey.ac.uk,c.saelim@surrey.ac.uk,m.wolf@surrey.ac.uk}

	\preprint{DMUS--MP--25/03}

	\title{Extremal Black Holes from Homotopy Algebras}

	\author[a]{Jan~Gutowski\,\orcidlink{0000-0001-8807-3818}\,}
	\author[a]{Chettha~Saelim\,\orcidlink{0009-0006-8249-1334}\,}
	\author[a]{Martin~Wolf\,\orcidlink{0009-0002-8192-3124}\,}

	\affil[a]{School of Mathematics and Physics,\\ University of Surrey, Guildford GU2 7XH, United Kingdom}

	\abstract{The uniqueness and rigidity of black holes remain central themes in gravitational research. In this work, we investigate the construction of all extremal black hole solutions to the Einstein equation for a given near-horizon geometry, employing the homotopy algebraic perspective, a powerful and increasingly influential framework in both classical and quantum field theory. Utilising Gau{\ss}ian null coordinates, we recast the deformation problem as an analysis of the homotopy Maurer--Cartan equation associated with an $L_\infty$-algebra. Through homological perturbation theory, we systematically solve this equation order by order in directions transverse to the near-horizon geometry. As a concrete application of this formalism, we examine the deformations of the extremal Kerr horizon. Notably, this homotopy-theoretic approach enables us to characterise the moduli space of deformations by studying only the lowest-order solutions, offering a systematic way to understand the landscape of extremal black hole geometries.}

	\acknowledgements{We thank Christian Saemann and Alessandro Torrielli for useful discussions.}

	\declarations{
		\textbf{Funding.}
		C.S.~has been supported by a Royal Thai Government Doctoral Studentship.\\[5pt]
		\textbf{Conflict of interest.}
		The authors have no relevant financial or non-financial interests to disclose.\\[5pt]
		\textbf{Data statement.}
		No additional research data beyond the data presented and cited in this work are needed to validate the research findings in this work.\\[5pt]
		\textbf{Licence statement.}
		For the purpose of open access, the authors have applied a Creative Commons Attribution (CC-BY) license to any author-accepted manuscript version arising.
	}

	\begin{body}

		\section{Introduction and conclusions}

		There is significant interest in exploring aspects of black hole uniqueness, as well as black hole rigidity. Particularly strong uniqueness theorems hold for stationary, asymptotically flat black holes in four dimensions~\cite{Robinson:1975bv,Mazur:1982db,Carter:1971zc,Bunting:1987uls,Israel:1967wq}. These uniqueness theorems were formulated initially for non-extremal black holes, but have also been extended to the extremal cases~\cite{Amsel:2009et}. It is known, however, that such uniqueness theorems break down in higher dimensions, as they exploit properties of curvature which are specific to the fact that the spatial cross-sections of the geometry are three-dimensional. This is explicitly evidenced by the construction of stationary and asymptotically flat five-dimensional black ring solutions~\cite{Emparan:2001wn}. In particular, there exist examples of such black rings which have the same conserved charges as certain Myers--Perry black holes~\cite{Myers:1986un}. Other even more notable examples of black hole non-uniqueness are provided by the asymptotically flat five-dimensional bubbling solutions constructed in~\cite{Horowitz:2017fyg}, which all have the same near-horizon geometry, and conserved charges, as the Breckenridge--Myers--Peet--Vafa (BMPV) black hole~\cite{Breckenridge:1996is}, but have a non-trivial topology outside the horizon. However, there are examples of higher-dimensional uniqueness theorems for asymptotically flat static spacetimes~\cite{Gibbons:2002av}.

		Furthermore, the uniqueness theorems are also not generically formulated when there is a cosmological constant. In the case of higher-dimensional solutions, in~\cite{Gibbons:2002av}, if the assumption of asymptotic flatness is dropped, then non-uniqueness manifests via a construction of an infinite family of regular black holes. In terms of four-dimensional black hole uniqueness, for the special case of the uniqueness theorem for static asymptotically flat solutions constructed in~\cite{Israel:1967wq}, asymptotic flatness is used to prove that a certain harmonic function constructed from the geometry must be constant. We remark that the dimensionality of the geometry also plays a critical role in the proof of~\cite{Israel:1967wq} as it utilises the Gau{\ss}--Bonnet theorem. There has been further progress for non-asymptotically flat extremal black hole solutions in the special case when the near-horizon geometry is static. It has been established in~\cite{Katona:2023vtq} that any such vacuum solution in four or more dimensions for which the spatial cross-section of the event horizon is maximally symmetric, and compact without boundary must be isometric to the extremal Schwarzschild de Sitter solution (or its near-horizon geometry). This uniqueness theorem utilises a systematic order-by-order expansion of the Einstein equations written in Gau{\ss}ian null coordinates.

		The issue of rigidity also plays an important role in understanding the structure of black holes. Indeed, in four dimensions, the existence of an axisymmetric Killing vector field plays an important role in formulating the known uniqueness theorems. Initially, in four dimensions, rigidity was established for non-extremal black holes~\cite{Hawking:1971vc,Hawking:1973uf}. This was then extended~\cite{Hollands:2006rj} for non-extremal black holes in more than four dimensions, and also including a negative cosmological constant. A further extension, again for non-extremal solutions, was constructed in~\cite{Hollands:2022ajj}, incorporating higher derivative corrections to general relativity. The rigidity of extremal, asymptotically flat black holes was established in~\cite{Hollands:2008wn}, subject to a certain additional `diophantine condition'. Another approach to establishing rigidity theorems for extremal black holes in more than four dimensions without making any assumption regarding the asymptotic geometry is to first consider the event horizon of the geometry, and establish a `near-horizon' rigidity theorem and then attempt to extend the rigidity away from the near-horizon region. In terms of the first step, horizon rigidity theorems have been established for supersymmetric solutions in many supergravity theories~\cite{Gutowski:2013kma,Gran:2013wca,Gran:2014fsa,Gutowski:2016gkg} by utilising general Lichnerowicz type theorems to establish supersymmetry enhancement at the horizon, which produces additional isometries. Horizon rigidity has also been established for (non-supersymmetric) vacuum solutions with zero cosmological constant in four or more dimensions provided that a certain one-form which forms part of the near-horizon geometry is not closed~\cite{Dunajski:2023xrd}. This theorem also holds when there is a cosmological constant~\cite{Colling:2024usk}. The issue of whether such horizon rigidity theorems can be extended into the bulk geometry is an open question.

		In this work, we propose an alternative approach to the study of extremal black hole solutions by employing the so-called `homotopy algebraic perspective' on classical and quantum field theories. This framework, rooted in the deep structural parallels between homotopical algebra and field theory, offers one of the most general mathematical formalisms for analysing such theories. Its origins lie in the Batalin--Vilkovisky (BV) formalism~\cite{Batalin:1977pb,Batalin:1981jr,Batalin:1984jr,Batalin:1984ss,Batalin:1985qj,Schwarz:1992nx}, where the central object, called the BV complex, is a differential graded commutative algebra. This complex can be identified with the Chevalley--Eilenberg algebra of a cyclic $L_\infty$-algebra~\cite{Alexandrov:1995kv,Stasheff:1997iz,Zeitlin:2007fp,Jurco:2018sby}. These $L_\infty$-algebras, which are special instances of homotopy algebras and whose origin is in closed string field theory~\cite{Zwiebach:1992ie}, generalise metric differential graded Lie algebras by relaxing the Jacobi identities up to coherent homotopies.

		Crucially, the BV field content naturally organises into the graded vector space underlying the $L_\infty$-algebra, and the kinematic structure is encoded by differentials that endow this graded vector space with the structure of a cochain complex. In turn, its cohomology captures the space of on-shell free fields modulo gauge transformations. Moreover, interactions are governed by the higher products of the $L_\infty$-algebra. The BV anti-bracket then induces a compatible inner product on the $L_\infty$-algebra also called a cyclic structure. This inner product then allows for the reformulation of the BV action as the homotopy Maurer--Cartan action for the $L_\infty$-algebra. Consequently, any variational field theory can be reformulated as the homotopy Maurer--Cartan theory of a cyclic $L_\infty$-algebra. For details on the BV formalism and $L_\infty$-algebras, see~\cite{Kajiura:2003ax,Doubek:2017naz,Hohm:2017pnh,Jurco:2018sby,Jurco:2019bvp,Borsten:2024gox}. Moreover, when considering field theories on manifolds with boundaries, one needs to generalise the notion of cyclic $L_\infty$-algebras to so-called cyclic `relative' $L_\infty$-algebras~\cite{Alfonsi:2024utl}. Essentially, these are pairs of $L_\infty$-algebras, one in the bulk and one in the boundary, and with a morphism between them; see~\cite{Chiaffrino:2023wxk} for a different approach to dealing with boundaries.

		It is important to realise that this connection between field theory and homotopical algebra is significantly deeper than the consideration of equations of motion and actions, revealing an emerging dictionary between quantities in homotopical algebra and quantities in field theory, see~\cite[Section 1]{Borsten:2023ned} for a summary. A key ingredient is the notion of quasi-isomorphism in homotopy algebras, which generalises the familiar concept from cochain complexes. In the setting of $L_\infty$-algebras, quasi-isomorphisms reflect `semi-classical equivalence' between field theories or, put differently, they preserve the tree-level scattering amplitudes. See~\cite[Section 3.4]{Borsten:2021gyl} for more details on notions of equivalence in this context. This insight allows one to classify field theories not merely by their actions or symmetries, but by the homotopy type of their algebraic structures.

		A particularly powerful feature of homotopy algebras is the existence of minimal models: canonical representatives within each quasi-isomorphism class, defined on the cohomology of the original cochain complex. These minimal models are unique up to isomorphism and, in the context of field theory, they encode the essential tree-level dynamics. Their construction is facilitated by the homological perturbation lemma~\cite{brown1967twisted,Gugenheim1989:aa,gugenheim1991perturbation,Crainic:0403266}, which systematically organises the perturbative expansion essentially mirroring the structure of tree-level Feynman diagrams. Such an expansion naturally leads to recursive formulations such as the Berends--Giele relations~\cite{Macrelli:2019afx} and perturbiners~\cite{Lopez-Arcos:2019hvg} for any field theory, offering a homotopical reinterpretation~\cite{Arvanitakis:2019ald} of classical scattering theory.

		The homological perturbation lemma is not just restricted to the construction of the minimal model but rather it can be used to transfer the $L_\infty$-structure from one cochain complex to another such that the two $L_\infty$-structures are quasi-isomorphic. This process is called homotopy transfer~\cite{Loday:2012aa}, and it often allows us to construct a simpler or more physically meaningful $L_\infty$-structure. In the language of field theory, homotopy transfer provides a rigorous framework for reformulating a theory on different field spaces, whilst preserving its perturbative properties. When the target field space is a subspace of the original, the transfer procedure acquires a familiar physical interpretation: it corresponds to integrating out degrees of freedom. This perspective is deeply embedded in the BV formalism, where effective actions arise naturally through such reductions. The process has been explored from various angles in the literature~\cite{Dresse:1990dj,Henneaux:1989ua,Barnich:2004cr}, and continues to inform developments in effective field theory and string theory~\cite{Arvanitakis:2020rrk,Arvanitakis:2021ecw}. However, not all quasi-isomorphisms between $L_\infty$-algebras, that is, not all semi-classical equivalences between field theories, are captured by homotopy transfer. Nevertheless, any quasi-isomorphism between $L_\infty$-algebras can be lifted to a span of $L_\infty$-algebras in which the two quasi-isomorphic $L_\infty$-algebras are, in fact, obtained from a correspondence $L_\infty$-algebra by homotopy transfer~\cite{JalaliFarahani:2023sfq}.

		Whilst in this work we shall exclusively work at the classical level, it should be pointed out that the homotopy-theoretic framework is not confined to tree-level phenomena. As originally suggested in~\cite{Zwiebach:1992ie}, and further developed in~\cite{Markl:1997bj,Doubek:2017naz,Jurco:2019yfd,Saemann:2020oyz,Borsten:2025hrn} as well as in~\cite{Costello:2016vjw,Costello:2021jvx}, many of these algebraic structures extend to loop-level quantum corrections.

		\paragraph{Goals and outline of the paper.}
		The main objective of this work is to perturbatively construct extremal black hole solutions for a given near-horizon geometry using the homotopy algebraic perspective. We shall exemplify the general construction by focussing on the deformations of the extremal Kerr horizon.

		This work generalises~\cite{Li:2015wsa}, which focuses on transverse deformations of the near-horizon geometry. Our approach is more general as it allows us to deduce the moduli space of deformations by considering only the lowest-order solutions. We find that the finiteness of the moduli space dimension of deformation can be extended to each order in the transverse direction ($r$), that is, the dimension increases by a finite number for each order in $r$. Specifically, for the deformation problem of the extremal Kerr horizon, the number of dimensions of solutions up to order $r^n$ is $0$ for $n=1$, $2$ for $n=2$, and at most $2k-2$ for $n=k$. These results generalise the finiteness theorem of general near-horizon deformations and the uniqueness theorem for extremal Kerr horizon deformations given in~\cite{Li:2015wsa}.

		In \cref{sec:ExtremalBlackHoleSolutions}, we briefly recap the definition of the near-horizon geometry of an extremal black hole, which is naturally defined in Gau{\ss}ian null coordinates (\cref{sec:GaussianNullCoordinates}). We then discuss the isometries of near-horizon geometries in \cref{sec:IsometriesOfNear-HorizonSolutions}. The section concludes with an explicit example demonstrating how to express the extremal Kerr black hole solution in Gau{\ss}ian null coordinates and extract its near-horizon geometry (\cref{sec:GaussianNullKerr}). We remark that it would also be interesting to examine the effect of different types of asymptotic geometry on the deformation theory. However, the Gau{\ss}ian null coordinates are in general defined only in a neighbourhood of the event horizon, and consequently asymptotic properties of the geometry cannot be straightforwardly encoded in these coordinates. In terms of the analysis of black hole uniqueness theorems, for example~\cite{Amsel:2009et}, after axisymmetry is established, the metric can be written in Weyl--Papapetrou coordinates which are globally well-defined on the domain of outer communication and the asymptotic properties can be considered in these coordinates. Furthermore, in higher dimensions, it has also been established that uniqueness theorems can be formulated for non-asymptotically flat solutions, provided that there are sufficiently large numbers of isometries,~\cite{Alaee:2019qhj}.

		In \cref{sec:DeformingNear-HorizonGeometries}, we begin by setting up the deformation problem for an arbitrary near-horizon geometry in \cref{sec:deformationSetting}, along with the basis that will be used throughout the paper. In \cref{sec:bianchiIdentityEinsteinEquation}, the (contracted) Bianchi identity is used to extract a set of independent equations from the Einstein equation. In the second half of \cref{sec:DeformingNear-HorizonGeometries}, we focus on analysing the lowest-order deformation. In \cref{sec:firstOrderDeformation}, we derive the lowest-order independent Einstein equation and present it as a linear operator $\mu_1$ acting on the deformation. The lowest-order infinitesimal transformation is then discussed and fixed. We proceed to compute the Green function of $\mu_1$ in terms of the Green function of a linear operator~\eqref{eq:spatialCrossSectionGreenFunctionOrderN} on the spatial cross section (co-dimension two). Towards the end of \cref{sec:firstOrderDeformation}, the Einstein equation is further simplified in a particular gauge. In \cref{sec:extremalKerrFirstOrder}, an example of the deformation of the extremal Kerr horizon is discussed, and the lowest-order Einstein equation is presented and solved along with the corresponding Green function.

		In \cref{sec:homotopyAlgebras}, we provide a brief review of homotopy algebras. An $L_\infty$-algebra, which encodes all information about a deformation problem, is defined in \cref{sec:LInftyAlgebras}. In homotopy algebraic formalism, the equation of motion is expressed as the homotopy Maurer--Cartan equation, as presented in \cref{sec:HomotopyMaurer-CartanTheory}. In \cref{sec:homologicalPerturbations}, we review of how the homotopy Maurer--Cartan equation can be solved with the help of the homological perturbation lemma.

		In \cref{sec:lowestOrderDeformationsRevisited}, we reformulate all the ingredients from \cref{sec:firstOrderDeformation} in the language of homotopy algebras. From this perspective, it is clear that the moduli space of deformations is parametrised by the lowest-order solution. We give an example of the deformation of the extremal Kerr horizon, explaining how one can determine the position in the moduli space of deformations for a given `full' solution. We then consider the next-to-lowest-order equation and explain how one could compute the solutions in \cref{sub:nextToLowestOrderDeformations}. Again in \cref{sec:GaussianNullKerr}, as an example, we apply the formalism to the deformation problem of the extremal Kerr horizon and compute the next-to-lowest order solutions explicitly.

		\section{Extremal black hole solutions}\label{sec:ExtremalBlackHoleSolutions}

		\subsection{Gau{\ss}ian null coordinates}\label{sec:GaussianNullCoordinates}

		Being central to our discussion, we shall start off by summarising the construction of \uline{Gau{\ss}ian null coordinates} following~\cite{Moncrief:1983xua}. These coordinates are much simpler to handle than other coordinates as the metric is fully determined by a scalar, a one-form, and a symmetric rank-2 tensor all in codimension two. Furthermore, the geometry close to the horizon of any extremal black hole can be transformed into this coordinate system. We end this section by defining a frame basis that we will use through out this paper. In the following, $\nabla$ will always denote the Levi-Civita connection for a given metric. Furthermore, for $x$ a local coordinate, we shall write $\partial_x\coloneqq\parder{x}$.

		\paragraph{General construction.}
		Let $M$ be a $d$-dimensional manifold with a Lorentzian metric $g$. Suppose that $M$ admits a null hypersurface $\Sigma\hookrightarrow M$, called the \uline{horizon}, that is, its normal vector field $N\in\Gamma(\Sigma,TM)$ satisfies $g(N,N)|_\Sigma=0$.\footnote{Note that this implies that in $\Sigma$, the integral curves of $N$ are geodesics. To see this, let us choose a function $f\in\scC^\infty(M)$ such that $f$ is constant on $\Sigma$ and define a vector field $N'\in TM$ by $g(N',X)=X(f)$ for all $X\in TM$. Then, for all points $p\in\Sigma$, $N_p\propto N'_p$. Therefore, one only needs to show that $\nabla_{N'}N'=\alpha N'$ on $\Sigma$ for some $\alpha\in\scC^\infty(\Sigma)$. This follows immediately from the fact that $N'$ is null on $\Sigma$, that is, $Y\big(g(N',N')\big)=\beta Y(f)=\beta g(N',X) $ on $\Sigma$ for all $Y\in\Gamma(\Sigma,TM)$ and some $\beta\in\scC^{\infty}(\Sigma)$. Note that we need to make use of the identity $g(\nabla_X N',N')-g(\nabla_{N'}N',X)=\big[X\big(N'(f)\big)-g(N',\nabla_XN')\big]-\big[N'\big(X(f)\big)-g(N',\nabla_{N'}X)\big]=[X,N'](f)-g(N',[X,N'])=0$ for all $X\in TM$ to write $Y\big(g(N',N')\big)$ as $2g(\nabla_{N'}N',Y)$.} We shall also assume that $N$ is future-directed. Moreover, suppose that $\Sigma$ is foliated by closed space-like hypersurfaces $S$, called \uline{spatial cross sections}, with the leaf space generated by $N$. We coordinatise $S$ by $y^i$ for $i,j,\ldots=1,\ldots,d-2$, and we may extend these local coordinates to local coordinates $(u,y^i)$ on a tubular neighbourhood in $\Sigma$ by requiring that a point in $\Sigma$ given by $(u,y^i)$ is a point on the integral curve of $N$ with parameter value $u$ that passes through the point in $S$ given by $y^i$. Hence, $N=\partial_u$. Furthermore, there is a unique past-directed vector field $P\in\Gamma(\Sigma,TM)$ such that
		\begin{equation}\label{eq:defOfP}
			\begin{gathered}
				g(P,P)|_\Sigma\ =\ 0~,
				\quad
				g(P,N)|_\Sigma\ =\ 1~,
				\\
				g(P,\partial_i)|_\Sigma\ =\ 0
				\ewith
				\partial_i\ \coloneqq\ \partial_{y^i}
				\eforall
				i\ =\ 1,\ldots,d-2~,
			\end{gathered}
		\end{equation}
		and which we extend to a tubular neighbourhood $U\subseteq M$ by requiring that its integral curves are affinely parametrised geodesics.\footnote{By a slight abuse of notation, we shall not make a notational distinction between vector fields on $\Sigma$ and their extensions to $M$.} This now allows us to coordinatise $U$ by $(r,u,y^i)$ with $r$ the affine parameter of the geodesic generated by $P$ and passing through the point in $\Sigma$ given by $(u,y^i)$. In these coordinates, we have
		\begin{equation}\label{eq:PandN}
			P\ =\ \partial_r
			\eand
			N\ =\ \partial_u~.
		\end{equation}
		Note that in writing this, we have also extended $N$ and $\partial_i$ from $\Sigma$ to $U$ by means of the push-forward with respect to the one-parameter subgroup of geodesics of $P$. Note also that the first constraint of~\eqref{eq:defOfP} holds on $U$ because of $Pg(P,P)=2g(P,\nabla_PP)=0$. Then, because of $0=Ng(P,P)=2g(P,\nabla_NP)=2g(P,\nabla_PN)=2Pg(P,N)$, we conclude that $g(P,N)=1$ also holds on $U$. Likewise, it also follows that the third constraint in~\eqref{eq:defOfP} must hold on $U$ as well.

		In conclusion, in the local coordinates $(r,u,y^i)$, the metric $g$ takes the form\footnote{Here, `$\odot$' denotes the symmetric tensor product with $\alpha\odot\beta\coloneqq\alpha\otimes\beta+\beta\otimes\alpha$ for any two one-forms $\alpha$ and $\beta$.}
		\begin{equation}\label{eq:metricInGaussianNullCoordinates}
			g\ =\ \rmd u\odot\big[\rmd r+r\alpha_i(r,u,y)\rmd y^i-\tfrac12rB(r,u,y)\rmd u\big]+\tfrac12\gamma_{ij}(r,u,y)\rmd y^i\odot\rmd y^j~.
		\end{equation}
		They are referred to as the \uline{Gau{\ss}ian null coordinates}.

		\paragraph{Extremal case.}
		A \uline{Killing horizon} is a null hypersurface $\Sigma$ in a Lorentzian manifold $(M,g)$ defined by having a Killing vector field $K$ as its normal vector field such that $g(K,K)|_\Sigma=0$. The \uline{surface gravity}, denoted by $\kappa\in\scC^\infty(\Sigma)$, of a Killing horizon is then given by
		\begin{equation}
			(\nabla_KK)_p\ =\ \kappa(p)K_p
			\eforall
			p\ \in\ \Sigma~.
		\end{equation}
		Using the Killing property of $K$, it is not difficult to see that
		\begin{equation}\label{eq:surfaceGravityIdentity}
			Xg(K,K)\ =\ -2g(X,\nabla_KK)
			\quad\Rightarrow\quad
			Xg(K,K)|_\Sigma\ =\ -2\kappa g(X,K)|_\Sigma
		\end{equation}
		for all $X\in\Gamma(M,TM)$.

		Below, we shall only be interested in \uline{extremal black holes} which are black hole solutions to the Einstein equation that admit Killing horizons with vanishing surface gravity. Such Killing horizons are called \uline{degenerate}. To make contact with our previous discussion about the Gau{\ss}ian null coordinates, we now assume that $N$ in~\eqref{eq:PandN} is the Killing vector field $K$ defining the Killing horizon $\Sigma$ located at $r=0$. Indeed, we may always do this since for a general vector field
		\begin{subequations}
			\begin{equation}
				K\ =\ K^r\partial_r+K^u\partial_u+K^i\partial_i
			\end{equation}
			with the boundary conditions
			\begin{equation}\label{eq:NKillingBoundaryConditions}
				K^r|_{r=0}\ =\ N^r|_{r=0}\ =\ 0~,
				\quad
				K^u|_{r=0}\ =\ N^u|_{r=0}\ =\ 1~,
				\eand
				K^i|_{r=0}\ =\ N^i|_{r=0}\ =\ 0~,
			\end{equation}
		\end{subequations}
		it is not too difficult to see that upon imposing the Killing property on $K$ for~\eqref{eq:metricInGaussianNullCoordinates}, the $rr$, $ri$, and $ru$ components of $\caL_Kg=0$, with $\caL$ the Lie derivative, directly imply that $K^r=0$, $K^u=1$, and $K^i=0$ and so, $K=N=\partial_u$ on all of the tubular neighbourhood $U$. Consequently, the Killing property of $N$ is indeed compatible with its extension property~\eqref{eq:PandN}. Furthermore, the $uu$, $ui$, and $ij$ components of $\caL_Ng=0$ imply that $\alpha_i$, $B$, and $\gamma_{ij}$ in~\eqref{eq:metricInGaussianNullCoordinates} must be independent of $u$. Next, because of~\eqref{eq:surfaceGravityIdentity}, we have $Xg(K,K)|_\Sigma=0$ for extremal black holes and so, from~\eqref{eq:metricInGaussianNullCoordinates}, we obtain
		\begin{equation}
			\partial_r|_{r=0}\big(rB(r,y)\big)\ =\ 0~.
		\end{equation}
		Hence, we can write $rB(r,y)=r^2\beta(r,y)$ for $\beta$ some other function. In conclusion, the metric~\eqref{eq:metricInGaussianNullCoordinates} simplifies in the extremal case to
		\begin{equation}\label{eq:metricInGaussianNullCoordinatesExtremal}
			g\ =\ \rmd u\odot\big[\rmd r+r\alpha_i(r,y)\rmd y^i-\tfrac12r^2\beta(r,y)\rmd u\big]+\tfrac12\gamma_{ij}(r,y)\rmd y^i\odot\rmd y^j~.
		\end{equation}
		This form of the metric is the starting point of our discussion.

		\subsection{Near-horizon geometries}\label{sec:nearHorizonGeometries}

		We now have all of the necessary ingredients to define near-horizon geometries of extremal black holes which are also solutions to the Einstein equations. Therefore, it is appropriate to use them as backgrounds for deformations. We will conclude this section by providing some comments on isometries of near-horizon geometries.

		\paragraph{Near-horizon limit.}
		Consider the one-parameter family of (local) diffeomorphisms
		\begin{equation}\label{eq:diffeo}
			(u,r,y^i)\ \rightarrow\ (u/\eps,r\eps,y^i)
			\eforall
			\eps\ >\ 0~.
		\end{equation}
		Then, from~\eqref{eq:metricInGaussianNullCoordinatesExtremal} we obtain the one-parameter family
		\begin{equation}\label{eq:oneParameterFamily}
			g_\eps\ \coloneqq\ \rmd u\odot\big[\rmd r+r\alpha_i(\eps r,y)\rmd y^i-\tfrac12r^2\beta(\eps r,y)\rmd u\big]+\tfrac12\gamma_{ij}(\eps r,y)\rmd y^i\odot\rmd y^j
		\end{equation}
		of metrics. The limit $\eps\to 0$ is called the \uline{near-horizon limit}. The geometry in this limit is called the \uline{near-horizon geometry}, and in this limit, the metric~\eqref{eq:oneParameterFamily} becomes
		\begin{subequations}\label{eq:nearHorizonMetric}
			\begin{equation}
				\mathring g\ =\ \rmd u\odot\big[\rmd r+r\mathring\alpha_i(y)\rmd y^i-\tfrac12r^2\mathring\beta(y)\rmd u\big]+\tfrac12\mathring\gamma_{ij}(y)\rmd y^i\odot\rmd y^j~,
			\end{equation}
			where
			\begin{equation}
				\mathring\alpha_i(y)\ \coloneqq\ \alpha_i(0,y)~,
				\quad
				\mathring\beta(y)\ \coloneqq\ \beta(0,y)~,
				\eand
				\mathring\gamma_{ij}(y)\ \coloneqq\ \gamma_{ij}(0,y)~.
			\end{equation}
		\end{subequations}
		Note that we will always use the diacritic `$\circ$' to indicate that an object constitutes a near-horizon datum, and we shall refer to $\mathring g$ as the \uline{near-horizon metric}.

		\paragraph{Adapted frame basis.}
		We shall make use of two different bases. The first one is the one we have already discussed, the coordinate basis given by the Gau{\ss}ian null coordinates. We shall use the collective coordinate index $I\sim(r,u,i)$.

		The second basis is the \uline{null orthonormal basis} with respected to the near-horizon geometry. We label this basis by $A\sim(+,-,a)$, and it is defined by
		\begin{subequations}\label{eq:defAdaptedFrameII}
			\begin{equation}
				\mathring e^+\ \coloneqq\ \rmd u~,
				\quad
				\mathring e^-\ \coloneqq\ \rmd r+r\mathring\alpha_i\rmd y^i-\tfrac12r^2\mathring\beta\rmd u~,
				\eand
				\mathring e^a\ \coloneqq\ \rmd y^i\mathring e_i{}^a~,
			\end{equation}
			where
			\begin{equation}
				\mathring e_i{}^a\mathring e_j{}^b\delta_{ab}\ =\ \mathring\gamma_{ij}
			\end{equation}
		\end{subequations}
		which define the basis coefficients $\mathring e_I{}^A$. Here, we have suppressed the explicit dependence on $y^i$. In this basis, the metric~\eqref{eq:metricInGaussianNullCoordinatesExtremal} takes the form
		\begin{equation}\label{eq:metricInGaussianNullCoordinatesExtremalAdaptedFrameII}
			g\ =\ \tfrac12g_{AB}\mathring e^A\odot\mathring e^B\ =\ \mathring e^+\odot\big[\mathring e^-+r(\alpha_a-\mathring\alpha_a)\mathring e^a-\tfrac12r^2(\beta-\mathring\beta)\mathring e^+\big]+\tfrac12\gamma_{ab}\mathring e^a\odot\mathring e^b~,
		\end{equation}
		where $\alpha_a\coloneqq\mathring E_a{}^i\alpha_i$, etc.~with $\mathring E_a{}^i$ the inverse of $\mathring e_i{}^a$.

		\subsection{Isometries of near-horizon solutions}\label{sec:IsometriesOfNear-HorizonSolutions}

		Next, let us discuss the isometries of near-horizon solutions. In particular, we summarise and extend the near-horizon rigidity theorem established in~\cite{Dunajski:2023xrd} about the existence of isometries.

		\paragraph{Decomposing $\mathring\alpha$.}
		A key result in establishing this rigidity is~\cite[Lemma 0]{Lucietti:2012sf} which states that given any near-horizon geometry, there exists a positive function $\Gamma$ and a one-form $\mathring V$ on the spatial cross section $\mathring S$ at $r=0$ such that $\mathring\alpha$ in~\eqref{eq:nearHorizonMetric} can be written as
		\begin{subequations}\label{eq:alphaAndV}
			\begin{equation}\label{eq:alphdecomp}
				\mathring\alpha\ =\ \tfrac1\Gamma(\mathring V-\rmd\Gamma)~,
			\end{equation}
			and moreover, the one-form $\mathring V$ satisfies
			\begin{equation}\label{eq:divergfreeV}
				\mathring{\tilde\nabla}^i\mathring V_i\ =\ 0~,
			\end{equation}
		\end{subequations}
		where $\mathring{\tilde\nabla}_i$ is the Levi-Civita connection with respect to $\mathring\gamma_{ij}$. Indeed, this immediately follows via defining the following elliptic operators
		\begin{equation}\label{eq:definitionDAndDAdjoint}
			\sfD\ \coloneqq\ \mathring{\tilde\nabla}^i\mathring{\tilde\nabla}_i-\mathring\alpha^i\mathring{\tilde\nabla}_i
			\eand
			\sfD^\dagger\ \coloneqq\ \mathring{\tilde\nabla}^i\mathring{\tilde\nabla}_i+\mathring\alpha^i\mathring{\tilde\nabla}_i+\mathring{\tilde\nabla}_i\mathring\alpha^i~.
		\end{equation}
		Here, $\sfD^\dagger$ is the adjoint with respect to the standard inner product
		\begin{equation}
			\inner{f}{g}\ \coloneqq\ \int\rmd^{d-2}y\sqrt{\det(\mathring\gamma(y))}\,f(y)g(y)
		\end{equation}
		on the vector space of smooth functions $\scC^\infty(\mathring S)$ on $\mathring S$. Focussing on $\sfD^\dagger$, it is known from~\cite{Andersson:2007fh} that such an elliptic operator has a real principal eigenvalue $\sigma_0$, and that there exists a unique (up to scaling) positive eigenfunction $\Gamma\in\scC^\infty(\mathring S)$ satisfying $\sfD^\dagger\Gamma=\sigma_0\Gamma$. Then,
		\begin{equation}
			\sigma_0\int\rmd^{d-2}y\sqrt{\det(\mathring\gamma(y))}\,\Gamma(y)\ =\ \int\rmd^{d-2}y\sqrt{\det(\mathring\gamma(y))}\,\sfD^\dagger\Gamma(y)\ =\ 0
		\end{equation}
		on using Stokes' Theorem. As $\Gamma>0$, it follows that $\int\rmd^{d-2}y\sqrt{\det(\mathring\gamma(y))}\,\Gamma(y)>0$, and hence $\sigma_0=0$. Therefore, there exists a positive function $\Gamma$ satisfying $\sfD^\dagger\Gamma=0$. Having established the existence of such a $\Gamma$, we may now define $\mathring V$ by means of~\eqref{eq:alphdecomp}, and the condition $\sfD^\dagger\Gamma=0$ implies~\eqref{eq:divergfreeV}.

		\paragraph{Imposing the Einstein condition.}
		The result~\eqref{eq:alphaAndV} is true for all near-horizon geometries. In order to proceed further, it is however necessary to assume the form of the stress-energy tensor. In~\cite{Dunajski:2023xrd}, it was assumed that that the $d$-dimensional spacetime is Einstein, with \uline{cosmological constant} $\Lambda$, and the components of the \uline{Einstein equation}, when reduced to $\mathring S$, are equivalent to
		\begin{equation}\label{eq:backgroundEinsteinEquations}
			\begin{aligned}
				\tfrac{2}{d-2}\Lambda\ &=\ -\mathring\beta+\tfrac12\mathring\gamma^{ij}\mathring{\tilde\nabla}_i\mathring\alpha_j-\tfrac12\mathring\gamma^{ij}\mathring\alpha_i\mathring\alpha_j~,
				\\
				\tfrac{2}{d-2}\Lambda\mathring\gamma_{ij}\ &=\ \mathring{\tilde R}_{ij}+\mathring{\tilde\nabla}_{(i}\mathring\alpha_{j)}-\tfrac12\mathring\alpha_i\mathring\alpha_j~.
			\end{aligned}
		\end{equation}
		Here, as before, $\mathring{\tilde\nabla}_i$ is the Levi-Civita connection with respect to $\mathring\gamma_{ij}$ and $\mathring{\tilde R}_{ij}$ the associated \uline{Ricci tensor}, respectively.

		On making use of the decomposition~\eqref{eq:alphdecomp}, the second equation in~\eqref{eq:backgroundEinsteinEquations} can be rewritten as
		\begin{equation}
			\mathring{\tilde R}_{ij}\ =\ \tfrac1{2\Gamma^2}\mathring V_i\mathring V_j-\tfrac1{2\Gamma^2}\mathring{\tilde\nabla}_i\Gamma\mathring{\tilde\nabla}_j\Gamma-\tfrac1\Gamma\mathring{\tilde\nabla}_{(i}\mathring V_{j)}+\tfrac1\Gamma\mathring{\tilde\nabla}_i\mathring{\tilde\nabla}_j\Gamma+\tfrac2{d-2}\mathring\gamma_{ij}~.
		\end{equation}
		Next, we take the divergence of this expression, use the (contracted) Bianchi identity, and simplify the result using the analysis of~\cite[Appendix A]{Dunajski:2023xrd}. After some calculation, we find the condition
		\begin{subequations}
			\begin{equation}\label{eq:keyisoeq1}
				\mathring{\tilde\nabla}_{(i}\mathring V_{j)}\mathring{\tilde\nabla}^{(i}\mathring V^{j)}\ =\ \mathring{\tilde\nabla}^i\mathring W_i+\mathring U\mathring{\tilde\nabla}^i\mathring V_i~,
			\end{equation}
			with
			\begin{equation}\label{eq:keyisoeq2}
				\begin{aligned}
					\mathring W_i\ &\coloneqq\ \mathring V^j\mathring{\tilde\nabla}_{(i}\mathring V_{j)}-\big(\tfrac12\mathring{\tilde\nabla}^j\mathring{\tilde\nabla}_j\Gamma+\tfrac12\mathring{\tilde\nabla}^j\mathring V_j+\tfrac2{d-2}\Lambda\Gamma\big)\mathring V_i~,
					\\
					\mathring U\ &\coloneqq\ -\tfrac1{2\Gamma}\mathring V_i\mathring V^i+\tfrac12\mathring{\tilde\nabla}^i\mathring{\tilde\nabla}_i\Gamma+\tfrac12\mathring{\tilde\nabla}^i\mathring V_i+\tfrac1{2\Gamma}\mathring V^i\mathring{\tilde\nabla}_i\Gamma+\tfrac2{d-2}\Lambda\Gamma~.
				\end{aligned}
			\end{equation}
		\end{subequations}
		We stress that in deriving this condition, we have not used~\eqref{eq:divergfreeV} and $\sfD^\dagger\Gamma=0$. Then, on substituting the condition~\eqref{eq:divergfreeV} into~\eqref{eq:keyisoeq1}, integrating over $\mathring S$ and using Stokes' Theorem, we obtain
		\begin{equation}
			\int\rmd^{d-2}y\sqrt{\det(\mathring\gamma(y))}\,\mathring{\tilde\nabla}_{(i}{\mathring V}_{j)}\mathring{\tilde\nabla}^{(i}{\mathring V}^{j)}\ =\ 0~,
		\end{equation}
		from which it follows that
		\begin{equation}\label{eq:LieDerivativeVGamma}
			\mathring{\tilde\nabla}_{(i}\mathring V_{j)}\ =\ 0
			\quad\Leftrightarrow\quad
			\caL_{\mathring V}\mathring\gamma\ =\ 0~.
		\end{equation}

		Having established~\eqref{eq:LieDerivativeVGamma}, it remains to determine necessary and sufficient conditions for the Lie derivative of the remaining near-horizon data with respect to $\mathring V$ to vanish. We remark that
		\begin{equation}\label{eq:lienhsuff}
			\caL_{\mathring V}\mathring\alpha\ =\ -\tfrac1{\Gamma^2}(\caL_{\mathring V}\Gamma)(\mathring V-\rmd\Gamma)-\tfrac1\Gamma\rmd(\caL_{\mathring V}\Gamma)\ =\ -\tfrac1{\Gamma^2} (\caL_{\mathring V}\Gamma)\mathring V-\rmd\big(\tfrac1\Gamma\caL_{\mathring V}\Gamma\big)\,,
		\end{equation}
		so clearly, if $\caL_{\mathring V}\Gamma=0$ then $\caL_{\mathring V}\mathring\alpha=0$. We now claim that also the converse holds. Indeed, suppose that $\caL_{\mathring V}\mathring\alpha=0$, then on taking the divergence of~\eqref{eq:lienhsuff}, we find the condition
		\begin{equation}
			\mathring{\tilde\nabla}^i\mathring{\tilde\nabla}_i\big(\tfrac1\Gamma\caL_{\mathring V}\Gamma\big)+\mathring\alpha^i\mathring{\tilde\nabla}_i\big(\tfrac1\Gamma\caL_{\mathring V}\Gamma\big)-\tfrac2\Gamma\big(\tfrac1\Gamma\caL_{\mathring V}\Gamma\big)^2\ =\ 0~,
		\end{equation}
		which is equivalent to
		\begin{equation}
			\mathring V^i\mathring{\tilde\nabla}_i\big( \mathring{\tilde\nabla}^j\mathring{\tilde\nabla}_j\log\Gamma+\caL_{\mathring V}\log\Gamma\big)\ =\ \tfrac2\Gamma(\caL_{\mathring V}\log\Gamma)^2~.
		\end{equation}
		Upon integrating this condition over $\mathring S$ and using Stokes's Theorem together with~\eqref{eq:divergfreeV}, we find
		\begin{equation}
			\int\rmd^{d-2}y\sqrt{\det(\mathring\gamma(y))}\,\tfrac2\Gamma(\caL_{\mathring V}\log\Gamma)^2\ =\ 0~,
		\end{equation}
		and hence, $\caL_{\mathring V}\Gamma=0$. Consequently, $\caL_{\mathring V}\mathring\alpha=0$ if and only if $\caL_{\mathring V}\Gamma=0$.

		Furthermore, if $\caL_{\mathring V}\Gamma=0$ then $\caL_{\mathring V}\mathring\alpha=0$, and it follows from~\eqref{eq:backgroundEinsteinEquations} that also $\caL_{\mathring V}\mathring\beta=0$. Consequently, $\caL_{\mathring V}\Gamma=0$ is a necessary and sufficient condition for the Lie derivative of \uline{all} the near-horizon data with respect to $\mathring V$ to vanish.

		It is straightforward to see that $\caL_{\mathring V} \Gamma=0$ holds. This is because on taking the trace of the Ricci tensor given in~\eqref{eq:keyisoeq2}, one obtains
		\begin{equation}
			\mathring{\tilde R}-2\Lambda\ =\ \tfrac1{2\Gamma^2}\mathring V_i\mathring V^i-\tfrac1{2\Gamma^2}\mathring{\tilde\nabla}^i\Gamma\mathring{\tilde\nabla}_i\Gamma+\tfrac1\Gamma\mathring{\tilde\nabla}^i\mathring{\tilde\nabla}_i\Gamma~.
		\end{equation}
		Upon taking the Lie derivative of this expression with respect to $\mathring V$ and recalling~\eqref{eq:LieDerivativeVGamma}, we obtain
		\begin{equation}\label{eq:liescalar}
			\begin{aligned}
				&-\tfrac1{\Gamma^3}(\caL_{\mathring V}\Gamma)\mathring V_i\mathring V^i+\tfrac1{\Gamma^3}(\caL_{\mathring V}\Gamma)\mathring{\tilde\nabla}^i\Gamma\mathring{\tilde\nabla}_i\Gamma
				\\
				&\kern1cm-\tfrac1{\Gamma^2}\mathring{\tilde\nabla}^i\Gamma\mathring{\tilde\nabla}_i(\caL_{\mathring V}\Gamma)-\tfrac1{\Gamma^2}(\caL_{\mathring V}\Gamma)\mathring{\tilde\nabla}^i\mathring{\tilde\nabla}_i\Gamma+\tfrac1\Gamma \mathring{\tilde\nabla}^i\mathring{\tilde\nabla}_i(\caL_{\mathring V}\Gamma)\ =\ 0~.
			\end{aligned}
		\end{equation}
		Next, we multiply this expression by $\caL_{\mathring V}\Gamma$ and integrate over $\mathring S$. We obtain
		\begin{equation}
			\begin{aligned}
				\int\rmd^{d-2}y\sqrt{\det(\mathring\gamma(y))}\,&\big\{\tfrac1{\Gamma^3}\mathring V_i\mathring V^i(\caL_{\mathring V}\Gamma)^2+\tfrac1{\Gamma}\mathring{\tilde\nabla}^i\big(\caL_{\mathring V}\log\Gamma\big)\mathring{\tilde\nabla}_i\big(\caL_{\mathring V}\log\Gamma\big)\big\}\ =\ 0~,
			\end{aligned}
		\end{equation}
		where we have integrated by parts in order to eliminate the two Laplacian terms in the final two terms of~\eqref{eq:liescalar}. As the integrand is a sum of two non-negative terms, we immediately arrive at
		\begin{equation}
			\caL_{\mathring V}\Gamma\ =\ 0~.
		\end{equation}
		In conclusion, it follows that the Lie derivative of \uline{all} the near-horizon data with respect to $\mathring V$ vanishes.

		The above summarises the result of~\cite{Dunajski:2023xrd}, where $\caL_{\mathring V}\Gamma=0$ was established for $\Lambda\leq 0$ for any value of $d$, and for all values of $\Lambda$ when $d=4$. This was then extended, using a slightly different method to that given above, to include the case of $\Lambda>0$ in $d>4$ in \cite{Colling:2024usk}. We further recall that the rigidity theorem was extended to the case of Einstein--Maxwell theory in $d=4$ in~\cite{Colling:2024txz}.

		\subsection{Example: extremal Kerr}\label{sec:GaussianNullKerr}

		The goal of this section is the construction of the extremal Kerr black hole in the Gau{\ss}ian null coordinates order-by-order in $r$, the affine parameter of the null geodesics generated by $P$ as discussed in \cref{sec:GaussianNullCoordinates}.

		\paragraph{Metric in Kerr coordinates and Killing horizon.}
		Recall that in the \uline{Kerr coordinates}, the extremal Kerr black hole metric is given by \cite{Kerr:1963ud}
		\begin{equation}\label{eq:extremalKerrMetricInKerrCoordinates}
			\begin{aligned}
				g_{\rm eK}\ &\coloneqq\ \frac{m^2\sin^2\theta-(\rho-m)^2}{2(m^2\cos^2\theta+\rho^2)}\,\rmd v\odot\rmd v+\frac12(m^2\cos^2\theta+\rho^2)\,\rmd\theta\odot\rmd\theta
				\\
				&\kern1cm+\frac{\sin^2\theta[(m^2+\rho^2)^2-m^2 (\rho-m)^2\sin^2\theta]}{2(m^2\cos^2\theta+\rho^2)}\,\rmd\phi\odot\rmd\phi
				\\
				&\kern1cm+\rmd v\odot\rmd\rho-\frac{2m^2\rho\sin^2\theta}{m^2\cos^2\theta+\rho^2}\,\rmd v\odot\rmd\phi-m\sin^2\theta\,\rmd\rho\odot\rmd\phi~,
			\end{aligned}
		\end{equation}
		where the parameter $m>0$ is the mass and the coordinate ranges are $\rho>0$, $v\in\IR$, $\theta\in(0,\pi)$, and $\phi\in(0,2\pi)$.

		It is not too difficult to see that apart from $\partial_\phi$, the metric~\eqref{eq:extremalKerrMetricInKerrCoordinates} admits another Killing vector
		\begin{subequations}
			\begin{equation}
				N\ \coloneqq\ \partial_v+\tfrac{1}{2m}\partial_\phi
			\end{equation}
			and
			\begin{equation}
				g_{\rm eK}(N,X)|_{\rho=m}\ =\ 0
				\eforall
				X\ \in\ \{N,\partial_v,\partial_\theta,\partial_\phi\}~.
			\end{equation}
		\end{subequations}
		We thus have a Killing horizon at $\rho=m$. Furthermore, one also checks that it has indeed vanishing surface gravity.

		\paragraph{Construction of Gau{\ss}ian null coordinates.}
		We now wish to change the coordinates $(\rho,v,\theta,\phi)$ to the Gau{\ss}ian null coordinates $(r,u,y^1\coloneqq x,y^2\coloneqq\varphi)$ with $r\geq0$, $u\in\IR$, $x\in(-1,1)$ and $\varphi\in(0,2\pi)$, and our goal will be to construct this change of coordinates so that we obtain $\alpha_i$, $\beta$, and $\gamma_{ij}$ in~\eqref{eq:metricInGaussianNullCoordinatesExtremal} to second order in $r$, the affine parameter of the geodesics generated by $P$.

		To this end, we assume that
		\begin{equation}\label{eq:VectorFieldsOnSigmaKerr}
			\begin{gathered}
				\partial_u|_{r=0}\ =\ N|_{\rho=m}~,
				\\
				\partial_1|_{r=0}\ =\ f'(\cos\theta)\partial_v|_{\rho=m}-\tfrac{1}{\sin\theta}\partial_\theta|_{\rho=m}-\tfrac{1}{2m }f'(\cos\theta)\partial_\phi|_{\rho=m}~,
				\\
				\partial_2|_{r=0}\ =\ \partial_\phi|_{\rho=m}~,
			\end{gathered}
		\end{equation}
		where here and in the following, $f$ is an arbitrary function of $\cos\theta$ and the prime indicates the derivative with respect to the argument. We note that we can choose any three vector fields on the Killing horizon as long as they are commuting and linearly independent, and $\partial_u|_{r=0}$ generates the Killing horizon. This reflects the freedom in choosing local coordinates on the Killing horizon. Our choice for $\partial_2|_{r=0}$ in~\eqref{eq:VectorFieldsOnSigmaKerr} is natural in that it is preferable to have $\partial_2$ as a Killing vector. Furthermore, the components of the near-horizon metric~\eqref{eq:nearHorizonMetric} will turn out to be independent of $f$: this function will play the role of the residual gauge transformations that we will discuss in \cref{sec:firstOrderDeformation}.

		To zeroth order in $r$, the change of coordinates can be deduced from~\eqref{eq:VectorFieldsOnSigmaKerr} as
		\begin{equation}\label{eq:extremalKerrInGaussianNullCoordinatesZerothOrder}
			\begin{aligned}
				\rho\ &=\ m+\caO(r)~,
				\\
				v\ &=\ u+\tfrac12f(x)+\caO(r)~,
				\\
				\theta\ &=\ \arccos x +\caO(r)~,
				\\
				\phi\ &=\ \varphi+\tfrac{1}{2m}u+\tfrac{1}{4m}f(x)+\caO(r)~.
			\end{aligned}
		\end{equation}

		Next, to extend~\eqref{eq:extremalKerrInGaussianNullCoordinatesZerothOrder} beyond the leading order, following the constructions in \cref{sec:GaussianNullCoordinates}, we need to solve the geodesic equation $\nabla_PP=0$ on the tubular region $U$ subject to the boundary conditions~\eqref{eq:defOfP}. The latter result in
		\begin{subequations}\label{eq:extremalKerrPHorizon}
			\begin{equation}
				P\ =\ P^\rho\partial_\rho+P^v\partial_v+P^\theta\partial_\theta+P^\phi\partial_\phi
			\end{equation}
			with
			\begin{equation}
				\begin{gathered}
					P^\rho|_{\rho=m}\ =\ \frac{2}{1+x^2}~,
					\quad
					P^v|_{\rho=m}\ =\ \frac{(1-x^2)[4m^2-f'^2(x)]}{8m^2(1+x^2)}~,
					\\
					P^\theta|_{\rho=m}\ =\ \frac{\sqrt{1-x^2}f'(x)}{2m^2(1+x^2)}~,
					\quad
					P^\phi|_{\rho=m}\ =\ \frac{4m^2(3+x^2)-(1-x^2)f'^2(x)}{16m^3(1+x^2)}~.
				\end{gathered}
			\end{equation}
		\end{subequations}
		Consequently, with
		\begin{equation}\label{eq:extremalKerrIntegralCurvesP}
			\dder[\rho]{r}\ =\ P^\rho~,
			\quad
			\dder[v]{r}\ =\ P^v~,
			\quad
			\dder[\theta]{r}\ =\ P^\theta~,
			\eand
			\dder[\phi]{r}\ =\ P^\phi
		\end{equation}
		the conditions~\eqref{eq:extremalKerrPHorizon} extend the change of coordinates~\eqref{eq:extremalKerrInGaussianNullCoordinatesZerothOrder} to first order in $r$ as
		\begin{equation}\label{eq:extremalKerrGaussianNullCoordinatesFirstOrder}
			\begin{aligned}
				\rho\ &=\ m+\frac{2}{1+x^2}r+\caO(r^2)~,
				\\
				v\ &=\ u+\frac12f(x)+\frac{(1-x^2)[4m^2-f'^2(x)]}{8m^2(1+x^2)}r+\caO(r^2)~,
				\\
				\theta\ &=\ \arccos x+\frac{\sqrt{1-x^2}f'(x)}{2m^2(1+x^2)}r+\caO(r^2)~,
				\\
				\phi\ &=\ \varphi+\frac{1}{2m}u+\frac{1}{4m}f(x)+\frac{4m^2(3+x^2)-(1-x^2)f'^2(x)}{16m^3(1+x^2)}r+\caO(r^2)~.
			\end{aligned}
		\end{equation}
		
		\pagebreak

		Next, since $g(P,N)|_{\rho=m}=1$ and $g(P,\partial_\phi)|_{\rho=m}=0$ and since $N$ and $\partial_\phi$ are Killing vectors, we immediately have that $g(P,N)=1$ and $g(P,\partial_\phi)=0$ on the tubular neighbourhood $U$.\footnote{Recall that for $c$ an affinely parametrised geodesic with parameter $t\in I\subseteq\IR$ and $K$ a Killing vector for a metric $g$, the quantity $g_{c(t)}(K_{c(t)},\partial_t c(t))$ is constant for all $t\in I$.} Upon solving these two equations, we can express $P^v$ and $P^\phi$ in terms of $P^\rho$ as
		\begin{equation}\label{eq:extremalKerrPvPphi}
			\begin{aligned}
				P^v\ &=\ \frac{\rho^2+m^2}{(\rho-m)^2}\left(P^\rho-\frac{2m\rho}{m^2\cos^2\theta+\rho^2}\right)-1~,
				\\
				P^\phi\ &=\ \frac{m}{(\rho-m)^2}\left(P^\rho-\frac{2m\rho}{m^2\cos^2\theta+\rho^2}\right).
			\end{aligned}
		\end{equation}
		Hence, once we have determined $P^\rho$ and $P^\theta$, we automatically obtain $P^v$ and $P^\phi$ via these equations. To determine $P^\rho$ and $P^\theta$, we note that since $g(P,P)|_{\rho=m}=0$, the geodesic equation implies that $g(P,P)=0$ on the tubular neighbourhood $U$. Explicitly, this equation reads as
		\begin{equation}\label{eq:extremalKerrPdotP}
			(P^\rho)^2+(\rho-m)^2(P^\theta)^2\ =\ \frac{m^2(m^2\cos^2\theta+\rho^2)+m^2\rho^2\cos^2\theta+2m^3\rho\sin^2\theta+\rho^4}{(m^2\cos^2\theta+\rho^2)^2}~.
		\end{equation}
		We shall use this equation together with the $\rho$ component of the geodesic equation $\nabla_PP=0$,
		\begin{equation}\label{eq:extremalKerrGeodesicEquationRhoComponent}
			\begin{aligned}
				& \dder[P^\rho]{r}+\frac{\rho[(P^\rho)^2-(\rho-m)^2(P^\theta)^2]-m^2P^\theta P^\rho\sin(2\theta)}{m^2\cos^2\theta+\rho^2}-\frac{(P^\rho)^2}{\rho-m}
				\\
				&\kern3cm+\frac{2m\rho^2(\rho^2+m^2)}{(m^2\cos^2\theta+\rho^2)^3}+\frac{m[m^3+\rho(m^2+3m\rho-\rho^2)]}{(\rho-m)(m^2\cos^2\theta+\rho^2)^2}\ =\ 0~,
			\end{aligned}
		\end{equation}
		to determine $P^\rho$ and $P^\theta$ and thus, the change of coordinates~\eqref{eq:extremalKerrGaussianNullCoordinatesFirstOrder} via~\eqref{eq:extremalKerrIntegralCurvesP} to higher orders in $r$. Note that in deriving~\eqref{eq:extremalKerrGeodesicEquationRhoComponent}, we have made use of the algebraic expressions~\eqref{eq:extremalKerrPvPphi}.

		In particular, the second-order contribution to $\rho$ follows from the zeroth order piece of~\eqref{eq:extremalKerrGeodesicEquationRhoComponent}. Likewise, we can find the second-order contributions to $v$, $\theta$, and $\phi$; however, a quicker route to finding the second-order contributions to the latter coordinates is to note that in the Gau{\ss}ian null coordinates, the $ri$ and $rr$ components of the metric are zero. Finally, we also need the third-order contribution to $\rho$ which follows from the second-order piece of~\eqref{eq:extremalKerrPdotP}. Putting everything together, we arrive at
		\begin{equation}\label{eq:extremalKerrChangeOfCoordinates}
			\begin{aligned}
				\rho\ &=\ m+\frac{2}{1+x^2}r-\frac{(1-x^2)[2m-xf'(x)]}{m^2(1+x^2)^3}r^2
				\\
				&\kern1cm+\frac{(1-x^2)[4m^2(5-4x^2-x^4)-8mx(3-x^2)f'(x)-(1-4x^2+3x^4)f'^2(x)]}{4m^4(1+x^2)^5}r^3
				\\
				&\kern1cm+\caO(r^4)~,
				\\
				v\ &=\ u+\frac12f(x)+\frac{(1-x^2)[4m^2-f'^2(x)]}{8m^2(1+x^2)}r
				\\
				&\kern1cm-\frac{(1-x^2)[16m^3-4m^2x(3+x^2)f'(x)-4mf'^2(x)+x(1-x^2)f'^3(x)]}{16m^4(1+x^2)^3}r^2+\caO(r^3)~,
				\\
				\theta\ &=\ \arccos x+\frac{\sqrt{1-x^2}f'(x)}{2m^2(1+x^2)}r
				\\
				&\kern1cm-\frac{\sqrt{1-x^2}[2m^2x(1+x^2)+4mf'(x)-x(1-x^2)f'^2(x)]}{4m^4(1+x^2)^3}r^2+\caO(r^3)~,
				\\
				\phi\ &=\ \varphi+\frac{1}{2m}u+\frac{1}{4m}f(x)+\frac{4m^2(3+x^2)-(1-x^2)f'^2(x)}{16m^3(1+x^2)}r
				\\
				&\kern1cm-\frac{1}{{32m^5(1+x^2)^3}}[8m^3(5+2x^2+x^4)-4m^2x(3-2x^2-x^4)f'(x)
				\\
				&\kern1cm-2m(3-2x^2-x^4)f'^2(x)+x(1-x^2)^2f'^3(x)]r^2+\caO(r^3)~.
			\end{aligned}
		\end{equation}
		As we shall show next, these orders are sufficient to determine coefficients $\alpha_i$, $\beta$, and $\gamma_{ij}$ in the metric~\eqref{eq:metricInGaussianNullCoordinatesExtremal} to second order in $r$.

		\paragraph{Metric in Gau{\ss}ian null coordinates.}
		Indeed, we implement the change of coordinates~\eqref{eq:extremalKerrChangeOfCoordinates} into~\eqref{eq:extremalKerrMetricInKerrCoordinates} to arrive at
		\begin{subequations}\label{eq:extremalKerrMetricGaussianNullCoordinates}
			\begin{equation}
				\begin{aligned}
					\alpha_x\ &=\ -\frac{2x}{1+x^2}+\frac{8mx(2-x^2)+2(2-5x^2-x^4)f'(x)+x(1-x^4)f''(x)}{2m^2(1+x^2)^3}r
					\\
					&\kern1cm-\frac{1}{2m^4(1+x^2)^5}\{2mx[2m(15-22x^2+3x^4)+(3-x^2-3x^4+x^6)f''(x)]
					\\
					&\kern1cm+f'(x)[16m(1-6x^2+3x^4)-x(15-22x^2+3x^4)f'(x)
					\\
					&\kern1cm+(1-3x^2-x^4+3x^6)f''(x)]\}r^2+\caO(r^3)~,
					\\
					\alpha_\varphi\ &=\ \frac{4(1-x^2)}{(1+x^2)^2}-\frac{2(1-x^2)[m(3-10x^2-x^4)-x(5-x^2)f'(x)]}{m^2(1+x^2)^4}r
					\\
					&\kern1cm+\frac{(1-x^2)}{2m^4(1+x^2)^6}[4m^2(11-55x^2+17x^4+3x^6)-8mx(15-22x^2+3x^4)f'(x)
					\\
					&\kern1cm-(5-41x^2+31x^4-3x^6)f'^2(x)]r^2+\caO(r^3)
				\end{aligned}
			\end{equation}
			and
			\begin{equation}
				\begin{aligned}
					\beta\ &=\ -\frac{3-6x^2-x^4}{m^2(1+x^2)^3}+\frac{2m(5-21x^2+7x^4+x^6)-4x(3-4x^2+x^4)f'(x)}{m^4(1+x^2)^5}r
					\\
					&\kern1cm-\frac{1}{2m^6(1+x^2)^7}[4m^2(19-121x^2+97x^4-7x^6-4x^8)
					\\
					&\kern1cm-2mx(95-224x^2+146x^4-16x^6-x^8)f'(x)
					\\
					&\kern1cm-3(2-21x^2+35x^4-19x^6+3x^8)f'^2(x)]r^2+\caO(r^3)
				\end{aligned}
			\end{equation}
			and
			\begin{equation}
				\begin{aligned}
					\gamma_{xx}\ &=\ \frac{m^2(1+x^2)}{1-x^2}+\frac{4m+2x^3f'(x)-(1-x^4)f''(x)}{1- x^4}r
					\\
					&\kern1cm+\frac{1}{4m^2(1-x^2)(1+x^2)^3}\{4m^2(1+13x^2-11x^4+x^6)-8m(1-x^4)f''(x)
					\\
					&\kern1cm+2f'(x)[4mx(5-5x^2+2x^4)+(2-7x^2+5x^4+2x^6)f'(x)
					\\
					&\kern1cm-2x^3(1-x^4)f''(x)]+(1-x^4)^2f''^2(x)\}r^2+\caO(r^3)~,
					\\
					\gamma_{x\varphi}\ &=\ \frac{2(1-x^2)[2mx+f'(x)]}{(1+x^2)^2}r-\frac{(1-x^2)}{2m^2(1+x^2)^4}\{2mx[2m(7-5x^2)+(1-x^4)f''(x)]
					\\
					&\kern1cm+f'(x)[4m(2-11x^2-x^4)-x(11-x^2)f'(x)+(1-x^4)f''(x)]\}r^2+\caO(r^3)~,
					\\
					\gamma_{\varphi\varphi}\ &=\ \frac{4m^2(1-x^2)}{1+x^2}+\frac{8x(1-x^2)[2mx+f'(x)]}{(1+x^2)^3}r
					\\
					&\kern1cm+\frac{2(1-x^2)}{m^2(1+x^2)^5}[2m^2(3-11x^2+11x^4+x^6)-4mx(4-7x^2+x^4)f'(x)
					\\
					&\kern1cm-(1-7x^2+4x^4)f'^2(x)]r^2+\caO(r^3)
				\end{aligned}
			\end{equation}
		\end{subequations}
		for the coefficients $\alpha_i$, $\beta$, and $\gamma_{ij}$ in the metric~\eqref{eq:metricInGaussianNullCoordinatesExtremal}.

		\paragraph{Near-horizon limit.}
		Next, the near-horizon metric~\eqref{eq:nearHorizonMetric} obtained by rescaling $r\to\eps r$ and $u\to u/\eps$ following from~\eqref{eq:extremalKerrMetricGaussianNullCoordinates} and taking the limit $\eps\to 0$ \footnote{This is equivalent to the zeroth-order in $r$ of~\eqref{eq:extremalKerrMetricGaussianNullCoordinates}.} is given by
		\begin{subequations}\label{eq:extremalKerrNearHorizonMetricGaussianNullCoordinates}
			\begin{equation}
				\mathring g_{\rm eK}\ =\ \rmd u\odot\big[\rmd r+r\mathring\alpha_i(y)\rmd y^i-\tfrac12r^2\mathring\beta(y)\rmd u\big]+\tfrac12\mathring\gamma_{ij}(y)\rmd y^i\odot\rmd y^j~,
			\end{equation}
			where
			\begin{equation}
				\begin{aligned}
					\mathring\alpha_i(y)\rmd y^i\ &=\ \frac{4(1-x^2)}{(1+x^2)^2}\rmd\varphi -\frac{2x}{1+x^2}\rmd x~,
					\\
					\mathring\beta(y)\ &=\ -\ \frac{3-6x^2-x^4}{m^2(1+x^2)^3}~,
					\\
					\tfrac12\mathring\gamma_{ij}(y)\rmd y^i\odot\rmd y^j\ &=\ \frac{m^2(1+x^2)}{2(1-x^2)}\rmd x\odot\rmd x+2m^2\frac{1-x^2}{1+x^2}\rmd\varphi\odot\rmd\varphi
				\end{aligned}
			\end{equation}
		\end{subequations}
		for all $x\in(-1,1)$ and $\varphi\in(0,2\pi)$. We stress that for the purpose of finding the near-horizon metric, one does not actually need to solve the geodesic equation since one only needs the zeroth-order-in-$r$ part of $\alpha_i$, $\beta$, and $\gamma_{ij}$. Finally, the non-vanishing coefficients $\mathring e_i{}^a$ of the basis~\eqref{eq:defAdaptedFrameII} are given by
		\begin{equation}\label{eq:defAdaptedFrameIIExtremalKerr}
			\mathring e_x{}^1\ =\ m\sqrt{\frac{1+x^2}{1-x^2}}
			\eand
			\mathring e_\varphi{}^2\ =\ 2m\sqrt{\frac{1-x^2}{1+x^2}}~.
		\end{equation}

		\section{Deforming near-horizon geometries}\label{sec:DeformingNear-HorizonGeometries}

		In this work, we shall be interested in extremal black hole solutions with a fixed near-horizon geometry. In particular, given a near-horizon geometry in Gau{\ss}ian null coordinates, we shall construct extremal black hole solutions by means of deformation theory. Subject of this section is to first set up the precise problem we wish to study. We shall then discuss general first-order deformations which we then exemplify in the context of the extremal Kerr black hole.

		\subsection{Setting}\label{sec:deformationSetting}

		Our starting point is a near-horizon solution,
		\begin{equation}\label{eq:backgroundMetric}
			\mathring g\ =\ \rmd u\odot\big[\rmd r+r\mathring\alpha_i(y)\rmd y^i-\tfrac12r^2\mathring\beta(y)\rmd u\big]+\tfrac12\mathring\gamma_{ij}(y)\rmd y^i\odot\rmd y^j~,
		\end{equation}
		to the Einstein equation~\eqref{eq:backgroundEinsteinEquations}.

		\paragraph{Deformation problem.}
		We now wish to deform the near-horizon metric~\eqref{eq:backgroundMetric} away from the near-horizon limit to construct a new solution to the full Einstein equation such that the near-horizon metric remains fixed and the full metric is in the Gau{\ss}ian-null-coordinate form~\eqref{eq:metricInGaussianNullCoordinatesExtremal}. In particular, we consider the deformed metric
		\begin{subequations}\label{eq:deformedMetric}
			\begin{equation}
				g\ \coloneqq\ \rmd u\odot\big[\rmd r+r\alpha_i(r,y)\rmd y^i-\tfrac12r^2\beta( r,y)\rmd u\big]+\tfrac12\gamma_{ij}(r,y)\rmd y^i\odot\rmd y^j~,
			\end{equation}
			where
			\begin{equation}\label{eq:deformations}
				\begin{aligned}
					\alpha_i(r,y)\ &\coloneqq\ \mathring\alpha_i(y)+\kappa h_i(r,y)~,
					\\
					\beta( r,y)\ &\coloneqq\ \mathring\beta(y)+\kappa h(r,y)~,
					\\
					\gamma_{ij}(r,y)\ &\coloneqq\ \mathring\gamma_{ij}(y)+\kappa h_{ij}(r,y)~,
				\end{aligned}
			\end{equation}
			with $\kappa$ \uline{Einstein's gravitational constant} and with the \uline{deformations} $h_i$, $h$, and $h_{ij}$ satisfying
			\begin{equation}\label{eq:deformationsBoundaryConditions}
				h_i|_{r=0}\ =\ 0~,
				\quad
				h|_{r=0}\ =\ 0~,
				\eand
				h_{ij}|_{r=0}\ =\ 0~.
			\end{equation}
		\end{subequations}
		These boundary conditions ensure that we only deform the metric away from the near-horizon geometry; evidently the near-horizon limit, see \cref{sec:nearHorizonGeometries}, of~\eqref{eq:deformedMetric} is~\eqref{eq:backgroundMetric}. We stress that, because of our discussion in \cref{sec:GaussianNullCoordinates}, the type of deformations~\eqref{eq:deformedMetric} does not constrain the possible extremal black hole solutions to the Einstein equation as for such solutions one can always bring the metric in the Gau{\ss}ian-null-coordinate form~\eqref{eq:metricInGaussianNullCoordinatesExtremal}. We remark that we do constrain our deformations to remain stationary; a class of $u$-dependent non-stationary deformations was constructed in \cite{Kapec:2023ruw} to examine quantum corrections to the black hole entropy. 

		Below, we shall mostly make use of the basis~\eqref{eq:defAdaptedFrameII}. In this basis, the deformed metric~\eqref{eq:deformedMetric} becomes
		\begin{subequations}\label{eq:deformedMetricAdaptedFrameII}
			\begin{equation}
				g\ =\ \tfrac12g_{AB}\mathring e^A\odot\mathring e^B\ =\ \mathring e^+\odot\big[\mathring e^-+\kappa rh_a\mathring e^a-\tfrac12\kappa r^2h\mathring e^+\big]+\tfrac12\underbrace{(\delta_{ab}+\kappa h_{ab})}_{\eqqcolon\,\gamma_{ab}}\mathring e^a\odot\mathring e^b~,
			\end{equation}
			where
			\begin{equation}
				h_a\ \coloneqq\ \mathring E_a{}^ih_i
				\eand
				h_{ab}\ \coloneqq\ \mathring E_a{}^i\mathring E_b{}^jh_{ij}~.
			\end{equation}
		\end{subequations}

		\subsection{Bianchi identity and Einstein equation}\label{sec:bianchiIdentityEinsteinEquation}

		It is well-known that due to the (contracted) Bianchi identity, not all the components of the Einstein equation constitute independent equations. In this section, we shall make use of these identities to extract an independent set of equations which our deformed metric~\eqref{eq:deformedMetric} will have to satisfy, under the assumption that the Einstein equation~\eqref{eq:backgroundEinsteinEquations} for the near-horizon metric~\eqref{eq:backgroundMetric} holds.

		\paragraph{Independent equations.}
		We denote the components of the deformed metric~\eqref{eq:deformedMetric} in the basis~\eqref{eq:defAdaptedFrameII} by $g_{AB}$. In this basis, the components of the Einstein equation are given by
		\begin{equation}\label{eq:EinsteinEquationAdaptedFrameI}
			G^{AB}+\Lambda g^{AB}\ =\ 0
			\ewith
			G^{AB}\ \coloneqq\ R^{AB}-\tfrac12Rg^{AB}
		\end{equation}
		with, as before, $\Lambda$ the cosmological constant, $G^{AB}$ the components of the \uline{Einstein tensor}, $R^{AB}$ the components of the Ricci tensor, and $R$ the curvature scalar for the Levi-Civita connection for the metric $g_{AB}$. We now claim that the independent set of equations arising from~\eqref{eq:EinsteinEquationAdaptedFrameI} is given by the $++$, $+a$, and $ab$ components. Put differently, for the deformation problem defined in \cref{sec:deformationSetting}, it is enough to consider the equations
		\begin{equation}\label{eq:independentEinsteinEquationAdaptedFrameI}
			\begin{gathered}
				G^{++}\ =\ R^{++}\ =\ 0~,
				\quad
				G^{a+}\ =\ R^{a+}\ =\ 0~,
				\\
				G^{ab}+\Lambda\gamma^{ab}\ =\ R^{ab}-\big(R^{+-}+\tfrac12R_c{}^c-\Lambda\big)\gamma^{ab}\ =\ 0
			\end{gathered}
		\end{equation}
		and to solve for the deformations~\eqref{eq:deformations} under the assumptions~\eqref{eq:deformationsBoundaryConditions}.

		Indeed, to verify this claim, we first note that
		\begin{equation}\label{eq:EinsteinEquationBoundaryConditionsAdaptedFrameI}
			\begin{gathered}
				\big(G^{--}+\Lambda g^{--}\big)\big|_{r=0}\ =\ 0~,
				\quad
				\big(G^{a-}+\Lambda g^{a-}\big)\big|_{r=0}\ =\ 0~,
				\\
				\big(G^{+-}+\Lambda\big)\big|_{r=0}\ =\ -\tfrac12\mathring\gamma^{ij}\big(\mathring{\tilde R}_{ij}+\mathring{\tilde\nabla}_i\mathring\alpha_j-\tfrac12\mathring\alpha_i\mathring\alpha_j\big)+\Lambda\ =\ 0~,
			\end{gathered}
		\end{equation}
		where we have made use of the explicit form of the components of Ricci tensor provided in \cref{app:RicciTensor} and assumed~\eqref{eq:backgroundEinsteinEquations}.

		Next, by using the fact that
		\begin{equation}
			\omega_{A-}{}^+\ =\ \omega_{-A}{}^+\ =\ \omega_{--}{}^A\ =\ 0\,,
		\end{equation}
		where $\omega_{AB}{}^C$ is the connection one-form, of the Levi-Civita connection, $\nabla_A$ (see also \cref{app:Vielbein}), it follows that the $+$ component of the \uline{contracted Bianchi identity},
		\begin{equation}\label{eq:contractedBianchiIdentityAdaptedFrameI}
			\nabla_A G^{AB}\ =\ 0
			\quad\Leftrightarrow\quad
			\nabla_A\big(G^{AB}+\Lambda g^{AB}\big)\ =\ 0~,
		\end{equation}
		is given by
		\begin{equation}
			\begin{aligned}
				-\big(\partial_r+\omega_{a-}{}^a\big)\big(G^{+-}+\Lambda\big)\ &=\ \big(\mathring{E}_++2\omega_{++}{}^++\omega_{-+}{}^-+\omega_{a+}{}^a\big)G^{++}
				\\
				&\kern1cm+\big(\mathring{E}_a+2\omega_{+a}{}^++\omega_{-a}{}^-+\omega_{ba}{}^b+\omega_{a+}{}^+\big)G^{a+}
				\\
				&\kern1cm+\omega_{ab}{}^+\big(G^{ab}+\Lambda\gamma^{ab}\big)\,.
			\end{aligned}
		\end{equation}
		Upon solving this differential equation supplemented by the boundary conditions~\eqref{eq:EinsteinEquationBoundaryConditionsAdaptedFrameI}, we can express the component $G^{+-}+\Lambda$ in terms of the components $G^{++}$, $G^{a+}$, and $G^{ab}+\Lambda \gamma^{ab}$. Likewise, the $a$ components of~\eqref{eq:contractedBianchiIdentityAdaptedFrameI} can be written as
		\begin{equation}
			\begin{aligned}
				&-\big[\delta^a{}_b\big(\partial_r+\omega_{c-}{}^c\big)+\omega_{-b}{}^a+\omega_{b-}{}^a\big]\big(G^{b-}+\Lambda g^{b-}\big)
				\\
				&\kern1cm=\ \omega_{++}{}^aG^{++}+\big(\omega_{+-}{}^a+\omega_{-+}{}^a\big)\big(G^{+-}+\Lambda\big)
				\\
				&\kern2cm+\big[\delta^a{}_b\big(\mathring{E}_++\omega_{++}{}^++\omega_{-+}{}^-+\omega_{c+}{}^c\big)+\omega_{b+}{}^a+\omega_{+b}{}^a\big]G^{b+}
				\\
				&\kern2cm+\big[\delta^a{}_c\big(\mathring{E}_b+\omega_{+b}{}^++\omega_{-b}{}^-+\omega_{db}{}^d\big)+\omega_{bc}{}^a\big]\big(G^{bc}+\Lambda\gamma^{bc}\big)\,.
				\end{aligned}
		\end{equation}
		Hence, using this, the boundary conditions~\eqref{eq:EinsteinEquationBoundaryConditionsAdaptedFrameI}, and our above result that the $G^{+-}+\Lambda $ components can be expressed in terms of $G^{++}$, $G^{a+}$, and $G^{ab}+\Lambda \gamma^{ab}$, we conclude that also $G^{a-}+\Lambda g^{a-}$ can be expressed in terms of these components. Finally, the $-$ component of~\eqref{eq:contractedBianchiIdentityAdaptedFrameI} can be written as
		\begin{equation}
			\begin{aligned}
				&-\big(\partial_r+\omega_{a-}{}^a\big)\big(G^{--}+\Lambda g^{--}\big)
				\\
				&\kern1cm=\ \omega_{++}{}^-G^{++}+\big(\mathring{E}_++\omega_{++}{}^++2\omega_{-+}{}^-+\omega_{a+}{}^a+\omega_{+-}{}^-\big)\big(G^{+-}+\Lambda\big)
				\\
				&\kern2cm+\big(\omega_{a+}{}^-+\omega_{+a}{}^-\big)G^{+a}+\big(\mathring{E}_a +\omega_{+a}{}^++2\omega_{-a}{}^-+\omega_{ba}{}^b+\omega_{a-}{}^-\big)\big(G^{a-}+\Lambda g^{a-}\big)
				\\
				&\kern2cm+\omega_{ab}{}^-\big(G^{ab}+\Lambda\gamma^{ab}\big)\,,
			\end{aligned}
		\end{equation}
		and, again, we can express the $G^{--}+\Lambda g^{--}$ component in terms of $G^{++}$, $G^{a+}$, and $G^{ab}+\Lambda \gamma^{ab}$.

		In conclusion, this verifies our claim, and it is indeed enough to consider the $++$, $a+$, and $ab$ components of the Einstein equation, that is, the equations~\eqref{eq:independentEinsteinEquationAdaptedFrameI}. Hence, we may use
		\begin{equation}\label{eq:independentEinsteinEquationAdaptedFrameII}
			G^{++}\ =\ 0~,
			\quad
			G^{+a}\ =\ 0~,
			\eand
			G^{ab}+\Lambda\gamma^{ab}\ =\ 0
		\end{equation}
		and solve for the deformations~\eqref{eq:deformations}, or as in~\eqref{eq:deformedMetricAdaptedFrameII} when written in the basis~\eqref{eq:defAdaptedFrameII}, under the assumptions~\eqref{eq:deformationsBoundaryConditions}.

		The equations~\eqref{eq:independentEinsteinEquationAdaptedFrameII} also follow from the \uline{Einstein--Hilbert action},
		\begin{equation}
			S_{\rm EH}\ \coloneqq\ -\frac{1}{\kappa^2}\int{\rm vol}_M\,(R-2\Lambda)~,
		\end{equation}
		where $\kappa$ is again Einstein's gravitational constant, upon assuming the metric~\eqref{eq:deformedMetricAdaptedFrameII} and varying with respect to $h$, $h_a$, and $h_{ab}$ respectively. Indeed, for general variations of $S_{\rm EH}$ we have
		\begin{equation}\label{eq:variationEinsteinHilbertAction}
			\delta S_{\rm EH}\ =\ \frac{1}{\kappa^2}\int{\rm vol}_M\,(G^{IJ}+\Lambda g^{IJ})\delta g_{IJ}~.
		\end{equation}
		From~\eqref{eq:deformedMetricAdaptedFrameII}, it now follows that
		\begin{equation}
			\delta g_{AB}\ =\ \mathring E_A{}^I\mathring E_B{}^J\delta g_{IJ}\ =\
			\begin{cases}
				-\kappa r^2\delta h & \efor (A,B)\ =\ (+,+)
				\\
				\kappa r\delta h_a & \efor (A,B)\ =\ (a,+)
				\\
				\kappa \delta h_{ab} & \efor (A,B)\ =\ (a,b)
				\\
				0 & \eelse
			\end{cases}
		\end{equation}
		and so,~\eqref{eq:variationEinsteinHilbertAction} becomes
		\begin{equation}
		   \delta S_{\rm EH}\ =\ \frac{1}{\kappa}\int{\rm vol}_M\,\big[2rG^{a+}\delta h_a-r^2G^{++}\delta h+(G^{ab}+\Lambda\gamma^{ab})\delta h_{ab}\big]\,.
		\end{equation}
		Hence, the equations of motion are
		\begin{equation}\label{eq:independentEinsteinEquationAdaptedFrameIIModified}
			\tfrac2\kappa rG^{a+}\ =\ 0~,
			\quad
			-\tfrac1\kappa r^2G^{++}\ =\ 0~,
			\eand
			\tfrac1\kappa(G^{ab}+\Lambda\gamma^{ab})\ =\ 0~.
		\end{equation}
		\pagebreak
		Evidently, the first two equations are satisfied for $r=0$, and for $r>0$, we get $G^{a+}=0$ and $G^{++}=0$. Assumed continuity of the solutions then implies that $G^{a+}=0$ and $G^{++}=0$ also hold for $r=0$. Altogether, we recover~\eqref{eq:independentEinsteinEquationAdaptedFrameII}. For the sake of convenience when reformulating everything in terms of homotopy algebras, we shall work with the equations~\eqref{eq:independentEinsteinEquationAdaptedFrameIIModified} in the following.

		\subsection{First-order deformations}\label{sec:firstOrderDeformation}

		As explained in \cref{sec:bianchiIdentityEinsteinEquation}, for metrics of the form~\eqref{eq:deformedMetricAdaptedFrameII}, it is sufficient to solve the equations~\eqref{eq:independentEinsteinEquationAdaptedFrameIIModified} in order to solve the Einstein equation. Upon inserting the metric~\eqref{eq:deformedMetricAdaptedFrameII} and expanding in powers of $\kappa$, we may write these equations formally as
		\begin{equation}\label{eq:homotopyMaurerCartanActionMotivation}
			\mu_1(\Theta)+\tfrac\kappa2\mu_2(\Theta,\Theta)+\tfrac{\kappa^2}{3!}\mu_3(\Theta,\Theta,\Theta)+\cdots\ =\ 0
			\ewith
			\Theta\ \coloneqq\
			\begin{pmatrix}
				h_a
				\\
				h
				\\
				h_{ab}
			\end{pmatrix},
		\end{equation}
		where $\mu_1$ is a linear differential operator depending on the near-horizon metric, and $\mu_2$, $\mu_3$, $\ldots$ can be understood as interaction terms amongst the deformations $\Theta$ and which also depend on the near-horizon metric. Note that the $\kappa$ dependence in~\eqref{eq:homotopyMaurerCartanActionMotivation} has been made explicit. The general objective now is to solve~\eqref{eq:homotopyMaurerCartanActionMotivation} perturbatively in powers of $\kappa$ and depending on what interaction terms are included. As a warm up, we shall now analyse~\eqref{eq:homotopyMaurerCartanActionMotivation} to lowest order, that is,
		\begin{equation}\label{eq:zerothOrderEinsteinEquationMotivation}
			\mu_1(\Theta)\ =\ 0~.
		\end{equation}

		\paragraph{Lowest-order Einstein equation.}
		As detailed in \cref{app:perturbativeExpansionEinsteinEquation}, the lowest-order equation~\eqref{eq:zerothOrderEinsteinEquationMotivation} is explicitly given by
		\begin{subequations}\label{eq:zerothOrderEinsteinEquation}
			\begin{equation}\label{eq:zerothOrderEinsteinEquation:a}
				\underbrace{
					\begin{pmatrix}
						\delta_a{}^c\big(2r\partial_r+r^2\partial_r^2\big) & 0 & \sfd_a{}^{cd}
						\\
						0 & 0 & \frac12\delta^{cd}r^2\partial_r^2
						\\
						\sfd_{ab}{}^c & \delta_{ab}\big(1+2r\partial_r+\frac12r^2\partial_r^2\big) & \sfd_{ab}{}^{cd}
					\end{pmatrix}
				}_{\coloneqq\,\mu_1}
				\begin{pmatrix}
					h_c
					\\
					h
					\\
					h_{cd}
				\end{pmatrix}
				\ =\ 0~,
			\end{equation}
			where
			\begin{equation}\label{eq:zerothOrderEinsteinEquation:b}
				\begin{aligned}
					\sfd_a{}^{cd}\ &\coloneqq\ r\big[\delta_a{}^{(c}\big(\mathring{\tilde\nabla}^{d)}-\mathring\alpha^{d)}\big)-\delta^{cd}\big(\mathring{\tilde\nabla}_a-\tfrac12\mathring\alpha_a\big)\big]\partial_r+r^2\big(\mathring\alpha_a\delta^{cd}-\mathring\alpha^{(c}\delta_a{}^{d)}\big)\partial_r^2~,
					\\
					\sfd_{ab}{}^c\ &\coloneqq\ \delta_{(a}{}^c\big(\mathring{\tilde\nabla}_{b)}-\mathring\alpha_{b)}\big)-\delta_{ab}\big(\mathring{\tilde\nabla}^c-\tfrac32\mathring\alpha^c\big)
					\\
					&\kern1cm+r\big[\delta_{(a}{}^c\big(\mathring{\tilde\nabla}_{b)}-3\mathring\alpha_{b)}\big)-\delta_{ab}\big(\mathring{\tilde\nabla}^c-\tfrac72\mathring\alpha^c\big)\big]\partial_r
					\\
					&\kern1cm+r^2\big(\delta_{ab}\mathring\alpha^c-\delta_{(a}{}^c\mathring\alpha_{b)}\big)\partial_r^2~,
					\\
					\sfd_{ab}{}^{cd}\ &\coloneqq\ \delta_{(a}{}^{(c}\big\{\big(\mathring{\tilde\nabla}^{d)}-\mathring\alpha^{d)}\big)\mathring{\tilde\nabla}_{b)}-\delta_{b)}{}^{d)}\big[\tfrac{2}{d-2}\Lambda+\tfrac12\big(\mathring{\tilde\nabla}_e-\mathring\alpha_e\big)\mathring{\tilde\nabla}^e\big]\big\}-\tfrac12\delta^{cd}\mathring{\tilde\nabla}_a\mathring{\tilde\nabla}_b
					\\
					&\kern1cm+\delta_{ab}\big\{\delta^{cd}\big[\tfrac{1}{d-2}\Lambda+\tfrac12\big(\mathring{\tilde\nabla}^e-\mathring\alpha^e\big)\mathring{\tilde\nabla}_e\big]-\big(\tfrac12\mathring{\tilde\nabla}^{(c}-\mathring\alpha^{(c}\big)\mathring{\tilde\nabla}^{d)}+\tfrac12\big(\mathring{\tilde\nabla}^{(c}-\mathring\alpha^{(c}\big)\mathring\alpha^{d)}\big\}
					\\
					&\kern1cm+r\big\{
					\delta_{(a}{}^{(d}\big[2\mathring\alpha^{c)}\mathring\alpha_{b)}-\mathring{\tilde\nabla}^{c)}\mathring\alpha_{b)}-\mathring\alpha_{b)}\mathring{\tilde\nabla}^{c)}-\mathring\alpha^{c)}\mathring{\tilde\nabla}_{b)}
					\\
					&\kern1cm-\delta_{b)}{}^{c)}\big(\mathring\beta+\mathring\alpha_e\mathring\alpha^{e}-\mathring\alpha^e\mathring{\tilde\nabla}_e-\tfrac12\mathring{\tilde\nabla}^e\mathring\alpha_e\big)\big]+\delta^{cd}\big(\mathring\alpha_{(a}\mathring{\tilde\nabla}_{b)}+\tfrac12\mathring{\tilde\nabla}_{(a}\mathring\alpha_{b)}-\tfrac12\mathring\alpha_a\mathring\alpha_b\big)
					\\
					&\kern1cm+\delta_{ab}\big[\mathring\alpha^{(c}\mathring{\tilde\nabla}^{d)}+\tfrac12\mathring{\tilde\nabla}^{(c}\mathring\alpha^{d)}-\tfrac32\mathring\alpha^{(c}\mathring\alpha^{d)}+\delta^{cd}\big(\mathring\beta+\mathring\alpha^e\mathring\alpha_{e}-\mathring\alpha^e\mathring{\tilde\nabla}_e-\tfrac12\mathring{\tilde\nabla}^e\mathring\alpha_e\big)\big]\big\}\partial_r
					\\
					&\kern1cm+r^2\big\{\delta_{(a}{}^{(c}\big[\mathring\alpha_{b)}\mathring\alpha^{d)}-\tfrac12\big(\mathring\alpha_e\mathring\alpha^e+\mathring\beta\big)\delta_{b)}{}^{d)}\big]-\tfrac12\mathring\alpha_a\mathring\alpha_b\delta^{cd}
					\\
					&\kern1cm+\tfrac12\delta_{ab}\big[\delta^{cd}\big(\mathring\beta+\mathring\alpha^e\mathring\alpha_e\big)-\mathring\alpha^c\mathring\alpha^d\big]\big\}\partial^2_r~.
				\end{aligned}
			\end{equation}
		\end{subequations}
		As before, $\mathring{\tilde\nabla}_a$ is the Levi-Civita connection with respect to $\mathring\gamma_{ab}=\delta_{ab}$ and indices are raised and lowered by $\mathring\gamma^{ab}$ and $\mathring\gamma_{ab}$.

		\paragraph{Lowest-order infinitesimal gauge transformations.}
		The Einstein equation is invariant under diffeomorphisms. Infinitesimally, such gauge transformations are given by
		\begin{equation}\label{eq:infinitesimalDiffeomorphismActionOnMetric}
			g\ \mapsto\ g+\kappa\caL_Xg
		\end{equation}
		for $X$ some vector field and $\caL$ the Lie derivative. Given that we work with metrics in the Gau{\ss}ian-null-coordinate form~\eqref{eq:deformedMetricAdaptedFrameII}, we wish to find the residual gauge transformations that preserve this form. Put differently, we wish to find the resulting gauge transformations
		\begin{equation}
			h_a\ \mapsto\ h_a+\delta h_a~,
			\quad
			h\ \mapsto\ h+\delta h~,
			\eand
			h_{ab}\ \mapsto\ h_{ab}+\delta h_{ab}
		\end{equation}
		of the deformations satisfying \eqref{eq:deformationsBoundaryConditions}. Let us now discuss these transformations to lowest order, that is, the gauge redundancy of~\eqref{eq:zerothOrderEinsteinEquation}.

		To this order, the gauge transformations~\eqref{eq:infinitesimalDiffeomorphismActionOnMetric} reduce to
		\begin{subequations}\label{eq:generalFirstOrderGaugeTransformations}
			\begin{equation}
				g\ \mapsto\ g+\kappa\caL_X\mathring g
			\end{equation}
			and with $\delta g_{AB}\coloneqq\kappa(\caL_X\mathring g)_{AB}=2\kappa\mathring\nabla_{(A}X_{B)}$ explicitly given as
			\begin{equation}
				\begin{aligned}
					\delta g_{+-}\ &=\ \kappa\partial_rX_++\tfrac12\kappa r^2\mathring\beta\partial_rX_-+\kappa r\mathring\beta X_--\kappa\mathring\alpha_aX^a~,
					\\
					\delta g_{--}\ &=\ 2\kappa\partial_rX_-~,
					\\
					\delta g_{a-}\ &=\ -\kappa r\mathring\alpha_a\partial_rX_-+\kappa\partial_rX_a+\kappa\mathring{\tilde{\nabla}}_aX_-~,
					\\
					\delta g_{++}\ &=\ \kappa r^2\mathring\beta\partial_r X_+-2\kappa r\mathring\beta X_++\kappa r^2\big(\mathring\beta\mathring\alpha_a-\mathring{\tilde\nabla}_a\mathring\beta\big)X^a~,
					\\
					\delta g_{a+}\ &=\ -\kappa r\mathring\alpha_a\partial_rX_++\tfrac12\kappa r^2\mathring\beta\partial_rX_a+\kappa\mathring{\tilde\nabla}_a X_++\kappa\mathring\alpha_aX_+-2\kappa r\mathring{\tilde\nabla}_{[a}\mathring\alpha_{b]}X^b
					\\
					&\kern1cm-\tfrac12\kappa r^2\big(\mathring\beta\mathring\alpha_a-\mathring{\tilde\nabla}_a\mathring\beta\big)X_-~,
					\\
					\delta g_{ab}\ &=\ 2\kappa\mathring{\tilde\nabla}_{(a}X_{b)}-2\kappa r\mathring\alpha_{(a}\partial_rX_{b)}~.
				\end{aligned}
			\end{equation}
		\end{subequations}
		Here, as before, $\mathring{\tilde\nabla}_a$ is the Levi-Civita connection with respect to $\mathring\gamma_{ab}=\delta_{ab}$. To retain the Gau{\ss}ian-null-coordinate form~\eqref{eq:deformedMetricAdaptedFrameII}, we now need to impose the conditions
		\begin{subequations}\label{eq:conditionsFirstOrderGaugeTransformations}
			\begin{equation}
				\delta g_{A-}\ =\ 0~,
				\quad
				\delta g_{++}\ =\ -\kappa r^2\delta h~,
				\quad
				\delta g_{a+}\ =\ \kappa r\delta h_a~,
				\eand
				\delta g_{ab}\ =\ \kappa\delta h_{ab}
			\end{equation}
			with
			\begin{equation}
				\delta h_a|_{r=0}\ =\ 0~,
				\quad
				\delta h|_{r=0}\ =\ 0~,
				\eand
				\delta h_{ab}|_{r=0}\ =\ 0~.
			\end{equation}
		\end{subequations}
		Using the explicit formul{\ae}~\eqref{eq:generalFirstOrderGaugeTransformations}, it is not too difficult to see that
		\begin{equation}
			X^+\ =\ -\tfrac12c~,
			\quad
			X^-\ =\ \tfrac14r^2\big(\mathring\beta+\mathring\alpha^a\mathring{\tilde\nabla}_a\big)c~,
			\eand
			X^a\ =\ \tfrac12r\mathring{\tilde\nabla}^ac
		\end{equation}
		is the most general vector field satisfying the conditions~\eqref{eq:conditionsFirstOrderGaugeTransformations}.\footnote{Here, we ignore the isometries of the background metric ($\caL_X\mathring g=0$) since infinitesimal gauge transformations associated with those do not affect the metric at lowest order.} Here, $c$ is an arbitrary smooth function on the spatial cross section $\mathring S$ at $r=0$. Consequently,
		\begin{equation}\label{eq:firstOrderGaugeTransformations}
			\begin{aligned}
				\delta h_a\ &=\ \tfrac14r\big[\mathring\alpha^b\big(\mathring{\tilde\nabla}_a-\mathring\alpha_a\big)-\mathring{\tilde\nabla}_a\mathring\alpha^b+2\mathring{\tilde\nabla}^b\mathring\alpha_a+2\mathring\beta\delta_a{}^b\big]\mathring{\tilde\nabla}_bc~,
				\\
				\delta h\ &=\ \tfrac12r\big(\mathring{\tilde\nabla}^a\mathring\beta-\mathring\beta\mathring\alpha^a\big)\mathring{\tilde\nabla}_ac~,
				\\
				\delta h_{ab}\ &=\ r\big(\mathring{\tilde\nabla}_{(a}-\mathring\alpha_{(a}\big)\mathring{\tilde\nabla}_{b)}c~.
			\end{aligned}
		\end{equation}

		\paragraph{Gauge fixing.}
		In the previous paragraph, we have derived the residual gauge transformations that preserve the Gau{\ss}ian-null-coordinate form to lowest order. We may gauge-fix these transformation following~\cite{Fontanella:2016lzo,Dunajski:2023xrd}.

		In particular, using~\eqref{eq:firstOrderGaugeTransformations}, it is not too difficult to see that for
		\begin{subequations}
			\begin{equation}
				h'_{ab}\ \coloneqq\ h_{ab}+\delta h_{ab}
			\end{equation}
			we have
			\begin{equation}\label{eq:gaugeTransformHab}
				\delta^{ab}\partial_rh'_{ab}|_{r=0}\ =\ \delta^{ab}\partial_rh_{ab}|_{r=0}+\sfD c
			\end{equation}
			with
			\begin{equation}\label{eq:definitionDAndDAdjointAdaptedBasisII}
				\sfD\ \coloneqq\ \mathring{\tilde\nabla}^a\mathring{\tilde\nabla}_a-\mathring\alpha^a\mathring{\tilde\nabla}_a
				\eand
				\sfD^\dagger\ \coloneqq\ \mathring{\tilde\nabla}^a\mathring{\tilde\nabla}_a+\mathring\alpha^a\mathring{\tilde\nabla}_a+\mathring{\tilde\nabla}_a\mathring\alpha^a~.
			\end{equation}
		\end{subequations}
		which are the operators~\eqref{eq:definitionDAndDAdjoint} written in the adapted basis. Furthermore, we have the identifications
		\begin{equation}
			\scC^\omega(\mathring S)\ \cong\ \im(\sfD)\oplus(\im(\sfD))^\perp
			\eand
			(\im(\sfD))^\perp\ \cong\ \ker(\sfD^\dagger)~,
		\end{equation}
		where $\scC^\omega(\mathring S)$ are the real analytic functions on $\mathring S$. Using these identifications, the transformation~\eqref{eq:gaugeTransformHab} can be written as
		\begin{subequations}
			\begin{equation}
				\delta^{ab}\partial_rh'_{ab}|_{r=0}\ =\ \Gamma
				\ewith
				\Gamma\ \coloneqq\ \big(\delta^{ab}\partial_rh_{ab}|_{r=0}\big)\big|_{\ker(\sfD^\dagger)}~,
			\end{equation}
			where we have fixed the gauge parameter $c$ such
			\begin{equation}
				\sfD c\ =\ -\big(\delta^{ab}\partial_rh_{ab}|_{r=0}\big)\big|_{\im(\sfD)}~.
			\end{equation}
		\end{subequations}
		Note that this does not completely fix $c$ since we can still have residual gauge transformations with $\sfD c=0$. However, there is a maximum principle argument that shows that the only solution to this equation must be constant~\cite{PUCCI20041}.

		In conclusion, we may always work in a gauge in which
		\begin{equation}\label{eq:gaugeFixingCondition}
			\delta^{ab}\partial_rh_{ab}|_{r=0}\ =\ \Gamma
			\ewith
			\sfD^\dagger\Gamma\ =\ 0~.
		\end{equation}
		Importantly, as shown in~\cite{Dunajski:2023xrd}, solutions to the differential equation $\sfD^\dagger\Gamma=0$ are unique up to a multiplicative constant.

		\paragraph{Green's function.}
		In the following, we are interested in solutions to the differential equation $\mu_1(\Theta)=\varrho$ with $\mu_1$ as defined in~\eqref{eq:zerothOrderEinsteinEquation} and $\varrho$ a general source term. As we have seen in the preceding paragraph, for homogeneous solutions, we can always fix the gauge~\eqref{eq:gaugeFixingCondition}. In addition, homogeneous solutions have to satisfy the boundary conditions~\eqref{eq:deformationsBoundaryConditions}. Consequently, for particular solutions, and without loss of generality, we may require
		\begin{equation}\label{eq:deformationsBoundaryConditionsParticularSolutions}
			h_a^{(p)}|_{r=0}\ =\ 0~,
			\quad
			h^{(p)}|_{r=0}\ =\ 0~,
			\quad
			h_{ab}^{(p)}|_{r=0}\ =\ 0~,
			\eand
			\delta^{ab}\partial_rh_{ab}^{(p)}|_{r=0}\ =\ 0~,
		\end{equation}
		where the superscript `$(p)$' stands for particular. Put differently, we shall now find the Green function for $\mu_1$ that produces particular solutions that obey these boundary conditions.

		To construct this Green function, we first bring $\mu_1$ into an equivalent upper triangular form. In particular, upon writing
		\begin{equation}
			(\mu_1(\Theta))(r,y)\ =\ \int\rmd r'\int\rmd^{d-2}y'\,\sqrt{\det(\mathring\gamma(y'))}\underbrace{\delta(r-r')\frac{\delta^{(d-2)}(y-y')}{\sqrt{\det(\mathring\gamma(y))}}\mu_1(r',y')}_{\eqqcolon\,\mu_1(r,y;r',y')}\Theta(r',y')
		\end{equation}
		we define
		\begin{subequations}\label{eq:operatorT}
			\begin{equation}
				\tilde\mu_1\ \coloneqq\ \sft\circ\mu_1
			\end{equation}
			with an injective\footnote{See \cref{app:injectivity} for details on the injectivity.} operator $\sft$ given by
			\begin{equation}
				\begin{aligned}
					(\tilde\mu_1(\Theta))(r,y)\ &=\ \int\rmd r'\int\rmd^{d-2}y'\,\sqrt{\det(\mathring\gamma(y'))}
					\\
					&\kern1cm\times\underbrace{\int\rmd r''\int\rmd^{d-2}y''\,\sqrt{\det(\mathring\gamma(y''))}\sft(r,y;r'',y'')\mu_1(r'',y'';r',y')}_{\eqqcolon\,\tilde\mu_1(r,y;r',y')}\Theta(r',y')
				\end{aligned}
			\end{equation}
			and
			\begin{equation}\label{eq:defsft}
				\sft(r,y;r',y')\ \coloneqq\ \frac{\delta^{(d-2)}(y-y')}{\sqrt{\det(\mathring\gamma(y'))}}
				\begin{pmatrix}
					\frac{1}{r'}\delta_a{}^c\theta(r-r') & \sft_a{}^0 & 0
					\\
					\sft_0{}^c & \sft_0{}^0 & \sft_0{}^{cd}
					\\
					\sft_{ab}{}^c & \sft_{ab}{}^0 & \sft_{ab}{}^{cd}
				\end{pmatrix},
			\end{equation}
			and
			\begin{equation}
				\begin{aligned}
					\sft_a{}^{0}\ &\coloneqq\ -\tfrac{1}{r'}\theta(r-r')\big(\tfrac{2(d-3)}{d-2}\mathring{\tilde\nabla}_a+\mathring\alpha_a\big)+\tfrac{r}{r'^2}\theta (r-r')\big(\tfrac{2(d-3)}{d-2}\mathring{\tilde\nabla}_a-\tfrac{d-4}{d-2}\mathring\alpha_a\big)\,,
					\\
					\sft_0{}^c\ &\coloneqq\ \tfrac{1}{r'}\theta(r-r')\big(\tfrac{d-3}{d-2}\mathring{\tilde\nabla}^c-\tfrac{3d-8}{2(d-2)}\mathring\alpha^c\big)-\delta(r-r')\tfrac{d-3}{d-2}\mathring\alpha^c~,
					\\
					\sft_0{}^0\ &\coloneqq\ \tfrac{1}{r'}\theta(r-r')\big(-\tfrac{d-3}{d-2}\mathring{\tilde\nabla}_a\mathring{\tilde\nabla}^a+\tfrac{d-3}{d-2}\mathring\alpha_a\mathring{\tilde\nabla}^a-\tfrac{d-4}{d-2}\mathring{\tilde\nabla}_a\mathring\alpha^a+\tfrac{3d-10}{2(d-2)}\mathring\alpha_a\mathring\alpha^a+\tfrac{2(d-4)}{(d-2)^2}\Lambda\big)
					\\
					&\kern1cm+\tfrac{r}{r'^2}\theta(r-r')\big(\tfrac{d-3}{d-2}\mathring{\tilde\nabla}_a\mathring{\tilde\nabla}^a-\tfrac{3(d-3)}{d-2}\mathring\alpha_a\mathring{\tilde\nabla}^a-\mathring{\tilde\nabla}_a\mathring\alpha^a+\tfrac{3d-8}{2(d-2)}\mathring\alpha_a\mathring\alpha^a+\tfrac{2}{d-2}\Lambda\big)
					\\
					&\kern1cm+\delta(r-r')\big(\tfrac{d-3}{d-2}\mathring\alpha_a\mathring\alpha^a-\tfrac{d-3}{d-2}\mathring\beta\big)\,,
					\\
					\sft_0{}^{cd}\ &\coloneqq\ \delta(r-r')\tfrac{1}{d-2}\delta^{cd}~,
					\\
					\sft_{ab}{}^c\ &\coloneqq\ -\tfrac{1}{r'}\theta(r-r')\big[-\delta_{(a}{}^c\mathring{\tilde\nabla}_{b)}+\delta_{(a}{}^c\mathring\alpha_{b)}+\tfrac{\delta_{ab}}{d-2}\big(\mathring{\tilde\nabla}^c-\mathring\alpha^c\big)\big]-\delta(r-r')\big(\delta_{(a}{}^c\mathring\alpha_{b)}-\tfrac{\delta_{ab}}{d-2}\mathring\alpha^c\big)\,,
					\\
					\sft_{ab}{}^0\ &\coloneqq\ -\tfrac{1}{r'}\theta(r-r')\big[\mathring{\tilde\nabla}_a\mathring{\tilde\nabla}_b+\mathring{\tilde\nabla}_{(a}\mathring\alpha_{b)}-\mathring\alpha_{(a}\mathring{\tilde\nabla}_{b)}-\mathring\alpha_a\mathring\alpha_b
					\\
					&\kern1cm+\tfrac{\delta_{ab}}{d-2}\big(-\mathring{\tilde\nabla}_c\mathring{\tilde\nabla}^c-\mathring{\tilde\nabla}_c\mathring\alpha^c+\mathring\alpha_c\mathring{\tilde\nabla}^c+\mathring\alpha_c\mathring\alpha^c\big)\big]
					\\
					&\kern1cm-\tfrac{r}{r'^2}\theta(r-r') \big[-\mathring{\tilde\nabla}_a\mathring{\tilde\nabla}_b+3\mathring\alpha_{(a}\mathring{\tilde\nabla}_{b)}-\mathring\alpha_a\mathring\alpha_b+\tfrac{\delta_{ab}}{d-2}\big(\mathring{\tilde\nabla}_c\mathring{\tilde\nabla}^c-3\mathring\alpha_c\mathring{\tilde\nabla}^c+\mathring\alpha_c\mathring\alpha^c\big)\big]
					\\
					&\kern1cm+\delta(r-r')\big(\mathring\alpha_a\mathring\alpha_b-\tfrac{\delta_{ab}}{d-2}\mathring\alpha_c\mathring\alpha^c\big)+\tfrac{2}{d-2}\tfrac{r-r'}{r'^2}\theta(r-r')\delta_{ab}~,
					\\
					\sft_{ab}{}^{cd}\ &\coloneqq\ -\delta(r-r')\big[\delta_{(a}{}^c\delta_{b)}{}^d-\tfrac{1}{d-2}\delta_{ab}\delta^{cd}\big]\,,
				\end{aligned}
			\end{equation}
		\end{subequations}
		and where $\theta(r)$ is the Heaviside step function. A short calculation shows that $\tilde\mu_1$ is again a local differential operator when acting on functions satisfying~\eqref{eq:deformationsBoundaryConditionsParticularSolutions}. Explicitly, we have
		\begin{subequations}\label{eq:upperTriangularMu1}
			\begin{equation}
				\tilde\mu_1\ =\
				\begin{pmatrix}
					\delta_a{}^c(1+r\partial_r) & 0 & \bar\sfd_a{}^{cd}
					\\
					0 & 1+2r\partial_r+\tfrac12r^2\partial_r^2 & \bar\sfd^{cd}
					\\
					0 & 0 & \bar\sfd_{ab}{}^{cd}+\tfrac{1}{d-2}\delta_{ab}\delta^{cd}
				\end{pmatrix}
			\end{equation}
			with
			\begin{equation}\label{eq:upperTriangularMu1B}
				\begin{aligned}
					\bar\sfd_a{}^{cd}\ &\coloneqq\ \delta_a{}^{(c}\mathring{\tilde\nabla}^{d)}-\tfrac{1}{d-2}\delta^{cd}\mathring{\tilde\nabla}_a-r\big(\delta_a{}^{(c}\mathring\alpha{}^{d)}-\tfrac{1}{d-2}\delta^{cd}\mathring\alpha_a\big)\partial_r~,
					\\
					\bar\sfd^{cd}\ &\coloneqq\ \tfrac12\mathring{\tilde\nabla}^{(c}\mathring\alpha^{d)}+\tfrac12\mathring{\tilde\nabla}^{(c}\mathring{\tilde\nabla}^{d)}-\tfrac12\mathring\alpha^{(c}\mathring{\tilde\nabla}^{d)}-\tfrac12\mathring\alpha^c\mathring\alpha^d-\tfrac{1}{2d-4}\delta^{cd}\big[\big(\mathring{\tilde\nabla}^e-\mathring\alpha^e\big)\mathring{\tilde\nabla}_e+\big(\mathring{\tilde\nabla}^e-\mathring\alpha^e\big)\mathring\alpha_e\big]
					\\
					&\kern1cm+r\big[\mathring\alpha^c\mathring\alpha^d-\mathring\alpha^{(c}\mathring{\tilde\nabla}^{d)}-\tfrac12\mathring{\tilde\nabla}^{(c}\mathring\alpha^{d)}-\tfrac{1}{d-2}\delta^{cd}\big(\mathring\alpha_e\mathring\alpha^e-\mathring\alpha_e{}\mathring{\tilde\nabla}^e-\tfrac12\mathring{\tilde\nabla}_{e}\mathring\alpha^e\big)\big]\partial_r
					\\
					&\kern1cm+\tfrac12r^2\big(\mathring\alpha^c\mathring\alpha^d-\tfrac{1}{d-2}\delta^{cd}\mathring\alpha^e\mathring\alpha_e\big)\partial_r^2~,
					\\
					\bar\sfd_{ab}{}^{cd}\ &\coloneqq\ \delta_{(b}{}^{(c}\big\{\mathring\alpha^{d)}\mathring{\tilde\nabla}_{a)}-\mathring\alpha_{a)}\mathring{\tilde\nabla}^{d)}-\mathring{\tilde\nabla}^{d)}\mathring{\tilde\nabla}_{a)}+\mathring{\tilde\nabla}_{a)}\mathring{\tilde\nabla}^{d)}+\delta_{a)}{}^{d)}\big[\tfrac12\big(\mathring{\tilde\nabla}_e-{\alpha}_e\big)\mathring{\tilde\nabla}^e+\tfrac{2}{d-2}\Lambda\big]\big\}
					\\
					&\kern1cm-\tfrac{1}{d-2}\delta_{ab}\delta^{cd}\big[\tfrac12\big(\mathring{\tilde\nabla}_e-{\alpha}_e\big)\mathring{\tilde\nabla}^e+\tfrac{2}{d-2}\Lambda\big]
					\\
					&\kern1cm+r\big\{\delta_{(b}{}^{(c}\big[\mathring{\tilde\nabla}^{d)}\mathring\alpha_{a)}-\mathring{\tilde\nabla}_{a)}\mathring\alpha^{d)}+\delta_{a)}{}^{d)}\big(\tfrac12\mathring\alpha_e\mathring\alpha^e-\mathring\alpha^e\mathring{\tilde\nabla}_e-\tfrac{2}{d-2}\Lambda\big)\big]
					\\
					&\kern1cm-\tfrac{1}{d-2}\delta_{ab}\delta^{cd}\big(\tfrac12\mathring\alpha_e\mathring\alpha^e-\mathring\alpha^e\mathring{\tilde\nabla}_e-\tfrac{2}{d-2}\Lambda\big)\big\}\partial_r
					\\
					&\kern1cm+r^2\big(\delta_{(a}{}^c\delta_{b)}{}^d-\tfrac{1}{d-2}\delta_{ab}\delta^{cd}\big)\big(\tfrac14\mathring\alpha^e\mathring\alpha_e+\tfrac14\mathring{\tilde\nabla}^e\mathring\alpha_e-\tfrac{1}{d-2}\Lambda\big)\partial_r^2~.
				\end{aligned}
			\end{equation}
		\end{subequations}
		Notice that $\bar\sfd_a{}^{cd}$, $\bar\sfd^{cd}$, and $\bar\sfd_{ab}{}^{cd}$ are all traceless over the $cd$ indices, and $\bar\sfd_{ab}{}^{cd}$ is also traceless over the $ab$ indices. This implies that the $\delta^{ab}h_{ab}$ part of $h_{ab}$ is decoupled.

		The above now implies that the Green functions $\sfg$ of $\mu_1$ and $\tilde\sfg$ of $\tilde\mu_1$ are related by
		\begin{equation}\label{eq:fullGreenFunction}
			\sfg(r,y;r',y')\ =\ \int\rmd r''\int\rmd^{d-2}y''\,\sqrt{\det(\mathring\gamma(y''))}\,\wick{\tilde\sfg(r,y;r'',\c y'')\,\sft(r'',\c y'';r',y')}~,
		\end{equation}
		and all that remains is finding $\tilde\sfg$; here, $\wick{\tilde\sfg(\ldots,\c y'')\sft(\ldots,\c y'')}$ refers to the derivatives with respect to $y''$ in $\sft$ to act on $\tilde\sfg$.

		To this end, we introduce the Green functions
		\begin{subequations}\label{eq:GreenFunction}
			\begin{equation}\label{eq:GreenFunction:a}
				\begin{aligned}
					\partial_r^\ell\sfg_\ell(r,y;r',y')\ &=\ \delta(r-r')\frac{\delta^{(d-2)}(y-y')}{\sqrt{\det(\mathring\gamma(y))}}
					\efor
					\ell\ \in\ \IN~,
					\\
					\bar\sfd_{ab}{}^{ef}(r,y)\bar\sfg_{ef}{}^{cd}(r,y;r',y')\ &=\ \big(\delta_{(a}{}^c\delta_{b)}{}^d-\tfrac{1}{d-2}\delta_{ab}\delta^{cd}\big)\delta(r-r')\frac{\delta^{(d-2)}(y-y')}{\sqrt{\det(\mathring\gamma(y))}}~.
				\end{aligned}
			\end{equation}
			Consequently, the Green function $\tilde\sfg$ is given by
			\begin{equation}\label{eq:GreenFunction:b}
				\tilde\sfg\ =\
				\begin{pmatrix}
					\delta_a{}^c\frac1r\sfg_1 & 0 & -\frac1r\sfg_1\circ\bar\sfd_a{}^{ef}\circ\bar\sfg_{ef}{}^{cd}
					\\
					0 & \frac{2}{r^2}\sfg_2 & -\frac{2}{r^2}\sfg_2\circ\bar\sfd^{ef}\circ\bar\sfg_{ef}{}^{cd}
					\\
					0 & 0 & \bar\sfg_{ab}{}^{cd}+\tfrac{1}{d-2}\delta_{ab}\delta^{cd}\delta(r-r')\frac{\delta^{(d-2)}(y-y')}{\sqrt{\det(\mathring\gamma(y))}}
				\end{pmatrix}
			\end{equation}
			with
			\begin{equation}\label{eq:GreenFunction:c}
				\begin{aligned}
					&\big(\sfg_1\circ\bar\sfd_a{}^{ef}\circ\bar\sfg_{ef}{}^{cd}\big)(r,y;r',y')
					\\
					&\kern1cm=\ \int\rmd r''\int\rmd^{d-2}y''\,\sqrt{\det(\mathring\gamma(y''))}\,\sfg_1(r,y;r'',y'')\bar\sfd_a{}^{ef}(r'',y'')\bar\sfg_{ef}{}^{cd}(r'',y'';r',y')~,
					\\
					&\big(\sfg_2\circ\bar\sfd{}^{ef}\circ\bar\sfg_{ef}{}^{cd}\big)(r,y;r',y')
					\\
					&\kern1cm=\ \int\rmd r''\int\rmd^{d-2}y''\,\sqrt{\det(\mathring\gamma(y''))}\,\sfg_2(r,y;r'',y'')\bar\sfd{}^{ef}(r'',y'')\bar\sfg_{ef}{}^{cd}(r'',y'';r',y')
				\end{aligned}
			\end{equation}
		\end{subequations}
		and so,
		\begin{equation}
			\tilde\mu_1(r,y)\tilde\sfg(r,y;r',y')\ =\
			\begin{pmatrix}
				\delta_a{}^c & 0 & 0
				\\
				0 & 1 & 0
				\\
				0 & 0 & \delta_{(a}{}^c\delta_{b)}{}^d
			\end{pmatrix}
			\delta(r-r')\frac{\delta^{(d-2)}(y-y')}{\sqrt{\det(\mathring\gamma(y))}}~.
		\end{equation}
		Hence, to find $\tilde\sfg$, we only need to find the Green functions~\eqref{eq:GreenFunction:a}.

		The Green functions $\sfg_1$ and $\sfg_2$ are easily constructed
		\begin{equation}
			\begin{aligned}
				\sfg_1(r,y;r',y')\ &=\ \theta(r-r')\frac{\delta^{(d-2)}(y-y')}{\sqrt{\det(\mathring\gamma(y))}}~,
				\\
				\sfg_2(r,y;r',y')\ &=\ (r-r')\theta(r-r')\frac{\delta^{(d-2)}(y-y')}{\sqrt{\det(\mathring\gamma(y))}}~.
			\end{aligned}
		\end{equation}
		To find the Green function $\bar\sfg_{ab}{}^{cd}$, we note that $\bar\sfd_{ab}{}^{cd}$ in~\eqref{eq:upperTriangularMu1} is of the form
		\begin{equation}\label{eq:defsfasfbsfc}
			\bar\sfd_{ab}{}^{cd}\ =\ r^2\bar\sfa_{ab}{}^{cd}(y)\partial_r^2+r\bar\sfb_{ab}{}^{cd}(y)\partial_r+\bar\sfc_{ab}{}^{cd}(y)
		\end{equation}
		with $\bar\sfa_{ab}{}^{cd}$ an invertible zeroth-order differential operator, and $\bar\sfb_{ab}{}^{cd}$ and $\bar\sfc_{ab}{}^{cd}$ first- and second-order differential operators, respectively. For fixed $y$, we thus obtain Euler's differential equation in $r$ which, in turn, allows us to separate out the $r$ and $r'$ dependence in $\bar\sfg_{ab}{}^{cd}(r,y;r',y')$ straightforwardly by constructing the Green function for the resulting partial differential equation. In particular, it is not too difficult to see that upon considering a function $\varrho_{ab}(r,y)=\sum_{n>0}\frac{r^n}{n!}\varrho^{(n)}_{ab}(y)$, we find
		\begin{subequations}
			\begin{equation}
				\begin{aligned}
					&\int\rmd r'\int\rmd^{d-2}y'\sqrt{\det(\mathring\gamma(y'))}\,\bar\sfg_{ab}{}^{cd}(r,y;r',y')\varrho_{cd}(r',y')
					\\
					&\kern1cm=\ \sum_{n>0}\frac{r^n}{n!}\int\rmd^{d-2}y'\sqrt{\det(\mathring\gamma(y'))}\,\bar\sfg^{(n)}_{ab}{}^{cd}(y,y')\varrho^{(n)}_{cd}(y')
				\end{aligned}
			\end{equation}
			with
			\begin{equation}\label{eq:spatialCrossSectionGreenFunctionOrderN}
				\begin{aligned}
					&\big[n(n-1)\bar\sfa_{ab}{}^{cd}(y)+n\bar\sfb_{ab}{}^{cd}(y)+\bar\sfc_{ab}{}^{cd}(y)\big]\bar\sfg^{(n)}_{ab}{}^{cd}(y;y')
					\\
					&\kern1cm=\ \big(\delta_{(a}{}^c\delta_{b)}{}^d-\tfrac{1}{d-2}\delta_{ab}\delta^{cd}\big)\frac{\delta^{(d-2)}(y-y')}{\sqrt{\det(\mathring\gamma(y))}}~.
				\end{aligned}
			\end{equation}
		\end{subequations}
		The explicit form of the Green function $\bar\sfg^{(n)}_{ab}{}^{cd}(y;y')$ now depends on the chosen near-horizon geometry. We will construct it explicitly in our example in \cref{sec:extremalKerrFirstOrder}.

		\paragraph{Lowest-order Einstein equations simplified.}
		With the gauge fixing condition~\eqref{eq:gaugeFixingCondition} and the operator $\sft$ defined in~\eqref{eq:operatorT}, we can further simplify the lowest-order Einstein equations~\eqref{eq:zerothOrderEinsteinEquation}. We first note that in this gauge, the boundary conditions are
		\begin{equation}\label{eq:boundaryAndGaugeFixingConditions}
			h_a|_{r=0}\ =\ 0~,
			\quad
			h|_{r=0}\ =\ 0~,
			\quad
			h_{ab}|_{r=0}\ =\ 0~,
			\eand
			\delta^{ab}\partial_rh_{ab}|_{r=0}\ =\ \Gamma
		\end{equation}
		with $\sfD^\dagger\Gamma=0$ and $\sfD^\dagger$ as defined in~\eqref{eq:definitionDAndDAdjointAdaptedBasisII}. Under these conditions, using the operator $\sft$ defined in~\eqref{eq:operatorT}, the equation~\eqref{eq:zerothOrderEinsteinEquation} can be transformed into the equivalent equation
		 \begin{subequations}\label{eq:simplifiedLowestOrderEinstein}
			\begin{equation}\label{eq:simplifiedLowestOrderEinsteinA}
				\tilde\mu_1(\Theta)\ =\ \rho
				\ewith
				\Theta\ \coloneqq\
				\begin{pmatrix}
					h_a
					\\
					h
					\\
					h_{ab}
				\end{pmatrix}
				\eand
				\rho\ \coloneqq\
				\begin{pmatrix}
					\varrho_a
					\\
					\varrho
					\\
					\varrho_{ab}
				\end{pmatrix},
			\end{equation}
			where $\tilde\mu_1$ is as defined in~\eqref{eq:upperTriangularMu1} and
			\begin{equation} \label{eq:Defvarrho}
				\begin{aligned}
					\varrho_a\ &\coloneqq\ r\big(\tfrac{d-3}{d-2}\mathring{\tilde\nabla}_a\Gamma-\tfrac{d-4}{2(d-2)}\mathring\alpha_a\Gamma\big)\,,
					\\
					\varrho\ &\coloneqq\ r\big(\tfrac{1}{d-2}\Lambda\Gamma-\tfrac{2d-5}{2(d-2)}\mathring{\tilde\nabla}_a\mathring\alpha^a\Gamma-\tfrac{2(d-3)}{d-2}\mathring\alpha^a\mathring{\tilde\nabla}_a\Gamma+\tfrac{3d-8}{4(d-2)}\mathring\alpha_a\mathring\alpha^a\Gamma\big)\,,
					\\
					\varrho_{ab}\ &\coloneqq\ -r\big[\tfrac12\mathring{\tilde\nabla}_{(a}\mathring\alpha_{b)}\Gamma+2\mathring\alpha_{(a}\mathring{\tilde\nabla}_{b)}\Gamma-\tfrac12\mathring\alpha_{(a}\mathring\alpha_{b)}\Gamma
					\\
					&\kern1cm-\tfrac{1}{d-2}\delta_{ab}\big(\tfrac12\mathring{\tilde\nabla}^c\mathring\alpha_c\Gamma+2\mathring\alpha^c\mathring{\tilde\nabla}_c\Gamma-\tfrac12\mathring\alpha_c\mathring\alpha^c\Gamma+\Gamma\big)\big]\,.
				\end{aligned}
			\end{equation}
		\end{subequations}

		Notice that the first two rows and the trace of the last row of~\eqref{eq:simplifiedLowestOrderEinsteinA} can be uniquely solved to obtain $h_a$, $h$ and $\delta^{ab}h_{ab}$ in terms of $\Gamma$ and $\bar h_{ab}\coloneqq h_{ab}-\tfrac{1}{d-2}\delta_{ab}\delta^{cd}h_{cd}$. By writing $\bar h_{ab}$ as $\bar h_{ab}=\sum_{n>0}\tfrac{r^n}{n!}\bar h_{ab}^{(n)}$, the traceless part of the last row of~\eqref{eq:simplifiedLowestOrderEinsteinA} can be written as
		\begin{subequations}\label{eq:simplifiedLowestOrderEinsteinTracelessComponents}
			\begin{equation}
				[\bar\sfb_{ab}{}^{cd}(y)+\bar\sfc_{ab}{}^{cd}(y)\big]\bar h_{cd}^{(1)}(y)\ =\ \tfrac{1}{r}\big(\varrho_{ab}-\tfrac{1}{d-2}\delta_{ab}\delta^{cd}\varrho_{cd}\big)
			\end{equation}
			and
			\begin{equation}
				[n(n-1)\bar\sfa_{ab}{}^{cd}(y)+n\bar\sfb_{ab}{}^{cd}(y)+\bar\sfc_{ab}{}^{cd}(y)\big]\bar h_{cd}^{(n)}(y)\ =\ 0
			\end{equation}
		\end{subequations}
		for $n>1$, where $\bar\sfa_{ab}{}^{cd},\bar\sfb_{ab}{}^{cd}$ and $\bar\sfc_{ab}{}^{cd}$ are defined in~\eqref{eq:defsfasfbsfc}. By examining the form of $\bar\sfd_{ab}{}^{cd}$ in~\eqref{eq:upperTriangularMu1B}, one can deduce that~\eqref{eq:simplifiedLowestOrderEinsteinTracelessComponents} are elliptic equations on the spatial cross section. From standard Fredholm theory, it follows that the space of solutions of $h_{ab}^{(n)}$ is finite-dimensional. Therefore, at each order in $r$ the moduli space of deformations is finite-dimensional.

		\subsection{Example: extremal Kerr}\label{sec:extremalKerrFirstOrder}

		We shall now apply the above to the example of the extremal Kerr black hole.

		\paragraph{Lowest-order Einstein equations.}
		Upon specialising to the near-horizon extremal Kerr metric~\eqref{eq:extremalKerrNearHorizonMetricGaussianNullCoordinates}, we shall simplify
		\begin{equation}\label{eq:simplifiedLowestOrderEinsteinExtremalKerr}
			\tilde\mu_1(\Theta)\ =\ \rho
			\ewith
			\Theta\ \coloneqq\
			\begin{pmatrix}
				h_a
				\\
				h
				\\
				h_{ab}
			\end{pmatrix}
			\eand
			\rho\ \coloneqq\
			\begin{pmatrix}
				\varrho_a
				\\
				\varrho
				\\
				\varrho_{ab}
			\end{pmatrix},
		\end{equation}
		given in~\eqref{eq:simplifiedLowestOrderEinstein}, where $\varrho$ depends on $\Gamma$. Recall that $\Gamma$ is a solution of $\sfD^\dagger\Gamma=0$, where $\sfD^\dagger$ is defined~\eqref{eq:definitionDAndDAdjointAdaptedBasisII}. It is not too difficult to see that this is solved (uniquely) by
		\begin{equation}
			\Gamma\ =\ \tfrac{1}{m}A(1+x^2)~,
		\end{equation}
		where $A$ is an arbitrary constant. Hence,
		\begin{equation}
			\begin{gathered}
				\varrho_1\ =\ A\frac{x}{m^2}\sqrt{\frac{1-x^2}{1+x^2}}r~,
				\quad
				\varrho_2\ =\ 0~,
				\\
				\varrho\ =\ A\frac{7-10x^2-9x^4}{2m^3(1+x^2)^2}r~,
				\\
				\varrho_{11}-\tfrac12\Gamma r\ =\ -\varrho_{22}+\tfrac12\Gamma r\ =\ -A\frac{1-10x^2+9x^4}{2m^3(1+x^2)^2}r~,
				\\
				\varrho_{12}\ =\ \varrho_{21}\ =\ -A\frac{5x(1-x^2)}{m^3(1+x^2)^2}r~.
			\end{gathered}
		\end{equation}

		We shall now make the assumption of \uline{axis-symmetry}, that is, we require that there is no explicit $\varphi$-dependence. The properties of non-axisymmetric first order deformations were considered in \cite{Jezierski:2012lzy}, where a Fourier decomposition was used. It was shown that for all but a finite number of Fourier modes, regular $\varphi$ dependent first-order deformations do not exist. The remaining Fourier modes were analysed in~\cite{Chrusciel:2017vie}, where a numerical analysis suggests that the solutions associated to these modes also cannot be regular. In this setting, the operators $\bar\sfd_a{}^{cd}$, $\bar\sfd^{cd}$, and $\bar\sfd_{ab}{}^{cd}$ featuring in $\tilde\mu_1$ in~\eqref{eq:upperTriangularMu1} are thus given by
		\begin{subequations}\label{eq:differentialExtremalKerrAxisSymmetric}
			\begin{equation}
				\begin{aligned}
					\bar\sfd_1{}^{11}\ &=\ \bar\sfd_2{}^{12}\ =\ \frac{1}{m}\sqrt{\frac{1-x^2}{1+x^2}}\bigg(\frac{1}{2}\partial_x-\frac{2x}{1-x^4}+r\frac{x}{1+x^2}\partial_r\bigg)~,
					\\
					\bar\sfd_1{}^{12}\ &=\ -\bar\sfd_2{}^{11}\ =\ -\frac{r}{m}\sqrt{\frac{1-x^2}{1+x^2}}\frac{1}{1+x^2}\partial_r~,
					\\
					\bar\sfd^{11}\ &=\ -\frac{1}{2m^2}\bigg\{-\frac{1-x^2}{2(1+x^2)}\partial^2_x+\frac{x(3+x^2)}{(1+x^2)^2}\partial_x+\frac{1-x^2}{(1+x^2)^2}
					\\
					&\kern1cm+r\bigg[-\frac{2x(1-x^2)}{(1+x^2)^2}\partial_x+\frac{3+2x^2+3x^4}{(1+x^2)^3}\bigg]\partial_r
					\\
					&\kern1cm+r^2\frac{2(1-x^2)^2}{(1+x^2)^3}\partial^2_r\bigg\}\,,
					\\
					\bar \sfd^{12}\ &=\ -\frac{1}{2m^2}\bigg\{\frac{1-x^2}{(1+x^2)^2}\partial_x+\frac{2x(-3+x^2)}{(1+x^2)^3}+r\bigg[\frac{2(1-x^2)}{(1+x^2)^2}\partial_x-\frac{2x(1+3x^2)}{(1+x^2)^3}\bigg]\partial_r
					\\
					&\kern1cm+r^2\frac{4x(1-x^2)}{(1+x^2)^3}\partial^2_r\bigg\}\,,
					\\
					\bar\sfd_{ab}{}^{cd}\ &=\ \bar\sfa_{ab}{}^{cd}r^2\partial^2_r+\bar\sfb_{ab}{}^{cd}r\partial_r+\bar\sfc_{ab}{}^{cd}~,
				\end{aligned}
			\end{equation}
			where
			\begin{equation}\label{eq:sourcesExtremalKerrAxisSymmetric}
				\begin{gathered}
					\bar\sfa_{11}{}^{11}\ =\ \bar\sfa_{12}{}^{12}\ =\ \frac{1}{m^2}\frac{1+6x^2-3x^4}{4(1+x^2)^3}~,
					\quad
					\bar\sfa_{11}{}^{12}\ =\ -\bar\sfa_{12}{}^{11}\ =\ 0~,
					\\
					\bar\sfb_{11}{}^{11}\ =\ \bar\sfb_{12}{}^{12}\ =\ \frac{1}{m^2}\bigg[\frac{x(1-x^2)}{(1+x^2)^2}\partial_x+ \frac{1-x^2}{(1+x^2)^2}\bigg]\,,
					\\
					\bar\sfb_{11}{}^{12}\ =\ -\bar\sfb_{12}{}^{11}\ =\ \frac{1}{m^2}\bigg[\frac{2x(1-x^2)}{(1+x^2)^3}\bigg]\,,
					\\
					\bar\sfc_{11}{}^{11}\ =\ \bar\sfc_{12}{}^{12}\ =\ \frac{1}{m^2}\bigg[\frac{1-x^2}{4(1+x^2)}\partial^2_x-\frac{x}{2(1+x^2)}\partial_x -\frac{2(1+x^4)}{(1-x^2)(1+x^2)^3}\bigg]\,,
					\\
					\bar\sfc_{11}{}^{12}\ =\ -\bar\sfc_{12}{}^{11}\ =\ \frac{1}{m^2}\bigg[\frac{1-x^2}{(1+x^2)^2}\partial_x - \frac{2x}{(1+x^2)^3}\bigg]\,.
				\end{gathered}
			\end{equation}
		\end{subequations}
		The remaining components not displayed here follow straightforwardly from the tracelessness condition of $\bar\sfd_a{}^{cd}$, $\bar\sfd^{cd}$, and $\bar\sfd_{ab}{}^{cd}$, respectively.

		\paragraph{Smoothness conditions.}
		Generally, the metric must be smooth but it is not obvious that this will be the case in our given chart as there are coordinate singularities for $x=\pm1$; see e.g.~the metric~\eqref{eq:extremalKerrNearHorizonMetricGaussianNullCoordinates} or the operator~\eqref{eq:differentialExtremalKerrAxisSymmetric}. Therefore, we need conditions on our deformations as $x\to\pm1$ such that the resulting metric can be smoothly extended to $x=\pm1$.

		To this end, following~\cite{Kunduri:2013gce}, consider a general deformation on the spatial cross section
		\begin{equation}
			\tfrac12h_{ij}\rmd y^i\odot\rmd y^j\ =\ \tfrac12h_{xx}\rmd x\odot\rmd x+h_{x\varphi}\rmd x\odot\rmd\varphi+\tfrac12h_{\varphi\varphi}\rmd\varphi\odot\rmd\varphi
		\end{equation}
		and the change of coordinates
		\begin{equation}
			(x,\varphi)\ \mapsto\ \big(\sqrt{1\mp x}\sin\varphi,\sqrt{1\mp x}\cos\varphi\big)\,.
		\end{equation}
		Upon performing this change of coordinates in $h$ and requiring that $h$ can be smoothly extended to $x=\pm1$, it follows that
		\begin{equation}
			h_{x\varphi}\ =\ \caO(1\mp x)~,
			\quad
			h_{\varphi\varphi}\ =\ \caO(1\mp x)~,
			\eand
			h_{xx}\ =\ \frac{h_{\varphi\varphi}}{4(1\mp x)^2}+\cdots
		\end{equation}
		as $x\to\pm1$ and where the ellipsis denotes terms that are smooth as $x\to\pm1$. In the basis~\eqref{eq:defAdaptedFrameII} (see also~\eqref{eq:defAdaptedFrameIIExtremalKerr}), these conditions amount to
		\begin{equation}\label{eq:smoothnessCondition}
			\bar h_{ab}|_{x\to\pm1}\ =\ 0
			\ewith
			\bar h_{ab}\ \coloneqq\ h_{ab}-\tfrac12\delta_{ab}\delta^{cd}h_{cd}~.
		\end{equation}

		One can repeat the same analysis and show that for $h_a$ to be smoothly extended to $x=\pm1$ provided that
		\begin{equation}\label{eq:smoothnessConditionOneForm}
			h_a\ =\ \caO(\sqrt{1\mp x})
		\end{equation}
		as $x\to\pm1$. However, using the Green function~\eqref{eq:GreenFunction:b}, we obtain
		\begin{subequations}\label{eq:SolhaInTermsOfhab}
			\begin{equation}
				h_a(r,y)\ =\ \frac{1}{r}\int\rmd r'\int\rmd^{d-2}y'\,\sqrt{\det(\mathring\gamma(y'))}\,\sfg_1(r,y;r',y')\big[\varrho_a(r',y')-\bar\sfd_a{}^{cd}(r',y')\bar h_{cd}(r',y')\big]
			\end{equation}
			with
			\begin{equation}
				\bar{h}_{ab}(r,y)\ =\ \int\rmd r'\int\rmd^{d-2}y''\,\sqrt{\det(\mathring\gamma(y'))}\,\bar\sfg_{ab}{}^{cd}(r,y;r',y')\varrho_{cd}(r',y')
			\end{equation}
		\end{subequations}
		for $h_a$ as a solution to~\eqref{eq:simplifiedLowestOrderEinsteinExtremalKerr}. Since every operator acting on $\varrho_a$ and $\bar h_{ab}$ in~\eqref{eq:SolhaInTermsOfhab} are smooth covariant operators with respect to the spatial cross section (the spatial part of $\sfg_1$ is the delta function), the $h_a$ provided in~\eqref{eq:SolhaInTermsOfhab} is smooth if both $\bar h_{ab}$ and $\varrho_a$ are smooth. Hence, we do not need to impose~\eqref{eq:smoothnessConditionOneForm} as an extra condition on the solution since it follows from the smoothness condition~\eqref{eq:smoothnessCondition} on $h_{ab}$. Finally, we also note that there are no extra smoothness conditions on the deformation $h$, since it can always be smoothly extended to $x=\pm1$.

		In conclusion, we shall augment the boundary conditions~\eqref{eq:boundaryAndGaugeFixingConditions} to
		\begin{equation}\label{eq:augmentedBoundaryAndGaugeFixingConditions}
			h_a|_{r=0}\ =\ 0~,
			\quad
			h|_{r=0}\ =\ 0~,
			\quad
			h_{ab}|_{r=0}\ =\ 0~,
			\quad
			\bar h_{ab}|_{x\to\pm1}\ =\ 0~,
			\eand
			\delta^{ab}\partial_rh_{ab}|_{r=0}\ =\ \Gamma~.
		\end{equation}

		\paragraph{Lowest-order solutions.}
		To construct solutions to~\eqref{eq:simplifiedLowestOrderEinsteinExtremalKerr} in the near-horizon extremal Kerr setting with axis-symmetry subject to the boundary conditions~\eqref{eq:augmentedBoundaryAndGaugeFixingConditions}, we note that the trace part of the last row~\eqref{eq:simplifiedLowestOrderEinsteinExtremalKerr} reduces to the algebraic condition
		\begin{equation}\label{eq:extremalKerrLowestOrderSolutionTracePart}
			\delta^{ab}h_{ab}\ =\ \Gamma r\ =\ \tfrac1mA(1+x^2)r~.
		\end{equation}
		Consequently, the conditions $\delta^{ab}h_{ab}|_{r=0}=0$ as well as the gaug-fixing condition in~\eqref{eq:augmentedBoundaryAndGaugeFixingConditions} are already built in. Hence, the trace part of $h_{ab}$ is fixed. Therefore, we only need to solve for $(h_a,h,\bar h_{ab})$ with $\bar h_{ab}$ defined in~\eqref{eq:smoothnessCondition} subject to
		\begin{equation}\label{eq:reducedAugmentedBoundaryAndGaugeFixingConditions}
			h_a|_{r=0}\ =\ 0~,
			\quad
			h|_{r=0}\ =\ 0~,
			\quad
			\bar h_{ab}|_{r=0}\ =\ 0~,
			\eand
			\bar h_{ab}|_{x\to\pm1}\ =\ 0~.
		\end{equation}

		We first construct the most general homogeneous solution $(h_a^{(h)},h^{(h)},\bar h_{ab}^{(h)})$. In particular, the last row of~\eqref{eq:simplifiedLowestOrderEinsteinExtremalKerr} reads as
		\begin{equation}\label{eq:simplifiedLinearEqKerrHorizonHab}
			\bar\sfd_{ab}{}^{cd}\bar h^{(h)}_{cd}\ =\ 0
		\end{equation}
		with $\bar\sfd_{ab}{}^{cd}$ as given in~\eqref{eq:differentialExtremalKerrAxisSymmetric}. Because of the boundary condition $\bar h^{(h)}_{ab}|_{r=0}=0$ in~\eqref{eq:reducedAugmentedBoundaryAndGaugeFixingConditions}, the Taylor expansion of $\bar h^{(h)}_{ab}$ is of the form
		\begin{equation}
			\bar h^{(h)}_{ab}\ =\ \sum_{n>0}\frac{r^n}{n!}\bar h^{(h,n)}_{ab}
		\end{equation}
		with $\bar h^{(h,n)}_{ab}=\bar h^{(h,n)}_{ab}(x)$ and so,~\eqref{eq:simplifiedLinearEqKerrHorizonHab} becomes
		\begin{equation}
			\big[n(n-1)\bar\sfa_{ab}{}^{cd}+n\bar\sfb_{ab}{}^{cd}+\bar\sfc_{ab}{}^{cd}\big]\bar h^{(h,n)}_{ab}\ =\ 0~.
		\end{equation}
		This is generally solved by
		\begin{subequations}\label{eq:extremalKerrLowestOrderSolutionBarHAB}
			\begin{equation}\label{eq:extremalKerrLowestOrderSolutionBarHAB:a}
				\begin{aligned}
					\bar h^{(h,1)}_{11}(x)\ &=\ \frac{1}{m}\left[\frac{3K^{(1)}_1+K^{(1)}_2x(3-x^2)}{3(1+x^2)^2}-\frac{2x(K^{(1)}_4x(3-x^2)+3 K^{(1)}_3)}{3(1-x^2)(1+x^2)^2}\right],
					\\
					\bar h^{(h,1)}_{12}(x)\ &=\ \frac{1}{m}\left[\frac{2x(K^{(1)}_2x(3-x^2)+3K^{(1)}_1)}{3 (1-x^2)(1+x^2)^2}+\frac{3 K^{(1)}_3+K^{(1)}_4x(3-x^2)}{3(1+x^2)^2}\right]
					\end{aligned}
			\end{equation}
			and
			\begin{equation}\label{eq:extremalKerrLowestOrderSolutionBarHAB:b}
				\begin{aligned}
					\bar h^{(h,n)}_{11}(x)\ &=\ \frac{(K^{(n)}_1(1-x^2)-2K^{(n)}_3 x)\sfP_n^2(x)+(K^{(n)}_2(1-x^2)-2K^{(n)}_4x)\sfQ_n^2(x)}{m^n(1+x^2)^{n+1}}~,
					\\
					\bar h^{(h,n)}_{12}(x)\ &=\ \frac{(2K^{(n)}_1x+K^{(n)}_3(1-x^2))\sfP_n^2(x)+(2K^{(n)}_2x+K^{(n)}_4(1-x^2))\sfQ_n^2(x)}{m^n(1+x^2)^{n+1}}
				\end{aligned}
			\end{equation}
		\end{subequations}
		for all $n>1$. Here, $K^{(n)}_{1,\ldots,4}$ are arbitrary constants and $\sfP_n^2$ and $\sfQ_n^2$ are the \uline{associated Legendre functions of the first and second kind}, respectively. Again, the remaining components not displayed follow from the tracelessness condition. Upon imposing the smoothness condition $\bar h^{(h)}_{ab}\big|_{x\to\pm1}=0$ from~\eqref{eq:reducedAugmentedBoundaryAndGaugeFixingConditions}, the general solution~\eqref{eq:extremalKerrLowestOrderSolutionBarHAB} reduces to
		\begin{equation}\label{eq:extremalKerrSmoothLowestOrderSolutionBarHAB}
			\begin{gathered}
			   \bar h^{(h,1)}_{11}(x)\ =\ \bar h^{(h,1)}_{12}(x)\ =\ 0~,
			   \\
			   \bar h^{(h,n)}_{11}\ =\ \frac{(K^{(n)}_1(1-x^2)-2K^{(n)}_3 x)\sfP_n^2(x)}{m^n(1+x^2)^{n+1}}~,
			   \quad
			   \bar h^{(h,n)}_{12}\ =\ \frac{(2K^{(n)}_1x+K^{(n)}_3(1-x^2))\sfP_n^2(x)}{m^n(1+x^2)^{n+1}}
		   \end{gathered}
		\end{equation}
		for all $n>1$. The homogeneous solutions $h^{(h)}_a$ and $h^{(h)}$ now follow immediately as one can simply integrate the first and second row of the homogeneous part of~\eqref{eq:simplifiedLowestOrderEinsteinExtremalKerr}. Concretely, we have
		\begin{subequations}
			\begin{equation}
				h^{(h)}_a\ =\ \sum_{n>0}\frac{r^n}{n!}h^{(h,n)}_a
				\eand
				h^{(h)}\ =\ \sum_{n>0}\frac{r^n}{n!}h^{(h,n)}
			\end{equation}
			with $h^{(h,n)}_a=h^{(h,n)}_a(x)$ and $h^{(h,n)}=h^{(h,n)}(x)$ and to accommodate the boundary conditions $h^{(h)}_a\big|_{r=0}=0$ and $h^{(h)}\big|_{r=0}=0$ in~\eqref{eq:reducedAugmentedBoundaryAndGaugeFixingConditions}. Using the explicit smooth solution~\eqref{eq:extremalKerrSmoothLowestOrderSolutionBarHAB} for $\bar h^{(h)}_{ab}$, a short calculation shows that these coefficients are given by
			\begin{equation}
				h^{(h,1)}_a(x)\ =\ 0
				\eand
				h^{(h,1)}(x)\ =\ 0~,
			\end{equation}
			and
			\begin{equation}
				\begin{aligned}
					h^{(h,n)}_1(x)\ &=\ \frac{(1+x^2)^{-n-3}}{m^{n+1}(n+1)}\sqrt{\frac{1+x^2}{1-x^2}}\big\{(n-1)(1+x^2)\sfP_{n+1}^2(x)[(1-x^2)K_1^{(n)}-2xK_3^{(n)}]
					\\
					&\kern1cm+\sfP_n^2(x)\big[x(1-x^2)K_1^{(n)}(3n+7-(n+1)x^2)
					\\
					&\kern1cm+2K_3^{(n)}(n+1-(n+5)x^2+2(n+1) x^4)\big]\big\}\,,
					\\
					h^{(h,n)}_2(x)\ &=\ \frac{(1+x^2)^{-n-3}}{m^{n+1}(n+1)}\sqrt{\frac{1+x^2}{1-x^2}}\big\{(n-1)(1+x^2)\sfP_{n+1}^2(x)[2xK_1^{(n)}+(1-x^2)K_3^{(n)}]
					\\
					&\kern1cm+\sfP_n^2(x)\big[x(1-x^2)K_3^{(n)}(3n+7-(n+1)x^2)
					\\
					&\kern1cm-2K_1^{(n)}(n+1-(n+5)x^2+2(n+1) x^4)\big]\big\}\,,
					\\
					h^{(h,n)}(x)\ &=\ -\frac{(1+x^2)^{-n-4}}{m^{n+2} (n+1)(n+2)(1-x^2)}
					\\
					&\kern1cm\times\big\{2K_3^{(n)}\big((n-1)(1+x^2)[(2n+3-(2n+7)x^2+2(2n+3)x^4)\sfP_{n+1}^2(x)
					\\
					&\kern1cm-nx(1+x^2)\sfP_{n+2}^2(x)]+x[2(n+2)(n+5)-(7n^2+34n+43)x^2
					\\
					&\kern1cm+2 (n+5) (2 n+3) x^4-3 (n+1)^2 x^6]\sfP_n^2(x)\big)
					\\
					&\kern1cm+(1-x^2)K_1^{(n)}\big(\sfP_n^2(x)[-4n^2-11n-9+(n^2+35n+60) x^2
					\\
					&\kern1cm-(10n^2+33n+29)x^4+(n+2)(n-1)x^6]
					\\
					&\kern1cm+(n-1)(1+x^2)[x(6n+13-(2n+3)x^2)\sfP_{n+1}^2(x)
					\\
					&\kern1cm+n(1+x^2)\sfP_{n+2}^2(x)]\big)\big\}
				\end{aligned}
			\end{equation}
		\end{subequations}
		for all $n>1$.

		Next, to construct a particular solution $(h_a^{(p)},h^{(p)},\bar h_{ab}^{(p)})$ to~\eqref{eq:simplifiedLowestOrderEinsteinExtremalKerr}, we need $\tilde\sfg$ defined in~\eqref{eq:GreenFunction:b}. As discussed, the non-trivial part for constructing $\tilde\sfg$ is finding the Green function $\bar\sfg^{(n)}_{ab}{}^{cd}$ given by~\eqref{eq:spatialCrossSectionGreenFunctionOrderN}. Notice that since $\varrho_a$, $\varrho$, and $\varrho_{ab}$ only depend linearly on $r$, we only need $\bar\sfg^{(1)}_{ab}{}^{cd}$. This is simply the Green function for a boundary value problem of~\eqref{eq:simplifiedLinearEqKerrHorizonHab} since we require $h_{ab}^{(p)}$ to vanish as $x\to\pm1$ by virtue of~\eqref{eq:reducedAugmentedBoundaryAndGaugeFixingConditions}. With the help of the Heaviside step function, we can construct the Green function.\footnote{\label{foot:GreenFunction}Concretely, one needs to diagonalise~\eqref{eq:simplifiedLinearEqKerrHorizonHab}, which leads to two independent uncoupled ordinary differential equations. Then, for a boundary value problem for $x\in[-1,1]$, the Green function of the operator $a(x)\partial^2_x+b(x)\partial_x+c(x)$ is given by $\tfrac{1}{a(x')W(x')}[\theta(x'-x)f_1(x)f_2(x')+\theta(x-x')f_2(x)f_1(x')]$ where $f_1$ and $f_2$ are two independent homogeneous solutions satisfying the boundary conditions at $x=-1$ and $x=1$, respectively, and $W$ is the \uline{Wronskian} of $f_1$ and $f_2$.} It is not too difficult to show that
		\begin{subequations}\label{eq:spatialGreenFunctionExtremalKerrN=1}
			\begin{equation}
				\begin{aligned}
					\bar\sfg^{(1)}_{11}{}^{11}(x,\varphi;x',\varphi')\ &=\ -\frac{(1+x'^2)[1-x'^2+4xx'-x^2(1-x'^2)]}{24(1-x^2)(1-x'^2)(1+x^2)^2}\delta(\varphi-\varphi')
					\\
					&\kern1cm\times[(2-x)(1+x)^2(2+x')(1-x')^2\theta(x'-x)
					\\
					&\kern1cm+(1-x)^2(2+x)(2-x')(1+x')^2\theta(x-x')]~,
					\\
					\bar\sfg^{(1)}_{22}{}^{12}(x,\varphi;x',\varphi')\ &=\ -\frac{(1+x'^2)(x-x')(1+xx')}{12(1-x^2)(1-x'^2)(1+x^2)^2}\delta(\varphi-\varphi')
					\\
					&\kern1cm\times[(2-x)(1+x)^2(2+x')(1-x')^2\theta(x'-x)
					\\
					&\kern1cm+(1-x)^2(2+x)(2-x')(1+x')^2\theta(x-x')]
				\end{aligned}
			\end{equation}
			as well as
			\begin{equation}
				\bar\sfg^{(1)}_{11}{}^{11}(x,\varphi;x',\varphi')\ =\ \bar\sfg^{(1)}_{12}{}^{12}(x,\varphi;x',\varphi')
				\eand
				\bar\sfg^{(1)}_{11}{}^{12}(x,\varphi;x',\varphi')\ =\ -\bar\sfg^{(1)}_{12}{}^{11}(x,\varphi;x',\varphi')
			\end{equation}
		\end{subequations}
		when acting on sources that vanish as $x\to\pm1$.\footnote{Our sources in~\eqref{eq:sourcesExtremalKerrAxisSymmetric} do satisfy this conditions.} Again, the remaining components can be recovered from traceless and symmetric properties.

		We now have all the ingredients to solve~\eqref{eq:simplifiedLowestOrderEinsteinExtremalKerr}. In particular, using the above, the most general solution to~\eqref{eq:simplifiedLowestOrderEinsteinExtremalKerr} subject to the boundary conditions~\eqref{eq:augmentedBoundaryAndGaugeFixingConditions} is given by
		\begin{subequations}\label{eq:generalLowestOrderSolutionExtremalKerr}
			\begin{equation}
				h_a\ =\ \sum_{n>0}\frac{r^n}{n!}h^{(n)}_a~,
				\quad
				h\ =\ \sum_{n>0}\frac{r^n}{n!}h^{(n)}~,
				\eand
				h_{ab}\ =\ \sum_{n>0}\frac{r^n}{n!}h^{(n)}_{ab}
			\end{equation}
			with
			\begin{equation}
				\begin{gathered}
					h^{(1)}_1(x)\ =\ \frac{Ax(1-x^2)(59-55x^2-23x^4-5x^6)}{10m^2(1+x^2)^4}\sqrt{\frac{1+x^2}{1-x^2}}~,
					\\
					h^{(1)}_2(x)\ =\ -\frac{A(1-x^2)(7-45x^2-3x^4+x^6)}{5m^2(1+x^2)^4}\sqrt{\frac{1+x^2}{1-x^2}}~,
					\\
					h^{(1)}(x)\ =\ \frac{2A(35-225x^2+135x^4+5x^6-6x^8)}{15m^3(1+x^2)^5}~,
					\\
					h^{(1)}_{11}(x)-\frac12\Gamma\ =\ -h^{(1)}_{22}(x)+\frac12\Gamma\ =\ \frac{A}{m}\frac{(1-x^2)(5-16x^2-5x^4)}{10(1+x^2)^2}~,
					\\
					\bar{h}^{(1)}_{12}(x)\ =\ \frac{A}{m}\frac{x(1-x^2)(9+x^2)}{5(1+x^2)^2}
				\end{gathered}
			\end{equation}
			and
			\begin{equation}
				\begin{aligned}
					h^{(n)}_1(x)\ &=\ \frac{(1+x^2)^{-n-3}}{m^{n+1}(n+1)}\sqrt{\frac{1+x^2}{1-x^2}}\big\{(n-1)(1+x^2)\sfP_{n+1}^2(x)[(1-x^2)K_1^{(n)}-2xK_3^{(n)}]
					\\
					&\kern1cm+\sfP_n^2(x)\big[x(1-x^2)K_1^{(n)}(3n+7-(n+1)x^2)
					\\
					&\kern1cm+2K_3^{(n)}(n+1-(n+5)x^2+2(n+1)x^4)\big]\big\}\,,
					\\
					h^{(n)}_2(x)\ &=\ \frac{(1+x^2)^{-n-3}}{m^{n+1}(n+1)}\sqrt{\frac{1+x^2}{1-x^2}}\big\{(n-1)(1+x^2)\sfP_{n+1}^2(x)[2 xK_1^{(n)}+(1-x^2)K_3^{(n)}]
					\\
					&\kern1cm+\sfP_n^2(x)\big[x(1-x^2)K_3^{(n)}(3n+7-(n+1)x^2)
					\\
					&\kern1cm-2K_1^{(n)}(n+1-(n+5)x^2+2(n+1)x^4)\big]\big\}\,,
					\\
					h^{(n)}(x)\ &=\ -\frac{(1+x^2)^{-n-4}}{m^{n+2}(n+1)(n+2)(1-x^2)}
					\\
					&\kern1cm\times\big\{2K_3^{(n)}\big((n-1)(1+x^2)[(2n+3-(2n+7)x^2+2(2n+3)x^4)\sfP_{n+1}^2(x)
					\\
					&\kern1cm-nx(1+x^2)\sfP_{n+2}^2(x)]+x[2(n+2)(n+5)-(7n^2+34n+43)x^2
					\\
					&\kern1cm+2(n+5)(2n+3)x^4-3(n+1)^2x^6]\sfP_n^2(x)\big)
					\\
					&\kern1cm+(1-x^2)K_1^{(n)}\big(\sfP_n^2(x)[-4n^2-11n-9+(n^2+35n+60) x^2
					\\
					&\kern1cm-(10n^2+33n+29)x^4+(n+2)(n-1)x^6]
					\\
					&\kern1cm+(n-1)(1+x^2)[x(6n+13-(2n+3)x^2)\sfP_{n+1}^2(x)
					\\
					&\kern1cm+n(1+x^2)\sfP_{n+2}^2(x)]\big)\big\}\,,
					\\
					\bar h^{(n)}_{11}(x)\ &=\ \frac{(K^{(n)}_1(1- x^2)-2K^{(n)}_3 x)\sfP_n^2(x)}{m^n(1+x^2)^{n+1}}\,,
					\\
					\bar h^{(n)}_{12}\ &=\ \frac{(2K^{(n)}_1x+ K^{(n)}_3(1-x^2))\sfP_n^2(x)}{m^n(1+x^2)^{n+1}}
				\end{aligned}
			\end{equation}
		\end{subequations}
		for all $n>1$ and with the trace of $h_{ab}$ given in~\eqref{eq:extremalKerrLowestOrderSolutionTracePart}.

		We recover the first-order transverse deformation in the $r$-direction, as presented in~\cite{Li:2015wsa}, by examining the order-$r$ component of our solution~\eqref{eq:generalLowestOrderSolutionExtremalKerr}. This confirms that the first-order transverse deformation of the extremal Kerr horizon is unique up to the overall scaling factor $A$. Given the scaling symmetry $r \to\lambda r$ and $u\to u/\lambda$,~\cite{Li:2015wsa} concludes that the deformation is uniquely determined and corresponds precisely to the Kerr solution, a result we will later demonstrate explicitly. Our findings allow us to make a more general statement: the dimension of the moduli space for transverse deformations of the extremal Kerr horizon at order $r^n$ is at most $2n-2$. The reason for this upper bound will become clear in \cref{sec:higherOrderDeformations}.

		\paragraph{Extremal Kerr solution.}
		Let us now make contact with the extremal Kerr metric~\eqref{eq:extremalKerrMetricGaussianNullCoordinates}. In particular, in \cref{sec:GaussianNullKerr}, we have shown that there is a family of coordinate transformations~\eqref{eq:extremalKerrChangeOfCoordinates} parametrised by a function $f$ that transform the extremal Kerr metric in the Kerr coordinates~\eqref{eq:extremalKerrMetricInKerrCoordinates} into the Gau{\ss}ian-null-coordinate form~\eqref{eq:extremalKerrMetricGaussianNullCoordinates}, and the near-horizon geometry~\eqref{eq:extremalKerrNearHorizonMetricGaussianNullCoordinates} is independent of $f$. Furthermore, upon inspecting the Ricci tensor~\eqref{eq:RicciTensorAdaptedFrameII}, the pieces of order $r$ in~\eqref{eq:extremalKerrMetricGaussianNullCoordinates} will satisfy the lowest-order Einstein equation\footnote{For more details, see also \cref{sec:lowestOrderDeformationsRevisited}.} and so, we can compare these pieces with our general solution~\eqref{eq:generalLowestOrderSolutionExtremalKerr}. To fix the function $f$, we make use of the gauge transformation~\eqref{eq:firstOrderGaugeTransformations} to arrive at the gauge-fixing condition $\delta^{ab}\partial_rh_{ab}|_{r=0}=\Gamma$ from~\eqref{eq:gaugeFixingCondition} for~\eqref{eq:extremalKerrMetricGaussianNullCoordinates}; recall that this condition is part of the boundary conditions~\eqref{eq:augmentedBoundaryAndGaugeFixingConditions} and which our general solution~\eqref{eq:generalLowestOrderSolutionExtremalKerr} satisfies.

		In particular, the gauge-fixing condition $\delta^{ab}\partial_rh_{ab}|_{r=0}=\Gamma$ for~\eqref{eq:extremalKerrMetricGaussianNullCoordinates} amounts to
		\begin{equation}
			\frac{1}{m^2(1+x^2)}[(-1+x^2)f''(x)+2xf'(x)]+\frac{4 }{m(1+x^2)}\ =\ \frac1m A(1+x^2)~,
		\end{equation}
		and which is generally solved by
		\begin{equation}
			\begin{aligned}
				f(x)\ &=\ \tfrac{1}{60}Amx^2(26+3x^2)
				\\
				&\kern1cm+\tfrac{m}{30}\log(1-x)(28A-60+K_1)
				\\
				&\kern1cm+\tfrac{m}{30}\log(1+x)(28A-60-K_1)+K_2~,
			\end{aligned}
		\end{equation}
		where $K_1$ and $K_2$ are arbitrary constants. For $f$ to be smooth as $x\to\pm1$, we require
		\begin{equation}\label{eq:extremalKerrConstantA}
			K_1\ =\ 0
			\eand
			A\ =\ \tfrac{15}{7}~.
		\end{equation}
		Furthermore, without loss of generality, $K_2=0$ since the metric~\eqref{eq:extremalKerrMetricGaussianNullCoordinates} depends only through derivatives of $f$. Therefore, to order $r$, the metric~\eqref{eq:extremalKerrMetricGaussianNullCoordinates} is given by
		\begin{equation}\label{eq:alphabetagammaExtreamlKerrGAugeFixed}
			\begin{aligned}
				\alpha_x\ &=\ -\frac{2x}{1+x^2}+\frac{3x(59-55x^2-23x^4-5x^6)}{14m(1+x^2)^3}r+\caO(r^2)~,
				\\
				\alpha_\varphi\ &=\ \frac{4(1-x^2)}{(1+x^2)^2}-\frac{6(1-x^2)(7-45 x^2-3 x^4+x^6)}{7m(1+x^2)^4}r+\caO(r^2)~,
				\\
				\beta\ &=\ -\frac{3-6x^2-x^4}{m^2(1+x^2)^3}+\frac{2(35-225 x^2+135x^4+5 x^6-6x^8)}{7m^3(1+x^2)^5}r+\caO(r^2)~,
				\\
				\gamma_{xx}\ &=\ \frac{m^2(1+x^2)}{1-x^2}+\frac{3m(5-3x^2+13x^4+5x^6)}{7(1-x^2)(1+x^2)} r+\caO(r^2)~,
				\\
				\gamma_{x\varphi}\ &=\ \frac{6mx(1-x^2)(9+x^2)}{7(1+x^2)^2}r+\caO(r^2)~,
				\\
				\gamma_{\varphi\varphi}\ &=\ \frac{4m^2(1-x^2)}{1+x^2}+\frac{24mx^2(1-x^2)(9+x^2)}{7(1+x^2)^3}r+\caO(r^2)~.
				\end{aligned}
		\end{equation}
		Upon converting this into the basis~\eqref{eq:defAdaptedFrameII} (see also~\eqref{eq:defAdaptedFrameIIExtremalKerr}), we arrive at deformations $(h^{(1)}_a,h^{(1)},h^{(1)}_{ab})$ listed in~\eqref{eq:generalLowestOrderSolutionExtremalKerr} with $A$ as in~\eqref{eq:extremalKerrConstantA}. In the above formul{\ae}, we have set $\kappa=1$.

		\paragraph{Green's function.}
		In \cref{sec:firstOrderDeformation}, we have introduced the general Green function in~\eqref{eq:fullGreenFunction} and reduced it to finding the Green function $\bar\sfg^{(n)}_{ab}{}^{cd}$ in~\eqref{eq:spatialCrossSectionGreenFunctionOrderN}. In the previous paragraph, we have already completed this task for $n=1$ in~\eqref{eq:spatialGreenFunctionExtremalKerrN=1}. It remains to construct $\bar\sfg^{(n)}_{ab}{}^{cd}$ for all $n>1$. However, this is somewhat more difficult when compared to the $n=1$ case since from~\eqref{eq:extremalKerrLowestOrderSolutionBarHAB:b} it follows that the homogeneous solutions $\bar h^{(h)}_{ab}$ either satisfy the smoothness condition $\bar h^{(h)}_{ab}|_{x\to\pm1}=0$ from~\eqref{eq:reducedAugmentedBoundaryAndGaugeFixingConditions} at both $x\to\pm 1$ or not at all. Therefore, we cannot use the same method as in~\eqref{eq:spatialGreenFunctionExtremalKerrN=1} to construct $\bar\sfg^{(n)}_{ab}{}^{cd}$ such that it gives a particular solution $h^{(p)}_{ab}$ that always satisfies the smoothness condition.\footnote{See \cref{foot:GreenFunction} on \cpageref{foot:GreenFunction}.} However, if we relax this condition say at $x=1$, one can then show that one of the Green functions $\bar\sfg^{(n)}_{ab}{}^{cd}$ is given by
		\begin{subequations}\label{eq:spatialGreenFunctionExtremalKerrN>1}
			\begin{equation}
				\begin{aligned}
					\bar\sfg^{(n)}_{11}{}^{11}(x,\varphi;x',\varphi')\ &=\ \frac{(1+x^2)^{-n-1}(1+x'^2)^n (1-x+x'(1+x))(1-x'+x (1+x'))}{2(n-1)n(n+1)(n+2)}
					\\
					&\kern1cm\times[\theta(x-x')\sfP_n^2(x')\sfQ_n^2(x)+\theta(x'-x)\sfP_n^2(x)\sfQ_n^2(x')]\delta(\varphi-\varphi')~,
					\\
					\bar\sfg^{(n)}_{22}{}^{12}(x,\varphi;x',\varphi')\ &=\ \frac{(1+x^2)^{-n-1}(1+x'^2)^n (x-x')(1+x x')}{(n-1) n(n+1)(n+2)}
					\\
					&\kern1cm\times[\theta(x-x')\sfP_n^2(x')\sfQ_n^2(x)+\theta(x'-x)\sfP_n^2(x)\sfQ_n^2(x')]\delta(\varphi-\varphi')
				\end{aligned}
			\end{equation}
			and
			\begin{equation}
				\bar\sfg^{(n)}_{11}{}^{11}(x,\varphi;x',\varphi')\ =\ \bar\sfg^{(n)}_{12}{}^{12}(x,\varphi;x',\varphi')
				\eand
				\bar\sfg^{(n)}_{11}{}^{12}(x,\varphi;x',\varphi')\ =\ -\bar\sfg^{(n)}_{12}{}^{11}(x,\varphi;x',\varphi')
			\end{equation}
		\end{subequations}
		for all $n>1$.

		One has to be careful when working with this Green function since the resulting particular solution might not satisfy the smoothness condition at $x=1$. However, we claim that
		\begin{subequations}\label{eq:criterionForSmoothSolutions}
			\begin{equation}\label{eq:ParticularSolutionBoundaryAtx=1}
				\lim_{x\rightarrow 1}\int\rmd^2y'\sqrt{\det(\mathring\gamma(y'))}\,\bar\sfg^{(n)}_{ab}{}^{cd}(y;y')\varrho^{(n)}_{cd}(y')\ =\ 0
			\end{equation}
			with $\bar\sfg^{(n)}_{ab}{}^{cd}(y;y')$ from~\eqref{eq:spatialGreenFunctionExtremalKerrN>1} for all $n>1$ if and only if\footnote{see~\eqref{eq:spatialCrossSectionGreenFunctionOrderN}}
			\begin{equation}\label{eq:SpatialEquationOrdernWithSource}
				[n(n-1)\bar\sfa_{ab}{}^{cd}(y)+n\bar\sfb_{ab}{}^{cd}(y)+\bar\sfc_{ab}{}^{cd}(y)]\bar h^{(n)}_{cd}(y)\ =\ \varrho^{(n)}_{ab}(y)
			\end{equation}
		\end{subequations}
		has a smooth solution for all $n>1$. Indeed, if~\eqref{eq:ParticularSolutionBoundaryAtx=1} holds, it is evident that then the equation~\eqref{eq:SpatialEquationOrdernWithSource} admits the smooth solution $\bar h^{(n)}_{ab}(y)=\int\rmd^2y'\sqrt{\det(\mathring\gamma(y'))}\,\bar\sfg^{(n)}_{ab}{}^{cd}(y;y')\varrho^{(n)}_{cd}(y')$. Conversely, suppose that the equation~\eqref{eq:SpatialEquationOrdernWithSource} admits a smooth solution but, for a contradiction, that~\eqref{eq:ParticularSolutionBoundaryAtx=1} does not hold. Then, to have $\bar h_{ab}|_{x\to\pm1}=0$ with $\bar h_{ab}=\bar h_{ab}^{(h)}+\bar h_{ab}^{(p)}$, there must be a homogeneous solution $\bar h_{ab}^{(h)}$ such that
		\begin{equation}
			\bar h_{ab}^{(h)}\big|_{x\to-1}=0
			\eand
			\bar h_{ab}^{(h,n)}\big|_{x\to1}\ =\ -\lim_{x\rightarrow 1}\int\rmd^2y'\sqrt{\det(\mathring\gamma(y'))}\,\bar\sfg^{(n)}_{ab}{}^{cd}(y;y')\varrho^{(n)}_{cd}(y')~.
		\end{equation}
		However, upon inspecting the general homogenous solutions~\eqref{eq:extremalKerrLowestOrderSolutionBarHAB:b}, this is not possible without affecting the behaviour of $\bar h_{ab}^{(h,n)}$ as $x\to-1$ since
		\begin{equation}
			\lim_{x\to\pm1}\sfP_n^2(x)\ =\ 0
			\eand
			\lim_{x\to\pm1}(1-x^2)\sfQ_n^2(x)\ =\ 2(\pm 1)^{n+1}~.
		\end{equation}
		Therefore, we conclude that on the image of $\tilde\mu_1$ (and thus on the image of $\mu_1$) on smooth tensor fields, we can use the Green function given by~\eqref{eq:spatialGreenFunctionExtremalKerrN>1}.

		\section{Homotopy algebras: a brief recap}\label{sec:homotopyAlgebras}

		\subsection{\texorpdfstring{$L_\infty$}{L infinity}-algebras}\label{sec:LInftyAlgebras}

		In this section, we briefly review the basic facts about $L_\infty$-algebras and the associated homotopy Maurer--Cartan theory. For more details and the conventions we follow here, see e.g.~\cite{Jurco:2018sby,Jurco:2019bvp} and~\cite[Appendix A]{Borsten:2021hua}.

		\paragraph{$L_\infty$-algebras.}
		An \uline{$L_\infty$-algebra} or \uline{strongly homotopy Lie algebra} extends the notion of a differential graded Lie algebra. Concretely, it consists of a $\IZ$-graded vector space $V=\bigoplus_{k\in\IZ}V_k$ together with graded anti-symmetric $i$-linear maps $\mu_i:V\times\cdots\times V\rightarrow V$ of degree $2-i$. These maps satisfy the so-called \uline{homotopy Jacobi identities},
		\begin{subequations}
			\begin{equation}\label{eq:homotopyJacobiIdentities}
				\sum_{i_1+i_2=i}\sum_{\sigma\in{\rm\overline{Sh}}(i_1;i)}(-1)^{i_2}\chi(\sigma;v_1,\ldots,v_i)\mu_{i_2+1}(\mu_{i_1}(v_{\sigma(1)},\ldots,v_{\sigma(i_1)}),v_{\sigma(i_1+1)},\ldots,v_{\sigma(i)})\ =\ 0
			\end{equation}
			for all homogeneous $v_1,\ldots,v_i\in V$ and $i\in\IN$ where the sum is understood to be taken over all \uline{unshuffles}. Recall that these are permutations $\sigma$ of $\{1,\ldots,i\}$ with $\sigma(1)<\cdots<\sigma(i_1)$ and $\sigma(i_1)<\cdots<\sigma(i)$. Furthermore, $\chi(\sigma;v_1,\ldots,v_i)$ is a sign factor called the \uline{Koszul sign}, and it is defined by
			\begin{equation}\label{eq:KoszulSign}
				v_1\wedge\ldots\wedge v_i\ =\ \chi(\sigma;v_1,\ldots,v_i)v_{\sigma(1)}\wedge\ldots\wedge v_{\sigma(i)}~.
			\end{equation}
		\end{subequations}
		Here and in the following, we shall denote the \uline{degree} of a homogeneous element $v\in V$ by $|v|$. We shall also refer to the $\mu_i$ as (higher) products. Explicitly, the lowest few homotopy Jacobi identities~\eqref{eq:homotopyJacobiIdentities} are given by
		\begin{equation}\label{eq:lowestHomotopyJacobiIdentities}
			\begin{gathered}
				\mu_1(\mu_1(v_1))\ =\ 0~,
				\\
				\mu_1(\mu_2(v_1,v_2))\ =\ \mu_2(\mu_1(v_1),v_2)+(-1)^{|v_1|}\mu_2(v_1,\mu_1(v_2))~,
				\\
				\mu_2(\mu_2(v_1,v_2),v_3)+(-1)^{|v_1|\,|v_2|}\mu_2(v_2,\mu_2(v_1,v_3))-\mu_2(v_1,\mu_2(v_2,v_3))
				\\
				\kern2cm=\ \mu_1(\mu_3(v_1,v_2,v_3))+\mu_3(\mu_1(v_1),v_2,v_3)+(-1)^{|v_1|}\mu_3(v_1,\mu_1(v_2),v_3)
				\\
				\kern5cm+\,(-1)^{|v_1|+|v_2|}\mu_3(v_1,v_2,\mu_1(v_3))~,
				\\
				\vdots
			\end{gathered}
		\end{equation}
		In particular, the first relation says that $\mu_1$ is a differential and so, any $L_\infty$-algebra $(V,\mu_i)$ has an underlying cochain complex $(V,\mu_1)$. Furthermore, the second relation says that $\mu_1$ a derivation with respect to $\mu_2$, and the third relation says that $\mu_3$ captures the failure of $\mu_2$ to satisfy the standard Jacobi identity.

		\paragraph{Cyclic structure.}
		A \uline{cyclic $L_\infty$-algebra} extends the notion of a metric differential graded Lie algebra. Concretely, given an $L_\infty$-algebra $(V,\mu_i)$, a \uline{cyclic structure} on $(V,\mu_i)$ is a non-degenerate graded symmetric bilinear form $\inner{-}{-}:V\times V\rightarrow\IR$ such that
		\begin{equation}\label{eq:cyclicity}
			\inner{v_1}{\mu_i(v_2,\ldots,v_{i+1})}\ =\ (-1)^{i+i(|v_1|+|v_{i+1}|)+|v_{i+1}|\sum_{j=1}^{i}|v_j|}\inner{v_{i+1}}{\mu_i(v_1,\ldots,v_{i})}
		\end{equation}
		for all homogeneous $v_1,\ldots,v_i\in V$.

		\paragraph{$L_\infty$-morphisms.}
		Morphisms of Lie algebras are maps that preserve the Lie bracket. In the context of $L_\infty$-algebras, this notion generalises as follows. An \uline{$L_\infty$-morphism} $\phi:(V,\mu_i)\rightarrow(V',\mu'_i)$ of $L_\infty$-algebras $(V,\mu_i)$ and $(V',\mu'_i)$ is a collection of graded anti-symmetric $i$-linear maps $\phi_i:V\times\cdots\times V\rightarrow V'$ of degree $1-i$ such that
		\begin{subequations}\label{eq:LInfinityMorphism}
			\begin{equation}
				\begin{aligned}
					&\sum_{i_1+i_2=i}\sum_{\sigma\in{\rm\overline{Sh}}(i_1;i)}(-1)^{i_2}\chi(\sigma;v_1,\ldots,v_i)\phi_{i_2+1}(\mu_{i_1}(v_{\sigma(1)},\ldots,v_{\sigma(i_1)}),v_{\sigma(i_1+1)},\ldots,v_{\sigma(i)})
					\\
					&=\ \sum_{j\geq1}\frac{1}{j!}\sum_{k_1+\cdots+k_j=i}\sum_{\sigma\in{\rm \overline{Sh}}(k_1,\ldots,k_{j-1};i)}\chi(\sigma;v_1,\ldots,v_i)\zeta(\sigma;v_1,\ldots,v_i)
					\\
					&\kern1cm\times\mu'_j\Big(\phi_{k_1}\big(v_{\sigma(1)},\ldots,v_{\sigma(k_1)}\big),\ldots,\phi_{k_j}\big(v_{\sigma(k_1+\cdots+k_{j-1}+1)},\ldots,v_{\sigma(i)}\big)\Big)
				\end{aligned}
			\end{equation}
			for all homogeneous $v_1,\ldots,v_i\in V$ and $i\in\IN$. Here, $\chi(\sigma;v_1,\ldots,v_i)$ is again the Koszul sign~\eqref{eq:KoszulSign} and $\zeta(\sigma;v_1,\ldots,v_i)$ is another sign factor given by
			\begin{equation}\label{eq:zetaSign}
				\zeta(\sigma;v_1,\ldots,v_i)\ \coloneqq\ (-1)^{\sum_{1\leq m<n\leq j}k_mk_n+\sum_{m=1}^{j-1}k_m(j-m)+\sum_{m=2}^j(1-k_m)\sum_{k=1}^{k_1+\cdots+k_{m-1}}|v_{\sigma(k)}|}~.
			\end{equation}
		\end{subequations}
		Explicitly, the lowest few relations of~\eqref{eq:LInfinityMorphism} read as
		\begin{equation}
			\begin{gathered}
				\phi_1(\mu_1(v_1))\ =\ \mu'_1(\phi_1(v_1))~,
				\\
				\phi_1(\mu_2(v_1,v_2))-\phi_2(\mu_1(v_1),v_2)+(-1)^{|v_1||v_2|}\phi_2(\mu_1(v_2),v_1)
				\\
				=\ \mu'_1(\phi_2(v_1,v_2))+\mu'_2(\phi_1(v_1),\phi_1(v_2))
				\\
				\vdots
			\end{gathered}
		\end{equation}
		In particular, the first relation says that $\phi_1$ is a morphism of cochain complexes.

		We call an $L_\infty$-morphism $\phi:(V,\mu_i)\rightarrow(V',\mu'_i)$ an \uline{$L_\infty$-quasi-isomorphism} whenever $\phi_1$ induces an isomorphism on the cohomologies of the cochain complexes $(V,\mu_1)$ and $(V',\mu'_1)$. Furthermore, it is called an \uline{$L_\infty$-isomorphism} whenever $\phi_1$ is invertible. Moreover, if we are given inner products $\inner{-}{-}$ on $(V,\mu_i)$ and $\inner{-}{-}'$ on $(V',\mu'_i)$, then an $L_\infty$-morphism $\phi:(V,\mu_i)\rightarrow(V',\mu'_i)$ between cyclic $L_\infty$-algebras is an $L_\infty$-morphism that satisfies
		\begin{subequations}\label{eq:cyclicLInfinityMorphism}
			\begin{equation}
				\inner{v_1}{v_2}\ =\ \inner{\phi_1(v_1)}{\phi_1(v_2)}'
			\end{equation}
			for all $v_1,v_2\in V$ and for all $i\geq3$ and $v_1,\ldots,v_i\in V$ as well as
			\begin{equation}
				\sum_{\substack{i_1+i_2=i\\i_1,i_2\geq1}}\inner{\phi_{i_1}(v_1,\ldots,v_{i_1})}{\phi_{i_2}(v_{i_1+1},\ldots,v_{i})}'\ =\ 0~.
			\end{equation}
		\end{subequations}
		Note that the $L_\infty$-morphisms of cyclic $L_\infty$-algebras require $\phi_1$ to be injective. Indeed, for arbitrary $v_2,v_3\in V$ suppose that $\phi_1(v_2)=\phi_1(v_3)$. Then, for arbitrary $v_1\in V$ we have that $\inner{v_1}{v_2}=\inner{\phi_1(v_1)}{\phi_1(v_2)}'=\inner{\phi_1(v_1)}{\phi_1(v_3)}'=\inner{v_1}{v_3}$. Hence, $\inner{v_1}{v_2-v_3}=0$ and so, from the non-degeneracy of $\inner{-}{-}$, it follows that $v_2=v_3$.

		\paragraph{Structural theorems.}
		An $L_\infty$-algebra $(V,\mu_i)$ is called \uline{minimal} whenever $\mu_1=0$ and \uline{strict} whenever $\mu_{i>2}=0$. Furthermore, it is called \uline{linearly contractible} whenever $\mu_{i>1}=0$ and its underlying cochain complex has trivial cohomology. We now have the following structural theorems:
		\begin{center}
			\begin{tabularx}{\textwidth}{rX}
				\uline{Strictification theorem:} & every $L_\infty$-algebra is $L_\infty$-quasi-isomorphic to a strict $L_\infty$-algebra~\cite{igor1995,Berger:0512576}.
				\\
				\uline{Decomposition theorem:} & every $L_\infty$-algebra is $L_\infty$-isomorphic to the direct sum of a minimal and a linearly contractible $L_\infty$-algebra~\cite{Kajiura:2003ax}.
				\\
				\uline{Minimal model theorem:} & every $L_\infty$-algebra is $L_\infty$-quasi-isomorphic to a minimal $L_\infty$-algebra~\cite{kadeishvili1982algebraic,Kajiura:2003ax}; this is a direct consequence of the decomposition theorem.
			\end{tabularx}
		\end{center}

		\subsection{Homotopy Maurer--Cartan theory}\label{sec:HomotopyMaurer-CartanTheory}

		Given an $L_\infty$-algebra $(V,\mu_i)$, we have naturally associated with it \uline{homotopy Maurer--Cartan theory} which generalises the standard Maurer--Cartan theory for Lie algebras.

		\paragraph{Homotopy Maurer--Cartan equation.}
		Concretely, a \uline{gauge potential} is an element $a\in V_1$ and its \uline{curvature} is
		\begin{equation}\label{eq:hMCCurvature}
			f\ \coloneqq\ \sum_{i\geq1}\frac1{i!}\mu_i(a,\ldots,a)\ \in\ V_2
		\end{equation}
		and which obeys the \uline{Bianchi identity},
		\begin{equation}
			\sum_{i\geq0}\frac1{i!}\mu_{i+1}(a,\ldots,a,f)\ =\ 0~,
		\end{equation}
		as a direct consequence of the homotopy Jacobi identities~\eqref{eq:homotopyJacobiIdentities} for elements of degree one. Furthermore, a \uline{homotopy Maurer--Cartan element} is a gauge potential whose curvature vanishes. Provided that $(V,\mu_i)$ comes with a cyclic structure $\inner{-}{-}$ of degree $-3$, homotopy Maurer--Cartan elements are the extrema of the \uline{homotopy Maurer--Cartan action} that is given by
		\begin{equation}\label{eq:hMCAction}
			S\ \coloneqq\ \sum_{i\geq1}\frac1{(i+1)!}\inner{a}{\mu_i(a,\ldots,a)}~.
		\end{equation}
		This follows again from the homotopy Jacobi identities~\eqref{eq:homotopyJacobiIdentities} as well as the cyclicity condition~\eqref{eq:cyclicity}. Note that the equation of motion
		\begin{equation}\label{eq:hMCEquation}
			f\ =\ 0
		\end{equation}
		is also called the \uline{homotopy Maurer--Cartan equation}.

		The action~\eqref{eq:hMCAction} is invariant under the infinitesimal \uline{gauge transformations}
		\begin{equation}\label{eq:hMCGaugeTransformations}
			\delta_{c_0}a\ \coloneqq\ \sum_{i\geq0}\frac1{i!}\mu_{i+1}(a,\ldots,a,c_0)
		\end{equation}
		which are parametrised by $c_0\in V_0$. Correspondingly, the curvature~\eqref{eq:hMCCurvature} transforms as
		\begin{equation}
			\delta_{c_0}f\ =\ \sum_{i\geq0}\frac1{i!}\mu_{i+2}(a,\ldots,a,f,c_0)~.
		\end{equation}
		We also have infinitesimal \uline{higher gauge transformations} that are recursively given by
		\begin{equation}\label{eq:hMCHigherGaugeTransformations}
			\delta_{c_{-k-1}}c_{-k}\ \coloneqq\ \sum_{i\geq0}\frac1{i!}\mu_{i+1}(a,\ldots,a,c_{-k-1})
		\end{equation}
		for $c_{-k}\in V_{-k}$. It is not too difficult to see that these invariance and covariance statements under these (higher) gauge transformation are again a direct consequence of the homotopy Jacobi identities~\eqref{eq:homotopyJacobiIdentities}.

		\paragraph{$L_\infty$-morphisms.}
		Furthermore, suppose that we are given an $L_\infty$-morphism $\phi:(V,\mu_i)\rightarrow (V',\mu'_i)$; see~\eqref{eq:LInfinityMorphism}. Upon setting
		\begin{equation}\label{eq:gaugePotentialUnderLInfinityMorphism}
			a'\ \coloneqq\ \sum_{i\geq1}\frac1{i!}\phi_i(a,\ldots,a)~,
		\end{equation}
		it follows from~\eqref{eq:homotopyJacobiIdentities} that the curvature $f'$ is given by
		\begin{equation}
			f'\ =\ \sum_{i\geq1}\frac1{i!}\mu'_i(a',\ldots,a')\ =\ \sum_{i\geq0}\frac1{i!}\phi_{i+1}(a,\ldots,a,f)~.
		\end{equation}
		Hence, homotopy Maurer--Cartan elements are mapped to homotopy Maurer--Cartan elements under $L_\infty$-morphisms. More than that, one can show that gauge equivalence classes of homotopy Maurer--Cartan elements are mapped to gauge equivalence classes of homotopy Maurer--Cartan elements.\footnote{\uline{Warning:} it is not in general true that gauge equivalence classes of gauge potentials are mapped to gauge equivalence classes of gauge potentials.} Finally, if the $L_\infty$-morphism is also cyclic, that is, if also~\eqref{eq:cyclicLInfinityMorphism} holds and the inner products $\inner{-}{-}$ and $\inner{-}{-}'$ are both of degree $-3$, then
		\begin{equation}
			S\ =\ \sum_{i\geq1}\frac1{(i+1)!}\inner{a}{\mu_i(a,\ldots,a)}\ =\ \sum_{i\geq1}\frac1{(i+1)!}\inner{a'}{\mu'_i(a',\ldots,a')}\ =\ S'
		\end{equation}
		for the corresponding homotopy Maurer--Cartan actions.

		\paragraph{Field theories as homotopy Maurer--Cartan theories.}
		The crucial point now is that \uline{any} Batalin--Vilkovisky quantisable (Lagrangian) field theory can be reformulated as the homotopy Maurer--Cartan theory for a (cyclic) $L_\infty$-algebra~\cite{Kajiura:2003ax,Doubek:2017naz,Jurco:2018sby,Jurco:2019bvp}; see~\cite{Hohm:2017pnh} for a discussion at the level of the equation of motion. In addition, when considering field theories with boundaries, one needs to generalise the notion of (cyclic) $L_\infty$-algebras to (cyclic) \uline{relative} $L_\infty$-algebras~\cite{Alfonsi:2024utl} which are pairs of $L_\infty$-algebras, one in the bulk and one in the boundary, and with a $L_\infty$-morphism between then; see~\cite{Chiaffrino:2023wxk} for a different approach to dealing with boundaries. Below, we shall apply this \uline{homotopy algebraic perspective} to the discssion of general deformations of near-horizon geometries. This now also justifies or choice of notation in~\eqref{eq:homotopyMaurerCartanActionMotivation} and, more generally, in \cref{sec:firstOrderDeformation}.

		\subsection{Homological perturbations}\label{sec:homologicalPerturbations}

		In \cref{sec:firstOrderDeformation}, we have solved the deformation equation~\eqref{eq:homotopyMaurerCartanActionMotivation} to lowest order, and in the preceding section we have explained that this equation can, in fact, be understood as the homotopy Maurer--Cartan equation for an $L_\infty$-algebra. We shall now recap the general derivation of perturbative solutions to homotopy Mauer--Cartan equations by using homological perturbation theory.

		\paragraph{Special deformation retracts.}
		A \uline{deformation retract} (see e.g.~\cite{Loday:2012aa}) of cochain complexes $(V,\mu_1)$ and $(V',\mu'_1)$ of vector spaces constitutes of morphisms $\sfp$ and $\sfe$ of cochain complexes of degree~$0$ together with a morphism $\sfh$ of vector spaces of degree~$-1$, called a \uline{contracting homotopy}, such that
		\begin{subequations}\label{eq:deformationRetract}
			\begin{equation}
				\begin{tikzcd}
					\ar[loop,out=160,in=200,distance=20,"\sfh" left] (V,\mu_1)\arrow[r,shift left]{}{\sfp} & (V',\mu'_1)\arrow[l,shift left]{}{\sfe}
				\end{tikzcd}
			\end{equation}
			and
			\begin{equation}
				\sfid\ =\ \sfe\circ\sfp+\mu_1\circ\sfh+\sfh\circ\mu_1
				\eand
				\sfp\circ\sfe\ =\ \sfid~.
			\end{equation}
		\end{subequations}
		Thus, $\sfp$ is a surjection and $\sfe$ an injection.

		A deformation retract is called a \uline{special deformation retract} whenever the \uline{side conditions},
		\begin{equation}\label{eq:sideConditions}
			\sfp\circ\sfh\ =\ 0~,
			\quad
			\sfh\circ\sfe\ =\ 0~,
			\eand
			\sfh\circ\sfh\ =\ 0~,
		\end{equation}
		are satisfied as well. Importantly, the side conditions can be assumed without loss of generality, and we shall do so in the following, since we may always turn a deformation retract into a special one, see e.g.~\cite{Crainic:0403266,Loday:2012aa}, by means of
		\begin{equation}
			\sfh\ \rightarrow\ (\sfid-\,\sfe\circ\sfp)\circ\sfh\circ(\sfid-\,\sfe\circ\sfp)\circ\mu_1\circ(\sfid-\,\sfe\circ\sfp)\circ\sfh\circ(\sfid-\,\sfe\circ\sfp)~.
		\end{equation}

		\paragraph{Hodge--Kodaira decomposition.}
		Consider a special deformation retract, that is,~\eqref{eq:deformationRetract} together with~\eqref{eq:sideConditions}, in the special case when $V'$ is the cohomology $V^\circ\coloneqq H^\bullet(V)$ of the cochain complex $(V,\mu_1)$ and with $\mu'_1=0$.\footnote{The pair $(V^\circ,0)$ is trivially a cochain complex.} Evidently, in this case $\sfp$ and $\sfe$ are quasi-isomorphisms of cochain complexes. Note that for cochain complexes of vector spaces, such a special deformation retract always exist~\cite{Weibel:1994aa} as short exact sequences of vector spaces always split.

		We then also have\footnote{The second equation holds for general special deformation retracts.}
		\begin{equation}\label{eq:SDRAdditionalProperties}
			\mu_1\ =\ \mu_1\circ\sfh\circ\mu_1
			\eand
			\sfh\ =\ \sfh\circ\mu_1\circ\sfh~.
		\end{equation}
		In addition, $\sfe\circ\sfp$, $\mu_1\circ\sfh$, and $\sfh\circ\mu_1$ are all projectors and because of~\eqref{eq:deformationRetract}, we have the decomposition
		\begin{subequations}\label{eq:HodgeKodairaDecomposition}
			\begin{equation}
				V\ \cong\ V_{\rm harm}\oplus V_{\rm ex}\oplus V_{\rm coex}~,
			\end{equation}
			where
			\begin{equation}
				V_{\rm harm}\ \coloneqq\ \im(\sfe\circ\sfp)~,
				\quad
				V_{\rm ex}\ \coloneqq\ \im(\mu_1\circ\sfh)~,
				\quad
				V_{\rm coex}\ \coloneqq\ \im(\sfh\circ\mu_1)~,
			\end{equation}
		\end{subequations}
		and with the identification $V_{\rm harm}\cong V^\circ$. This is known as the \uline{Hodge--Kodaira decomposition}. It is now not too difficult to see that we have the identifications
		\begin{equation}
			\begin{gathered}
				V_{\rm harm}\ \cong\ \im(\sfe)~,
				\quad
				V_{\rm ex}\ \cong\ \im(\mu_1)~,
				\quad
				V_{\rm coex}\ \cong\ \im(\sfh)~,
				\\
				V_{\rm harm}\oplus V_{\rm ex}\ \cong\ \ker(\mu_1)~,
				\quad
				V_{\rm ex}\oplus V_{\rm coex}\ \cong\ \ker(\sfp)~,
				\quad
				V_{\rm harm}\oplus V_{\rm coex}\ \cong\ \ker(\sfh)~.
			\end{gathered}
		\end{equation}
		Note that in this context, the contracting homotopy $\sfh$ is also called the \uline{propagator}.

		\paragraph{Example.}
		In view of our applications in \cref{sec:higherOrderDeformations}, consider a cochain complex $(V,\mu_1)$ that is concentrated in degrees one and two, that is, $V=V_1\oplus V_2$. We then have
		\begin{subequations}\label{eq:twoTermsHodgeDecomposition}
			\begin{equation}
				\begin{tikzcd}
					& V_1\ar[dl,"\sfid|_{V_1}-\,\epsilon_1\circ\pi_1",shift right,swap]\ar[dr,"\pi_1",shift left]\ar[rrr,"\mu_1"] &&& V_2\ar[dl,"\sfid|_{V_2}-\,\epsilon_2\circ\pi_2",shift right,swap]\ar[dr,"\pi_2",shift left]
					\\
					H^1(V)\ar[ur,"\iota_1",shift right, swap] && V_1/\ker(\mu_1)\ar[ul,"\epsilon_1",shift left]\ar[r,"\hat\mu_1"] & \im(\mu_1)\ar[ur,"\iota_2",shift right, swap] && H^2(V)\ar[ul,"\epsilon_2",shift left]
				\end{tikzcd}
			\end{equation}
			where $H^1(V)=\ker(\mu_1)$ and $H^2(V)=V_2/\im(\mu_1)$ with $\pi_{1,2}$ the canonical quotient projections and $\epsilon_{1,2}$ choices of right-inverses, $\iota_{1,2}$ the inclusions, and $\hat\mu_1$ the canonical isomorphism given by the first isomorphism theorem. We may write
			\begin{equation}
				\hat\mu_1\ =\ (\sfid|_{V_2}-\,\epsilon_2\circ\pi_2)\circ\mu_1\circ\epsilon_1\ =\ \mu_1\circ\epsilon_1~.
			\end{equation}
		\end{subequations}
		Note that the combination $\mu_1\circ\epsilon_1$ is independent of the choice of $\epsilon_1$.

		We then have a special deformation retract, see~\eqref{eq:deformationRetract} and~\eqref{eq:sideConditions}, given by
		\begin{subequations}\label{eq:exampleSpecialDeformationRetract}
			\begin{equation}\label{eq:exampleSpecialDeformationRetract:a}
				\begin{tikzcd}
					\ar[loop,out=160,in=200,distance=20,"\sfh" left] (V,\mu_1)\arrow[r,shift left]{}{\sfp} & (V^\circ,0)\arrow[l,shift left]{}{\sfe}
				\end{tikzcd}
			\end{equation}
			with $V^\circ\coloneqq H^\bullet(V)$ and
			\begin{equation}\label{eq:exampleSpecialDeformationRetract:b}
				\begin{gathered}
					\sfp|_{V_1}\ \coloneqq\ \sfid|_{V_1}-\,\epsilon_1\circ\pi_1~,
					\quad
					\sfp|_{V_2}\ \coloneqq\ \pi_2~,
					\quad
					\sfe|_{H^1(V)}\ \coloneqq\ \iota_1~,
					\quad
					\sfe|_{H^2(V)}\ \coloneqq\ \epsilon_2~,
					\\
					\sfh\ \coloneqq\ \epsilon_1\circ\hat\mu_1^{-1}\circ(\sfid|_{V_2}-\,\epsilon_2\circ\pi_2)~.
				\end{gathered}
			\end{equation}
		\end{subequations}
		Indeed, using~\eqref{eq:exampleSpecialDeformationRetract:b}, it immediately follows that $\sfp\circ\sfe=\sfid$. It also follows that
		\begin{subequations}\label{eq:checkConditionSDR}
			\begin{equation}
				\begin{aligned}
					\sfe|_{H^1(V)}\circ\sfp|_{V_1}+\sfh\circ\mu_1\ &=\ \iota_1\circ(\sfid|_{V_1}-\epsilon_1\circ\pi_1)+\epsilon_1\circ\hat\mu_1^{-1}\circ(\sfid|_{V_2}-\,\epsilon_2\circ\pi_2)\circ\mu_1
					\\
					&=\ (\sfid|_{V_1}-\epsilon_1\circ\pi_1)+\epsilon_1\circ\hat\mu_1^{-1}\circ\hat\mu_1\circ\pi_1
					\\
					&=\ \sfid|_{V_1}
				\end{aligned}
			\end{equation}
			as well as
			\begin{equation}
				\begin{aligned}
					\sfe|_{H^2(V)}\circ\sfp|_{V_2}+\mu_1\circ\sfh\ &=\ \epsilon_2\circ\pi_2+\mu_1\circ\epsilon_1\circ\hat\mu_1^{-1}\circ(\sfid|_{V_2}-\,\epsilon_2\circ\pi_2)
					\\
					&=\ \epsilon_2\circ\pi_2+(\sfid|_{V_2}-\,\epsilon_2\circ\pi_2)
					\\
					&=\ \sfid|_{V_2}~
				\end{aligned}
			\end{equation}
		\end{subequations}
		and so, the conditions~\eqref{eq:deformationRetract} for a deformation retract are satisfied. Furthermore, it is also easy to see that $\sfh$ satisfies the side conditions~\eqref{eq:sideConditions} as well so that~\eqref{eq:exampleSpecialDeformationRetract} is indeed a special deformation retract.

		\paragraph{Homological perturbation lemma.}
		Given an $L_\infty$-algebra $(V,\mu_i)$ and a special deformation retract onto its underlying cohomology $(V^\circ,0)$ as discussed above, the \uline{homological perturbation lemma} now states that the $L_\infty$-algebra structure $\mu_i$ can be transferred to an $L_\infty$-algebra structure $\mu^\circ_i$ on $V^\circ$~\cite{Kajiura:2003ax,Crainic:0403266}. Concretely, the quasi-isomorphism $\sfe$ of cochain complexes lifts to an $L_\infty$-quasi-isomorphism $\sfE:(V^\circ,\mu_i^\circ)\rightarrow(V,\mu_i)$, see~\eqref{eq:LInfinityMorphism}, with the component maps recursively given by
		\begin{subequations}\label{eq:homotopyTransfer}
			\begin{equation}\label{eq:LInftyQuasiIsoMinimalModel}
				\begin{aligned}
					\sfE_1(v^\circ_1)\ &\coloneqq\ \sfe(v^\circ_1)~,
					\\
					\sfE_2(v^\circ_1,v^\circ_2)\ &\coloneqq\ -\,\sfh(\mu_2(\sfE_1(v^\circ_1),\sfE_1(v^\circ_2)))~,
					\\
					&~~\vdots
					\\
					\sfE_i(v^\circ_1,\ldots,v^\circ_i)\ &\coloneqq\ -\sum_{j=2}^i\frac1{j!} \sum_{k_1+\cdots+k_j=i}\sum_{\sigma\in\overline{\rm Sh}(k_1,\ldots,k_{j-1};i)}\chi(\sigma;v^\circ_1,\ldots,v^\circ_i)\zeta(\sigma;v^\circ_1,\ldots,v^\circ_i)
					\\
					&\kern0.5cm\times\sfh\left\{\mu_j\Big(\sfE_{k_1}\big(v^\circ_{\sigma(1)},\ldots,v^\circ_{\sigma(k_1)}\big),\ldots,\sfE_{k_j}\big(v^\circ_{\sigma(k_1+\cdots+k_{j-1}+1)},\ldots,v^\circ_{\sigma(i)}\big)\Big)\right\}
				\end{aligned}
			\end{equation}
			for all homogeneous $v^\circ_1,\ldots,v^\circ_i\in V^\circ$ and $i\in\IN$ with $\chi(\sigma;v^\circ_1,\ldots,v^\circ_i)$ again the Koszul sign defined in~\eqref{eq:KoszulSign} and $\zeta(\sigma;v^\circ_1,\ldots,v^\circ_i)$ the sign defined in~\eqref{eq:zetaSign}, respectively. Furthermore, the higher products $\mu^\circ_{i>1}$ induced on $V^\circ$ are given by
			\begin{equation}\label{eq:minimalModelHigherProducts}
				\begin{aligned}
					\mu^\circ_2(v^\circ_1,v^\circ_2)\ &\coloneqq\ \sfp(\mu_2(\sfE_1(v^\circ_1),\sfE_1(v^\circ_2))~,
					\\
					&~~\vdots
					\\
					\mu^\circ_i(v^\circ_1,\ldots,v^\circ_i)\ &\coloneqq\ \sum_{j=2}^i\frac1{j!}\sum_{k_1+\cdots+k_j=i}\sum_{\sigma\in\overline{\rm Sh}(k_1,\ldots,k_{j-1};i)}\chi(\sigma;v^\circ_1,\ldots,v^\circ_i)\zeta(\sigma;v^\circ_1,\ldots,v^\circ_i)
					\\
					&\kern0.5cm\times\sfp\left\{\mu_j\Big(\sfE_{k_1}\big(v^\circ_{\sigma(1)},\ldots,v^\circ_{\sigma(k_1)}\big),\ldots,\sfE_{k_j}\big(v^\circ_{\sigma(k_1+\cdots+k_{j-1}+1)},\ldots,v^\circ_{\sigma(i)}\big)\Big)\right\}
				\end{aligned}
			\end{equation}
		\end{subequations}
		for all homogeneous $v^\circ_1,\ldots,v^\circ_i\in V^\circ$ and $i\in\IN$. These formul{\ae} also extend to the cyclic setting~\cite{Doubek:2017naz}. Note that the above constitutes, in fact, the aforementioned minimal model theorem for $L_\infty$-algebras.

		\paragraph{Solving the homotopy Maurer--Cartan equation.}
		Given an $L_\infty$-algebra $(V,\mu_i)$ and a special deformation retract onto its underlying cohomology, upon recalling~\eqref{eq:gaugePotentialUnderLInfinityMorphism}, we obtain the general perturbative solution
		\begin{subequations}\label{eq:MCElement}
			\begin{equation}\label{eq:MCElement:a}
				a\ =\ a_{\rm harm}+a_{\rm ex}+a_{\rm coex}
				\ewith
				\sfh(a)\ =\ 0
			\end{equation}
			to the homotopy Maurer--Cartan equation~\eqref{eq:hMCEquation} under the Hodge--Kodaira decomposition~\eqref{eq:HodgeKodairaDecomposition} by means of
			\begin{equation}\label{eq:MCElement:b}
				a_{\rm harm}\ =\ \sfE_1(a^\circ)\ =\ \sfe(a^\circ)~,
				\quad
				a_{\rm ex}\ =\ 0~,
				\eand
				a_{\rm coex}\ =\ \sum_{i\geq2}\frac1{i!}\sfE_i(a^\circ,\ldots,a^\circ)~.
			\end{equation}
			Here, the $\sfE_i$ are given by~\eqref{eq:LInftyQuasiIsoMinimalModel} and $a^\circ\in H^1(V)$ satisfies the \uline{minimal model Maurer--Cartan equation}
			\begin{equation}\label{eq:minimalModelMaurerCartanEquation}
				\sum_{i\geq2}\frac1{i!}\mu^\circ_i(a^\circ,\ldots,a^\circ)\ =\ 0
			\end{equation}
		\end{subequations}
		with the $\mu^\circ_i$ given by~\eqref{eq:minimalModelHigherProducts}. The condition $\sfh(a)=0$ holds in~\eqref{eq:MCElement:b} because of the side conditions~\eqref{eq:sideConditions}. Note, however, that $\sfh(a)=0$ is \uline{not} a restriction on the solutions as it can always be assumed without loss of generality; it constitutes a gauge generalising the well-known Lorenz gauge. Indeed, when $V_0$ is trivial then $\sfh(a)=0$ holds trivially and when $V_0$ is non-trivial, when the $\mu_{i>1}=0$, the infinitesimal gauge transformations~\eqref{eq:hMCGaugeTransformations} are $a'=a+\mu_1(c_0)$ and with $c_0\coloneqq-\sfh(a)$, we immediately get $\sfh(a')=\sfh(a)-(\sfh\circ\mu_1\circ\sfh)(a)=\sfh(a)-\sfh(a)=0$ because of~\eqref{eq:SDRAdditionalProperties}. This can then be extended to when the higher products are non-vanishing by recursive means as explained in \cref{app:contractingHomotopyGaugeFixing}.

		We note that the recursion relations~\eqref{eq:LInftyQuasiIsoMinimalModel} are, in fact, the \uline{Berend--Giele recursions}~\cite{Macrelli:2019afx,Jurco:2019yfd}, and in~\cite{Lopez-Arcos:2019hvg} this construction was related to the \uline{perturbiner approach} of constructing perturbative solutions to the equations of motion of some theory.

		\section{Higher-order deformations}\label{sec:higherOrderDeformations}

		\paragraph{Motivation.}
		In this section, we wish to revisit the deformation equation~\eqref{eq:homotopyMaurerCartanActionMotivation}, that is,
		\begin{equation}\label{eq:homotopyMaurerCartanActionMotivationRepeated}
			\mu_1(\Theta)+\tfrac\kappa2\mu_2(\Theta,\Theta)+\tfrac{\kappa^2}{3!}\mu_3(\Theta,\Theta,\Theta)+\cdots\ =\ \sum_{i\geq1}\frac{\kappa^{i-1}}{i!}\mu_i(\Theta,\ldots,\Theta)\ =\ 0~.
		\end{equation}
		The standard approach of constructing perturbative solutions to this equation is to substitute the Taylor series
		\begin{equation}\label{eq:standardPerturbativeSolution}
			\Theta\ =\ \Theta^{(0)}+\kappa\Theta^{(1)}+\tfrac{\kappa^2}2\Theta^{(2)}+\cdots\ =\ \sum_{i\geq0}\frac{\kappa^i}{i!}\Theta^{(i)}~,
		\end{equation}
		to obtain a recursive set of equations for the coefficients $\Theta^{(i)}$ which, in turn, can then be solved. Here, however, we wish to proceed differently using the homotopy-algebra formalism since we are already given the general perturbative solution~\eqref{eq:MCElement} via the homotopy transfer to the minimal model of the underlying $L_\infty$-algebra. In particular, general perturbative solutions are of the form
		\begin{subequations}
			\begin{equation}\label{eq:minmalModelExpansionRepeated}
				\Theta\ =\ \sfE_1(\Theta^\circ)+\tfrac\kappa2\sfE_2(\Theta^\circ,\Theta^\circ)+\tfrac{\kappa^2}{3!}\sfE_3(\Theta^\circ,\Theta^\circ,\Theta^\circ)+\cdots\ =\ \sum_{i\geq1}\frac{\kappa^{i-1}}{i!}\sfE_i(\Theta^\circ,\ldots,\Theta^\circ)
			\end{equation}
			with the $\sfE_i$ given by~\eqref{eq:LInftyQuasiIsoMinimalModel} and $\Theta^\circ$ an element in the first cohomology group of the cochain complex underlying the construction and subject to
			\begin{equation}\label{eq:minimalModelMaurerCartanEquationRepeated}
				\sum_{i\geq2}\frac{\kappa^{i-1}}{i!}\mu^\circ_i(\Theta^\circ,\ldots,\Theta^\circ)\ =\ 0
			\end{equation}
		\end{subequations}
		with the $\mu^\circ_i$ given by~\eqref{eq:minimalModelHigherProducts}. However, we cannot yet identify the coefficients $\Theta^{(i)}$ in~\eqref{eq:standardPerturbativeSolution} with $\sfE_{i+1}(\Theta^\circ,\ldots,\Theta^\circ)$ since $\Theta^\circ$ has an expansion in terms $\kappa$. Hence, upon writing
		\begin{equation}
			\Theta^\circ\ =\ \Theta^{\circ\,(0)}+\kappa\Theta^{\circ\,(1)}+\tfrac1{2!}\kappa^2 \Theta^{\circ\,(2)}+\cdots\ =\ \sum_{i\geq0}\frac{\kappa^i}{i!}\Theta^{\circ\,(i)}
		\end{equation}
		with $\Theta^{\circ\,(i)}\in H^1(V)$ for all $i\in\IN_0$. Upon inserting this expansion into~\eqref{eq:minimalModelMaurerCartanEquationRepeated}, the lowest few equations are
		\begin{equation}\label{eq:lowestMinimalModelMaurerCartanEquationRepeated}
			\begin{aligned}
				\mu^\circ_2(\Theta^{\circ\,(0)},\Theta^{\circ\,(0)})\ &=\ 0~,
				\\
				\mu^\circ_2(\Theta^{\circ\,(0)},\Theta^{\circ\,(1)})+\tfrac1{3!}\mu^\circ_3(\Theta^{\circ\,(0)},\Theta^{\circ\,(0)},\Theta^{\circ\,(0)})\ &=\ 0~,
				\\
				&~\,\vdots
			\end{aligned}
		\end{equation}
		Likewise,~\eqref{eq:minmalModelExpansionRepeated} is then given by
		\begin{equation}\label{eq:firstOrderKappaSolution}
			\Theta\ =\ \underbrace{\sfE_1(\Theta^{\circ\,(0)})}_{=\,\Theta^{(0)}}+\kappa\underbrace{\big[\sfE_1(\Theta^{\circ\,(1)})+\tfrac12\sfE_2(\Theta^{\circ\,(0)},\Theta^{\circ\,(0)})\big]}_{=\,\Theta^{(1)}}+\cdots~.
		\end{equation}
		Note that if one is only interested in solutions to some fixed order $\kappa^n$ with $n>0$, then there is always the `gauge freedom' $\Theta\mapsto\Theta'\coloneqq\Theta+\kappa^n\sfE_1(X^\circ)$ with $X^\circ\in H^1(V)$ as $\Theta'$ will again satisfy the $n$-th-order equations of motion. Below, we shall be interested in solutions to order $\kappa$ and so, we may drop $\sfE_1(\Theta^{\circ\,(1)})$ in~\eqref{eq:firstOrderKappaSolution} and instead consider
		\begin{equation}\label{eq:firstOrderKappaSolutionReduced}
			\Theta\ =\ \sfE_1(\Theta^{\circ\,(0)})+\tfrac12\kappa\sfE_2(\Theta^{\circ\,(0)},\Theta^{\circ\,(0)})~.
		\end{equation}

		\subsection{Lowest-order deformations revisited}\label{sec:lowestOrderDeformationsRevisited}

		Let us recall the results from \cref{sec:firstOrderDeformation} for~\eqref{eq:homotopyMaurerCartanActionMotivationRepeated} to lowest order,
		\begin{equation}
			\mu_1(\Theta)\ =\ 0~,
		\end{equation}
		and formulate them in the language of homotopy algebras as outlined in \cref{sec:homologicalPerturbations}.

		\paragraph{Deformation complex.}
		We first note that instead of incorporating the gauge transformation in the degree zero of the $L_\infty$-algebra, we fix the gauge beforehand by~\eqref{eq:gaugeFixingCondition}.
		Hence, the underlying cochain complex is simply
		  \begin{equation}\label{eq:cochainDeformationNH}
			 \begin{tikzcd}
				\underbrace{\Omega^1_r(S)\oplus\scC^\infty_r(S)\oplus\scS^2_r(S)|_\text{gf}}_{\eqqcolon\,V_1}\ar[r,"\mu_1"] & \underbrace{\Omega^1_r(S)\oplus\scC^\infty_r(S)\oplus\scS^2_r(S)}_{\eqqcolon\,V_2}
			\end{tikzcd}
		\end{equation}
		where $\Omega^1_r(S)$, $\scC^\infty_r(S)$, and $\scS^2_r(S)$ are one-parameter families of smooth one-forms, functions, and symmetric tensor fields on the spatial cross section $S$ with parameter $r\geq0$ and which vanish at $r=0$. Furthermore, $|_\text{gf}$ indicates that we restrict to the elements of $\scS^2_r(S)$ that satisfy the gauge-fixing condition~\eqref{eq:gaugeFixingCondition}. In addition, $\mu_1$ was defined in~\eqref{eq:zerothOrderEinsteinEquation}. Note that the complex~\eqref{eq:cochainDeformationNH} follows directly from the full Batalin--Vilkovisky complex involving ghosts, anti-fields, and trivial pairs upon imposing the gauge-fixing condition leading to Gau{\ss}ian null coordinates and~\eqref{eq:gaugeFixingCondition} via a gauge-fixing fermion. For details, see e.g.~\cite{Jurco:2018sby,Borsten:2021hua}.

		\paragraph{Hodge--Kodaira decomposition.}
		Given the cochain complex~\eqref{eq:cochainDeformationNH}, we may now consider the Hodge--Kodaira decomposition from \cref{sec:homologicalPerturbations}. In particular, we have
		\begin{equation}
			\begin{tikzcd}
				& V_1\ar[dl,"\sfid|_{V_1}-\,\epsilon_1\circ\pi_1",shift right,swap]\ar[dr,"\pi_1",shift left]\ar[rrr,"\mu_1"] &&& V_2\ar[dl,"\sfid|_{V_2}-\,\epsilon_2\circ\pi_2",shift right,swap]\ar[dr,"\pi_2",shift left]
				\\
				H^1(V)\ar[ur,"\iota_1",shift right, swap] && V_1/\ker(\mu_1)\ar[ul,"\epsilon_1",shift left]\ar[r,"\hat\mu_1"] & \im(\mu_1)\ar[ur,"\iota_2",shift right, swap] && H^2(V)\ar[ul,"\epsilon_2",shift left]
			\end{tikzcd}
		\end{equation}
		from~\eqref{eq:twoTermsHodgeDecomposition}.

		Furthermore, the Green function $\sfg(r,y;r',y')$ defined in~\eqref{eq:fullGreenFunction} is related to the maps $\epsilon_1$ and $\hat\mu_1$ as
		\begin{equation}
			\int\rmd r'\int\rmd^{d-2}y'\,\sqrt{\det(\mathring\gamma(y'))}\,\sfg(r,y;r',y')\rho(r',y')\ =\ ((\epsilon_1\circ\hat\mu_1^{-1})(\rho))(r,y)
		\end{equation}
		for all $\rho\in\im(\mu_1)$. Hence, the choice of $\sfg(r,y;r',y')$ implies the choice of $\epsilon_1$. Thus, the special deformation retract between the cochain complex~\eqref{eq:cochainDeformationNH} and its cohomology is the one given in~\eqref{eq:exampleSpecialDeformationRetract} with the contracting homotopy written as
		\begin{equation}\label{eq:actionOfPropagatorDeformation}
			(\sfh(\rho))(r,y)\ =\ \int\rmd r'\int\rmd^{d-2}y'\,\sqrt{\det(\mathring\gamma(y'))}\,\sfg(r,y;r',y')((\sfid|_{V_2}-\,\epsilon_2\circ\pi_2)(\rho))(r',y')
		\end{equation}
		for all $\rho\in V_2$. We shall see below that when constructing explicit solutions, $\sfh$ will always act on elements of $\im(\mu_1)$ only and so, $\sfid|_{V_2}-\,\epsilon_2\circ\pi_2$ acts as the identity and is independent of $\epsilon_2$.

		\paragraph{Comparing to extremal Kerr.}
		Upon inspecting the formula for the Ricci tensor in \cref{app:RicciTensor}, one can see that in the equations of motion~\eqref{eq:independentEinsteinEquationAdaptedFrameIIModified} every $r$-derivative is always accompanied by an explicit multiplication by $r$. Consequently, $\frac{\kappa^{n-1}}{n!}\mu_n(\Theta,\ldots,\Theta)$ and $\frac{\kappa^{n-1}}{n!}\sfE_n(\Theta^\circ,\ldots,\Theta^\circ)$ are at least order $r^n$ for $n>0$. Therefore, when one performs the construction of solutions via homotopy transfer to order $\kappa^{n-1}$ as outlined at the beginning of this section, terms of order less than or equal to $r^n$ of those solutions will get no correction from the next level of the recursive construction. Hence, we can compare the solutions from the recursive construction of order $\kappa^{n-1}$ to a known full solution (e.g. extremal Kerr black hole) up to order $r^n$ and fix some of the degrees of freedom. However, we can do better than that since we have defined what we mean by the lowest-order solution, that is, elements of $H^1(V)$, along with the projection. Hence, for any known all-order solution, we can simply project it onto $H^1(V)$ and read off all values of degrees of freedom regardless of how the higher-order parts of the solutions look. Concretely, from~\eqref{eq:HodgeKodairaDecomposition} with~\eqref{eq:checkConditionSDR}, the projection is
		\begin{equation}
			\sfe|_{H^1(V)}\circ\sfp|_{V_1}\ =\ \sfid|_{V_1}-\sfh\circ\mu_1~,
		\end{equation}
		and so, for a given all-order solution $\Theta$, its lowest-order solution is
		\begin{equation}\label{eq:ProjectionToH^1(V)}
			\Theta^{(0)}\ =\ (\sfid|_{V_1}-\sfh\circ\mu_1)(\Theta)~.
		\end{equation}

		As an example, for deformations of the extremal Kerr horizon, we can use this method to deduce all the constants $A, K_1^{(n)}$, and $K_3^{(n)}$ appearing in~\eqref{eq:generalLowestOrderSolutionExtremalKerr} specific for the extremal Kerr metric~\eqref{eq:extremalKerrMetricGaussianNullCoordinates} without the need to consider higher-order solutions. In particular, in this case, with~\eqref{eq:actionOfPropagatorDeformation}~\eqref{eq:ProjectionToH^1(V)} becomes
		\begin{equation}\label{eq:projectOntoH^1Kerr}
			\Theta^{(0)}_\text{Kerr}(r,y)\ =\ \Theta_\text{Kerr}(r,y)-\int\rmd r'\int\rmd^{2}y'\,\sqrt{\det(\mathring\gamma(y'))}\,\sfg(r,y;r',y')(\mu_1(\Theta_\text{Kerr}))(r',y')
		\end{equation}
		since $\pi_2\circ\mu_1=0$ and, as before, $\sfg$ is defined in~\eqref{eq:fullGreenFunction} with $\bar\sfg_{ab}^{(n)}{}^{cd}$ given by~\eqref{eq:spatialGreenFunctionExtremalKerrN=1} and~\eqref{eq:spatialGreenFunctionExtremalKerrN>1}, and $\Theta_\text{Kerr}$ is the deformation part of the Kerr metric~\eqref{eq:alphabetagammaExtreamlKerrGAugeFixed} in the basis~\eqref{eq:defAdaptedFrameII} (see also~\eqref{eq:defAdaptedFrameIIExtremalKerr}). We can obtain $\Theta_\text{Kerr}$ to order $r^n$ by computing~\eqref{eq:alphabetagammaExtreamlKerrGAugeFixed} to order $r^n$. In other words, one needs to write the extremal Kerr metric in the Gau{\ss}ian-null-coordinate form to order $r^n$, which can be done by repeating the construction in \cref{sec:GaussianNullKerr}. Upon combining this with~\eqref{eq:projectOntoH^1Kerr}, to order $r^3$, the symmetric tensor part of $\Theta^{(0)}_\text{Kerr}$, denoted by $\big(\Theta^{(0)}_\text{Kerr}\big)_{ab}$, is given by
		\begin{equation}
			\begin{aligned}
				\big(\Theta^{(0)}_{\text{Kerr}}\big)_{ab}\ &=\ r
				\begin{pmatrix}
					\frac{3(5-3x^2+13x^4+5x^6)}{7m(1+x^2)^2} & \frac{3x(1-x^2)(9+x^2)}{7m(1+x^2)^2}
					\\
					\frac{3x(1-x^2)(9+x^2)}{7m(1+x^2)^2} & \frac{6x^2(9+x^2)}{7m(1+x^2)^2}
				\end{pmatrix}
				\\
				&\kern1cm+\frac{r^2}{2}
				\begin{pmatrix}
					\frac{3(1-x^2)(277-46x-277x^2)}{49m^2(x^2+1)^3} & \frac{3(1-x^2)(23+554 x-23x^2)}{49m^2(x^2+1)^3}
					\\
					\frac{3(1-x^2)(23+554x-23x^2)}{49m^2(x^2+1)^3} & -\frac{3(1-x^2)(277-46 x-277x^2)}{49m^2(x^2+1)^3}
					\end{pmatrix}
					\\
				&\kern1cm+\frac{r^3}{3!}
				\begin{pmatrix}
					\frac{20x(1-x^2)(671+381x-671x^2)}{343m^3(1+x^2)^4} & -\frac{10x(1-x^2)(381-2684x-381x^2)}{343m^3(1+x^2)^4}
					\\
					-\frac{10x(1-x^2)(381-2684x-381x^2)}{343m^3(1+x^2)^4} & -\frac{20x(1-x^2)(671+381x-671x^2)}{343m^3(1+x^2)^4}
				\end{pmatrix}
				\\
				&\kern1cm+\caO(r^4)~.
				\end{aligned}
		\end{equation}
		By comparing this to the lowest-order solution~\eqref{eq:generalLowestOrderSolutionExtremalKerr}, we can deduce the values of $A$, $K_1^{(n)}$, and $K_3^{(n)}$ for $n=2$ and $3$ of the extremal Kerr black hole. We obtain
		\begin{equation}\label{eq:valuesOfAK1K3}
			A\ =\ \tfrac{15}{7}~,
			\quad
			K_{1}^{(2)}\ =\ \tfrac{277}{49}~,
			\quad
			K_{3}^{(2)}\ =\ \tfrac{23}{49}~,
			\quad
			K_1^{(3)}\ =\ \tfrac{2684}{1029}~,
			\quad
			K_3^{(3)}\ =\ -\tfrac{254}{343}~.
		\end{equation}
		For consistency, one can also check that this is also true for the scalar and the one-form parts of $\Theta^{(0)}_{\text{Kerr}}$. The result for $A$ agrees with~\eqref{eq:extremalKerrConstantA} derived before with a different method. Let us remark that to fix $K_1^{(n)}$ and $K_3^{(n)}$ one needs determine the terms of order $r^n$ for the extremal Kerr metric~\eqref{eq:extremalKerrMetricGaussianNullCoordinates} in Gau{\ss}ian null coordinates following the calculation in \cref{sec:GaussianNullKerr}. As before, in the above formul{\ae} we have set $\kappa=1$.

		\subsection{Next-to-lowest-order deformations}\label{sub:nextToLowestOrderDeformations}

		Having formulated the cochain complex and the special deformation retract onto its cohomology for our deformation problem, let us now move on an introduce the first non-trivial non-linearities. In particular, let us now study~\eqref{eq:homotopyMaurerCartanActionMotivationRepeated} to next-to-lowest order,
		\begin{equation}\label{eq:NTLOMCEquation}
			\mu_1(\Theta)+\tfrac\kappa2\mu_2(\Theta,\Theta)\ =\ 0
		\end{equation}
		and define the underlying strict $L_\infty$-algebra, that is, the higher product $\mu_2$.

		\paragraph{$L_\infty$-algebra.}
		To satisfy the homotopy Jacobi identities~\eqref{eq:homotopyJacobiIdentities} (see also~\eqref{eq:lowestHomotopyJacobiIdentities}), we require $\mu_2$ to be non-trivial only between elements of $V_1$ in~\eqref{eq:cochainDeformationNH}. Upon expanding the independent Einstein equations~\eqref{eq:independentEinsteinEquationAdaptedFrameIIModified} to the next-to-lowest-order, we obtain
		\begin{subequations}\label{eq:DefOfMu2NH}
			\begin{equation}
				\mu_2(\Theta,\Theta)\ \coloneqq\
				\begin{pmatrix}
					\mu_2(\Theta,\Theta)_a
					\\
					\mu_2(\Theta,\Theta)_0
					\\
					\mu_2(\Theta,\Theta)_{ab}
				\end{pmatrix}
				\ewith
				\Theta\ =\
				\begin{pmatrix}
					h_a\\ h\\ h_{ab}
				\end{pmatrix}
				\ \in\ V_1~,
			\end{equation}
			where
			\begin{equation}
				\begin{aligned}
					\mu_2(\Theta,\Theta)_a\ &\coloneqq\ r\big(-2h^b\partial_rh_{ab}+h_a\partial_rh_b{}^b-4h_{ab}\partial_rh^b+2\mathring\alpha_ch^{bc}\partial_rh_{ab}+2h_{ab}\mathring\alpha_c\partial_rh^{bc}
					\\
					&\kern1cm-\mathring\alpha_ah^{bc}\partial_rh_{bc}-h_{ab}\mathring\alpha^b\partial_rh_c{}^c+\partial_rh^{bc}\mathring{\tilde\nabla}_ah_{bc}+\partial_rh_{ab}\mathring{\tilde\nabla}^bh_c{}^c-2\partial_rh_a{}^b\mathring{\tilde\nabla}^ch_{bc}
					\\
					&\kern1cm+2\mathring{\tilde\nabla}_a\partial_rh_{bc}h^{bc}+2h_{ab}\mathring{\tilde\nabla}^b\partial_rh_c{}^c-2h^{bc}\mathring{\tilde\nabla}_b\partial_rh_{ac}-2h_{ab}\mathring{\tilde\nabla}_c\partial_rh^{bc}\big)
					\\
					&\kern1cm+r^2\big[2h_a\partial_r^2h_b{}^b-2 h^b\partial_r^2h_{ab}-2\partial_rh^b\partial_rh_{ab}+\partial_rh_a\partial_rh_b{}^b-2\partial_r^2h^bh_{ab}
					\\
					&\kern1cm-\mathring\alpha_a\partial_rh^{bc}\partial_rh_{bc}+\mathring\alpha_b\big(2\partial_rh^{bc}\partial_rh_{ac}-\partial_rh_c{}^c\partial_rh_a{}^b\big)-2\mathring\alpha_a\partial_r^2h_{bc}h^{bc}
					\\
					&\kern1cm-2\mathring\alpha^b\partial_r^2h_c{}^ch_{ab}+2\mathring\alpha_b\partial_r^2h_{ac}h^{bc}+2\mathring\alpha_b\partial_r^2h^{bc}h_{ac}\big]\,,
					\\
					\mu_2(\Theta,\Theta)_0\ &\coloneqq\ -r^2\big(\partial_r^2h_{ab}h^{ab}+\tfrac12\partial_rh_{ab}\partial_rh^{ab}\big)\,,
					\\
					\mu_2(\Theta,\Theta)_{ab}\ &\coloneqq\ \tfrac{8}{d-2}\Lambda h_{c(a}h_{b)}{}^c+R^{(2)}_{ab}-4R^{(1)}_{c(a}h_{b)}{}^c+R^{(1)}h_{ab}-\tfrac12R^{(2)}\delta_{ab}-4h_{(a}R_{b)-}^{(1)}~,
				\end{aligned}
			\end{equation}
			where, as before, indices are raised and lowered with $\mathring\gamma_{ab}=\delta_{ab}$ and
			\begin{equation}
				\begin{aligned}
					R^{(1)}_{a-}\ &=\ \partial_rh_a-\tfrac12\mathring\alpha_b\partial_r h_a {}^b+\tfrac14\mathring\alpha_a\partial_rh_b{}^b-\tfrac12\mathring{\tilde\nabla}_a\partial_rh_b{}^b+\tfrac12\mathring{\tilde\nabla}_b\partial_rh_a{}^b
					\\
					&\kern1cm+r\big(\tfrac12\partial_r^2 h_a+\tfrac12\mathring\alpha_a \partial_r^2h_b{}^b-\tfrac12\mathring\alpha_b\partial_r^2h_a{}^b\big)
				\end{aligned}
			\end{equation}
			and
			\begin{equation}
				\begin{aligned}
					R^{(1)}_{ab}\ &=\ \mathring{\tilde\nabla}^c\mathring{\tilde\nabla}_{(a}h_{b)c}-\tfrac12\mathring{\tilde\nabla}_a\mathring{\tilde\nabla}_bh_c {}^c
					\\
					&\kern1cm-\tfrac12\mathring{\tilde\nabla}^c\mathring{\tilde\nabla}_ch_{ab}+\mathring{\tilde\nabla}_{(a} h_{b)}-\mathring\alpha^c\mathring{\tilde\nabla}_{(a}h_{b)c}+\tfrac12\mathring\alpha^c\mathring{\tilde\nabla}_ch_{ab}-h_{(a}\mathring\alpha_{b)}
					\\
					&\kern1cm+r\big[\mathring{\tilde\nabla}_{(a}\partial_rh_{b)}-\tfrac32\big(\partial_rh_a\mathring\alpha_b+\partial_rh_b\mathring\alpha_a \big)+\big(2\mathring\alpha_c\mathring\alpha_{(b}-\mathring{\tilde\nabla}_c\mathring\alpha_{( b}\big)\partial_rh_{a)}{}^c
					\\
					&\kern1cm+\big(-\mathring\alpha_c\mathring\alpha^c-\mathring\beta+\tfrac12\mathring{\tilde\nabla}_c \mathring\alpha^c\big)\partial_rh_{ab}+\tfrac12\big(-\mathring\alpha_a\mathring\alpha_b+\mathring{\tilde\nabla}_{(a}\mathring\alpha_{b)}\big)\partial_rh_c{}^c
					\\
					&\kern1cm-\mathring\alpha_{(a|}\mathring{\tilde\nabla}^c\partial_rh_{c|b)}-\mathring\alpha^c\mathring{\tilde\nabla}_{(a}\partial_rh_{b) c}+\mathring\alpha^c\mathring{\tilde\nabla}_c\partial_rh_{a b}+\mathring\alpha_{(a}\mathring{\tilde\nabla}_{b)}\partial_rh_c{}^c\big]
					\\
					&\kern1cm+r^2\big(\mathring\alpha^c\mathring\alpha_{(a}\partial_r^2h_{b)c}-\partial_r^2h_{(a}\mathring\alpha_{b)}-\tfrac12\mathring\alpha_a\mathring\alpha_b\partial_r^2h_c {}^c-\tfrac12(\mathring\alpha^c\mathring\alpha_c+\mathring\beta)\partial_r^2h_{ab}\big)
				\end{aligned}
			\end{equation}
			and
			\begin{equation}
				\begin{aligned}
					R_{ab}^{(2)}\ &=\ -2h^{cd}\mathring{\tilde\nabla}_c\mathring{\tilde\nabla}_{(a}h_{b)d}+h^{cd}\mathring{\tilde\nabla}_c\mathring{\tilde\nabla}_dh_{ab}+h^{cd}\mathring{\tilde\nabla}_{(a}\mathring{\tilde\nabla}_{b)}h_{cd}-h^{cd}\mathring\alpha_c\mathring{\tilde\nabla}_dh_{ab}+2h^{cd}\mathring\alpha_c\mathring{\tilde\nabla}_{(a}h_{b)d}
					\\
					&\kern1cm-2\mathring{\tilde\nabla}_ch^{cd}\mathring{\tilde\nabla}_{(a}h_{b)d}+\mathring{\tilde\nabla}_ch^{cd}\mathring{\tilde\nabla}_dh_{ab}+\tfrac12\mathring{\tilde\nabla}_ah_{cd}\mathring{\tilde\nabla}_bh^{cd}+\mathring{\tilde\nabla}^dh_c{}^c\mathring{\tilde\nabla}_{(a}h_{b)d}
					\\
					&\kern1cm-\tfrac12\mathring{\tilde\nabla}^dh_c{}^c\mathring{\tilde\nabla}_dh_{ab}-\mathring{\tilde\nabla}_ch_a{}^d\mathring{\tilde\nabla}_dh_b{}^c+\mathring{\tilde\nabla}^ch_{ad}\mathring{\tilde\nabla}_ch_b{}^d+h^c\mathring{\tilde\nabla}_ch_{ab}-2h^c\mathring{\tilde\nabla}_{(a}h_{b)c}-h_ah_b
					\\
					&\kern1cm+r\big[2\big(-2\mathring\alpha_c\mathring\alpha_{(a}+\mathring{\tilde\nabla}_c\mathring\alpha_{(a}\big)\partial_rh_{b)d}h^{cd}+\big(2\mathring\alpha_c\mathring\alpha_d-\mathring{\tilde\nabla}_c\mathring\alpha_d\big)h^{cd}\partial_rh_{ab}
					\\
					&\kern1cm+\big(\mathring\alpha_a\mathring\alpha_b-\mathring{\tilde\nabla}_{(a}\mathring\alpha_{b)}\big)h^{cd}\partial_rh_{cd}+2h_c{}^d\mathring\alpha_{(a|}\mathring{\tilde\nabla}{}^c\partial_rh_{|b)d}+2\mathring\alpha_ch^{cd}\mathring{\tilde\nabla}_{(a}\partial_rh_{b)d}
					\\
					&\kern1cm-2\mathring\alpha_ch^{cd}\mathring{\tilde\nabla}_d\partial_rh_{ab}-2\mathring\alpha_{(a|}h^{cd}\mathring{\tilde\nabla}_{|b)}\partial_rh_{cd}-2h\partial_rh_{ab}-2h_{(a}\partial_rh_{b)}+4\mathring\alpha_{(a|}h^c\partial_rh_{|b)c}
					\\
					&\kern1cm+2\mathring\alpha^ch_{(a}\partial_rh_{b)c}-4\mathring\alpha^ch_c\partial_rh_{ab}-\mathring\alpha_{(a}h_{b)}\partial_rh_c{}^c+\partial_rh^c\big(\mathring{\tilde\nabla}_ch_{ab}-2\mathring{\tilde\nabla}_{(a}h_{b)c}\big)
					\\
					&\kern1cm+\mathring{\tilde\nabla}^ch_c\partial_rh_{ab}-2\mathring{\tilde\nabla}^ch_{(a}\partial_rh_{b)c}+\mathring{\tilde\nabla}_{(a}h_{b)}\partial_rh_c{}^c-2h^c\mathring{\tilde\nabla}_{(a}\partial_rh_{b)c}+2h^c\mathring{\tilde\nabla}_c\partial_rh_{ab}
					\\
					&\kern1cm-2\mathring\alpha_c\partial_rh_{(a|}{}^d\mathring{\tilde\nabla}^ch_{|b)d}+2\mathring\alpha^c\partial_rh_{(a|}{}^d\mathring{\tilde\nabla}_dh_{|b)c}+\tfrac12\mathring\alpha^c\partial_rh_{ab}\mathring{\tilde\nabla}_ch_d{}^d-\mathring\alpha^c\partial_rh_{ab}\mathring{\tilde\nabla}^dh_{cd}
					\\
					&\kern1cm+\tfrac12\mathring\alpha^c\partial_rh_d{}^d\mathring{\tilde\nabla}_ch_{ab}-\mathring\alpha^c\partial_rh_d{}^d\mathring{\tilde\nabla}_{(a}h_{b)c}-\mathring\alpha^c\partial_rh_c{}^d\mathring{\tilde\nabla}_dh_{ab}+2\mathring\alpha^c\partial_rh_c{}^d\mathring{\tilde\nabla}_{(a}h_{b)d}
					\\
					&\kern1cm-\mathring\alpha_{(a|}\partial_rh^{cd}\mathring{\tilde\nabla}_{|b)}h_{cd}-\mathring\alpha_{(a}\partial_rh_{b)c}\mathring{\tilde\nabla}^ch_d{}^d+2\mathring\alpha_{(a}\partial_rh_{b)d}\mathring{\tilde\nabla}^ch_{dc}\big]
					\\
					&\kern1cm+r^2\big[-2\mathring\alpha_ch^{cd}\mathring\alpha_{(a}\partial_r^2h_{b)d}+\mathring\alpha_a\mathring\alpha_bh^{cd}\partial_r^2h_{cd}+\mathring\alpha_c\mathring\alpha_dh^{cd}\partial_r^2h_{ab}+2\mathring\alpha_{(a|}h^c\partial_r^2h_{|b)c}
					\\
					&\kern1cm-2\mathring\alpha^ch_c\partial_r^2h_{ab}-h\partial_r^2h_{ab}-\partial_rh_a\partial_rh_b-\big(\partial_rh+2\partial_rh^c\mathring\alpha_c\big)\partial_rh_{ab}+2\partial_rh^c\mathring\alpha_{(a}\partial_rh_{b)c}
					\\
					&\kern1cm+2\partial_rh_{(a|}\mathring\alpha^c\partial_rh_{|b)c}-\partial_rh_{(a}\mathring\alpha_{b)}\partial_rh_c{}^c-2\mathring\alpha_c\mathring\alpha_{(a}\partial_rh_{b)d}\partial_rh^{cd}+\mathring\alpha^c\mathring\alpha_{(a}\partial_rh_{b)c}\partial_rh_d{}^d
					\\
					&\kern1cm-\mathring\alpha_c\mathring\alpha_d\partial_rh_a{}^c\partial_rh_b{}^d+\big(\mathring\alpha_c\mathring\alpha^c+\mathring\beta\big)\partial_rh_a{}^d\partial_rh_{bd}+\tfrac12\mathring\alpha_a\mathring\alpha_b\partial_rh_{cd}\partial_rh^{cd}
					\\
					&\kern1cm+\mathring\alpha_c\mathring\alpha_d\partial_rh^{cd}\partial_rh_{ab}-\tfrac12\big(\mathring\alpha_c\mathring\alpha^c+\mathring\beta\big)\partial_rh_{ab}\partial_rh_d{}^d\big]
				\end{aligned}
			\end{equation}
			and
			\begin{equation}
				\begin{aligned}
					R^{(1)}\ &\coloneqq\ -\tfrac{2}{d-2}\Lambda h_a{}^a-\mathring{\tilde\nabla}_a\mathring\alpha_bh^{ab}+\mathring\alpha_a\mathring\alpha_bh^{ab}
					\\
					&\kern1cm-2h+2\mathring{\tilde\nabla}_ah^a-3\mathring\alpha_ah^a+\mathring{\tilde\nabla}_a\mathring{\tilde\nabla}_bh^{ab}-\mathring{\tilde\nabla}^a\mathring{\tilde\nabla}_ah_b{}^b-2\mathring\alpha_a\mathring{\tilde\nabla}_bh^{ab}+\mathring\alpha^a\mathring{\tilde\nabla}_ah_b{}^b
					\\
					&\kern1cm+r\big(-4\partial_rh-2\mathring\beta\partial_rh_a{}^a+2\mathring{\tilde\nabla}_a\partial_rh^a-7\mathring\alpha_a\partial_rh^a+3\mathring\alpha_a\mathring\alpha_b\partial_rh^{ab}-\mathring{\tilde\nabla}_a\mathring\alpha_b\partial_rh^{ab}
					\\
					&\kern1cm-2\mathring\alpha_a\mathring\alpha^a\partial_rh_b{}^b+\mathring{\tilde\nabla}_a\mathring\alpha^a\partial_rh_b{}^b-2\mathring\alpha_a\mathring{\tilde\nabla}_b\partial_rh^{ab}+2\mathring\alpha^a\mathring{\tilde\nabla}_a\partial_rh_b{}^b\big)
					\\
					&\kern1cm+r^2\big(-\partial_r^2h-2\mathring\alpha_a\partial_r^2h^a-\mathring\beta\partial_r^2h_a{}^a+\mathring\alpha_a\mathring\alpha_b\partial_r^2h^{ab}-\mathring\alpha_a\mathring\alpha^a\partial_r^2h_b{}^b\big)
				\end{aligned}
			\end{equation}
			and
			\begin{equation}
				\begin{aligned}
					R^{(2)}\ &\coloneqq\ \tfrac4{d-2}\Lambda h^{ab}h_{ab}+2\mathring{\tilde\nabla}_a\mathring\alpha_bh^{ac}h^b{}_c-2\mathring\alpha_a\mathring\alpha_bh^{ac}h^b{}_c-4h^{ab}\mathring{\tilde\nabla}_ah_b+6h^{ab}\mathring\alpha_ah_b
					\\
					&\kern1cm-2h^{ab}\mathring{\tilde\nabla}_a\mathring{\tilde\nabla}_ch_b{}^c+2h^{ab}\mathring{\tilde\nabla}{}^c\mathring{\tilde\nabla}_ch_{ab}+4h^{ab}\mathring\alpha^c\mathring{\tilde\nabla}_ah_{bc}-2h^{ab}\mathring\alpha^c\mathring{\tilde\nabla}_ch_{ab}-2h^{ab}\mathring{\tilde\nabla}^c\mathring{\tilde\nabla}_ah_{bc}
					\\
					&\kern1cm+2h^{ab}\mathring{\tilde\nabla}{}_a\mathring{\tilde\nabla}_bh_c{}^c+4h^{ab}\mathring\alpha_a\mathring{\tilde\nabla}^ch_{bc}-2h^{ab}\mathring\alpha_a\mathring{\tilde\nabla}_bh_c{}^c-3h_ah^a+2h^a\mathring{\tilde\nabla}_ah_b{}^b
					\\
					&\kern1cm-4 h^a\mathring{\tilde\nabla}^bh_{ab}-2\mathring{\tilde\nabla}^ah_{ab}\mathring{\tilde\nabla}_ch^{bc}+2\mathring{\tilde\nabla}^ah_{ab}\mathring{\tilde\nabla}^bh_c{}^c+\tfrac32\mathring{\tilde\nabla}^ah_{bc}\mathring{\tilde\nabla}_ah^{bc}-\tfrac12\mathring{\tilde\nabla}^ah_b{}^b\mathring{\tilde\nabla}_ah_c{}^c\\
					&\kern1cm-\mathring{\tilde\nabla}^ah^{bc}\mathring{\tilde\nabla}_bh_{ac}
					\\
					&\kern1cm+r\big(4\mathring\beta\partial_rh_{ab}h^{ab}-4h^{ab}\mathring{\tilde\nabla}_a\partial_rh_b+14h^{ab}\mathring\alpha_a\partial_rh_b-12h^{ab}\mathring\alpha_a\mathring\alpha^c\partial_rh_{bc}
					\\
					&\kern1cm+2h^{ab}\mathring{\tilde\nabla}^c\mathring\alpha_a\partial_rh_{bc}+4h^{ab}\mathring\alpha_a\mathring\alpha_b\partial_rh_c{}^c-2\mathring{\tilde\nabla}^a\mathring\alpha_ah^{bc}\partial_rh_{bc}+4h^{ab}\mathring\alpha_a\mathring{\tilde\nabla}^cb\partial_rh_{bc}
					\\
					&\kern1cm-4h^{ab}\mathring\alpha_a\mathring{\tilde\nabla}_b\partial_rh_c{}^c+2h^{ab}\mathring{\tilde\nabla}_a\mathring\alpha^c\partial_rh_{bc}+4\mathring\alpha_a\mathring\alpha^ah^{bc}\partial_rh_{bc}-2h^{ab}\mathring{\tilde\nabla}_a\mathring\alpha_b\partial_rh_c{}^c
					\\
					&\kern1cm+4h^{ab}\mathring\alpha^c\mathring{\tilde\nabla}_a\partial_rh_{bc}-4\mathring\alpha^ah^{bc}\mathring{\tilde\nabla}_a\partial_rh_{bc}-4 h\partial_rh_a{}^a-14\partial_rh_ah^a+12\mathring\alpha_ah_b\partial_rh^{ab}
					\\
					&\kern1cm-8\mathring\alpha_ah^a\partial_rh_b{}^b+2\partial_rh^a\mathring{\tilde\nabla}_ah_b{}^b-4\partial_rh^a\mathring{\tilde\nabla}^bh_{ab}+2\mathring{\tilde\nabla}_a h^a\partial_rh_b{}^b-2\mathring{\tilde\nabla}_a h_b\partial_rh^{ab}
					\\
					&\kern1cm-4 h^a\mathring{\tilde\nabla}^b\partial_rh_{ab}+4h^a\mathring{\tilde\nabla}_a\partial_rh_b{}^b-3\mathring\alpha_a\partial_rh_{bc}\mathring{\tilde\nabla}^ah^{bc}+2\mathring\alpha^a\partial_rh^{bc}\mathring{\tilde\nabla}_bh_{ac}
					\\
					&\kern1cm+\mathring\alpha_a\partial_rh_b{}^b\mathring{\tilde\nabla}^ah_c{}^c-2\mathring\alpha_a\partial_rh_b{}^b\mathring{\tilde\nabla}_ch^{ac}-2\mathring\alpha_a\partial_rh^{ab}\mathring{\tilde\nabla}_bh_c{}^c+4\mathring\alpha_a\partial_rh^{ab}\mathring{\tilde\nabla}^ch_{bc}\big)
					\\
					&\kern1cm+r^2\big(4h^{ab}\mathring\alpha_a\partial_r^2h_b+2h^{ab}\mathring\beta\partial_r^2h_{ab}-4h^{ab}\mathring\alpha_a\mathring\alpha^c\partial_r^2h_{bc}+2\mathring\alpha_a\mathring\alpha^ah^{bc}\partial_r^2h_{bc}
					\\
					&\kern1cm+2h^{ab}\mathring\alpha_a\mathring\alpha_b\partial_r^2h_c{}^c-2\partial_rh\partial_rh_a{}^a-2h\partial_r^2h_a{}^a+\tfrac32\mathring\beta\partial_rh_{ab}\partial_rh^{ab}-\tfrac12\mathring\beta\partial_rh_a{}^a\partial_rh_b{}^b
					\\
					&\kern1cm-3\partial_rh_a\partial_rh^a -4h_a\partial_r^2h^a +4\mathring\alpha^ah^b\partial_r^2h_{ab} -4\mathring\alpha_a h^a\partial_r^2h_b{}^b-4\mathring\alpha_a\partial_rh^a\partial_rh_b{}^b
					\\
					&\kern1cm+6\mathring\alpha_a\partial_rh_b\partial_rh^{ab} -3\mathring\alpha_a\mathring\alpha_b\partial_rh^{ac}\partial_rh^b{}_c+\tfrac32\mathring\alpha_a\mathring\alpha^a\partial_rh^{bc}\partial_rh_{bc}
					\\
					&\kern1cm+2\mathring\alpha_a\mathring\alpha_b\partial_rh^{ab}\partial_rh_c{}^c-\tfrac12\mathring\alpha_a\mathring\alpha^a\partial_rh_b{}^b\partial_rh_c{}^c\big)\,.
				\end{aligned}
			\end{equation}
		\end{subequations}
		See \cref{app:perturbativeExpansionEinsteinEquation} for details on the derivation. Note that we have used the background Einstein equation~\eqref{eq:backgroundEinsteinEquations}. The general expression for $\mu_2$ between any two elements of $V_1$ is obtained from polarising these formul{\ae}.

		\paragraph{Next-to-lowest-order solutions.}
		General next-to-lowest-order solutions will now be of the form~\eqref{eq:firstOrderKappaSolutionReduced},
		\begin{subequations}
			\begin{equation}\label{eq:SolutionOfNHDeformationNextToLowestOrder}
				\Theta\ =\ \sfE_1(\Theta^{\circ\,(0)})+\tfrac12\kappa\sfE_2(\Theta^{\circ\,(0)},\Theta^{\circ\,(0)})\ =\ \sfE_1(\Theta^{\circ\,(0)})-\tfrac12\kappa\sfh\big(\mu_2(\Theta^{\circ\,(0)},\Theta^{\circ\,(0)})\big)
			\end{equation}
			with $\Theta^{\circ\,(0)}\in H^1(V)$ subject to the minimal model Maurer--Cartan equation~\eqref{eq:lowestMinimalModelMaurerCartanEquationRepeated},
			\begin{equation}\label{eq:MinimalMCEquationOfNHDeformationNextToLowestOrder}
				0\ =\ \mu^\circ_2(\Theta^{\circ\,(0)},\Theta^{\circ\,(0)})\ =\ \sfp\big(\mu_2(\Theta^{\circ\,(0)},\Theta^{\circ\,(0)})\big)
				\quad\Leftrightarrow\quad
				\mu_2(\Theta^{\circ\,(0)},\Theta^{\circ\,(0)})\ \in\ \im(\mu_1)~.
			\end{equation}
		\end{subequations}
		Hence, in the solution~\eqref{eq:SolutionOfNHDeformationNextToLowestOrder}, the propagator $\sfh$ acts on elements of $\im(\mu_1)$ only and so, the $\epsilon_2\circ\pi_2$ term in~\eqref{eq:actionOfPropagatorDeformation} drops out as claimed.\footnote{Note that this remains true for higher-order solutions.} Hence, to write down general next-to-lowest-order solutions, all one needs to do is to solve~\eqref{eq:MinimalMCEquationOfNHDeformationNextToLowestOrder}. We shall now do this explicitly in the case of deformations of extremal Kerr horizon.

		\subsection{Example: extremal Kerr}\label{sec:NTLOKerrSolutions}

		We now have all of the ingredients in order to compute the next-to-lowest order deformations of the extremal Kerr horizon using homological perturbation theory.

		\paragraph{$L_\infty$-algebra.}
		We first substitute in the near horizon data~\eqref{eq:extremalKerrNearHorizonMetricGaussianNullCoordinates} into~\eqref{eq:DefOfMu2NH},
		\begin{subequations}\label{eq:mu2Kerr}
			\begin{equation}
				\begin{aligned}
					\mu_2(\Theta,\Theta)_1\ &=\ r\bigg[-4\partial_rh_1h_{11}-4\partial_rh_2h_{12}-h_1\partial_rh_{11}-2h_2\partial_rh_{12}+h_1\partial_rh_{22}
					\\
					&\kern1cm+\frac{1}{m}\sqrt{\frac{1-x^2}{1+x^2}}\bigg(\frac{4xh_{11}\partial_rh_{11}}{1-x^2}+\frac{4h_{11}\partial_rh_{12}}{1+x^2}-\frac{2xh_{11}\partial_rh_{22}}{1-x^2}+\frac{2h_{12}\partial_rh_{11}}{1+x^2}
					\\
					&\kern1cm+\frac{4xh_{12}\partial_rh_{12}}{1-x^2}+\frac{4h_{22}\partial_rh_{12}}{1+x^2}+\frac{2h_{12}\partial_rh_{22}}{1+x^2}-\frac{2xh_{22}\partial_rh_{22}}{1-x^2}
					\\
					&\kern1cm+\partial_xh_{22}\partial_rh_{11}+\partial_x h_{22}\partial_rh_{22}+2\partial_x\partial_rh_{22}h_{11}+2\partial_x\partial_rh_{22}h_{22}\bigg)\bigg]
					\\
					&\kern1cm+r^2\bigg[-2\partial_r^2h_1h_{11}-2\partial_r^2h_2h_{12}-2h_2\partial_r^2h_{12}+2h_1\partial_r^2h_{22}+\partial_rh_1\partial_rh_{22}
					\\
					&\kern1cm-\partial_rh_1\partial_rh_{11}-2\partial_rh_2\partial_rh_{12}+\frac{2}{m(1+x^2)}\sqrt{\frac{1-x^2}{1+x^2}}\big(2h_{11}\partial_r^2h_{12}
					\\
					&\kern1cm+2h_{22}\partial_r^2h_{12}+2xh_{11}\partial_r^2h_{22}+2xh_{22}\partial_r^2h_{22}+x\partial_rh_{22}\partial_rh_{22}
					\\
					&\kern1cm+x\partial_rh_{11}\partial_rh_{22}+\partial_rh_{12}\partial_rh_{22}+\partial_rh_{11}\partial_rh_{12}\big)\bigg]
				\end{aligned}
			\end{equation}
			and
			\begin{equation}
				\begin{aligned}
					\mu_2(\Theta,\Theta)_2\ &=\ r\bigg[-4\partial_rh_1h_{12}-4\partial_rh_2h_{22}+h_2\partial_rh_{11}-2h_1\partial_rh_{12}-h_2\partial_rh_{22}
					\\
					&\kern1cm+\frac{1}{m}\sqrt{\frac{1-x^2}{1+x^2}}\bigg(\frac{2xh_{12}\partial_rh_{11}}{1-x^2}+\frac{4h_{12}\partial_rh_{12}}{1+x^2}+\frac{2xh_{12}\partial_rh_{22}}{1-x^2}+\frac{4xh_{11}\partial_rh_{12}}{1-x^2}
					\\
					&\kern1cm+\frac{4xh_{22}\partial_rh_{12}}{1-x^2}+\frac{4h_{22}\partial_rh_{22}}{1+x^2}-\frac{2h_{11}\partial_rh_{11}}{1+x^2}-\frac{2h_{22}\partial_rh_{11}}{1+x^2}-\partial_xh_{11}\partial_rh_{12}
					\\
					&\kern1cm-2\partial_xh_{12}\partial_rh_{22}+\partial_xh_{22}\partial_rh_{12}-2\partial_x\partial_rh_{12}h_{11}-2\partial_x\partial_rh_{12} h_{22}\bigg)\bigg]
					\\
					&\kern1cm+r^2\bigg[-2\partial_r^2h_1h_{12}-2\partial_r^2h_2h_{22}+2h_2\partial_r^2h_{11}-2h_1\partial_r^2h_{12}+\partial_rh_2\partial_rh_{11}
					\\
					&\kern1cm-2\partial_rh_1\partial_rh_{12}-\partial_rh_2\partial_rh_{22}+\frac{2}{m(1+x^2)}\sqrt{\frac{1-x^2}{1+x^2}}\big(-2h_{11}\partial_r^2h_{11}
					\\
					&\kern1cm-2h_{22}\partial_r^2h_{11}-2xh_{11}\partial_r^2h_{12}-2xh_{22}\partial_r^2h_{12}-\partial_rh_{11}\partial_rh_{11}-x\partial_rh_{12}\partial_rh_{11}
					\\
					&\kern1cm-\partial_rh_{22}\partial_rh_{11}-x\partial_rh_{12}\partial_rh_{22}\big)\bigg]
				\end{aligned}
			\end{equation}
			and
			\begin{equation}
				\begin{aligned}
					\mu_2(\Theta,\Theta)_0\ &=\ -\frac12r^2(2h_{11}\partial_r^2h_{11}+4h_{12}\partial_r^2h_{12}+2h_{22}\partial_r^2h_{22}
					\\
					&\kern1cm+\partial_rh_{11}\partial_rh_{11}+2\partial_rh_{12}\partial_rh_{12}+\partial_rh_{22}\partial_rh_{22})
				\end{aligned}
			\end{equation}
			and
			\begin{equation}
				\begin{aligned}
					\mu_2(\Theta,\Theta)_{11}\ &=\ -2hh_{11}+\frac12h_1h_1+\frac32h_2h_2
					\\
					&\kern1cm-\frac{2}{m}\sqrt{\frac{1-x^2}{1+x^2}}\bigg(\frac{h_1h_{12}}{1+x^2}+\frac{2xh_1h_{11}}{1-x^2}+\frac{xh_2h_{12}}{1-x^2}+\frac{3h_2h_{11}}{1+x^2}+\frac{3h_2h_{22}}{1+x^2}\bigg)
					\\
					&\kern1cm+\frac{2}{m^2(1+x^2)^3}\big[4x(1+x^2)h_{11}h_{12}+2x(1+x^2)h_{12}h_{22}
					\\
					&\kern1cm-2x^2(3+x^2)h_{11}h_{11}+3(1-x^2)h_{22}h_{22}+3(1-x^2)h_{11}h_{22}
					\\
					&\kern1cm+(1-4x^2-x^4)h_{12}h_{12}\big]-\frac{1}{m}\sqrt{\frac{1-x^2}{1+x^2}}h_1\partial_xh_{22}
					\\
					&\kern1cm+\frac{2(1-x^2)}{m^2(1+x^2)^2}\big(\partial_xh_{22}h_{12}-2x\partial_xh_{22}h_{11}-x\partial_xh_{22}h_{22}\big)
					\\
					&\kern1cm+r\bigg\{-4\partial_rhh_{11}+2h\partial_rh_{22}+h_1\partial_rh_1+7h_2\partial_rh_2
					\\
					&\kern1cm+\frac{2}{m}\sqrt{\frac{1-x^2}{1+x^2}}\bigg(-\frac{\partial_rh_1h_{12}}{1+x^2}-\frac{2x\partial_rh_1h_{11}}{1-x^2}-\frac{x\partial_rh_2h_{12}}{1-x^2}-\frac{7\partial_rh_2h_{11}}{1+x^2}
					\\
					&\kern1cm-\frac{7\partial_rh_2h_{22}}{1+x^2}-\frac{2h_1\partial_rh_{12}}{1+x^2}-\frac{2xh_1\partial_rh_{22}}{1+x^2}-\frac{2h_2\partial_rh_{22}}{1+x^2}+\frac{x(1-2x^2)h_2\partial_rh_{12}}{1-x^4}\bigg)
					\\
					&\kern1cm+\frac{2}{m^2(1+x^2)^3}\big[4x(1+x^2)h_{11}\partial_rh_{12}+2x(1+x^2)h_{22}\partial_rh_{12}
					\\
					&\kern1cm+4(1-x^2)h_{12}\partial_rh_{12}+2x(3-x^2)h_{12}\partial_rh_{22}+(5-12x^2+3x^4)h_{11}\partial_rh_{22}
					\\
					&\kern1cm+(7-12x^2+x^4)h_{22}\partial_rh_{22}\big]+\frac{1}{m}\sqrt{\frac{1-x^2}{1+x^2}}\big(-\partial_rh_1 \partial_xh_{22}+\partial_xh_2\partial_rh_{12}\big)\bigg\}
					\\
					&\kern1cm+r^2\bigg[-\partial_r^2h h_{11}+h\partial_r^2h_{22}+\partial_rh\partial_rh_{22}+2h_2\partial_r^2h_2+\frac12\partial_rh_1\partial_rh_1
					\\
					&\kern1cm+\frac32\partial_rh_2\partial_rh_2+\frac{2}{m(1+x^2)}\sqrt{\frac{1-x^2}{1+x^2}}\big(-2\partial_r^2h_2h_{11}-2\partial_r^2h_2h_{22}
					\\
					&\kern1cm-\partial_rh_1\partial_rh_{12}-x\partial_rh_1\partial_rh_{22}+x\partial_rh_2\partial_rh_{12}-\partial_rh_2\partial_rh_{22}\big)
					\\
					&\kern1cm+\frac{3-6x^2-x^4}{2m^2(1+x^2)^3}\big(2h_{11}\partial_r^2h_{22}+2h_{22}\partial_r^2h_{22}+\partial_rh_{12}\partial_rh_{12}+\partial_rh_{22}\partial_rh_{22}\big)\bigg]
				\end{aligned}
			\end{equation}
			and
			\begin{equation}
				\begin{aligned}
					\mu_2(\Theta,\Theta)_{12}\ &=\ -2hh_{12}-h_1h_2+\frac{2}{m}\sqrt{\frac{1-x^2}{1+x^2}}\bigg(-\frac{x h_1h_{12}}{1-x^2}-\frac{h_2h_{12}}{1+x^2}+\frac{h_1h_{11}}{1+x^2}
					\\
					&\kern1cm+\frac{h_1h_{22}}{1+x^2}-\frac{x(2-x^2)h_2h_{11}}{1-x^4}-\frac{x(2-x^2)h_2h_{22}}{1-x^4}\bigg)
					\\
					&\kern1cm+\frac{2}{m^2(1+x^2)^3}\big[2x(1+x^2)h_{12}h_{12}-(1-x^2)h_{22}h_{12}
					\\
					&\kern1cm-(2+x^2+x^4)h_{11}h_{12}\big]-\frac{1}{m}\sqrt{\frac{1-x^2}{1+x^2}}\big(h_2\partial_x h_{22}+\partial_xh_2h_{11}+\partial_xh_2h_{22}\big)
					\\
					&\kern1cm+\frac{2(1-x^2)}{m^2(1+x^2)^2}\big(-x\partial_xh_{22}h_{12}+\partial_xh_{22}h_{11}+2\partial_xh_{22}h_{22}\big)
					\\
					&\kern1cm+r\bigg\{-4\partial_rh h_{12}-2h\partial_rh_{12}-3h_2\partial_rh_1-3h_1 \partial_rh_2
					\\
					&\kern1cm+\frac{2}{m}\sqrt{\frac{1-x^2}{1+x^2}}\bigg[-\frac{\partial_rh_1xh_{12}}{1-x^2}+\frac{3\partial_rh_1h_{11}}{1+x^2}+\frac{3\partial_rh_1h_{22}}{1+x^2}-\frac{\partial_rh_2h_{12}}{1+x^2}
					\\
					&\kern1cm-\frac{(4-3x^2)x\partial_rh_2h_{11}}{1-x^4}-\frac{(4-3x^2)x\partial_rh_2h_{22}}{1-x^4}+\frac{h_1\partial_rh_{11}}{1+x^2}
					\\
					&\kern1cm+\frac{h_1\partial_rh_{22}}{1+x^2}-\frac{(1-2x^2)xh_2\partial_rh_{11}}{2(1-x^4)}-\frac{(3-2x^2)xh_2\partial_rh_{22}}{2(1-x^4)}\bigg]
					\\
					&\kern1cm+\frac{2}{m^2(1+x^2)^3}\big[-2(1-x^2)x^2h_{12}\partial_rh_{22}-2x(1+x^2)h_{11}\partial_rh_{11}
					\\
					&\kern1cm-x(1+x^2)h_{22}\partial_rh_{11}-2(1-x^2)h_{12}\partial_rh_{11}-(3-6x^2-x^4)h_{11}\partial_rh_{12}
					\\
					&\kern1cm-(3-6x^2-x^4)h_{22}\partial_rh_{12}+x(5-3x^2)h_{11}\partial_rh_{22}
					\\
					&\kern1cm+x(10-6x^2)h_{22}\partial_rh_{22}\big]+\frac{1}{m}\sqrt{\frac{1-x^2}{1+x^2}}\big(-\partial_rh_2 \partial_xh_{22}-\frac12\partial_x h_2\partial_rh_{11}
					\\
					&\kern1cm+\frac12\partial_xh_2\partial_rh_{22}-\partial_x\partial_rh_2h_{11}-\partial_x\partial_rh_2h_{22}\big)\bigg\}
					\\
					&\kern1cm+r^2\bigg[-\partial_r^2hh_{12}-h\partial_r^2h_{12}-\partial_rh\partial_rh_{12}-h_2\partial_r^2h_1-h_1\partial_r^2h_2-\partial_rh_1\partial_rh_2
					\\
					&\kern1cm+\frac{1}{m(1+x^2)}\sqrt{\frac{1-x^2}{1+x^2}}\big(2\partial_r^2h_1h_{11}+2\partial_r^2h_1h_{22}-2x\partial_r^2h_2h_{11}-2x\partial_r^2h_2h_{22}
					\\
					&\kern1cm-x\partial_rh_2\partial_rh_{11}-x\partial_rh_2\partial_rh_{22}+\partial_rh_1\partial_rh_{11}+\partial_rh_1\partial_rh_{22}\big)
					\\
					&\kern1cm-\frac{3-6x^2-x^4}{2m^2(1+x^2)^3}\big(2h_{11}\partial_r^2h_{12}+2h_{22}\partial_r^2h_{12}+\partial_rh_{11}\partial_rh_{12}+\partial_rh_{12}\partial_rh_{22}\big)\bigg]
				\end{aligned}
			\end{equation}
			and
			\begin{equation}
				\begin{aligned}
					\mu_2(\Theta,\Theta)_{22}\ &=\ -2hh_{22}+\frac32h_1h_1+\frac12h_2h_2
					\\
					&\kern1cm+\frac{2}{m}\sqrt{\frac{1-x^2}{1+x^2}}\bigg(\frac{3xh_1h_{11}}{1+x^2}+\frac{3xh_1h_{22}}{1+x^2}-\frac{h_1h_{12}}{1+x^2}-\frac{xh_2h_{12}}{1-x^2}-\frac{2h_2h_{22}}{1+x^2}\bigg)
					\\
					&\kern1cm+\frac{2}{m^2(1+x^2)^3}\big[2x(1+x^2)h_{11}h_{12}+4x(1+x^2)h_{12}h_{22}+2(1-x^2)h_{22}h_{22}
					\\
					&\kern1cm+(2-5x^2-x^4)h_{11}h_{11}+(1-4x^2-x^4)h_{12}h_{12}+(2-5x^2-x^4)h_{11}h_{22}\big]
					\\
					&\kern1cm+\frac{1}{m}\sqrt{\frac{1-x^2}{1+x^2}}\big(h_1\partial_x h_{11}+2h_2\partial_xh_{12}+2\partial_xh_1h_{11}+2\partial_xh_1h_{22}\big)
					\\
					&\kern1cm+\frac{2(1-x^2)}{m^2(1+x^2)^2}\big(2x\partial_xh_{11}h_{11}+x\partial_xh_{11}h_{22}+2x\partial_xh_{12}h_{12}-\partial_xh_{11}h_{12}
					\\
					&\kern1cm-2\partial_xh_{12}h_{11}-4\partial_xh_{12}h_{22}\big)
					\\
					&\kern1cm+r\bigg\{-4\partial_rhh_{22}+2h\partial_rh_{11}+7h_1\partial_rh_1+h_2\partial_rh_2
					\\
					&\kern1cm+\frac{2}{m}\sqrt{\frac{1-x^2}{1+x^2}}\bigg[\frac{7x\partial_rh_1h_{11}}{1+x^2}-\frac{x\partial_rh_2h_{12}}{1-x^2}+\frac{7x\partial_rh_1h_{22}}{1+x^2}-\frac{\partial_rh_1h_{12}}{1+x^2}
					\\
					&\kern1cm-\frac{2\partial_rh_2h_{22}}{1+x^2}+\frac{2xh_1\partial_rh_{11}}{1+x^2}+\frac{2h_2\partial_rh_{11}}{1+x^2}-\frac{2h_1\partial_rh_{12}}{1+x^2}+\frac{x(3-2x^2)h_2\partial_rh_{12}}{1-x^4}\bigg]
					\\
					&\kern1cm+\frac{2}{m^2(1+x^2)^3}\big[4x^2(1-x^2)h_{12}\partial_rh_{12}-2x(3-x^2)h_{12}\partial_rh_{11}
					\\
					&\kern1cm-2x(5-3x^2)h_{11}\partial_rh_{12}-4x(5-3x^2)h_{22}\partial_rh_{12}+(1-5x^4)h_{11}\partial_rh_{11}
					\\
					&\kern1cm-(1+3x^4)h_{22}\partial_rh_{11}\big]+\frac{1}{m}\sqrt{\frac{1-x^2}{1+x^2}}\big(\partial_rh_1\partial_xh_{11}+2\partial_rh_2\partial_xh_{12}
					\\
					&\kern1cm-\partial_xh_2\partial_rh_{12}+2\partial_x\partial_rh_1h_{11}+2\partial_x\partial_rh_1h_{22}\big)\bigg\}
					\\
					&\kern1cm+r^2\bigg[-\partial_r^2hh_{22}+h\partial_r^2h_{11}+\partial_rh\partial_rh_{11}+2h_1\partial_r^2h_1+\frac32\partial_rh_1\partial_rh_1
					\\
					&\kern1cm+\frac12\partial_rh_2\partial_rh_2+\frac{2}{m(1+x^2)}\sqrt{\frac{1-x^2}{1+x^2}}\big(2x\partial_r^2h_1h_{11}+2x\partial_r^2h_1h_{22}
					\\
					&\kern1cm+x\partial_rh_1\partial_rh_{11}-\partial_rh_1\partial_rh_{12}+\partial_rh_2\partial_rh_{11}+x\partial_rh_2\partial_rh_{12}\big)
					\\
					&\kern1cm-\frac{3-6x^2-x^4}{2m^2(1+x^2)^3}\big(-2h_{11}\partial_r^2h_{11}-2h_{22}\partial_r^2h_{11}-\partial_rh_{11}^2-\partial_rh_{12}^2\big)\bigg]\,.
				\end{aligned}
			\end{equation}
		\end{subequations}

		\paragraph{Next-to-lowest-order solutions.}
		The only non-straightforward part of constructing the next-to-lowest-order solutions is to solve the minimal model Maurer--Cartan equation~\eqref{eq:MinimalMCEquationOfNHDeformationNextToLowestOrder}, that is,
		\begin{equation}
			\mu_2(\Theta^{\circ\,(0)},\Theta^{\circ\,(0)})\ \in\ \im(\mu_1)~.
		\end{equation}
		Put differently, we need to determine all those $\Theta^{\circ\,(0)}\in H^1(V)$ for which there is a $\Phi\in V_1$ with
		\begin{equation}\label{eq:nonStraightForwardEquation}
			\mu_1(\Phi)\ =\ \rho
			\ewith
			\rho\ \coloneqq\ \mu_2(\Theta^{\circ\,(0)},\Theta^{\circ\,(0)})~.
		\end{equation}
		Once we have found those $\Theta^{\circ\,(0)}\in H^1(V)$, we can then use~\eqref{eq:SolutionOfNHDeformationNextToLowestOrder} to write down the most general next-to-lowest-order solution. We now claim that~\eqref{eq:nonStraightForwardEquation} always holds, that is, for all $\Theta^{\circ\,(0)}\in H^1(V)$ there is such a $\Phi\in V_1$.

		To see this, recall that when constructing the Green function in the extremal Kerr horizon setting in \cref{sec:extremalKerrFirstOrder}, we found that if $\rho\in\im(\mu_1)$ with $\rho|_{r=0}=0$, then the Green function given in~\eqref{eq:fullGreenFunction} and~\eqref{eq:GreenFunction} with $\bar{\sfg}_{ab}{}^{cd}$ as in~\eqref{eq:spatialGreenFunctionExtremalKerrN=1} and~\eqref{eq:spatialGreenFunctionExtremalKerrN>1} always yields a smooth solution $\Phi$ to $\mu_1(\Phi)=\rho$ that vanishes at $r=0$ by means of
		\begin{equation}\label{eq:generalSmoothSolutionViaGreen}
			\Phi(r,y)\ =\ \int\rmd r'\int\rmd^{2}y'\,\sqrt{\det(\mathring\gamma(y'))}\sfg(r,y;r',y')\rho(r',y')~.
		\end{equation}
		Conversely, if~\eqref{eq:generalSmoothSolutionViaGreen} is smooth and vanishes at $r=0$, then $\rho\in\im(\mu_1)$ with $\rho|_{r=0}=0$.	Moreover, as also shown in \cref{sec:extremalKerrFirstOrder} around~\eqref{eq:criterionForSmoothSolutions}, this, in turn, is equivalent to saying that $\rho\in\im(\mu_1)$ with $\rho|_{r=0}=0$ if and only if
		\begin{subequations}
			\begin{equation}\label{eq:reducedMinimalMCofKerr}
				0\ =\ \lim_{x\to1}\int\rmd^2y'\,\sqrt{\det(\mathring\gamma(y'))}\bar{\sfg}^{(n)}_{ab}{}^{cd}(y;y')\varrho_{ab}^{(n)}(y)~
			\end{equation}
			for all $n\geq 2$, where
			\begin{equation}\label{eq:defRExpansionOftmu2}
				\varrho_{ab}(r,y)\ \coloneqq\ \int\rmd r'\int\rmd^2y'\,\sqrt{\det(\mathring\gamma(y'))}\big(\sft(r,y;r',y')\rho(r',y')\big)_{ab}\ =\ \sum_{n>0}\frac{r^n}{n!}\varrho_{ab}^{(n)}(y)
			\end{equation}
		\end{subequations}
		and $\sft$ as given in~\eqref{eq:operatorT}.

		Next, we use the criterion~\eqref{eq:reducedMinimalMCofKerr} with~\eqref{eq:defRExpansionOftmu2} for $\rho=\mu_2(\Theta^{\circ\,(0)},\Theta^{\circ\,(0)})$ and the general solution~\eqref{eq:generalLowestOrderSolutionExtremalKerr} for $\Theta^{\circ\,(0)}\in H^1(V)$ to check when~\eqref{eq:nonStraightForwardEquation} holds. Indeed, a lengthy calculation now shows that there are no constraints on the adminisable $\Theta^{\circ\,(0)}$, that is, for any solution~\eqref{eq:generalLowestOrderSolutionExtremalKerr}, the equation~\eqref{eq:reducedMinimalMCofKerr} holds. Consequently,~\eqref{eq:nonStraightForwardEquation} is always satisfied. See \cref{app:proofMinmimalMC} for details.

		Finally, upon combining~\eqref{eq:SolutionOfNHDeformationNextToLowestOrder} with~\eqref{eq:mu2Kerr}, and~\eqref{eq:actionOfPropagatorDeformation}, we arrive at the following solutions up to third order in $r$:
		\begin{subequations}\label{eq:NextToLowestOrderSolutionr^3}
			\begin{equation}
				\begin{aligned}
					h_1\ &=\ \frac{Arx(59-55x^2-23x^4-5x^6)}{10m^2(1+x^2)^3}\sqrt{\frac{1-x^2}{1+x^2}}
					\\
					&\kern1cm+\frac{r^2}{m^3(1+x^2)^4}\sqrt{\frac{1-x^2}{1+x^2}}\big[K_1^{(2)}x(9-8 x^2-x^4)+K_3^{(2)}(3-12 x^2+x^4)\big]
					\\
					&\kern1cm+\frac{5r^3}{8m^4(1+x^2)^5}\sqrt{\frac{1-x^2}{1+x^2}}\big[2 K_3^{(3)}x(5-14x^2+x^4)
					\\
					&\kern1cm-K_1^{(3)}(1-23x^2+19x^4+3x^6)\big]
					\\
					&\kern1cm-\frac{\kappa A^2r^2}{450m^3(1+x^2)^5}\sqrt{\frac{1-x^2}{1+x^2}}(138+10097x-414x^2-14848x^3-506x^4
					\\
					&\kern1cm-780x^5+46x^6+1602x^7+531x^9+54x^{11})
					\\
					&\kern1cm+\frac{\kappa r^3}{60m^4(1+x^2)^6}\sqrt{\frac{1-x^2}{1+x^2}}\big[AK_1^{(2)}(1-x^2)(70-1697x-1470x^2
					\\
					&\kern1cm+4758x^3-1750x^4+842x^5-210x^6+162x^7+15x^9)-AK_3^{(2)}(367+700x
					\\
					&\kern1cm-4342x^2-1260x^3+6048x^4-1820x^5-994x^6+140x^7-255x^8
					\\
					&\kern1cm-24x^{10})\big]+\caO(r^4,\kappa^3)
				\end{aligned}
			\end{equation}
			and
			\begin{equation}
				\begin{aligned}
					h_2\ &=\ -\frac{Ar(7-45x^2-3x^4+x^6)}{5m^2(1+x^2)^3}\sqrt{\frac{1-x^2}{1+x^2}}
					\\
					&\kern1cm-\frac{r^2}{m^3(1+x^2)^4}\sqrt{\frac{1-x^2}{1+x^2}}\big[K_1^{(2)}(3-12x^2+x^4)-K_3^{(2)}x(9-8x^2-x^4)\big]
					\\
					&\kern1cm-\frac{5r^3}{8m^4(1+x^2)^5}\sqrt{\frac{1-x^2}{1+x^2}}\big[2K_1^{(3)}x(5-14x^2+x^4)
					\\
					&\kern1cm+K_3^{(3)}(1-23x^2+19x^4+3x^6)\big]
					\\
					&\kern1cm+\frac{\kappa A^2r^2}{900m^3(1+x^2)^5}\sqrt{\frac{1-x^2}{1+x^2}}(5480-828x-32517x^2-92x^3+11179x^4
					\\
					&\kern1cm+828x^5+1122x^6+92x^7-2046x^8-45x^{10}+27x^{12})
					\\
					&\kern1cm+\frac{\kappa r^3}{60m^4(1+x^2)^6}\sqrt{\frac{1-x^2}{1+x^2}}\big[AK_1^{(2)}(427+700x-4162x^2-1260x^3+6072x^4
					\\
					&\kern1cm-1820x^5-1210x^6+140x^7-339x^8+12x^{10})
					\\
					&\kern1cm+2AK_3^{(2)}(1-x)(1-x^2)(35-803x-1538x^2+769x^3-106x^4
					\\
					&\kern1cm+132x^5+27x^6)\big]+\caO(r^4,\kappa^3)
				\end{aligned}
			\end{equation}
			and
			\begin{equation}
				\begin{aligned}
					h\ &=\ \frac{2Ar(35-225x^2+135x^4+5x^6-6x^8)}{15m^3(1+x^2)^5}
					\\
					&\kern1cm+\frac{r^2}{2m^4(1+x^2)^6}\big[K_1^{(2)}(13-84x^2+66x^4+4x^6+x^8)
					\\
					&\kern1cm-4K_3^{(2)}x(12-23x^2+6x^4+x^6)\big]
					\\
					&\kern1cm+\frac{r^3}{4m^5(1+x^2)^7}\big[5K_1^{(3)}x(13-52x^2+34x^4+4x^6+x^8)
					\\
					&\kern1cm+K_3^{(3)}(9-168x^2+262x^4-56x^6-15x^8)\big]
					\\
					&\kern1cm+\frac{\kappa A^2r^2}{450m^4(1+x^2)^7}(-7325+1104x+61687x^2-1012x^3-60983x^4-1564x^5
					\\
					&\kern1cm+12031x^6+644x^7+609x^8+92x^9-3111x^{10}+147x^{12}+81x^{14})
					\\
					&\kern1cm+\frac{\kappa r^3}{30m^5(1+x^2)^8}\big[-AK_1^{(2)}(1-x^2)(609+910x-6480x^2-1820x^3+8644x^4
					\\
					&\kern1cm-3080x^5-802x^6-420x^7-389x^8-70x^9+18x^{10})
					\\
					&\kern1cm-AK_3^{(2)}(126-2741x-2226x^2+12137x^3+1316x^4-14108x^5+2884x^6
					\\
					&\kern1cm+4302x^7-994x^8+951x^9-210x^{10}-87x^{11}-6x^{13})\big]+\caO(r^4,\kappa^3)
				\end{aligned}
			\end{equation}
			and
			\begin{equation}
				\begin{aligned}
					h_{11}\ &=\ \frac{Ar(5-3x^2+13x^4+5x^6)}{5m(1+x^2)^2}-\frac{3r^2(1-x^2)}{2m^2(1+x^2)^3}\big[-K_1^{(2)}(1-x^2)+2K_3^{(2)}x\big]
					\\
					&\kern1cm-\frac{5r^3x(1-x^2)}{2m^3(1+x^2)^4}\big[-K_1^{(3)}(1-x^2)+2K_3^{(3)}x\big]
					\\
					&\kern1cm+\frac{\kappa A^2r^2}{300m^2(1+x^2)^4}(-675+92x+2752x^2-1851x^4-92x^5-268x^6+747x^8
					\\
					&\kern1cm+420x^{10}+75x^{12})
					\\
					&\kern1cm-\frac{\kappa r^3(1-x^2)}{30m^3(1+x^2)^5}\big[10AK_1^{(2)}(4+14x-60x^2+41x^4-14x^5+12x^6+3x^8)
					\\
					&\kern1cm+AK_3^{(2)}x(-269-280x+721x^2-280x^3+69x^4+39x^6)\big]+\caO(r^4,\kappa^3)
				\end{aligned}
			\end{equation}
			and
			\begin{equation}
				\begin{aligned}
					h_{22}\ &=\ \frac{2Arx^2(9+x^2)}{5m(1+x^2)^2}+\frac{3r^2(1-x^2)}{2m^2(1+x^2)^3}\big[-K_1^{(2)}(1-x^2)+2K_3^{(2)}x\big]
					\\
					&\kern1cm+\frac{5r^3x(1-x^2)}{2m^3(1+x^2)^4}\big[-K_1^{(3)}(1-x^2)+2K_3^{(3)}x\big]
					\\
					&\kern1cm+\frac{\kappa A^2r^2}{150m^2(1+x^2)^4}(375-46x-1178x^2+1188x^4+46x^5+338x^6
					\\
					&\kern1cm
					-111x^8-12x^{10})
					\\
					&\kern1cm-\frac{\kappa r^3(1-x^2)}{30m^3(1+x^2)^5}\big[-10AK_1^{(2)}(7+14x-54x^2+41x^4-14x^5+6x^6)
					\\
					&\kern1cm+AK_3^{(2)}x(221+280x-769x^2+280x^3-21x^4+9x^6)\big]+\caO(r^4,\kappa^3)
				\end{aligned}
			\end{equation}
			and
			\begin{equation}
				\begin{aligned}
					h_{12}\ &=\ \frac{Arx(1-x^2)(9+x^2)}{5m(1+x^2)^2}+\frac{3r^2(1-x^2)}{2m^2(1+x^2)^3}\big[K_3^{(2)}(1-x^2)+2K_1^{(2)}x\big]
					\\
					&\kern1cm+\frac{5r^3x(1-x^2)}{2m^3(1+x^2)^4}\big[K_3^{(3)}(1-x^2)+2K_1^{(3)}x\big]
					\\
					&\kern1cm-\frac{\kappa A^2r^2 (1-x^2)^2}{300m^2(1+x^2)^4}(46+1925x+46x^2+902x^3+135x^5+6x^7)
					\\
					&\kern1cm-\frac{\kappa r^3(1-x)(1-x^2)}{6m^3(1+x^2)^5}\big[AK_1^{(2)}x(49+105x-44x^2+12x^3+3x^4+3x^5)
					\\
					&\kern1cm+AK_3^{(2)}(1+x)(11+28x-103x^2+28x^3-21x^4-3x^6)\big]+\caO(r^4,\kappa^3)
				\end{aligned}
			\end{equation}
		\end{subequations}
		For order $r^n$, the terms are of the form $(\ldots)K_p^{(j)}K_q^{(n-j)}$ where $(\ldots)$ is a rational function of $x$, and $p$ and $q$ are $1$ or $3$.

		\paragraph{Comparing to extremal Kerr.}
		Since we have the next-to-lowest-order solution, we can compare this solution to the order $r^2$ of the extremal Kerr solution to fix $K_1^{(2)}$ and $K_3^{(2)}$. This can be done by computing~\eqref{eq:alphabetagammaExtreamlKerrGAugeFixed} to order $r^2$. One can show that $K_1^{(2)}$ and $K_3^{(2)}$ are indeed those we have already found in~\eqref{eq:valuesOfAK1K3}.

		\appendix
		\addappheadtotoc
		\appendixpage

		\appendices

		\section{Vielbein formalism}\label{app:Vielbein}

		The following summarises our conventions for the vielbein formalism.

		\paragraph{Setting.}
		Let $(M,g)$ be a $d$-dimensional semi-Riemannian manifold with local coordinates $x^I$ with $I,J,\ldots=1,\ldots,d$. Then, $g=\frac12g_{IJ}\rmd x^I\odot\rmd x^J$. We introduce the vielbeins $E_A=E_A{}^I\partial_I$ for $A,B,\ldots=1,\ldots,d$ and $\partial_I\coloneqq\parder{x^I}$. Dually, we have $e^A=\rmd x^Ie_I{}^A$ with $E_A\intprod e^B=\delta_A{}^B$. Hence, the metric can be written as $g=\frac12g_{AB}e^A\odot e^B$ with $g_{AB}\coloneqq E_A{}^IE_B{}^Jg_{IJ}$. Furthermore, the \uline{structure functions} $C_{AB}{}^C$ are given by $[E_A,E_B]=C_{AB}{}^CE_C$ or, dually, $\rmd e^A=\frac12e^C\wedge e^BC_{BC}{}^A$.

		\paragraph{Torsion and curvature two-forms.}
		The \uline{torsion} and \uline{curvature two-forms} are defined by the \uline{Cartan structure equations}
		\begin{equation}
			\begin{gathered}
				T^A\ \coloneqq\ \tfrac12T_{BC}{}^Ae^B\wedge e^C\ \coloneqq\ \rmd e^A-e^B\wedge\omega_B{}^A~,
				\\
				R_A{}^B\ \coloneqq\ \tfrac12R_{CDA}{}^Be^C\wedge e^D\ \coloneqq\ \rmd\omega_A{}^B-\omega_A{}^C\wedge\omega_C{}^B~,
			\end{gathered}
		\end{equation}
		where $\omega_A{}^B=e^C\omega_{CA}{}^B$ is the \uline{connection one-form}. The \uline{Ricci tensor}, denoted by $R_{AB}$, and \uline{curvature scalar}, denoted by $R$, are defined by
		\begin{equation}\label{eq:RicciTensorCurvatureScalar}
			\begin{gathered}
				R_{AB}\ \coloneqq\ R_{CAB}{}^C\ =\ E_C\omega_{AB}{}^C-E_A\omega_{CB}{}^C-\omega_{CB}{}^E\omega_{AE}{}^C+\omega_{AB}{}^E\omega_{CE}{}^C-C_{CA}{}^E\omega_{EB}{}^C~,
				\\
				R\ \coloneqq\ g^{AB}R_{AB}~.
			\end{gathered}
		\end{equation}

		\paragraph{Levi-Civita connection.}
		The \uline{Levi-Civita connection} is obtained by imposing the \uline{torsion freeness},
		\begin{subequations}
			\begin{equation}
				T^A\ =\ 0
				\quad\Leftrightarrow\quad
				\omega_{[AB]}{}^{C}\ =\ \tfrac12C_{AB}{}^{C}
			\end{equation}
			and the \uline{metric compatibility},
			\begin{equation}
				\omega_{A(B}{}^{D}g_{C)D}\ =\ \tfrac12E_Ag_{BC}
			\end{equation}
		\end{subequations}
		Therefore,
		\begin{equation}\label{eq:connection1formformula}
			\omega_{AB}{}^C\ =\ \tfrac12\big[g^{CD}(E_Ag_{BD}+E_Bg_{AD}-E_Dg_{AB})+C^C{}_{AB}+C^C{}_{BA}+C_{AB}{}^C\big]\,,
		\end{equation}
		where indices are raised and lowered using $g_{AB}$.

		\paragraph{Adapted frame basis.}
		Next, we summarise some details about the non-coordinate basis~\eqref{eq:defAdaptedFrameII}. In particular, we have
		\begin{equation}
			\mathring e^+\ =\ \rmd u~,
			\quad
			\mathring e^-\ =\ \rmd r+r\mathring\alpha_i\rmd y^i-\tfrac12r^2\mathring\beta\rmd u~,
			\eand
			\mathring e^a\ =\ \rmd y^i\mathring e_i{}^\wha~,
		\end{equation}
		and which we collectively denote by $\mathring e^A$. Dually, we have
		\begin{equation}\label{eq:defDualAdaptedFrameII}
			\mathring E_+\ =\ \partial_u+\tfrac12r^2\mathring\beta\partial_r~,
			\quad
			\mathring E_-\ =\ \partial_r~,
			\eand
			\mathring E_a\ =\ \underbrace{\mathring E_a{}^i\partial_i}_{\eqqcolon\,\mathring{\tilde E}_a}-r\mathring\alpha_a\partial_r~,
		\end{equation}
		with $\mathring E_a{}^i$ the inverse of $\mathring e_i{}^a$ and $\mathring\alpha_a\coloneqq\mathring E_a{}^i\alpha_i$. We shall denote these vector fields collectively by $\mathring E_A$. The non-vanishing structure functions, denoted by $\mathring{C}_{AB}{}^C$ and defined by $[\mathring E_A,\mathring E_B]=\mathring{C}_{AB}{}^C\mathring E_C$, are then given by
		\begin{subequations}
			\begin{equation}\label{eq:structureFunctionsAdaptedFrameII}
				\begin{gathered}
					\mathring{C}_{+-}{}^-\ =\ -r\mathring\beta~,
					\quad
					\mathring{C}_{+a}{}^-\ =\ -\tfrac12r^2\big(\mathring{\tilde E}_a\mathring\beta-\mathring\alpha_a\mathring\beta\big)~,
					\\
					\mathring{C}_{-a}{}^-\ =\ -\mathring\alpha_a~,
					\quad
					\mathring{C}_{ab}{}^-\ =\ -r\big(\mathring{\tilde E}_a\mathring\alpha_b-\mathring{\tilde E}_b\mathring\alpha_a-\mathring{\tilde C}_{ab}{}^c\mathring\alpha_c\big)~,
					\quad
					\mathring{C}_{ab}{}^c\ =\ \mathring{\tilde C}_{ab}{}^c
				\end{gathered}
			\end{equation}
			where $\mathring{\tilde E}_a$ was defined in~\eqref{eq:defDualAdaptedFrameII} with
			\begin{equation}\label{eq:crossSectionStructureFunctionsAdaptedFrameII}
				[\mathring{\tilde E}_a,\mathring{\tilde E}_b]\ =\ \mathring{\tilde C}_{ab}{}^c\mathring{\tilde E}_c~.
			\end{equation}
		\end{subequations}

		\section{Curvatures in the adapted frame basis}\label{app:RicciTensor}

		We now compute the Ricci tensor and the curvature scalar of the metric~\eqref{eq:metricInGaussianNullCoordinatesExtremal} in the basis~\eqref{eq:defAdaptedFrameII}. In particular, from~\eqref{eq:metricInGaussianNullCoordinatesExtremalAdaptedFrameII}, we have
		\begin{equation}\label{eq:metricComponentsInGaussianNullCoordinatesExtremalAdaptedFrameII}
			(g_{AB})\ =\
			\begin{pmatrix}
				-r^2(\beta-\mathring\beta) & 1 & r(\alpha_a-\mathring\alpha_a)
				\\
				1 & 0 & 0
				\\
				r(\alpha_a-\mathring\alpha_a) & 0 & \gamma_{ab}
			\end{pmatrix}.
		\end{equation}

		\paragraph{Connection one-form.}
		Furthermore, upon inserting these metric components and the structure functions~\eqref{eq:structureFunctionsAdaptedFrameII} into the formula~\eqref{eq:connection1formformula} for the Levi-Civita connection, we obtain
		\begin{equation}\label{eq:connectionOneFormAdaptedFrameII}
			\begin{aligned}
				\omega_{ab}{}^c\ &=\ \underbrace{\tfrac12\big[\gamma^{cd}(\mathring{\tilde E}_a\gamma_{bd}+\mathring{\tilde E}_b\gamma_{ad}-\mathring{\tilde E}_d\gamma_{ab})+\mathring{\tilde C}^c{}_{ab}+\mathring{\tilde C}^c{}_{ba}+\mathring{\tilde C}_{ab}{}^c\big]}_{\eqqcolon\,\tilde\omega_{ab}{}^c}
				\\
				&\kern1cm -\tfrac{r}{2}\gamma^{cd}(\mathring{\alpha}_{a}\partial_r\gamma_{bd}+\mathring{\alpha}_{b}\partial_r\gamma_{ad}-\alpha_{d}\partial_r\gamma_{ab})\,,
			\end{aligned}
		\end{equation}
		In the following, we denote by $\tilde\nabla_a$ the covariant derivative with respect to $\tilde\omega_{ab}{}^c$.

		\paragraph{Ricci tensor and curvature scalar.}
		Upon combining~\eqref{eq:structureFunctionsAdaptedFrameII} and~\eqref{eq:RicciTensorCurvatureScalar}, we obtain for the components of the Ricci tensor
		\begin{subequations}\label{eq:RicciTensorAdaptedFrameII}
			\begin{equation}
				\begin{aligned}
					R_{++}\ &=\ r^2\big(\beta\alpha_a\alpha^a-\tfrac32\alpha^a\tilde\nabla_a\beta+\tilde\nabla_{[a}\alpha_{b]}\tilde\nabla^a\alpha^b-\tfrac12\beta\tilde\nabla^a\alpha_a+\tfrac12\tilde\nabla^a\tilde\nabla_a\beta\big)
					\\
					&\kern1cm+r^3\big[\tfrac14\partial_r\gamma_{ac}\gamma^{ac}\big(\beta\alpha_b\alpha^b-\alpha_b\tilde\nabla^b\beta\big)-\tfrac12\partial_r\gamma_{ab}\big(\tfrac12\beta\alpha^a\alpha^b-\alpha^a\tilde\nabla^b\beta\big)+\tfrac12\beta\tilde\nabla^a\partial_r\alpha_a
					\\
					&\kern1cm-\tfrac12\partial_r\beta\tilde\nabla^a\alpha_a+2\partial_r\beta\alpha^a\alpha_a-\alpha^a\tilde\nabla_a\partial_r\beta+\tfrac12\partial_r\alpha_a\tilde\nabla^a\beta-\beta\partial_r\alpha_a\alpha^a
					\\
					&\kern1cm-2\partial_r\alpha_a\alpha_b\tilde\nabla^{[b}\alpha^{a]}\big]
					\\
					&\kern1cm+\tfrac12r^4\big[\tfrac12\partial_r\gamma_{ac}\gamma^{ac}\big(\partial_r\beta\alpha_b\alpha^b-\beta\partial_r\alpha_b\alpha^b\big)+\partial_r\gamma_{ab}\big(\beta\partial_r\alpha_c\alpha^b\gamma^{ac}-\partial_r\beta\alpha^a\alpha^b\big)
					\\
					&\kern1cm-\tfrac14\beta^2\partial_r^2\gamma_{ab}\gamma^{ab}-\beta\partial_r^2\alpha_a\alpha^a+\partial_r^2\beta\alpha^a\alpha_a+\partial_r\alpha_a\alpha^b\big(\partial_r\alpha_c\alpha_b\gamma^{ac}-\partial_r\alpha_b\alpha^a\big)
					\\
					&\kern1cm+\tfrac18\beta^2\partial_r\gamma_{ab}\partial_r\gamma_{cd}\gamma^{ac}\gamma^{bd}\big]
					\\
					&\kern1cm-r^2(\beta-\mathring\beta)\big\{-\beta+\tfrac12\tilde\nabla^a \alpha_a-\tfrac12\alpha^a\alpha_a
					\\
					&\kern1cm-r\big(2\partial_r\beta+\tfrac12\beta \partial_r\gamma_{ab}\gamma^{ab}+2\partial_r\alpha_a\alpha^a-\tfrac12\alpha^a\alpha^b \partial_r\gamma_{ab}+\tfrac14\alpha_a\alpha^a \partial_r\gamma_{bc}\gamma^{bc}-\tfrac12\tilde\nabla^a\partial_r\alpha_a\big)
					\\
					&\kern1cm-\tfrac12r^2\big[\partial_r^2{\beta}+\partial_r^2{\alpha}_a\alpha^a+\tfrac12\beta \partial_r^2\gamma_{ab}\gamma^{ab}+\tfrac12\big(\partial_r\beta+\partial_r\alpha_b\alpha^b\big)\partial_r\gamma_{ac}\gamma^{ac}
					\\
					&\kern1cm-\partial_r\gamma_{ab}\partial_r\alpha_c\alpha^b\gamma^{ac}+\partial_r\alpha_a \partial_r\alpha_b\gamma^{ab}-\tfrac14\beta \partial_r\gamma_{ab}\partial_r\gamma_{cd}\gamma^{ac}\gamma^{bd}\big]\big\}
					\\
					&\kern1cm+\tfrac14r^4(\beta-\mathring\beta)^2\big(-\tfrac12 \partial_r^2\gamma_{ab}\gamma^{ab}+\tfrac14\partial_r\gamma_{ab}\partial_r\gamma_{cd}\gamma^{ac}\gamma^{bd}\big)
				\end{aligned}
			\end{equation}
			and
			\begin{equation}
				\begin{aligned}
					R_{--}\ =\ -\tfrac12 \partial_r^2\gamma_{ab}\gamma^{ab}+\tfrac14\partial_r\gamma_{ab}\partial_r\gamma_{cd}\gamma^{ac}\gamma^{bd}
				\end{aligned}
			\end{equation}
			and
			\begin{equation}
				\begin{aligned}
					R_{+-}\ &=\ -\beta+\tfrac12\tilde\nabla^a \alpha_a-\tfrac12\alpha^a\alpha_a
					\\
					&\kern1cm-r\big(2\partial_r\beta+\tfrac12\beta \partial_r\gamma_{ab}\gamma^{ab}+2\partial_r\alpha_a\alpha^a-\tfrac12\alpha^a\alpha^b \partial_r\gamma_{ab}+\tfrac14\alpha_a\alpha^a \partial_r\gamma_{bc}\gamma^{bc}-\tfrac12\tilde\nabla^a\partial_r\alpha_a\big)
					\\
					&\kern1cm-\tfrac12r^2\big[\partial_r^2{\beta}+\partial_r^2{\alpha}_a\alpha^a+\tfrac12\beta \partial_r^2\gamma_{ab}\gamma^{ab}+\tfrac12\big(\partial_r\beta+\partial_r\alpha_b\alpha^b\big)\partial_r\gamma_{ac}\gamma^{ac}
					\\
					&\kern1cm-\partial_r\gamma_{ab}\partial_r\alpha_c\alpha^b\gamma^{ac}+\partial_r\alpha_a \partial_r\alpha_b\gamma^{ab}-\tfrac14\beta \partial_r\gamma_{ab}\partial_r\gamma_{cd}\gamma^{ac}\gamma^{bd}\big]
					\\
					&\kern1cm-\tfrac12r^2(\beta-\mathring\beta)\big(-\tfrac12 \partial_r^2\gamma_{ab}\gamma^{ab}+\tfrac14\partial_r\gamma_{ab}\partial_r\gamma_{cd}\gamma^{ac}\gamma^{bd}\big)
				\end{aligned}
			\end{equation}
			and
			\begin{equation}
				\begin{aligned}
					R_{a-}\ &=\ \partial_r\alpha_a-\tfrac12\alpha^b \partial_r\gamma_{ab}+\tfrac14\alpha_a \partial_r\gamma_{bc}\gamma^{bc}-\tfrac12\tilde\nabla_a \partial_r\gamma_{bc}\gamma^{bc}+\tfrac12\tilde\nabla^b \partial_r\gamma_{ab}
					\\
					&\kern1cm+\tfrac12r\big[\partial_r^2\alpha_a+\alpha_a\partial_r^2\gamma_{bc}\gamma^{bc}-\alpha^b\partial_r^2\gamma_{ab}-\partial_r\alpha_b \partial_r\gamma_{ac}\gamma^{bc}+\tfrac12\partial_r\alpha_a \partial_r\gamma_{bc}\gamma^{bc}
					\\
					&\kern1cm-\tfrac12\alpha_a \partial_r\gamma_{bc}\partial_r\gamma_{de}\gamma^{bd}\gamma^{ce}+\alpha^b\big(\partial_r\gamma_{bc}\partial_r\gamma_{ad}\gamma^{cd}-\partial_r\gamma_{cd}\gamma^{cd} \partial_r\gamma_{ab}\big)\big]
					\\
					&\kern1cm+r(\alpha_a-\mathring\alpha_a)\big(-\tfrac12 \partial_r^2\gamma_{bc}\gamma^{bc}+\tfrac14\partial_r\gamma_{bc}\partial_r\gamma_{de}\gamma^{bd}\gamma^{ce}\big)
				\end{aligned}
			\end{equation}
			and
			\begin{equation}
				\begin{aligned}
					R_{a+}\ &=\ r\big(\beta\alpha_a -\tilde\nabla_a\beta-2\alpha^b\tilde\nabla_{[a}\alpha_{b]}+\tilde\nabla^b\tilde\nabla_{[a}\alpha_{b]}\big)
					\\
					&\kern1cm+r^2\big[-\tfrac12\beta\partial_r\alpha_a+2\partial_r\beta\alpha_a-\tfrac12\tilde\nabla_a\partial_r\beta-\tfrac32\alpha^b(\partial_r\alpha_a\alpha_b-\partial_r\alpha_b \alpha_a)+\alpha^b\tilde\nabla_b \partial_r\alpha_a
					\\
					&\kern1cm-\tfrac12\tilde\nabla_a (\partial_r\alpha_b\alpha^b)-\tfrac12\alpha_a\tilde\nabla^b\partial_r\alpha_b+\tfrac12\partial_r\alpha_a\tilde\nabla^b\alpha_b +\big(-\tfrac34\beta\alpha^b+\tfrac12\tilde\nabla^b\beta\big)\partial_r\gamma_{ab}
					\\
					&\kern1cm+\tfrac14\big(\tfrac32\beta\alpha_a-\tilde\nabla_a\beta\big)\partial_r\gamma_{bc}\gamma^{bc} +\alpha_b\tilde\nabla^{[c}\alpha^{b]}\partial_r\gamma_{ac}+\alpha^b\tilde\nabla_{[a}\alpha_{c]}\partial_r\gamma_{bd}\gamma^{cd}
					\\
					&\kern1cm-\tfrac12\alpha^b\tilde\nabla_{[a}\alpha_{b]}\partial_r\gamma_{cd}\gamma^{cd}-\tfrac14\beta\tilde\nabla_a \partial_r\gamma_{bc}\gamma^{bc}+\tfrac14\beta\tilde\nabla^b \partial_r\gamma_{ab}\big]
					\\
					&\kern1cm+r^3\big[-\tfrac14\beta\partial_r^2\alpha_a+\tfrac12\partial_r^2\beta\alpha_a+\tfrac12\alpha_a\alpha^b\partial_r^2\alpha_b-\tfrac12\alpha^b\alpha_b\partial_r^2\alpha_a-\tfrac12\partial_r\alpha_a\partial_r\alpha_b\alpha^b
					\\
					&\kern1cm+\tfrac12\partial_r\alpha_b \partial_r\alpha_c\alpha_a\gamma^{bc}-\tfrac14\beta\alpha^b \partial_r^2\gamma_{ab}+\tfrac14\beta \alpha_a\partial_r^2\gamma_{bc}\gamma^{bc}+\big(\tfrac14\beta\partial_r\alpha_c\gamma^{bc}-\tfrac12\partial_r\beta\alpha^b\big)\partial_r\gamma_{ab}
					\\
					&\kern1cm+\big(-\tfrac18\beta\partial_r\alpha_a+\tfrac14\partial_r\beta\alpha_a \big)\partial_r\gamma_{bc}\gamma^{bc}+\tfrac12\big(\partial_r\alpha_a\alpha^b\alpha^c-\partial_r\alpha_d\alpha^b\alpha_a\gamma^{dc}\big)\partial_r\gamma_{bc}
					\\
					&\kern1cm+\tfrac12\big(\partial_r\alpha_d\alpha^b\alpha_b\gamma^{cd}-\partial_r\alpha_b\alpha^b\alpha^c \big)\partial_r\gamma_{ac}+\tfrac14\big(\partial_r\alpha_b\alpha^b\alpha_a-\partial_r\alpha_a\alpha^b\alpha_b\big)\partial_r\gamma_{cd} \gamma^{cd}
					\\
					&\kern1cm-\tfrac18\beta\alpha_a \partial_r\gamma_{bc}\partial_r\gamma_{de}\gamma^{bd}\gamma^{ce}+\tfrac14\beta\alpha^b \big(\partial_r\gamma_{bd}\partial_r\gamma_{ac}\gamma^{dc}-\tfrac12\partial_r\gamma_{ab} \partial_r\gamma_{cd}\gamma^{cd}\big)\big]
					\\
					&\kern1cm+r(\alpha_a-\mathring\alpha_a)\big\{-\beta+\tfrac12\tilde\nabla^b \alpha_b-\tfrac12\alpha^b\alpha_b
					\\
					&\kern1cm-r\big(2\partial_r\beta+\tfrac12\beta \partial_r\gamma_{bc}\gamma^{bc}+2\partial_r\alpha_b\alpha^b-\tfrac12\alpha^b\alpha^c \partial_r\gamma_{bc}+\tfrac14\alpha_b\alpha^b \partial_r\gamma_{cd}\gamma^{cd}-\tfrac12\tilde\nabla^b\partial_r\alpha_b\big)
					\\
					&\kern1cm-\tfrac12r^2\big[\partial_r^2{\beta}+\partial_r^2{\alpha}_b\alpha^b+\tfrac12\beta \partial_r^2\gamma_{bc}\gamma^{bc}+\tfrac12\big(\partial_r\beta+\partial_r\alpha_c\alpha^c\big)\partial_r\gamma_{bd}\gamma^{bd}
					\\
					&\kern1cm-\partial_r\gamma_{bc}\partial_r\alpha_d\alpha^c\gamma^{bd}+\partial_r\alpha_b \partial_r\alpha_c\gamma^{bc}-\tfrac14\beta \partial_r\gamma_{bc}\partial_r\gamma_{de}\gamma^{bd}\gamma^{ce}\big]\big\}
					\\
					&\kern1cm -\tfrac12r^2(\beta-\mathring\beta)\big\{\partial_r\alpha_a-\tfrac12\alpha^b \partial_r\gamma_{ab}+\tfrac14\alpha_a \partial_r\gamma_{bc}\gamma^{bc}-\tfrac12\tilde\nabla_a \partial_r\gamma_{bc}\gamma^{bc}+\tfrac12\tilde\nabla^b \partial_r\gamma_{ab}
					\\
					&\kern1cm+\tfrac12r\big[\partial_r^2\alpha_a+\alpha_a\partial_r^2\gamma_{bc}\gamma^{bc}-\alpha^b\partial_r^2\gamma_{ab}-\partial_r\alpha_b \partial_r\gamma_{ac}\gamma^{bc}+\tfrac12\partial_r\alpha_a \partial_r\gamma_{bc}\gamma^{bc}
					\\
					&\kern1cm-\tfrac12\alpha_a \partial_r\gamma_{bc}\partial_r\gamma_{de}\gamma^{bd}\gamma^{ce}+\alpha^b\big(\partial_r\gamma_{bc}\partial_r\gamma_{ad}\gamma^{cd}-\partial_r\gamma_{cd}\gamma^{cd} \partial_r\gamma_{ab}\big)\big]\big\}
					\\
					&\kern1cm -\tfrac12r^3(\beta-\mathring\beta)(\alpha_a-\mathring\alpha_a)\big(-\tfrac12 \partial_r^2\gamma_{bc}\gamma^{bc}+\tfrac14\partial_r\gamma_{bc}\partial_r\gamma_{de}\gamma^{bd}\gamma^{ce}\big)
				\end{aligned}
			\end{equation}
			and
			\begin{equation}
				\begin{aligned}
					R_{ab}\ &=\ \tilde R_{ab}+\tilde\nabla_{(a}\alpha_{b)}-\tfrac12\alpha_a\alpha_b
					\\
					&\kern1cm+r\big[\tilde\nabla_{(a}\partial_r\alpha_{b)}-\tfrac32(\partial_r\alpha_a\alpha_b+\partial_r\alpha_b\alpha_a)+\big(\alpha^c\alpha_b-\tfrac12\tilde\nabla^c\alpha_b\big)\partial_r\gamma_{ac}
					\\
					&\kern1cm+\big(\alpha^c\alpha_a-\tfrac12\tilde\nabla^c\alpha_a\big)\partial_r\gamma_{bc}+\big(-\alpha^c\alpha_c -\beta+\tfrac12\tilde\nabla^c\alpha_c \big)\partial_r\gamma_{ab}
					\\
					&\kern1cm+\tfrac12\big(-\alpha_a\alpha_b+\tilde\nabla_{(a}\alpha_{b)}\big)\partial_r\gamma_{cd}\gamma^{cd}-\alpha_{(a|}\tilde\nabla^c \partial_r\gamma_{c|b)}-\alpha^c\tilde\nabla_{(a|}\partial_r\gamma_{c|b)}
					\\
					&\kern1cm+\alpha^c\tilde\nabla_c \partial_r\gamma_{ab}+\alpha_{(a}\tilde\nabla_{b)}\partial_r\gamma_{cd}\gamma^{cd}\big]
					\\
					&\kern1cm+r^2\big[\alpha_{(a|}\alpha^c\partial_r^2\gamma_{|b)c}-\partial_r^2\alpha_{(a}\alpha_{b)}-\tfrac12\alpha_a\alpha_b\partial_r^2\gamma_{cd}\gamma^{cd}-\tfrac12\big(\alpha_c\alpha^c+\beta\big)\partial_r^2\gamma_{ab}
					\\
					&\kern1cm-\tfrac12\partial_r\alpha_a \partial_r\alpha_b-\big(\tfrac12\partial_r\beta+\partial_r\alpha_c\alpha^c\big)\partial_r\gamma_{ab}+\partial_r\alpha_{(c}\alpha_{b)}\partial_r\gamma_{ad}\gamma^{cd}+\partial_r\alpha_{(c}\alpha_{a)}\partial_r\gamma_{bd}\gamma^{dc}
					\\
					&\kern1cm-\tfrac12\partial_r\alpha_{(a}\alpha_{b)}\partial_r\gamma_{cd}\gamma^{cd}-\alpha_{(a|}\alpha^d \partial_r\gamma_{|b)c}\partial_r\gamma_{e d}\gamma^{ce}+\tfrac12\alpha_{(a|}\alpha^c \partial_r\gamma_{|b)c} \partial_r\gamma_{de}\gamma^{de}
					\\
					&\kern1cm-\tfrac12\alpha^c\alpha^d \partial_r\gamma_{ac} \partial_r\gamma_{bd}+\tfrac12\big(\alpha_d\alpha^d+\beta\big)\partial_r\gamma_{ac}\partial_r\gamma_{be}\gamma^{ce}+\tfrac14\alpha_a\alpha_b \partial_r\gamma_{cd}\partial_r\gamma_{ef}\gamma^{ce}\gamma^{df}
					\\
					&\kern1cm+\tfrac12\alpha^c\alpha^d \partial_r\gamma_{cd}\partial_r\gamma_{ab}-\tfrac14\big(\alpha_c\alpha^c+\beta\big)\partial_r\gamma_{ab}\partial_r\gamma_{de}\gamma^{de} \big]
					\\
					&\kern1cm +2r(\alpha_{(b|}-\mathring\alpha_{(b|})\big\{\partial_r\alpha_{|a)}+\tfrac12\big(\tilde\nabla^c-\alpha^c\big) \partial_r\gamma_{|a)c}+\tfrac14\alpha_{|a)} \partial_r\gamma_{cd}\gamma^{cd}-\tfrac12\tilde\nabla_{|a)} \partial_r\gamma_{cd}\gamma^{cd}
					\\
					&\kern1cm+\tfrac12r\big[\partial_r^2\alpha_{|a)}+\alpha_{|a)}\partial_r^2\gamma_{cd}\gamma^{cd}-\alpha^c\partial_r^2\gamma_{|a)c}-\partial_r\alpha_c \partial_r\gamma_{|a)d}\gamma^{cd}+\tfrac12\partial_r\alpha_{|a)} \partial_r\gamma_{cd}\gamma^{cd}
					\\
					&\kern1cm-\tfrac12\alpha_{|a)} \partial_r\gamma_{cd}\partial_r\gamma_{ef}\gamma^{ce}\gamma^{df}+\alpha^d\big(\partial_r\gamma_{ed}\partial_r\gamma_{|a)c}\gamma^{ce}-\partial_r\gamma_{ce}\gamma^{ce} \partial_r\gamma_{|a)d}\big)\big]\big\}
					\\
					&\kern1cm+r^2(\alpha_a-\mathring\alpha_a)(\alpha_b-\mathring\alpha_b)\big(-\tfrac12 \partial_r^2\gamma_{cd}\gamma^{cd}+\tfrac14\partial_r\gamma_{cd}\partial_r\gamma_{ef}\gamma^{ce}\gamma^{df}\big)\,,
				\end{aligned}
			\end{equation}
		\end{subequations}
		where the indices are raised by $\gamma^{ab}$. Finally, the curvature scalar is given by
		\begin{equation}
			\begin{aligned}
				R\ &=\ \tilde R-2\beta+2\tilde\nabla^a \alpha_a-\tfrac32\alpha^a\alpha_a
				\\
				&\kern1cm r\big[-4\partial_r\beta-2\beta \partial_r\gamma_{ab}\gamma^{ab}-7\partial_r\alpha_a\alpha^a+3\alpha^a\alpha^b \partial_r\gamma_{ab}-2\alpha_a\alpha^a \partial_r\gamma_{bc}\gamma^{bc}+2\tilde\nabla^a\partial_r\alpha_a
				\\
				&\kern1cm-\tilde\nabla^b\alpha^a\partial_r\gamma_{ab}+\tilde\nabla^b\alpha_b \partial_r\gamma_{ac}\gamma^{ac}-2\alpha^a\tilde\nabla^b\partial_r\gamma_{ab}+2\alpha_b \tilde\nabla^b\partial_r\gamma_{ac}\gamma^{ac}\big]
				\\
				&\kern1cm+r^2\big[-\partial_r^2\beta-2\partial_r^2{\alpha}_a\alpha^a-\tfrac32\partial_r\alpha_a \partial_r\alpha_b\gamma^{ab}-\beta \partial_r^2\gamma_{ab}\gamma^{ab}+\alpha^a\alpha^b\partial_r^2\gamma_{ab}-\alpha_a\alpha^a\partial_r^2\gamma_{bc}\gamma^{bc}
				\\
				&\kern1cm-\partial_r\beta\partial_r\gamma_{ab}\gamma^{ab}-2\partial_r\alpha_b\alpha^b\partial_r\gamma_{ac}\gamma^{ac}+3\partial_r\alpha_c\alpha^b\partial_r\gamma_{ab}\gamma^{ca}+\tfrac34\beta \partial_r\gamma_{ab}\partial_r\gamma_{cd}\gamma^{ac}\gamma^{bd}
				\\
				&\kern1cm-\tfrac32\alpha^a\alpha^c \partial_r\gamma_{ad}\partial_r\gamma_{bc}\gamma^{bd}+\alpha^a\alpha^b \partial_r\gamma_{ab} \partial_r\gamma_{cd}\gamma^{cd}+\tfrac34\alpha_a\alpha^a\partial_r\gamma_{bc}\partial_r\gamma_{de}\gamma^{bd}\gamma^{ce}
				\\
				&\kern1cm-\tfrac14\big(\alpha_b\alpha^b+\beta\big)\partial_r\gamma_{ac}\gamma^{ac}\partial_r\gamma_{de}\gamma^{de}\big]\,.
			\end{aligned}
		\end{equation}

		\section{Perturbative expansion of the Einstein equation}\label{app:perturbativeExpansionEinsteinEquation}

		Consider the components~\eqref{eq:metricComponentsInGaussianNullCoordinatesExtremalAdaptedFrameII}. In light of~\eqref{eq:deformedMetricAdaptedFrameII}, we write
		\begin{subequations}\label{eq:deformedMetricAdaptedBasisII}
			\begin{equation}
				g_{AB}\ =\ \mathring g_{AB}+\kappa h_{AB}
			\end{equation}
			with
			\begin{equation}
				(\mathring g_{AB})\ \coloneqq\
				\begin{pmatrix}
					0 & 1 & 0
					\\
					1 & 0 & 0
					\\
					0 & 0 & \delta_{ab}
				\end{pmatrix}
				\eand
				(h_{AB})\ \coloneqq\
				\begin{pmatrix}
					-r^2h & 0 & rh_a
					\\
					0 & 0 & 0
					\\
					rh_a & 0 & h_{ab}
				\end{pmatrix}.
			\end{equation}
		\end{subequations}
		We shall now derive the Ricci tensor and scalar curvature as a series expansion in $\kappa$ by expanding the components~\eqref{eq:RicciTensorAdaptedFrameII} in powers of $\kappa$, and then also the Einstein equation.

		\subsection{Lowest-order Einstein equation}

		\paragraph{Connection one-form.}
		We have a series expansion of the connection one-form in $\kappa$, that is,
		\begin{equation}\label{eq:lowestOrderConnectionOneForm}
			\omega_{AB}{}^C\ =\ \mathring\omega_{AB}{}^C+\kappa\omega^{(1)}_{AB}{}^C+\caO(\kappa^2)~,
		\end{equation}
		where $\mathring\omega_{AB}{}^C$ is the connection one-form for the undeformed metric $\mathring g_{AB}$. Explicitly, combining the structure functions~\eqref{eq:structureFunctionsAdaptedFrameII} with the formula~\eqref{eq:connection1formformula} for the Levi-Civita connection, we obtain
		\begin{equation}\label{eq:backgroundConnectionOneFormAdaptedFrameII}
			\begin{gathered}
				\mathring\omega_{+-}{}^-\ =\ -r\mathring\beta\ =\ -\mathring\omega_{++}{}^+~,
				\\
				\mathring\omega_{a-}{}^-\ =\ \tfrac12\mathring\alpha_a\ =\ -\mathring\omega_{a+}{}^+\ =\ \delta_{ab}\,\mathring\omega_{-+}{}^b\ =\ -\mathring\omega_{-a}{}^-\ =\ \delta_{ab}\,\mathring\omega_{+-}{}^b\ =\ -\mathring\omega_{+a}{}^+~,
				\\
				\delta_{ab}\mathring\omega_{++}{}^b\ =\ -\tfrac12r^2\big(\mathring\beta\mathring\alpha_a-\mathring{\tilde E}_a\mathring\beta\big)\ =\ -\mathring\omega_{+a}{}^-~,
				\\
				\mathring\omega_{ab}{}^-\ =\ -\tfrac12r\big(\mathring{\tilde E}_a\mathring\alpha_b-\mathring{\tilde E}_b\mathring\alpha_a-\mathring{\tilde C}_{ab}{}^c\mathring\alpha_c\big)\ =\ -\mathring\omega_{a+}{}^c\delta_{bc}\ =\ -\mathring\omega_{+a}{}^c\delta_{bc}~,
				\\
				\mathring\omega_{ab}{}^c\ =\ \underbrace{\tfrac12\big(\mathring{\tilde C}^c{}_{ab}+\mathring{\tilde C}^c{}_{ba}+\mathring{\tilde C}_{ab}{}^c\big)}_{\eqqcolon\,\mathring{\tilde\omega}_{ab}{}^c}
			\end{gathered}
		\end{equation}
		for non-vanishing components for $\mathring\omega_{AB}{}^C$. In the following, we denote by $\mathring\nabla_A$ the covariant derivative with respect to $\mathring\omega_{AB}{}^C$ and by $\mathring{\tilde\nabla}_a$ the covariant derivative with respect to $\mathring{\tilde\omega}_{ab}{}^c$, respectively.

		Next, upon imposing the metric compatibility and the torsion-freeness on $\omega_{AB}{}^C$, it is a straightforward exercise to show that
		\begin{equation}\label{eq:lowestOrderConnectionOneFormFormula}
			\omega^{(1)}_{AB}{}^C\ =\ \tfrac12\mathring g^{CD}\big(\mathring\nabla_Ah_{BC}+\mathring\nabla_Bh_{AC}-\mathring\nabla_Ch_{AB}\big)\ =\ \mathring\nabla_{(A}h_{B)}{}^C-\tfrac12\mathring\nabla^Ch_{AB}
		\end{equation}
		from~\eqref{eq:lowestOrderConnectionOneForm}.

		\paragraph{Ricci tensor and curvature scalar.}
		Likewise, we have a series expansion of the Ricci tensor and the curvature scalar in $\kappa$,
		\begin{equation}\label{eq:lowestOrderRicciTensorCurvatureScalar}
			R_{AB}\ =\ \mathring R_{AB}+\kappa R^{(1)}_{AB}+\caO(\kappa^2)
			\eand
			R\ =\ \mathring R+\kappa R^{(1)}+\caO(\kappa^2)~,
		\end{equation}
		where $\mathring R_{AB}$ and $\mathring R$ are the Ricci tensor and the curvature scalar of $\mathring\omega_{AB}{}^C$, respectively. Explicitly, combining~\eqref{eq:structureFunctionsAdaptedFrameII} and~\eqref{eq:RicciTensorCurvatureScalar}, we obtain
		\begin{subequations}
			\begin{equation}
				\begin{aligned}
					\mathring R_{++}\ &=\ r^2\big(\mathring\beta\mathring\alpha_a\mathring\alpha^a-\tfrac32\mathring\alpha^a\mathring{\tilde\nabla}_a\mathring\beta+\mathring{\tilde\nabla}_{[a}\mathring\alpha_{b]}\mathring{\tilde\nabla}^a\mathring\alpha^b-\tfrac12\mathring\beta\mathring{\tilde\nabla}^a\mathring\alpha_a+\tfrac12\mathring{\tilde\nabla}^a\mathring{\tilde\nabla}_a\mathring\beta\big)\,,
					\\
					\mathring R_{+-}\ &=\ -\mathring\beta+\tfrac12\mathring{\tilde\nabla}^a\mathring\alpha_a-\tfrac12\mathring\alpha^a\mathring\alpha_a~,
					\\
					\mathring R_{a+}\ &=\ r\big(\mathring\beta\mathring\alpha_a-\mathring{\tilde\nabla}_a\mathring\beta-2\mathring\alpha^b\mathring{\tilde\nabla}_{[a}\mathring\alpha_{b]}+\mathring{\tilde\nabla}^b\mathring{\tilde\nabla}_{[a}\mathring\alpha_{b]}\big)\,,
					\\
					\mathring R_{ab}\ &=\ \mathring{\tilde R}_{ab}+\mathring{\tilde\nabla}_{(a}\mathring\alpha_{b)}-\tfrac12\mathring\alpha_a\mathring\alpha_b
				\end{aligned}
			\end{equation}
			for the non-vanishing components of the Ricci tensor. From this, the curvature scalar is
			\begin{equation}
				\mathring R\ =\ \mathring{\tilde R}-2\mathring\beta+2\mathring{\tilde\nabla}^a\mathring\alpha_a-\tfrac32\mathring\alpha^a\mathring\alpha_a~.
			\end{equation}
		\end{subequations}

		Next, upon substituting~\eqref{eq:lowestOrderConnectionOneForm} into~\eqref{eq:RicciTensorCurvatureScalar} and making use of~\eqref{eq:lowestOrderConnectionOneFormFormula}, we thus find
		\begin{equation}\label{eq:lowestOrderRicciTensorCurvatureScalarFormula}
			\begin{gathered}
				R^{(1)}_{AB}\ =\ \mathring\nabla_C\omega^{(1)}_{AB}{}^C-\mathring\nabla_A\omega^{(1)}_{CB}{}^C\ =\ \mathring\nabla_C\mathring\nabla_{(A}h_{B)}{}^C-\tfrac12\mathring\nabla_C\mathring\nabla^Ch_{AB}-\tfrac12\mathring\nabla_A\mathring\nabla_Bh_C{}^C~,
				\\
				R^{(1)}\ =\ \mathring\nabla^A\mathring\nabla^Bh_{AB}-\mathring\nabla_A\mathring\nabla^Ah_B{}^B-\mathring R_{AB}h^{AB}
			\end{gathered}
		\end{equation}
		for the terms $R^{(1)}_{AB}$ and $R^{(1)}$ from~\eqref{eq:lowestOrderRicciTensorCurvatureScalar}. Explicitly, using $\mathring g_{AB}$ and $h_{AB}$ from~\eqref{eq:deformedMetricAdaptedBasisII} and the components~\eqref{eq:backgroundConnectionOneFormAdaptedFrameII}, we obtain
		\begin{subequations}
			\begin{equation}
				\begin{aligned}
					R^{(1)}_{++}\ &=\ r^2\big[h\mathring\beta-\mathring\beta\mathring\alpha_a\mathring\alpha_bh^{ab}+\tfrac32\mathring\alpha_a\mathring{\tilde\nabla}_b\mathring\beta h^{ab}-\mathring{\tilde\nabla}_{[a}\mathring\alpha_{b]}\mathring{\tilde\nabla}_c\mathring\alpha^bh^{ac}-\mathring{\tilde\nabla}_{[a} \mathring\alpha_{b]}\mathring{\tilde\nabla}^a\mathring\alpha_ch^{bc}
					\\
					&\kern1cm+\tfrac12\mathring\beta\mathring{\tilde\nabla}_a\mathring\alpha_bh^{ab}-\tfrac12\mathring{\tilde\nabla}_a\mathring{\tilde\nabla}_b\mathring\beta h^{ab}+\tfrac32h\mathring\alpha^a \mathring\alpha_a+2\mathring\beta h^a\mathring\alpha_a-\tfrac32h^a\mathring{\tilde\nabla}_a\mathring\beta-\tfrac32\mathring\alpha^a \mathring{\tilde\nabla}_ah
					\\
					&\kern1cm+2\mathring{\tilde\nabla}^{[a}h^{b]}\mathring{\tilde\nabla}_{[a}\mathring\alpha_{b]}-h\mathring{\tilde\nabla}^a\mathring\alpha_a-\tfrac12\mathring\beta\mathring{\tilde\nabla}^ah_a+\tfrac12\mathring\beta\mathring\alpha_b\mathring{\tilde\nabla}_ah^{ab}-\tfrac14\mathring\beta\mathring\alpha^b\mathring{\tilde\nabla}_bh_a{}^a
					\\
					&\kern1cm+\tfrac12\mathring{\tilde\nabla}_a\mathring{\tilde\nabla}^ah-\tfrac12\mathring{\tilde\nabla}_b\mathring\beta\mathring{\tilde\nabla}_ah^{ab}+\tfrac14\mathring{\tilde\nabla}_b\mathring\beta\mathring{\tilde\nabla}^bh_a{}^a\big]
					\\
					&\kern1cm+r^3\big[\tfrac14\partial_rh_a{}^a\big(\mathring\beta\mathring\alpha^b\mathring\alpha_b-\mathring\alpha^b\mathring{\tilde\nabla}_b\mathring\beta\big)+\tfrac12\partial_rh^{ab}\big(-\tfrac12\mathring\beta\mathring\alpha_a\mathring\alpha_b+\mathring\alpha_a\mathring{\tilde\nabla}_b\mathring\beta\big)+\tfrac12\mathring\beta\mathring{\tilde\nabla}_a\partial_r h^a
					\\
					&\kern1cm-\tfrac12\partial_rh\mathring{\tilde\nabla}_a\mathring\alpha^a+2\partial_rh\mathring\alpha^a\mathring\alpha_a-\mathring\alpha^a\mathring{\tilde\nabla}_a\partial_rh+\tfrac12\partial_rh^a\mathring{\tilde\nabla}_a\mathring\beta-\mathring\beta\partial_rh^a\mathring\alpha_a
					\\
					&\kern1cm-2\partial_rh^a\mathring\alpha^b\mathring{\tilde\nabla}_{[b}\mathring\alpha_{a]}\big]
					\\
					&\kern1cm+r^4\big(-\tfrac18(\mathring\beta)^2\partial_r^2h_a{}^a-\tfrac12\mathring\beta\partial_r^2h^a\mathring\alpha_a+\tfrac12\partial_r^2h\mathring\alpha^a\mathring\alpha_a\big)
				\end{aligned}
			\end{equation}
			and
			\begin{equation}
				R^{(1)}_{--}\ =\ -\tfrac12\partial_r^2h_a{}^a
			\end{equation}
			and
			\begin{equation}
				\begin{aligned}
					R^{(1)}_{+-}\ &=\ -h-\tfrac12\mathring{\tilde\nabla}_a\mathring\alpha_bh^{ab}+\tfrac12\mathring\alpha_a\mathring\alpha_bh^{ab}+\tfrac12\mathring{\tilde\nabla}_ah^a-\tfrac12\mathring\alpha_c\mathring{\tilde\nabla}_ah^{ac}+\tfrac14\mathring\alpha_c\mathring{\tilde\nabla}^ch_a{}^a-h_a\mathring\alpha^a
					\\
					&\kern1cm+r\big(-2\partial_rh-\tfrac12\mathring\beta\partial_rh_a{}^a -2\partial_rh_a\mathring\alpha^a+\tfrac12\mathring\alpha_a\mathring\alpha_b\partial_rh^{ab}-\tfrac14\mathring\alpha^a\mathring\alpha_a\partial_rh_b{}^b+\tfrac12\mathring{\tilde\nabla}_a\partial_rh^a\big)
					\\
					&\kern1cm+r^2\big(-\tfrac12\partial_r^2h-\tfrac12\partial_r^2h_a\mathring\alpha^a-\tfrac14\mathring\beta\partial_r^2h_a{}^a\big)
				\end{aligned}
			\end{equation}
			and
			\begin{equation}
				\begin{aligned}
					R^{(1)}_{a-}\ &=\ \partial_rh_a-\tfrac12\mathring\alpha_b\partial_rh_a{}^b+\tfrac14\mathring\alpha_a\partial_rh_b{}^b-\tfrac12\mathring{\tilde\nabla}_a\partial_rh_b{}^b+\tfrac12\mathring{\tilde\nabla}_b\partial_rh_a{}^b
					\\
					&\kern1cm+r\big(\tfrac12\partial_r^2 h_a+\tfrac12\mathring\alpha_a\partial_r^2h_b{}^b-\tfrac12\mathring\alpha_b\partial_r^2h_a{}^b\big)
				\end{aligned}
			\end{equation}
			and
			\begin{equation}
				\begin{aligned}
					R^{(1)}_{a+}\ &=\ r\big[2\mathring{\tilde\nabla}_{[a}\mathring\alpha_{b]}\mathring\alpha_ch^{bc}-\mathring{\tilde\nabla}_b\mathring{\tilde\nabla}_{[a}\mathring\alpha_{c]}h^{bc}+\tfrac12h_a\mathring{\tilde\nabla}_b\mathring\alpha^b-\tfrac12h_a\mathring\alpha_b\mathring\alpha^b
					\\
					&\kern1cm +h\mathring\alpha_a-\mathring{\tilde\nabla}_ah-2\mathring\alpha^b\mathring{\tilde\nabla}_{[a}h_{b]}-2h^b\mathring{\tilde\nabla}_{[a}\mathring\alpha_{b]}+\mathring{\tilde\nabla}^b\mathring{\tilde\nabla}_{[a}h_{b]}
					\\
					&\kern1cm+\mathring{\tilde\nabla}_{[c}\mathring\alpha_{a]}\mathring{\tilde\nabla}_bh^{cb}+\mathring{\tilde\nabla}_{[b}\mathring\alpha_{c]}\mathring{\tilde\nabla}^bh_a{}^c+\tfrac12\mathring{\tilde\nabla}_{[a}\mathring\alpha_{c]}\mathring{\tilde\nabla}^ch_b{}^b\big]
					\\
					&\kern1cm+r^2\big[-\tfrac12\mathring\beta\partial_rh_a+2\partial_rh\mathring\alpha_a-\tfrac12\mathring{\tilde\nabla}_a\partial_rh-\tfrac32\mathring\alpha^b\big(\partial_rh_a\mathring\alpha_b-\partial_rh_b\mathring\alpha_a\big)+\mathring\alpha^b\mathring{\tilde\nabla}_b\partial_rh_a
					\\
					&\kern1cm-\tfrac12\mathring{\tilde\nabla}_a\big(\partial_rh^b\mathring\alpha_b\big)-\tfrac12\mathring\alpha_a\mathring{\tilde\nabla}_b\partial_rh^b+\tfrac12\partial_rh_a\mathring{\tilde\nabla}_b\mathring\alpha^b-\big(\tfrac34\mathring\beta\mathring\alpha^b-\tfrac12\mathring{\tilde\nabla}^b\mathring\beta\big)\partial_rh_{ab}
					\\
					&\kern1cm+\big(\tfrac38\mathring\beta\mathring\alpha_a-\tfrac14\mathring{\tilde\nabla}_a\mathring\beta\big)\partial_rh_b{}^b+\mathring\alpha^b\mathring{\tilde\nabla}_{[c}\mathring\alpha_{b]}\partial_rh_a{}^c+\mathring\alpha_b\mathring{\tilde\nabla}_{[a}\mathring\alpha_{c]}\partial_rh^{cb}
					\\
					&\kern1cm-\tfrac12\mathring\alpha^b\mathring{\tilde\nabla}_{[a}\mathring\alpha_{b]}\partial_r h_c {}^c-\tfrac12\mathring\beta\mathring{\tilde\nabla}_{[a}\partial_rh_{c]}{}^c\big]
					\\
					&\kern1cm+r^3\big(-\tfrac14\mathring\beta\partial_r^2h_a+\tfrac12\partial_r^2h\mathring\alpha_a+\mathring\alpha^b\mathring\alpha_{[a}\partial_r^2h_{b]}+\tfrac12\mathring\beta\mathring\alpha_{[a}\partial_r^2h_{b]}{}^b\big)
				\end{aligned}
			\end{equation}
			and
			\begin{equation}
				\begin{aligned}
					R^{(1)}_{ab}\ &=\ \mathring{\tilde\nabla}^c\mathring{\tilde\nabla}_{(a}h_{b)c}-\tfrac12\mathring{\tilde\nabla}_b\mathring{\tilde\nabla}_ah_c{}^c
					\\
					&\kern1cm-\tfrac12\mathring{\tilde\nabla}^c\mathring{\tilde\nabla}_ch_{ab}+\mathring{\tilde\nabla}_{(a} h_{b)}-\mathring\alpha^c\mathring{\tilde\nabla}_{(a}h_{b)c}+\tfrac12\mathring\alpha^c\mathring{\tilde\nabla}_ch_{ab}-h_{(a}\mathring\alpha_{b)}
					\\
					&\kern1cm+r\big[\mathring{\tilde\nabla}_{(a}\partial_rh_{b)}-\tfrac32\big(\partial_rh_a\mathring\alpha_b+\partial_rh_b\mathring\alpha_a\big)+\big(2\mathring\alpha_c\mathring\alpha_{(b}-\mathring{\tilde\nabla}_c\mathring\alpha_{(b}\big)\partial_rh_{a)}{}^c
					\\
					&\kern1cm+\big(-\mathring\alpha_c\mathring\alpha^c-\mathring\beta+\tfrac12\mathring{\tilde\nabla}_c\mathring\alpha^c\big)\partial_rh_{ab}+\tfrac12\big(-\mathring\alpha_a\mathring\alpha_b+\mathring{\tilde\nabla}_{(a}\mathring\alpha_{b)}\big)\partial_rh_c{}^c
					\\
					&\kern1cm-\mathring\alpha_{(a|}\mathring{\tilde\nabla}^c\partial_r h_{c|b)}-\mathring\alpha^c\mathring{\tilde\nabla}_{(a}\partial_rh_{b) c}+\mathring\alpha^c\mathring{\tilde\nabla}_c\partial_rh_{ab}+\mathring\alpha_{(a}\mathring{\tilde\nabla}_{b)}\partial_rh_c{}^c\big]
					\\
					&\kern1cm+r^2\big(\mathring\alpha^c\mathring\alpha_{(a}\partial_r^2h_{b)c}-\partial_r^2h_{(a}\mathring\alpha_{b)}-\tfrac12\mathring\alpha_a\mathring\alpha_b\partial_r^2h_c{}^c-\tfrac12(\mathring\alpha^c\mathring\alpha_c+\mathring\beta)\partial_r^2h_{ab}\big)
				\end{aligned}
			\end{equation}
			and
			\begin{equation}
				\begin{aligned}
					R^{(1)}\ &=\ -\mathring{\tilde{R}}_{ab}h^{ab}-2\mathring{\tilde\nabla}_a\mathring\alpha_bh^{ab}+\tfrac32\mathring\alpha_a\mathring\alpha_bh^{ab}
					\\
					&\kern1cm-2h+2\mathring{\tilde\nabla}_ah^a-3\mathring\alpha_ah^a+\mathring{\tilde\nabla}_a\mathring{\tilde\nabla}_bh^{ab}-\mathring{\tilde\nabla}^a\mathring{\tilde\nabla}_ah_b{}^b-2\mathring\alpha_a\mathring{\tilde\nabla}_bh^{ab}+\mathring\alpha^a\mathring{\tilde\nabla}_ah_b{}^b
					\\
					&\kern1cm+r\big(-4\partial_rh-2\mathring\beta\partial_rh_a{}^a+2\mathring{\tilde\nabla}_a\partial_rh^a-7\mathring\alpha_a\partial_rh^a+3\mathring\alpha_a\mathring\alpha_b\partial_rh^{ab}-\mathring{\tilde\nabla}_a\mathring\alpha_b\partial_rh^{ab}
					\\
					&\kern1cm-2\mathring\alpha_a\mathring\alpha^a\partial_rh_b{}^b+\mathring{\tilde\nabla}_a\mathring\alpha^a\partial_rh_b{}^b-2\mathring\alpha_a\mathring{\tilde\nabla}_b\partial_rh^{ab}+2\mathring\alpha^a\mathring{\tilde\nabla}_a\partial_rh_b{}^b\big)
					\\
					&\kern1cm+r^2\big(-\partial_r^2h-2\mathring\alpha_a\partial_r^2h^a-\mathring\beta\partial_r^2h_a{}^a+\mathring\alpha_a\mathring\alpha_b\partial_r^2h^{ab}-\mathring\alpha_a\mathring\alpha^a\partial_r^2h_b{}^b\big)\,.
				\end{aligned}
			\end{equation}
		\end{subequations}

		\paragraph{Einstein equation.}
		In \cref{sec:bianchiIdentityEinsteinEquation}, we have seen that because of the contracted Bianchi identity, the only independent components of the Einstein equation are the $++$, $a+$, and $ab$ components. Using the above and \eqref{eq:backgroundEinsteinEquations}, we arrive at the components listed in~\eqref{eq:zerothOrderEinsteinEquation}.

		\subsection{Next-to-lowest-order Einstein equation}

		\paragraph{Connection one-form.}
		We extend~\eqref{eq:lowestOrderConnectionOneForm} as
		\begin{equation}\label{eq:NTLOConnectionOneForm}
			\omega_{AB}{}^C\ =\ \mathring\omega_{AB}{}^C+\kappa\omega^{(1)}_{AB}{}^C+\tfrac{\kappa^2}2\omega^{(2)}_{AB}{}^C+\caO(\kappa^3)~.
		\end{equation}
		It is not to difficult to see that imposing the metric compatibility and torsion-freeness on $\omega_{AB}{}^C$ and using~\eqref{eq:lowestOrderConnectionOneFormFormula}, we obtain
		\begin{equation}\label{eq:NTLOConnectionOneFormFormula}
			\omega^{(2)}_{AB}{}^C\ =\ -2\mathring g^{CE}\omega^{(1)}_{AB}{}^Dh_{DE}\ =\ -2\omega^{(1)}_{AB}{}^Dh_D{}^C\ =\ -2\big(\mathring\nabla_{(A}h_{B)}{}^D-\tfrac12\mathring\nabla^Dh_{AB}\big)h_D{}^C~.
		\end{equation}

		\paragraph{Ricci tensor and curvature scalar.}
		Next, we extend~\eqref{eq:lowestOrderRicciTensorCurvatureScalar} as
		\begin{equation}\label{eq:NTLORicciTensorCurvatureScalar}
			\begin{aligned}
				R_{AB}\ &=\ \mathring R_{AB}+\kappa R^{(1)}_{AB}+\tfrac{\kappa^2}2R^{(2)}_{AB}+\caO(\kappa^3)~,
				\\
				R\ &=\ \mathring R+\kappa R^{(1)}+\tfrac{\kappa^2}2R^{(2)}+\caO(\kappa^3)~.
			\end{aligned}
		\end{equation}
		Upon substituting~\eqref{eq:NTLOConnectionOneForm} into~\eqref{eq:RicciTensorCurvatureScalar} and making use of~\eqref{eq:lowestOrderConnectionOneFormFormula} and~\eqref{eq:NTLOConnectionOneFormFormula}, we thus find
		\begin{subequations}\label{eq:NTLORicciTensorCurvatureScalarFormula}
			\begin{equation}
				\begin{aligned}
					R^{(2)}_{AB}\ &=\ \mathring\nabla_C\omega^{(2)}_{AB}{}^C-\mathring\nabla_A\omega^{(2)}_{CB}{}^C-2\omega^{(1)}_{CB}{}^E\omega^{(1)}_{AE}{}^C+2\omega^{(1)}_{AB}{}^E\omega^{(1)}_{CE}{}^C
					\\
					&=\ h^{CD}\big(-2\mathring\nabla_{C}\mathring\nabla_{(A}h_{B)D}+\mathring\nabla_{C}\mathring\nabla_{D}h_{AB}+\mathring\nabla_{A}\mathring\nabla_{B}h_{CD}\big)
					\\
					&\kern1cm-2\mathring\nabla_Ch^{CD}\mathring\nabla_{(A}h_{B)D}+\mathring\nabla_Ch^{CD}\mathring\nabla_Dh_{AB}+\tfrac12\mathring\nabla_Ah^{CD}\mathring\nabla_Bh_{CD}+\mathring\nabla^Ch_D{}^D\mathring\nabla_{(A}h_{B)C}
					\\
					&\kern1cm-\tfrac12\mathring\nabla^Ch_D{}^D\mathring\nabla_Ch_{AB}-\mathring\nabla^Ch_{AD}\mathring\nabla^Dh_{BC}+\mathring\nabla^Ch_{AD}\mathring\nabla_Ch_B{}^D
				\end{aligned}
			\end{equation}
			and
			\begin{equation}
				\begin{aligned}
					R^{(2)}\ &=\ h^{BC}\big(-4\mathring\nabla_{(C}\mathring\nabla_{A)}h^{A}{}_{B}+2\mathring\nabla_{C}\mathring\nabla_{B}h_{A}{}^{A}+2\mathring\nabla^{A}\mathring\nabla_{A}h_{BC}\big)
					\\
					&\kern1cm-2\mathring\nabla_Ch^{BC}\mathring\nabla^{A}h_{AB}+2\mathring\nabla_Ch^{BC}\mathring\nabla_Bh_{A}{}^{A}+\tfrac32\mathring\nabla^Ah^{BC}\mathring\nabla_Ah_{BC}
					\\
					&\kern1cm-\tfrac12\mathring\nabla^Ch_B{}^B\mathring\nabla_Ch_{A}{}^{A}-\mathring\nabla^Ch^{A}{}_{B}\mathring\nabla^Bh_{AC}+2\mathring R_{AB}h^{AC}h^{B}{}_C
				\end{aligned}
			\end{equation}
		\end{subequations}
		Explicitly, using $\mathring g_{AB}$ and $h_{AB}$ from~\eqref{eq:deformedMetricAdaptedBasisII} and the components~\eqref{eq:backgroundConnectionOneFormAdaptedFrameII}, we obtain
		\begin{subequations}
			\begin{equation}
				\begin{aligned}
					R_{++}^{(2)}\ &=\ r^2\big(2h^2+2\mathring\beta\mathring\alpha_a\mathring\alpha_bh^{ac}h^b{}_c-3\mathring\alpha_a\mathring{\tilde\nabla}_b\mathring\beta h^{ac}h^b{}_c+2\mathring{\tilde\nabla}_{[a}\mathring\alpha_{b]}\mathring{\tilde\nabla}_c\mathring\alpha_dh^{ac}h^{bd}
					\\
					&\kern1cm +2\mathring{\tilde\nabla}_{[a}\mathring\alpha_{b]}\mathring{\tilde\nabla}_c\mathring\alpha^bh^{ad}h^c{}_d+2\mathring{\tilde\nabla}_{[a}\mathring\alpha_{b]}\mathring{\tilde\nabla}^a\mathring\alpha_ch^{bd}h^c{}_d-\mathring\beta\mathring{\tilde\nabla}_a\mathring\alpha_bh^{ac}h^b{}_c
					\\
					&\kern1cm +\mathring{\tilde\nabla}_a\mathring{\tilde\nabla}_b\mathring\beta h^{ac}h^b{}_c-3h\mathring\alpha_a\mathring\alpha_bh^{ab}-4\mathring\beta h_a\mathring\alpha_bh^{ab}+3h_a\mathring{\tilde\nabla}_b\mathring\beta h^{ab}+3\mathring\alpha_a\mathring{\tilde\nabla}_bhh^{ab}
					\\
					&\kern1cm-4\mathring{\tilde\nabla}_{[a}h_{b]}\mathring{\tilde\nabla}_c\mathring\alpha^b h^{ac}-4\mathring{\tilde\nabla}_{[a}h_{b]}\mathring{\tilde\nabla}^a\mathring\alpha_ch^{bc}+\mathring\beta\mathring{\tilde\nabla}_ah_bh^{ab}+ 2h\mathring{\tilde\nabla}_a\mathring\alpha_bh^{ab}
					\\
					&\kern1cm+\tfrac12\mathring\beta \mathring\alpha^b\mathring{\tilde\nabla}_bh_{ac}h^{ac}+\tfrac12\mathring\beta \mathring\alpha_b\mathring{\tilde\nabla}_ch_a{}^ah^{bc}-\mathring\beta\mathring\alpha^b\mathring{\tilde\nabla}_ah_{bc}h^{ac}-\mathring\beta\mathring\alpha_b\mathring{\tilde\nabla}_ah_c{}^ah^{bc}
					\\
					&\kern1cm-\mathring{\tilde\nabla}_a\mathring{\tilde\nabla}_bhh^{ab}-\tfrac12\mathring{\tilde\nabla}^b\mathring\beta\mathring{\tilde\nabla}_bh_{ac}h^{ac}-\tfrac12\mathring{\tilde\nabla}_b\mathring\beta\mathring{\tilde\nabla}_ch_a{}^ah^{bc}+\mathring{\tilde\nabla}^b\mathring\beta\mathring{\tilde\nabla}_ah_{cb}h^{ac}
					\\
					&\kern1cm+\mathring{\tilde\nabla}_b\mathring\beta\mathring{\tilde\nabla}_ah^a{}_ch^{bc}+6hh_a\mathring\alpha^a+2\mathring\beta h_ah^a-3h^a\mathring{\tilde\nabla}_ah+2\mathring{\tilde\nabla}_{[a}h_{b]}\mathring{\tilde\nabla}^ah^b-2h\mathring{\tilde\nabla}^ah_a
					\\
					&\kern1cm-h\mathring\alpha^b\mathring{\tilde\nabla}_bh_a{}^a+2h\mathring\alpha^b\mathring{\tilde\nabla}_ah_b{}^a-\tfrac12\mathring\beta h^b\mathring{\tilde\nabla}_bh_a{}^a+\mathring\beta h^b\mathring{\tilde\nabla}_ah_b{}^a+\tfrac12\mathring{\tilde\nabla}^bh\mathring{\tilde\nabla}_bh_a{}^a
					\\
					&\kern1cm-\mathring{\tilde\nabla}^bh\mathring{\tilde\nabla}_ah_a{}^b\big)
					\\
					&\kern1cm+r^3\big[4h\partial_rh+\mathring\beta h\partial_rh_a{}^a-\tfrac12\partial_rh_{ac}h^{ac}\big(\mathring\beta\mathring\alpha_b\mathring\alpha^b-\mathring\alpha_b\mathring{\tilde\nabla}^b\mathring\beta\big)
					\\
					&\kern1cm-\tfrac12\partial_rh_a{}^ah^{bc}\big(\mathring\beta\mathring\alpha_b\mathring\alpha_c-\mathring\alpha_b\mathring{\tilde\nabla}_c\mathring\beta\big)+\partial_rh_a{}^bh^{ac}\mathring\beta\mathring\alpha_c\mathring\alpha_b-\partial_rh_a{}^bh^{ac}\mathring\alpha_c\mathring{\tilde\nabla}_b\mathring\beta
					\\
					&\kern1cm-\partial_rh^a{}_bh^{bc}\mathring\alpha_a\mathring{\tilde\nabla}_c\mathring\beta-\mathring\beta\mathring{\tilde\nabla}_a\partial_rh_bh^{ab}+\partial_rh\mathring{\tilde\nabla}_a\mathring\alpha_bh^{ab}-4\partial_rh\mathring\alpha_a\mathring\alpha_bh^{ab}
						\\
						&\kern1cm+2\mathring\alpha_a\mathring{\tilde\nabla}_b\partial_rhh^{ab}-\partial_rh_a\mathring{\tilde\nabla}_b\mathring\beta h^{ab}+2\mathring\beta\partial_rh_a\mathring\alpha_bh^{ab}+4\partial_rh_c\mathring\alpha^b\mathring{\tilde\nabla}_{[b}\mathring\alpha_{a]}h^{ac}
						\\
						&\kern1cm+4\partial_rh^a\mathring\alpha_c\mathring{\tilde\nabla}_{[b}\mathring\alpha_{a]}h^{bc}+\tfrac12\partial_rh_a{}^a\big(2h\mathring\alpha_b\mathring\alpha^b+2\mathring\beta h_b\mathring\alpha^b-h_b\mathring{\tilde\nabla}^b\mathring\beta-\mathring\alpha_b\mathring{\tilde\nabla}^bh\big)
					\\
					&\kern1cm-\partial_rh^{ab}\big(\tfrac32h\mathring\alpha_a\mathring\alpha_b+\mathring\beta h_a\mathring\alpha_b-h_a\mathring{\tilde\nabla}_b\mathring\beta-\mathring\alpha_a\mathring{\tilde\nabla}_bh\big)+\tfrac12\mathring\beta\partial_rh^b\mathring{\tilde\nabla}_bh_a{}^a-\mathring\beta\partial_rh^b\mathring{\tilde\nabla}_ah_b{}^a
					\\
					&\kern1cm-\partial_r h\mathring{\tilde\nabla}^a h_a-\tfrac12\partial_r h \mathring\alpha^b\mathring{\tilde\nabla}_b h_a{}^a+\partial_rh\mathring\alpha^b\mathring{\tilde\nabla}_a h_b{}^a+8\partial_rhh^a\mathring\alpha_a-2h^a\mathring{\tilde\nabla}_a\partial_rh
					\\
					&\kern1cm+\partial_rh^a\mathring{\tilde\nabla}_ah+2h\partial_rh^a\mathring\alpha_a-2\mathring\beta\partial_rh^ah_a-4\partial_rh^ah^b\mathring{\tilde\nabla}_{[b}\mathring\alpha_{a]}-4\partial_rh^a\mathring\alpha^b\mathring{\tilde\nabla}_{[b}h_{a]}\big]
					\\
					&\kern1cm+r^4\big[h\partial_r^2h+\tfrac14\mathring\beta^2\partial_r^2h_{ab}h^{ab}+\mathring\beta\partial_r^2h_a\alpha_bh^{ab}-\partial_r^2h\mathring\alpha_a\mathring\alpha_bh^{ab}
						\\
						&\kern1cm+\tfrac12\partial_rh_a{}^a\big(\partial_rh\mathring\alpha_b\mathring\alpha^b-\mathring\beta\partial_rh_b\mathring\alpha^b\big)+\partial_rh^{ab}\big(\mathring\beta\partial_rh_a\mathring\alpha_b-\partial_rh\mathring\alpha_a\mathring\alpha_b\big)
					\\
					&\kern1cm-\mathring\beta\partial_r^2h^ah_a+2\partial_r^2hh^a\mathring\alpha_a+\partial_rh^a\mathring\alpha^b\big(\partial_rh_a\mathring\alpha_b-\partial_rh_b\mathring\alpha_a\big)+\tfrac18\mathring\beta^2\partial_rh^{ab}\partial_rh_{ab}\big]
				\end{aligned}
			\end{equation}
			and
			\begin{equation}
				\begin{aligned}
					R_{--}^{(2)}\ &=\ \partial_r^2h_{ab}h^{ab}+\tfrac12\partial_rh_{ab}\partial_rh^{ab}
				\end{aligned}
			\end{equation}
			and
			\begin{equation}
				\begin{aligned}
					R_{+-}^{(2)}\ &=\ \mathring{\tilde\nabla}_a\mathring\alpha_bh^a{}_ch^{cb}-\mathring\alpha_a\mathring\alpha_bh^a{}_ch^{cb}-\mathring{\tilde\nabla}_ah_b
					h^{ab}-\tfrac12\mathring\alpha_b\mathring{\tilde\nabla}_ch_a{}^ah^{bc}-\tfrac12\mathring\alpha^a\mathring{\tilde\nabla}_ah_{bc}h^{bc}
					\\
					&\kern1cm+\mathring\alpha_b\mathring{\tilde\nabla}_ah_c{}^ah^{bc}+\mathring\alpha^b\mathring{\tilde\nabla}_ah_{bc}h^{ac}+2h_a\mathring\alpha_bh^{ab}+\tfrac12h^b\mathring{\tilde\nabla}_bh_a{}^a-h^b\mathring{\tilde\nabla}_ah_b{}^a- h^ah_a
					\\
					&\kern1cm+r\big(\mathring\beta \partial_rh_{ab}h^{ab}+4\partial_rh_a\mathring\alpha_b h^{ab}-2\mathring\alpha_a\mathring\alpha_b\partial_rh_c{}^bh^{ac}+\tfrac12\mathring\alpha_a\mathring\alpha_c\partial_rh_b{}^bh^{ac}
						\\
						&\kern1cm+\tfrac12\mathring\alpha_a\mathring\alpha^a\partial_rh_{bc}h^{bc}-\mathring{\tilde\nabla}_a\partial_rh_bh^{ab}-h\partial_rh_a{}^a-4\partial_rh^ah_a+2h_a\mathring\alpha_b\partial_rh^{ab}-h_a\alpha^a \partial_rh_b{}^b
						\\
						&\kern1cm+\tfrac12\partial_rh^b\mathring{\tilde\nabla}_bh_a{}^a-\partial_rh^b\mathring{\tilde\nabla}_ah_b{}^a\big)
					\\
					&\kern1cm+r^2\big[\partial_r^2{h}_a\mathring\alpha_bh^{ab}+\tfrac12\mathring\beta\partial_r^2h_{ab}h^{ab}-\partial_r^2{h}^a h_a-\tfrac12\big(\partial_rh+\partial_rh^b\mathring\alpha_b\big)\partial_rh_a{}^a
					\\
					&\kern1cm+\partial_rh^{ab}\partial_rh_a\mathring\alpha_b-\partial_rh^a\partial_rh_a+\tfrac14\mathring\beta\partial_rh^{ab}\partial_rh_{ab}\big]
				\end{aligned}
			\end{equation}
			and
			\begin{equation}
				\begin{aligned}
					R_{ab}^{(2)}\ &=\ -2h^{cd}\mathring{\tilde\nabla}_c\mathring{\tilde\nabla}_{(a}h_{b)d}+h^{cd}\mathring{\tilde\nabla}_c\mathring{\tilde\nabla}_dh_{ab}+h^{cd}\mathring{\tilde\nabla}_{(a}\mathring{\tilde\nabla}_{b)}h_{cd}-h^{cd}\mathring\alpha_c\mathring{\tilde\nabla}_dh_{ab}+2h^{cd}\mathring\alpha_c\mathring{\tilde\nabla}_{(a}h_{b)d}
					\\
					&\kern1cm-2\mathring{\tilde\nabla}_ch^{cd}\mathring{\tilde\nabla}_{(a}h_{b)d}+\mathring{\tilde\nabla}_ch^{cd}\mathring{\tilde\nabla}_dh_{ab}+\tfrac12\mathring{\tilde\nabla}_ah_{cd}\mathring{\tilde\nabla}_bh^{cd}+\mathring{\tilde\nabla}^dh_c{}^c\mathring{\tilde\nabla}_{(a}h_{b)d}
					\\
					&\kern1cm-\tfrac12\mathring{\tilde\nabla}^dh_c{}^c\mathring{\tilde\nabla}_dh_{ab}-\mathring{\tilde\nabla}_ch_a{}^d\mathring{\tilde\nabla}_dh_b{}^c+\mathring{\tilde\nabla}^ch_{ad}\mathring{\tilde\nabla}_ch_b{}^d+h^c\mathring{\tilde\nabla}_ch_{ab}-2h^c\mathring{\tilde\nabla}_{(a}h_{b)c}-h_ah_b
					\\
					&\kern1cm+r\big[2\big(-2\mathring\alpha_c\mathring\alpha_{(a}+\mathring{\tilde\nabla}_c\mathring\alpha_{(a}\big)\partial_rh_{b)d}h^{cd}+\big(2\mathring\alpha_c\mathring\alpha_d-\mathring{\tilde\nabla}_c\mathring\alpha_d\big)h^{cd}\partial_rh_{ab}
					\\
					&\kern1cm+\big(\mathring\alpha_a\mathring\alpha_b-\mathring{\tilde\nabla}_{(a}\mathring\alpha_{b)}\big)h^{cd}\partial_rh_{cd}+2h_c{}^d\mathring\alpha_{(a|}\mathring{\tilde\nabla}{}^c\partial_rh_{|b)d}+2\mathring\alpha_ch^{cd}\mathring{\tilde\nabla}_{(a}\partial_rh_{b)d}
					\\
					&\kern1cm-2\mathring\alpha_ch^{cd}\mathring{\tilde\nabla}_d\partial_rh_{ab}-2\mathring\alpha_{(a|}h^{cd}\mathring{\tilde\nabla}_{|b)}\partial_rh_{cd}-2h\partial_rh_{ab}-2h_{(a}\partial_rh_{b)}+4\mathring\alpha_{(a|}h^c\partial_rh_{|b)c}
					\\
					&\kern1cm+2\mathring\alpha^ch_{(a}\partial_rh_{b)c}-4\mathring\alpha^ch_c\partial_rh_{ab}-\mathring\alpha_{(a}h_{b)}\partial_rh_c{}^c+\partial_rh^c\big(\mathring{\tilde\nabla}_ch_{ab}-2\mathring{\tilde\nabla}_{(a}h_{b)c}\big)
					\\
					&\kern1cm+\mathring{\tilde\nabla}^ch_c\partial_rh_{ab}-2\mathring{\tilde\nabla}^ch_{(a}\partial_rh_{b)c}+\mathring{\tilde\nabla}_{(a}h_{b)}\partial_rh_c{}^c-2h^c\mathring{\tilde\nabla}_{(a}\partial_rh_{b)c}+2h^c\mathring{\tilde\nabla}_c\partial_rh_{ab}
					\\
					&\kern1cm-2\mathring\alpha_c\partial_rh_{(a|}{}^d\mathring{\tilde\nabla}^ch_{|b)d}+2\mathring\alpha^c\partial_rh_{(a|}{}^d\mathring{\tilde\nabla}_dh_{|b)c}+\tfrac12\mathring\alpha^c\partial_rh_{ab}\mathring{\tilde\nabla}_ch_d{}^d-\mathring\alpha^c\partial_rh_{ab}\mathring{\tilde\nabla}^dh_{cd}
					\\
					&\kern1cm+\tfrac12\mathring\alpha^c\partial_rh_d{}^d\mathring{\tilde\nabla}_ch_{ab}-\mathring\alpha^c\partial_rh_d{}^d\mathring{\tilde\nabla}_{(a}h_{b)c}-\mathring\alpha^c\partial_rh_c{}^d\mathring{\tilde\nabla}_dh_{ab}+2\mathring\alpha^c\partial_rh_c{}^d\mathring{\tilde\nabla}_{(a}h_{b)d}
					\\
					&\kern1cm-\mathring\alpha_{(a|}\partial_rh^{cd}\mathring{\tilde\nabla}_{|b)}h_{cd}-\mathring\alpha_{(a}\partial_rh_{b)c}\mathring{\tilde\nabla}^ch_d{}^d+2\mathring\alpha_{(a}\partial_rh_{b)d}\mathring{\tilde\nabla}^ch_{dc}\big]
					\\
					&\kern1cm+r^2\big[-2\mathring\alpha_ch^{cd}\mathring\alpha_{(a}\partial_r^2h_{b)d}+\mathring\alpha_a\mathring\alpha_bh^{cd}\partial_r^2h_{cd}+\mathring\alpha_c\mathring\alpha_dh^{cd}\partial_r^2h_{ab}+2\mathring\alpha_{(a|}h^c\partial_r^2h_{|b)c}
					\\
					&\kern1cm-2\mathring\alpha^ch_c\partial_r^2h_{ab}-h\partial_r^2h_{ab}-\partial_rh_a\partial_rh_b-\big(\partial_rh+2\partial_rh^c\mathring\alpha_c\big)\partial_rh_{ab}+2\partial_rh^c\mathring\alpha_{(a}\partial_rh_{b)c}
					\\
					&\kern1cm+2\partial_rh_{(a|}\mathring\alpha^c\partial_rh_{|b)c}-\partial_rh_{(a}\mathring\alpha_{b)}\partial_rh_c{}^c-2\mathring\alpha_c\mathring\alpha_{(a}\partial_rh_{b)d}\partial_rh^{cd}+\mathring\alpha^c\mathring\alpha_{(a}\partial_rh_{b)c}\partial_rh_d{}^d
					\\
					&\kern1cm-\mathring\alpha_c\mathring\alpha_d\partial_rh_a{}^c\partial_rh_b{}^d+\big(\mathring\alpha_c\mathring\alpha^c+\mathring\beta\big)\partial_rh_a{}^d\partial_rh_{bd}+\tfrac12\mathring\alpha_a\mathring\alpha_b\partial_rh_{cd}\partial_rh^{cd}
					\\
					&\kern1cm+\mathring\alpha_c\mathring\alpha_d\partial_rh^{cd}\partial_rh_{ab}-\tfrac12\big(\mathring\alpha_c\mathring\alpha^c+\mathring\beta\big)\partial_rh_{ab}\partial_rh_d{}^d\big]
				\end{aligned}
			\end{equation}
			and
			\begin{equation}
				\begin{aligned}
					R_{a+}^{(2)}\ &=\ r\big(-\mathring{\tilde\nabla}_b \mathring\alpha_c h_ah^{bc}+\mathring\alpha_b\mathring\alpha_ch_ah^{bc}+h_a\mathring{\tilde\nabla}_bh^b-\mathring\alpha_bh_a\mathring{\tilde\nabla}_ch^{bc}+\tfrac12\mathring\alpha_bh_a\mathring{\tilde\nabla}^bh_c{}^c-2h_ah_b\mathring\alpha^b
					\\
					&\kern1cm-4\mathring\alpha_c\mathring{\tilde\nabla}_{[a}\mathring\alpha_{b]}h^{bd}h_d{}^c+2\mathring{\tilde\nabla}_c\mathring{\tilde\nabla}_{[a}\mathring\alpha_{b]}h^{bd}h_d{}^c+4h_c\mathring{\tilde\nabla}_{[a}\mathring\alpha_{b]}h^{bc}+4\mathring\alpha_c\mathring{\tilde\nabla}_{[a}h_{b]}h^{bc}
					\\
					&\kern1cm-2\mathring{\tilde\nabla}_c\mathring{\tilde\nabla}_{[a}\mathring h_{b]}h^{bc}-2\mathring{\tilde\nabla}_{[c}\mathring\alpha_{a]}\mathring{\tilde\nabla}_bh^c{}_dh^{bd}-2\mathring{\tilde\nabla}_{[b}\mathring\alpha_{c]}\mathring{\tilde\nabla}_dh_a{}^ch^{bd}-\mathring{\tilde\nabla}_{[ a}\mathring\alpha_{c]}\mathring{\tilde\nabla}^ch_{bd}h^{bd}
					\\
					&\kern1cm-2\mathring{\tilde\nabla}_{[c}\mathring\alpha_{a]}\mathring{\tilde\nabla}_bh_d{}^bh^{cd}-2\mathring{\tilde\nabla}_{[b}\mathring\alpha_{c]}\mathring{\tilde\nabla}^bh_{ad}h^{cd}-\mathring{\tilde\nabla}_{[ a}\mathring\alpha_{c]}\mathring{\tilde\nabla}_dh_b{}^bh^{cd}-4h^b\mathring{\tilde\nabla}_{[a}h_{b]}
					\\
					&\kern1cm+2\mathring{\tilde\nabla}_{[c}h_{a]}\mathring{\tilde\nabla}_bh^{c b}+2\mathring{\tilde\nabla}_{[b}h_{c]}\mathring{\tilde\nabla}^bh_a{}^c+\mathring{\tilde\nabla}_{[a}h_{c]}\mathring{\tilde\nabla}^ch_b{}^b
					\\
					&\kern1cm+r^2\big[\mathring\alpha_b \mathring\alpha_ch_a\partial_rh^{bc}-\tfrac12\mathring\alpha^b \mathring\alpha_b h_a\partial_rh_c{}^c+3\mathring\alpha_c\big(\partial_rh_a\mathring\alpha_b-\partial_rh_b\mathring\alpha_c\big)h^{bc}
					\\
					&\kern1cm-2\mathring\alpha_b\mathring{\tilde\nabla}_c\partial_rh_ah^{bc}+\mathring{\tilde\nabla}_a(\partial_rh_b\mathring\alpha_ch^{bc})+\mathring\alpha_a\mathring{\tilde\nabla}_b\partial_rh_ch^{bc}-\partial_rh_a\mathring{\tilde\nabla}_b\mathring\alpha_ch^{bc}
					\\
					&\kern1cm+\big(\tfrac32\mathring\beta\mathring\alpha_b-\mathring{\tilde\nabla}_b\mathring\beta\big)\partial_rh_{ac}h^{bc}-\tfrac14\big(3\mathring\beta\mathring\alpha_a-2\mathring{\tilde\nabla}_a\mathring\beta\big)\partial_rh_{bc}h^{bc}+\tfrac12\mathring\beta\mathring{\tilde\nabla}_a\partial_rh_{bc}h^{bc}
					\\
					&\kern1cm-\tfrac12\mathring\beta\mathring{\tilde\nabla}_b\partial_rh_{ac}h^{bc}-2\mathring\alpha_d\mathring{\tilde\nabla}_{[c}\mathring\alpha_{b]}\partial_rh_a{}^ch^{bd}-2\mathring\alpha_d\mathring{\tilde\nabla}_{[a}\mathring\alpha_{c]}\partial_rh_b{}^ch^{bd}
					\\
					&\kern1cm-2\mathring\alpha^b\mathring{\tilde\nabla}_{[c}\mathring\alpha_{b]}\partial_rh_{ad}h^{cd}-2\mathring\alpha^b\mathring{\tilde\nabla}_{[a}\mathring\alpha_{c]}\partial_rh_{bd}h^{cd}+\mathring\alpha_d\mathring{\tilde\nabla}_{[a}\mathring\alpha_{b]}\partial_rh_c{}^ch^{bd}
					\\
					&\kern1cm+\mathring\alpha^b\mathring{\tilde\nabla}_{[a}\mathring\alpha_{b]}\partial_rh_{bd}h^{bd}-2h\partial_rh_a-6h^b\partial_rh_a\mathring\alpha_b+3h^b\partial_rh_b\mathring\alpha_a-\mathring\alpha^b\partial_rh_bh_a
					\\
					&\kern1cm+2h^b\mathring{\tilde\nabla}_b\partial_rh_a+\mathring\alpha^b\partial_rh^c\big(\mathring{\tilde\nabla}_ch_{ba}-\mathring{\tilde\nabla}_bh_{ca}-\mathring{\tilde\nabla}_ah_{bc}\big)-\mathring{\tilde\nabla}_a(\partial_rh^bh_b)
					\\
					&\kern1cm-\tfrac12\mathring\alpha_a\partial_rh^c\mathring{\tilde\nabla}_ch_b{}^b+\mathring\alpha_a\partial_rh^c\mathring{\tilde\nabla}_bh_c{}^b+\partial_rh_a\mathring{\tilde\nabla}^bh_b+\tfrac12\mathring\alpha^c\partial_rh_a\mathring{\tilde\nabla}_ch_b{}^b-\mathring\alpha^c\partial_rh_a\mathring{\tilde\nabla}_bh_c{}^b
					\\
					&\kern1cm-\big(h\mathring\alpha_b+\tfrac32\mathring\beta h_b-\mathring{\tilde\nabla}_bh\big)\partial_rh_a{}^b+\tfrac12\big(h\mathring\alpha_a-\tfrac12\mathring\beta h_a-\mathring{\tilde\nabla}_ah\big)\partial_rh_b{}^b+2h^b\mathring{\tilde\nabla}_{[c}\mathring\alpha_{b]}\partial_rh_a{}^c
					\\
					&\kern1cm+2\mathring\alpha^b\mathring{\tilde\nabla}_{[c}h_{b]}\partial_rh_a{}^c+2h^b\mathring{\tilde\nabla}_{[a}\mathring\alpha_{c]}\partial_rh_b{}^c+2\mathring\alpha^b\mathring{\tilde\nabla}_{[a}h_{c]}\partial_rh_b{}^c-h^b\mathring{\tilde\nabla}_{[a}\mathring\alpha_{b]}\partial_rh_c{}^c
					\\
					&\kern1cm-\mathring\alpha^b\mathring{\tilde\nabla}_{[a}h_{b]}\partial_rh_c{}^c+\tfrac14\mathring\beta\partial_rh^c{}_a\mathring{\tilde\nabla}_ch_b{}^b+\tfrac14\mathring\beta\partial_rh^{bc}\mathring{\tilde\nabla}_ah_{bc}-\tfrac12\mathring\beta\partial_rh_{ac}\mathring{\tilde\nabla}_bh^{bc}\big]
					\\
					&\kern1cm+r^3\big[-\mathring\alpha_a\mathring\alpha_b\partial_r^2h_ch^{bc}+\mathring\alpha_b\mathring\alpha_c\partial_r^2h_ah^{bc}+\tfrac12\mathring\beta\mathring\alpha_b \partial_r^2h_{ac}h^{bc}-\tfrac12\mathring\beta\mathring\alpha_a\partial_r^2h_{bc}h^{bc}
					\\
					&\kern1cm-h\partial_r^2h_a+\mathring\alpha_ah^b\partial_r^2h_b-2h^b\mathring\alpha_b\partial_r^2h_a-\partial_rh_a\partial_rh^b\mathring\alpha_b+\partial_rh^b\partial_rh_b\mathring\alpha_a-\tfrac12\mathring\beta h^b\partial_r^2h_{ab}
					\\
					&\kern1cm+\tfrac12\big(\mathring\beta\partial_r h^b-2\partial_rh\mathring\alpha^b\big)\partial_rh_{ab}-\tfrac14\big(\mathring\beta\partial_rh_a-2\partial_rh\mathring\alpha_a\big)\partial_rh_b{}^b
					\\
					&\kern1cm+\big(\partial_r h_a\mathring\alpha_b\mathring\alpha_c-\partial_rh_c\mathring\alpha_b\mathring\alpha_a\big)\partial_rh^{bc}+\big(\partial_rh^c\mathring\alpha^b\mathring\alpha_b-\partial_rh^b\mathring\alpha_b\mathring\alpha^c\big)\partial_rh_{ac}
					\\
					&\kern1cm+\tfrac12\big(\partial_rh^b\mathring\alpha_b\mathring\alpha_a-\partial_rh_a\mathring\alpha^b\mathring\alpha_b\big)\partial_rh_c{}^c-\tfrac14\mathring\beta\mathring\alpha_a\partial_rh^{bc}\partial_rh_{bc}
					\\
					&\kern1cm+\tfrac14\mathring\beta\mathring\alpha_b\big(2\partial_rh^{bc}\partial_rh_{ac}-\partial_rh_a{}^b\partial_rh_c{}^c\big)\big]
				\end{aligned}
			\end{equation}
			and
			\begin{equation}
				\begin{aligned}
					R_{a-}^{(2)}\ &=\ -h_b\partial_rh_a{}^b+\tfrac12h_a\partial_rh_b{}^b+\mathring\alpha_c\partial_rh_{ab}h^{cb}-\tfrac12\mathring\alpha_a\partial_rh_{bc}h^{bc}+\tfrac12\partial_rh^{bc}\mathring{\tilde\nabla}_ah_{bc}
					\\
					&\kern1cm+\tfrac12\partial_rh_{ca}\mathring{\tilde\nabla}^ch_b{}^b-\partial_r h_{ca}\mathring{\tilde\nabla}_b h^{bc}+\mathring{\tilde\nabla}_a\partial_rh_{bc}h^{bc}-\mathring{\tilde\nabla}_b\partial_rh_{ca}h^{bc}
					\\
					&\kern1cm+r\big[-h_b\partial_r^2h_a{}^b-\partial_rh_b\partial_rh_a{}^b+\tfrac12\partial_rh_a\partial_rh_b{}^b-\tfrac12\mathring\alpha_a\partial_rh^{bc}\partial_rh_{bc}
					\\
					&\kern1cm+\mathring\alpha_b\big(\partial_rh^{bc}\partial_rh_{ca}-\tfrac12\partial_rh_c{}^c\partial_rh_a{}^b\big)-\mathring\alpha_a\partial_r^2h_{bc}h^{bc}+\mathring\alpha_b\partial_r^2h_{ac}h^{bc}\big]
				\end{aligned}
			\end{equation}
			and
			\begin{equation}
				\begin{aligned}
					R^{(2)}\ &=\ 2\mathring{\tilde{R}}_{ab}h^{ac}h_c{}^b+4\mathring{\tilde\nabla}_a\mathring\alpha_bh^{ac}h^b{}_c-3\mathring\alpha_a\mathring\alpha_bh^{ac}h^b{}_c-4h^{ab}\mathring{\tilde\nabla}_ah_b+6h^{ab}\mathring\alpha_ah_b
					\\
					&\kern1cm-2h^{ab}\mathring{\tilde\nabla}_a\mathring{\tilde\nabla}_ch_b{}^c+2h^{ab}\mathring{\tilde\nabla}{}^c\mathring{\tilde\nabla}_ch_{ab}+4h^{ab}\mathring\alpha^c\mathring{\tilde\nabla}_ah_{bc}-2h^{ab}\mathring\alpha^c\mathring{\tilde\nabla}_ch_{ab}-2h^{ab}\mathring{\tilde\nabla}^c\mathring{\tilde\nabla}_ah_{bc}
					\\
					&\kern1cm+2h^{ab}\mathring{\tilde\nabla}{}_a\mathring{\tilde\nabla}_bh_c{}^c+4h^{ab}\mathring\alpha_a\mathring{\tilde\nabla}^ch_{bc}-2h^{ab}\mathring\alpha_a\mathring{\tilde\nabla}_bh_c{}^c-3h_ah^a+2h^a\mathring{\tilde\nabla}_ah_b{}^b
					\\
					&\kern1cm-4 h^a\mathring{\tilde\nabla}^bh_{ab}-2\mathring{\tilde\nabla}^ah_{ab}\mathring{\tilde\nabla}_ch^{bc}+2\mathring{\tilde\nabla}^ah_{ab}\mathring{\tilde\nabla}^bh_c{}^c+\tfrac32\mathring{\tilde\nabla}^ah_{bc}\mathring{\tilde\nabla}_ah^{bc}-\tfrac12\mathring{\tilde\nabla}^ah_b{}^b\mathring{\tilde\nabla}_ah_c{}^c\\
					&\kern1cm-\mathring{\tilde\nabla}^ah^{bc}\mathring{\tilde\nabla}_bh_{ac}
					\\
					&\kern1cm+r\big(4\mathring\beta\partial_rh_{ab}h^{ab}-4h^{ab}\mathring{\tilde\nabla}_a\partial_rh_b+14h^{ab}\mathring\alpha_a\partial_rh_b-12h^{ab}\mathring\alpha_a\mathring\alpha^c\partial_rh_{bc}
					\\
					&\kern1cm+2h^{ab}\mathring{\tilde\nabla}^c\mathring\alpha_a\partial_rh_{bc}+4h^{ab}\mathring\alpha_a\mathring\alpha_b\partial_rh_c{}^c-2\mathring{\tilde\nabla}^a\mathring\alpha_ah^{bc}\partial_rh_{bc}+4h^{ab}\mathring\alpha_a\mathring{\tilde\nabla}^cb\partial_rh_{bc}
					\\
					&\kern1cm-4h^{ab}\mathring\alpha_a\mathring{\tilde\nabla}_b\partial_rh_c{}^c+2h^{ab}\mathring{\tilde\nabla}_a\mathring\alpha^c\partial_rh_{bc}+4\mathring\alpha_a\mathring\alpha^ah^{bc}\partial_rh_{bc}-2h^{ab}\mathring{\tilde\nabla}_a\mathring\alpha_b\partial_rh_c{}^c
					\\
					&\kern1cm+4h^{ab}\mathring\alpha^c\mathring{\tilde\nabla}_a\partial_rh_{bc}-4\mathring\alpha^ah^{bc}\mathring{\tilde\nabla}_a\partial_rh_{bc}-4 h\partial_rh_a{}^a-14\partial_rh_ah^a+12\mathring\alpha_ah_b\partial_rh^{ab}
					\\
					&\kern1cm-8\mathring\alpha_ah^a\partial_rh_b{}^b+2\partial_rh^a\mathring{\tilde\nabla}_ah_b{}^b-4\partial_rh^a\mathring{\tilde\nabla}^bh_{ab}+2\mathring{\tilde\nabla}_a h^a\partial_rh_b{}^b-2\mathring{\tilde\nabla}_a h_b\partial_rh^{ab}
					\\
					&\kern1cm-4 h^a\mathring{\tilde\nabla}^b\partial_rh_{ab}+4h^a\mathring{\tilde\nabla}_a\partial_rh_b{}^b-3\mathring\alpha_a\partial_rh_{bc}\mathring{\tilde\nabla}^ah^{bc}+2\mathring\alpha^a\partial_rh^{bc}\mathring{\tilde\nabla}_bh_{ac}
					\\
					&\kern1cm+\mathring\alpha_a\partial_rh_b{}^b\mathring{\tilde\nabla}^ah_c{}^c-2\mathring\alpha_a\partial_rh_b{}^b\mathring{\tilde\nabla}_ch^{ac}-2\mathring\alpha_a\partial_rh^{ab}\mathring{\tilde\nabla}_bh_c{}^c+4\mathring\alpha_a\partial_rh^{ab}\mathring{\tilde\nabla}^ch_{bc}\big)
					\\
					&\kern1cm+r^2\big(4h^{ab}\mathring\alpha_a\partial_r^2h_b+2h^{ab}\mathring\beta\partial_r^2h_{ab}-4h^{ab}\mathring\alpha_a\mathring\alpha^c\partial_r^2h_{bc}+2\mathring\alpha_a\mathring\alpha^ah^{bc}\partial_r^2h_{bc}
					\\
					&\kern1cm+2h^{ab}\mathring\alpha_a\mathring\alpha_b\partial_r^2h_c{}^c-2\partial_rh\partial_rh_a{}^a-2h\partial_r^2h_a{}^a+\tfrac32\mathring\beta\partial_rh_{ab}\partial_rh^{ab}-\tfrac12\mathring\beta\partial_rh_a{}^a\partial_rh_b{}^b
					\\
					&\kern1cm-3\partial_rh_a\partial_rh^a -4h_a\partial_r^2h^a +4\mathring\alpha^ah^b\partial_r^2h_{ab} -4\mathring\alpha_a h^a\partial_r^2h_b{}^b-4\mathring\alpha_a\partial_rh^a\partial_rh_b{}^b
					\\
					&\kern1cm+6\mathring\alpha_a\partial_rh_b\partial_rh^{ab} -3\mathring\alpha_a\mathring\alpha_b\partial_rh^{ac}\partial_rh^b{}_c+\tfrac32\mathring\alpha_a\mathring\alpha^a\partial_rh^{bc}\partial_rh_{bc}
					\\
					&\kern1cm+2\mathring\alpha_a\mathring\alpha_b\partial_rh^{ab}\partial_rh_c{}^c-\tfrac12\mathring\alpha_a\mathring\alpha^a\partial_rh_b{}^b\partial_rh_c{}^c\big)\,.
				\end{aligned}
			\end{equation}
		\end{subequations}

		\paragraph{Einstein equation.}
		In \cref{sec:bianchiIdentityEinsteinEquation}, we have seen that because of the contracted Bianchi identity, the only independent components of the Einstein equation are the $++$, $a+$, and $ab$ components. Using the above and~\eqref{eq:backgroundEinsteinEquations}, we immediately arrive at~\eqref{eq:DefOfMu2NH}.

		\section{Injectivity}\label{app:injectivity}

		We shall now explain how the operator $\sft$ in~\eqref{eq:operatorT} und thus $\tilde\mu_1$ in~\eqref{eq:upperTriangularMu1} are constructed. It will become apparent that $\sft$ is injective, and $\tilde\mu_1$ is again a local differential operator when it acts on $\Theta^{(p)}=(h_a^{(p)},h^{(p)},h_{ab}^{(p)})^\sfT$ under the boundary conditions~\eqref{eq:deformationsBoundaryConditionsParticularSolutions}.

		We first start off with considering $\mu_1$ defined in~\eqref{eq:zerothOrderEinsteinEquation:a}
		\begin{equation}\label{eq:recallmu1H}
			\mu_1\ =\
			\begin{pmatrix}
				\delta_a{}^c\big(2r\partial_r+r^2\partial_r^2\big) & 0 & \sfd_a{}^{cd}
				\\
				0 & 0 & \frac12\delta^{cd}r^2\partial_r^2
				\\
				\sfd_{ab}{}^c & \delta_{ab}\big(1+2r\partial_r+\frac12r^2\partial_r^2\big) & \sfd_{ab}{}^{cd}
			\end{pmatrix}.
		\end{equation}
		The first line divided by $r$ is a total derivative in $r$, that is,
		\begin{equation}
			\begin{gathered}
				\delta_a{}^c\big(2\partial_r+r\partial_r^2\big)\ =\ \partial_r\circ\big[\delta_a{}^c\big(1+r\partial_r\big)]
				\\
				\tfrac1r\sfd_a{}^{cd} \ =\ \partial_r\circ
				\underbrace{\big[\delta_a{}^{(c}\mathring{\tilde\nabla}^{d)}-\delta^{cd}(\mathring{\tilde\nabla}_a+\tfrac12\mathring\alpha_a)+r(-\delta_a{}^{(c}\mathring\alpha^{d)}+\mathring\alpha_a\delta^{cd})\partial_r\big]}_{\eqqcolon\,\sfB_a{}^{cd}}\,.
			\end{gathered}
		\end{equation}
		Therefore, if we apply
		\begin{equation}
			\sfI\ \coloneqq\
			\begin{pmatrix}
				\delta_a{}^c\int_0^{r''}\rmd r\frac1r & 0 & 0
				\\
				0 & \int_0^{r''}\rmd r'\int_0^{r'}\rmd r\frac{2}{r^2} & 0
				\\
				0 & 0 & \delta_{(a}{}^c\delta_{b)}{}^d|_{r=r''}
			\end{pmatrix},
		\end{equation}
		with these integrations are seen as operators, to~\eqref{eq:recallmu1H}, we obtain
		\begin{equation}\label{eq:integratemu1H}
			\sfI\circ\mu_1\ =\
			\begin{pmatrix}
				\delta_a{}^c(1+r\partial_r) & 0 & \sfB_a{}^{cd}
				\\
				0 & 0 & \delta^{cd}
				\\
				\sfd_{ab}{}^c & \delta_{ab}(1+2r\partial_r +\tfrac12r^2\partial^2_r) & \sfd_{ab}{}^{cd}
			\end{pmatrix},
		\end{equation}
		where, we have made use of the boundary conditions~\eqref{eq:deformationsBoundaryConditionsParticularSolutions}.

		Next, we note that we can factor out $(1+r\partial_r)$ from $\sfd_{ab}{}^c$, that is,
		\begin{equation}
			\sfd_{ab}{}^c\ =\ \underbrace{\big[\delta_{(a}{}^c(\mathring{\tilde\nabla}_{b)}-\mathring\alpha_{b)})-\delta_{ab}(\mathring{\tilde\nabla}^c-\tfrac32\mathring\alpha^c)+r(-\delta_{(a}{}^c\alpha_{b)}+\delta_{ab}\alpha^c)\partial_r\big]}_{\eqqcolon\,\sfA_{ab}{}^c}\circ(1+r\partial_r)
		\end{equation}
		Therefore, we can use the second row of~\eqref{eq:integratemu1H} to get rid of the trace part of $cd$ in the remaining two rows of the last column and use the first row to eliminate the operator in the third row of the first column. This series of operations can be written in terms of matrices as
		\begin{equation}
			\begin{aligned}
				\sfI\sfI\ &\coloneqq\
				\begin{pmatrix}
					\delta_m{}^k & 0 & 0
					\\
					0 & 0 & \frac{1}{d-2}\delta^{kl}
					\\
					0 & \frac{1}{d-2}\delta_{mn} & -\delta_{(m}{}^k\delta_{n)}{}^l+\frac{1}{d-2}\delta^{kl}\delta_{mn}
				\end{pmatrix}
				\circ
				\begin{pmatrix}
					\delta_k{}^g & 0 & 0
					\\
					0 & 1 & 0
					\\
					-\sfA_{kl}{}^g & 0 & \delta_{(k}{}^g\delta_{l)}{}^h
				\end{pmatrix}
				\\
				&\kern1cm\circ
				\begin{pmatrix}
					\delta_g{}^e & 0 & 0
					\\
					0 & 1 & 0
					\\
					0 & -\sfd_{gh}{}^{pq}\delta_{pq} & \delta_{(g}{}^e\delta_{h)}{}^f
				\end{pmatrix}
				\circ
				\begin{pmatrix}
					\delta_e{}^a & -\sfB_e{}^{pq}\delta_{pq} & 0
					\\
					0 & 1 & 0
					\\
					0 & 0 & \delta_{(e}{}^a\delta_{f)}{}^b
				\end{pmatrix},
			\end{aligned}
		\end{equation}
		where the first matrix was added to make~\eqref{eq:integratemu1H} upper triangular when $\sfI\sfI$ is applied. It is then not too difficult to see that
		\begin{equation}
			\tilde\mu_1\ =\ \sft\circ\mu_1
			\ewith
			\sft\ =\ \sfI\sfI\circ\sfI
		\end{equation}
		where $\tilde\mu_1$ and $\sft$ as given in~\eqref{eq:upperTriangularMu1} and~\eqref{eq:operatorT}, respectively. Hence, we conclude that $\sft$ is injective since $\sfI$ is injective and $\sfI\sfI$ is invertible.

		\section{Contracting Homotopies}\label{app:contractingHomotopyGaugeFixing}

		We shall now explain that the condition $\sfh(a)=0$ in~\eqref{eq:MCElement:a} is indeed a gauge-fixing condition in the general case.

		\paragraph{Gauge transformations revisited.}
		Let $(V,\mu_i)$ be an $L_\infty$-algebra, and set $I\coloneqq[0,1]\subseteq\IR$. We can now construct a new $L_\infty$-algebra $(\Omega^\bullet(I,V),\mu_i^{\Omega^\bullet(I,V)})$ by setting
		\begin{subequations}
			\begin{equation}
				\Omega^\bullet(I,V)\ \coloneqq\ \bigoplus_{k\in\IZ}\Omega^\bullet_k(I,V)
				\ewith
				\Omega^\bullet_k(I,V)\ \coloneqq\ \bigoplus_{i+j=k}\Omega^i(I)\otimes V_j
			\end{equation}
			and
			\begin{equation}
				\begin{aligned}
					\mu_1^{\Omega^\bullet(I,V)}(\omega\otimes v)\ &\coloneqq\ \rmd\omega\otimes v+(-1)^{|\omega|}\omega\otimes\mu_1(v)~,
					\\[5pt]
					\mu_i^{\Omega^\bullet(I,V)}(\omega_1\otimes v_1,\ldots,\omega_i\otimes v_i)\ &\coloneqq\ (-1)^{i\sum_{j=1}^i|\omega_i|+\sum_{j=0}^{i-2}|\omega_{i-j}|\sum_{k=1}^{i-j-1}|v_k|}
					\\
					&\kern1cm\times(\omega_1\wedge\ldots\wedge\omega_i)\otimes\mu_i(v_1,\ldots v_i)
				\end{aligned}
			\end{equation}
		\end{subequations}
		for all homogeneous $\omega,\omega_1,\ldots,\omega_i\in\Omega^\bullet(I)$ and $v,v_1,\ldots,v_i\in V$. Then, $\Omega^\bullet_1(I,V)=\scC^\infty(I,V_1)\oplus\Omega^1(I,V_0)$ and so, elements $\sfa\in\Omega^\bullet_1(I,V)$ are of the form
		\begin{equation}
			\sfa(t)\ =\ a(t)+\rmd t\otimes c_0(t)
		\end{equation}
		for all $t\in I$ where $a\in\scC^\infty(I,V_1)$ and $c_0\in\scC^\infty(I,V_0)$, respectively. Likewise, since $\Omega^\bullet_2(I,V)=\scC^\infty(I,V_2)\oplus\Omega^1(I,V_1)$, the curvature $\sff\in\Omega^\bullet_2(I,V)$ of $\sfa$, see~\eqref{eq:hMCCurvature}, is given by
		\begin{equation}
			\begin{aligned}
				\sff(t)\ &=\ \sum_{i\geq1}\frac1{i!}\mu_i^{\Omega^\bullet(I,V)}(\sfa(t),\ldots,\sfa(t))
				\\
				&=\ f(t)+\rmd t\otimes\left\{\parder[a(t)]{t}-\sum_{i\geq0}\frac1{i!}\mu_{i+1}(a(t),\ldots,a(t),c_0(t))\right\}.
			\end{aligned}
		\end{equation}

		Upon imposing the partial flatness condition $\parder{t}\intprod\sff=0$, we obtain
		\begin{equation}\label{eq:partialFlatness}
			\parder[a(t)]{t}-\sum_{i\geq0}\frac1{i!}\mu_{i+1}(a(t),\ldots,a(t),c_0(t))\ =\ 0~.
		\end{equation}
		We now recover the infinitesimal gauge transformation~\eqref{eq:hMCGaugeTransformations} by means of
		\begin{equation}
			\delta_{c_0}a\ =\ \left.\parder[a(t)]{t}\right|_{t=0}\ =\ \sum_{i\geq0}\frac1{i!}\mu_{i+1}(a,\ldots,a,c_0)
			\ewith
			a\ \coloneqq\ a(0)
			\eand
			c_0\ \coloneqq\ c_0(0)~.
		\end{equation}
		Furthermore, upon solving the differential equation~\eqref{eq:partialFlatness} on all of $I$, we obtain finite gauge transformations between $a=a(0)$ and $a'=a(1)$. In conclusion, gauge transformations are given by \uline{partially flat homotopies}. This can also be generalised to higher gauge transformations~\eqref{eq:hMCHigherGaugeTransformations}; see e.g.~\cite[Section 4.1]{Jurco:2018sby} for more details.

		In the remainder of this section, we shall show that for a general initial condition $a=a(0)$, there is a perturbative solution to~\eqref{eq:partialFlatness} such that $\sfh(a')=0$ with $a'=a(1)$.

		\paragraph{Recursion relations.}
		In~\cite[Appendix A.3]{Borsten:2021hua}, it was shown that gauge transformations can be understood in terms of \uline{curved} $L_\infty$-morphisms. Those are generalisations of $L_\infty$-morphisms $\phi:(V,\mu_i)\rightarrow(V',\mu'_i)$ discussed in \cref{sec:LInftyAlgebras} by allowing also constant maps $\phi_0:\IR\rightarrow V'_1$. The defining relations~\eqref{eq:LInfinityMorphism} then only change in that $i=0$ is also allowed. Therefore, with $V'=V$ and $\mu'_i=\mu_i$, we may make the Ansatz
		\begin{equation}
			\begin{gathered}
				a(t)\ =\ \sum_{i\geq1}\frac1{i!}\phi_i(t)(a,\ldots,a)\ =\ \sum_{i\geq0}\sum_{n\geq0}\frac1{n!i!}t^n\phi_i^{(n)}(a,\ldots,a)~,
				\\
				c_0(t)\ =\ \sum_{i\geq1}\frac1{i!}\phi_{i+1}(t)(a,\ldots,a,c_0)\ =\ \sum_{i\geq0}\sum_{n\geq0}\frac1{n!i!}t^n\phi_{i+1}^{(n)}(a,\ldots,a,c_0)
			\end{gathered}
		\end{equation}
		to solve~\eqref{eq:partialFlatness}. We also assume that $c_0(t)$ is constant for all $t\in I$ and that $c_0=c_0(0)$ itself depends on $a=a(0)$. Therefore, these expansions can be rewritten as
		\begin{equation}\label{eq:aAndcExpansions}
			a(t)\ =\ \sum_{i\geq1}\sum_{n\geq0}\frac1{n!i!}t^n\alpha_i^{(n)}(a,\ldots,a)
			\eand
			c_0(t)\ =\ \sum_{i\geq1}\frac1{i!}\gamma_i(a,\ldots,a)
		\end{equation}
		for new coefficients $\alpha_i^{(n)}$ and $\gamma_i$ which are $i$-linear in $a=a(0)$.

		Upon inserting these expansions into the differential equation~\eqref{eq:partialFlatness} and suppressing the explicit dependence on $a$, we find
		\begin{subequations}\label{eq:aRecursion}
			\begin{equation}\label{eq:aRecursion:a}
				\begin{aligned}
					\alpha_i^{(n)}\ &=\ \sum_{j\geq0}\frac1{j!}\sum_{n_1+\cdots+n_j=n-1}\frac{(n-1)!}{n_1!\cdots n_j!}\sum_{k_1+\cdots+k_{j+1}=i}\frac{i!}{k_1!\cdots k_{j+1}!}\mu_{j+1}\big(\alpha_{k_1}^{(n_1)},\ldots,\alpha_{k_j}^{(n_j)},\gamma_{k_{j+1}}\big)
					\\
					&=\ \mu_1(\gamma_i)+\sum_{k_1+k_2=i}\frac{i!}{k_1!k_2!}\mu_2\big(\alpha_{k_1}^{(n-1)},\gamma_{k_2}\big)
					\\
					&\kern1cm+\frac12\sum_{n_1+n_2=n-1}\frac{(n-1)!}{n_1!n_2!}\sum_{k_1+k_2+k_3=i}\frac{i!}{k_1!k_2!k_3!}\mu_2\big(\alpha_{k_1}^{(n_1)},\alpha_{k_2}^{(n_2)},\gamma_{k_3}\big)+\cdots
				\end{aligned}
			\end{equation}
			for all $n,i\in\IN$. Note that the initial condition $a(0)=a$ amounts to
			\begin{equation}\label{eq:aRecursion:b}
				\alpha_i^{(0)}\ =\
				\begin{cases}
					a &\efor i\ =\ 1
					\\
					0 &\eelse
				\end{cases}.
			\end{equation}
		\end{subequations}
		Furthermore, using~\eqref{eq:aAndcExpansions} and~\eqref{eq:aRecursion}, the condition $\sfh(a(1))=0$ becomes
		\begin{subequations}\label{eq:recursionsH(a)}
			\begin{equation}\label{eq:recursionsH(a):a}
				\sfh(a)+\sfh(\mu_1(\gamma_1))\ =\ 0
			\end{equation}
			for $i=1$ and
			\begin{equation}\label{eq:recursionsH(a):b}
				\begin{aligned}
					&\sfh(\mu_1(\gamma_i))+\sum_{j\geq1}\frac1{j!}\sum_{n_1+\cdots+n_j=n-1}\frac{(n-1)!}{n_1!\cdots n_j!}
					\\
					&\kern2cm\times\sum_{k_1+\cdots+k_{j+1}=i}\frac{i!}{k_1!\cdots k_{j+1}!}\sfh\big(\mu_{j+1}\big(\alpha_{k_1}^{(n_1)},\ldots,\alpha_{k_j}^{(n_j)},\gamma_{k_{j+1}}\big)\big)\ =\ 0
				\end{aligned}
			\end{equation}
		\end{subequations}
		for all $i>1$. Hence,~\eqref{eq:recursionsH(a):a} is solved by
		\begin{subequations}\label{eq:cRecursion}
			\begin{equation}\label{eq:cRecursion:a}
				\gamma_1\ =\ -\sfh(a)~,
			\end{equation}
			which is what we have already obtained in the Abelian case around~\eqref{eq:minimalModelMaurerCartanEquation}. Like previously, here we have also made use of $\sfh=\sfh\circ\mu_1\circ\sfh$. This also yields
			\begin{equation}\label{eq:cRecursion:b}
				\begin{aligned}
					\gamma_i\ &=\ -\sum_{j\geq1}\frac1{j!}\sum_{n_1+\cdots+n_j=n-1}\frac{(n-1)!}{n_1!\cdots n_j!}
					\\
					&\kern1cm\times\sum_{k_1+\cdots+k_{j+1}=i}\frac{i!}{k_1!\cdots k_{j+1}!}\sfh\big(\mu_{j+1}\big(\alpha_{k_1}^{(n_1)},\ldots,\alpha_{k_j}^{(n_j)},\gamma_{k_{j+1}}\big)\big)\,.
				\end{aligned}
			\end{equation}
		\end{subequations}
		as a solution to~\eqref{eq:recursionsH(a):b}.

		Altogether, we have obtained a coupled set of recursion relations,~\eqref{eq:aRecursion:a} and~\eqref{eq:cRecursion:b}, together with the initial conditions~\eqref{eq:aRecursion:b} and~\eqref{eq:cRecursion:a}, respectively.

		\paragraph{Solution to recursion relations.}
		Upon iterating these recursion relations, it is not too difficult to see that the first few terms $\gamma_i$ and $\alpha_i^{(n)}$ are given by
		\begin{subequations}
			\begin{equation}
				\begin{aligned}
					\gamma_1(a)\ &=\ -\sfh(a)~,
					\\
					\gamma_2(a,a)\ &=\ 2\sfh(\mu_2(a,\sfh(a)))-\sfh(\mu_2(\mu_1(\sfh(a)),\sfh(a)))~,
					\\
					\gamma_3(a,a,a)\ &=\ 3\sfh(\mu_3(a,a,\sfh(a)))-3\sfh(\mu_3(a,\mu_1(\sfh(a)),\sfh(a)))
					\\
					&\kern1cm+\sfh(\mu_3(\mu_1(\sfh(a)),\mu_1(\sfh(a)),\sfh(a)))-6\sfh(\mu_2(a,\sfh(\mu_2(a,\sfh(a)))))
					\\
					&\kern1cm+3\sfh(\mu_2(a,\sfh(\mu_2(\mu_1(\sfh(a)),\sfh(a)))))+3\sfh(\mu_2(\mu_1(\sfh(a)),\sfh(\mu_2(a,\sfh(a)))))
					\\
					&\kern1cm-\tfrac{3}{2}\sfh(\mu_2(\mu_1(\sfh(a)),\sfh(\mu_2(\mu_1(\sfh(a)),\sfh(a)))))-3\sfh(\mu_2(\mu_2(a,\sfh(a)),\sfh(a)))
					\\
					&\kern1cm-3\sfh(\mu_2(\mu_1(\sfh(\mu_2(a,\sfh(a)))),\sfh(a)))+\sfh(\mu_2(\mu_2(\mu_1(\sfh(a)),\sfh(a)),\sfh(a)))
					\\
					&\kern1cm+\tfrac{3}{2}\sfh(\mu_2(\mu_1(\sfh(\mu_2(\mu_1(\sfh(a)),\sfh(a)))),\sfh(a)))
					\\
					&\kern6pt\vdots
				\end{aligned}
			\end{equation}
			and
			\begin{equation}
				\begin{aligned}
					\alpha_1^{(n)}(a)\ &=\
					\begin{cases}
						-\mu_1(\sfh(a)) & \efor n\ =\ 1
						\\
						0 &\eelse
					\end{cases},
					\\
					\alpha_2^{(n)}(a,a)\ &=\
					\begin{cases}
						\mu_1(\gamma_2(a,a))-2\mu_2(a,\sfh(a)) & \efor n\ =\ 1
						\\
						\mu_2(\mu_1(\sfh(a)),\sfh(a)) & \efor n\ =\ 2
						\\
						0 & \eelse
					\end{cases},
					\\
					\alpha_3^{(n)}(a,a,a)\ &=\
					\begin{cases}
						-3\mu_3(a,a,\sfh(a))+3\mu_2(a,\gamma_2(a,a))+6\mu_1(\gamma_3(a,a,a)) &\efor n\ =\ 1
						\\[6pt]
						\begin{aligned}
							&3\mu_3(a,\mu_1(\sfh(a)),\sfh(a))-\tfrac32\mu_2(\mu_1(\gamma_2(a,a)),\sfh(a))
							\\
							&+3\mu_2(\mu_2(a,\sfh(a)),\sfh(a))-\tfrac32\mu_2(\mu_1(\sfh(a)),\gamma_2(a,a))
						\end{aligned}
						&\efor n\ =\ 2
						\\[15pt]
						-\mu_3(\mu_1(\sfh(a)),\mu_1(\sfh(a)),\sfh(a))-\mu_2(\mu_2(\sfh(a),\sfh(a)),\sfh(a)) &\efor n\ =\ 3
						\\[6pt]
						0 &\eelse
					\end{cases}
					\\
					&\kern6pt\vdots
				\end{aligned}
			\end{equation}
		\end{subequations}
		for the solution to~\eqref{eq:partialFlatness} with $\sfh(a(1))=0$. In conclusion, this verifies explicitly the claim that the condition $\sfh(a)=0$ is indeed a gauge-fixing condition.

		\section{Next-to-lowest-order minimal model Maurer--Cartan equation}\label{app:proofMinmimalMC}

		We shall now provide details on the verification of our claim made in \cref{sec:NTLOKerrSolutions} that the equation~\eqref{eq:nonStraightForwardEquation} always holds by verifying~\eqref{eq:reducedMinimalMCofKerr} for $\rho$ as given in~\eqref{eq:nonStraightForwardEquation}.

		\paragraph{Evaluating the limit.}
		To do so, let us first evaluate the limit in~\eqref{eq:reducedMinimalMCofKerr} for an arbitrary symmetric tensor $\varrho^{(n)}_{ab}(y)$. In particular, using~\eqref{eq:spatialGreenFunctionExtremalKerrN>1} it is not to difficult to see that~\eqref{eq:reducedMinimalMCofKerr} is equivalent to
		\begin{equation}
			\begin{aligned}
				0\ &=\ \lim_{x\to1}(1+x^2)^{-n-1}\sfQ_n^2(x)\int_{-1}^x\rmd x'\,(1+x'^2)^n\sfP_n^2(x')
				\\
				&\kern1cm\times\big[(1-x+x'(1+x))(1-x'+x(1+x'))\bar\varrho^{(n)}_{11}(x',\varphi)
				\\
				&\kern1cm-2(x-x')(1+xx')\bar\varrho^{(n)}_{12}(x',\varphi)\big]
				\\
				0\ &=\ \lim_{x\to1}(1+x^2)^{-n-1}\sfQ_n^2(x)\int_{-1}^x\rmd x'\,(1+x'^2)^n\sfP_n^2(x')
				\\
				&\kern1cm\times\big[(1-x+x'(1+x))(1-x'+x(1+x'))\bar\varrho^{(n)}_{12}(x',\varphi)
				\\
				&\kern1cm+2(x-x')(1+xx')\bar\varrho^{(n)}_{11}(x',\varphi)\big]\,,
			\end{aligned}
		\end{equation}
		where $n\geq 2$, and $\bar\varrho_{ab} \coloneqq \varrho_{ab}-\tfrac{1}{2}\delta_{ab}\delta^{cd}\varrho_{cd}$. In deriving these equation, we have used the fact that $\lim_{x\to1}\int_{-1}^1\rmd x'\,\theta(x'-x)f(x',\varphi)=0$ for any bounded function $f(x,\varphi)$. Since $\sfQ_n^2(x)\sim\frac{1}{1-x}$ as $x\to1$, we can equivalently state that we require both integrals and their first derivatives to vanish at $x=1$. It turns out the only independent equations are
		\begin{subequations}\label{eq:reducedMinimalMCofKerrSimplified}
			\begin{eqnarray}
				0\! &=&\! \int_{-1}^1\rmd x\,(1+x^2)^n\sfP_n^2(x)\big[2x\bar\varrho^{(n)}_{11}(x,\varphi)-(1-x^2)\bar\varrho^{(n)}_{12}(x,\varphi)\big]\,,
				\\
				0\! &=&\! \int_{-1}^1\rmd x\,(1+x^2)^n\sfP_n^2(x)\big[2x\bar\varrho^{(n)}_{12}(x,\varphi)+(1-x^2)\bar\varrho^{(n)}_{11}(x,\varphi)\big]\,,
			\end{eqnarray}
		\end{subequations}
		where $n\geq 2$, and we have relabelled $x'$ by $x$. In our situation, $\bar\varrho^{(n)}_{ab}$ given in~\eqref{eq:defRExpansionOftmu2} will be independent of $\varphi$ because of our assumed axis-symmetry.

		\paragraph{Simplification.}
		Next, we substitute the expression for $\Theta^{\circ}$ given in~\eqref{eq:generalLowestOrderSolutionExtremalKerr} into the formula for $\varrho^{(n)}_{ab}$ given in~\eqref{eq:defRExpansionOftmu2}. After a lengthy but straightforward calculations, one can show that for $n=2$, both equations in~\eqref{eq:reducedMinimalMCofKerrSimplified} are satisfied, whilst for $n>2$, the right-hand sides of these equations are given by
		\begin{subequations}
			\begin{equation}\label{eq:reducedMinimalMCofKerrSimplifiedKerrA}
				\begin{aligned}
					\text{RHS(a)}\ &=\	\int_{-1}^1\rmd x\,\frac{\sfP^2_n(x)}{m^{n+2}}\bigg[nF_1(n,x)AK_1^{n-1}
					\\
					&\kern1cm+\sum_{j\geq2}^{n-2}\frac{n!}{j!(n-j)!}\frac{2(-jn+2j(j+1)+n^2)}{(j+1)n}F_2(n,j,x)K_1^jK_3^{n-j}\bigg]
				\end{aligned}
			\end{equation}
			and
			\begin{equation}\label{eq:reducedMinimalMCofKerrSimplifiedKerrB}
				\begin{aligned}
					\text{RHS(b)}\ &=\ \int_{-1}^1\rmd x\,\frac{\sfP^2_n(x)}{m^{n+2}}\bigg\{nF_1(n,x)AK_3^{n-1}
					\\
					&\kern1cm+\sum_{j\geq2}^{n-2}\frac{n!}{j!(n-j)!}F_2(n,j,x)\bigg[\frac{j^2-jn+2(n-1)n}{(n-1)n}K_3^{j}K_3^{n-j}
					\\
					&\kern1cm+\bigg(\frac{j^2}{(n-1)n}-\frac{j}{n-j+1}+\frac{j}{j+1}-\frac{j}{n-1}-\frac{n}{j+1}\bigg)K_1^{j}K_1^{n-j}\bigg]\bigg\}\,,
				\end{aligned}
			\end{equation}
			where
			\begin{equation}
				\begin{aligned}
					F_1(n,x)\ &\coloneqq\ -\frac{(2n-1)}{5n(n+2)(1+x^2)^3}\big[\big(-2nx^2(-51+75x^2+3x^4+9x^6+4x^8)
					\\
					&\kern1cm+2(-18+78x^2+51x^4+17x^6-5x^8-3x^{10})
					\\
					&\kern1cm+n^2(9-60x^2-18x^4+92x^6+49x^8+8 x^{10})\big)\sfP^2_n(x)
					\\
					&\kern1cm-2x\big(-30+41x^2+15x^4+11x^6+3x^8
					\\
					&\kern1cm+n^2(-9+7x^2+27x^4+13x^6+2x^8)
					\\
					&\kern1cm-n(-39+48x^2+42x^4+24x^6+5x^8)\big)\sfP_{n+1}^2(x)\big]\,
					\\
					F_2(n,j,x)\ &\coloneqq\ \frac{(j-1)(n-j-1)\sfP_{j+1}^2(x)\sfP_{n-j+1}^2(x)}{2(1+x^2)}
					\\
					&\kern1cm-\frac{(n-j-1)x\big(j-3+(j+1)x^2\big)\sfP_{j}^2(x)\sfP_{n-j+1}^2(x)}{2(1+x^2)^2}
					\\
					&\kern1cm-\frac{(j-1)x\big(n-j-3+(n-j+1)x^2\big)\sfP_{n+1}^2(x)\sfP_{n-j}^2(x)}{2(1 +x^2)^2}
					\\
					&\kern1cm+\frac{\big[j(n-j)(1+x^2)^2-x^2(3-x^2)\big(n-3+(1+n)x^2\big)\big]\sfP_j^2(x)\sfP_{n-j}^2(x)}{2(1+x^2)^3}~.
				\end{aligned}
			\end{equation}
		\end{subequations}
		In deriving these expressions, we have made use of the
		recursion relations of associated Legendre polynomials,
		\begin{subequations}
			\begin{eqnarray}
				(q-1)\sfP_{q+1}^2(x)\! &=&\! (2q+1)x\sfP_q^2(x)-(q+2)\sfP_{q-1}^2(x)\,,\label{eq:xP}
				\\
				(1-x^2)\partial_x\sfP_q^2(x)\! &=&\! \tfrac1{2q+1}\big[(q+1)(q+2)\sfP_{q-1}^2(x)-q(q-1)\sfP_{q+1}^2(x)\big]\label{eq:dP/dx}
			\end{eqnarray}
		\end{subequations}
		for all $q\in\mathbb{Z}$.

		Note that the reason for terms proportional to $AK_3^{n-1}$, $K_1^{j}K_1^{n-j}$, and $K_3^{j}K_3^{n-j}$ being absent in~\eqref{eq:reducedMinimalMCofKerrSimplifiedKerrA} is because they are multiplied by odd functions with respect to $x$ and so, the integrals vanishes. Terms not appearing in~\eqref{eq:reducedMinimalMCofKerrSimplifiedKerrB} vanish for the same reason as well.

		In remainder of proof, we will argue that
		\begin{subequations}\label{eq:intF12P}
			\begin{eqnarray}
				\int_{-1}^1\rmd x\,\sfP_n^2(x)F_1(n,x)\! &=&\! 0~,\label{eq:intF1P}
				\\
				\int_{-1}^1\rmd x\,\sfP_n^2(x)F_2(n,j,x)\! &=&\! 0\label{eq:intF2P}
			\end{eqnarray}
		\end{subequations}
		for all $n\geq 2$ and $2\leq j\leq n-2$ and which, in turn, verifies~\eqref{eq:reducedMinimalMCofKerrSimplified}
		and thus~\eqref{eq:nonStraightForwardEquation}, as claimed. This will occupy us for the remainder of this section.

		\paragraph{Notation and conventions.}
		To verify~\eqref{eq:intF12P}, we shall need some extra notation. Since the $P_n^2$ for $n\geq2$ form an orthogonal basis in the vector space of functions vanishing at $x=\pm1$, there is a unique decomposition
		\begin{equation}\label{eq:defOfCAsCoefficient}
			\begin{aligned}
				\frac{\sfP_n^2(x)}{(x^2-a^2)^p}\ =\ \sum_{q\geq2}C_p^{n,q}(a)\sfP_q^2(x)~,
			\end{aligned}
		\end{equation}
		where $a\in\rmi\IR$ with $\rmi$ is the imaginary unit. To be more precise, for all $n,q\geq2$ and $p\in\IZ$,
		\begin{subequations}\label{eq:defOfCp}
			\begin{equation}
				C_p^{n,q}(a)\ \coloneqq\ \frac1{N(q)}\int_{-1}^1\rmd x\,\frac1{(x^2-a^2)^p}\sfP^2_n(x)\sfP_q^2(x)\ =\ C_p^{q,n}(a)\frac{N(q)}{N(n)}~,
		   \end{equation}
		   where
		   \begin{equation}\label{eq:NormalisationOfIntPP}
				N(q)\ \coloneqq\ \int_{-1}^1\rmd x\,\sfP_q^2(x)\sfP_q^2(x)\ =\ \frac{2(q+2)(q+1)q(q-1)}{2q+1}~.
			\end{equation}
		\end{subequations}
		For convenience, we will also define $C_p^{n,q}(a)\coloneqq0$ for all $n$ or $q<2$.

		Upon multiplying~\eqref{eq:defOfCAsCoefficient} by $(x^2-a^2)^p\sfP_l^2(x)$, using~\eqref{eq:xP}, and performing the integrations on both sides, we can deduce recursion relations,
		\begin{itemize}
			\item $p=1$:
				\begin{subequations}\label{eq:recursionRelationPower-1}
				\begin{equation}\label{eq:CRecursionp=1}
					\delta^{nq}\ =\ f^-(q)C^{n,q-2}_1(a)+f^0(a,q)C^{n,q}_1(a)+f^+(q)C^{n,q+2}_1(a)~,
				\end{equation}
				where
				\begin{equation}
					\begin{gathered}
						f^-(q)\ \coloneqq\ \frac{(q-3)(q-2)}{(2q-3)(2q-1)}~,
						\quad
						f^+(q)\ \coloneqq\ \frac{(q+3)(q+4)}{(2q+3)(2q+5)}
						\\
						f^0(a,q)\ \coloneqq\ \frac{2q^2+2q-9}{(2q-1)(2q+3)}-a^2~,
					\end{gathered}
				\end{equation}
			\item $p=3$ and $a=\rmi$:
				\begin{equation}\label{eq:CRecursionp=3}
					\begin{aligned}
						\delta^{nq}\ &=\ g_1(q)C_3^{n,q-6}(\rmi)+g_2(q)C_3^{n,q-4}(\rmi)+g_3(q)C_3^{n,q-2}(\rmi)+g_4(q)C_3^{n,q}(\rmi)
						\\
						&\kern1cm+g_5(q)C_3^{n,q+2}(\rmi)+g_6(q)C_3^{n,q+4}(\rmi)+g_7(q)C_3^{n,q+6}(\rmi)~,
					\end{aligned}
				\end{equation}
				where
				\begin{equation}
					\begin{aligned}
						g_1(q)\ &\coloneqq\ \frac{(q-7)(q-6)(q-5)(q-4)(q-3)(q-2)}{(2q-11)(2 q-9)(2q-7)(2q-5)(2q-3)(2q-1)}~,
						\\
						g_2(q)\ &\coloneqq\ \frac{18(q-5)(q-4)(q-3)(q-2)(q^2-3q-8)}{(2q-9)(2q-7)(2q-5)(2q-3)(2q-1)(2q+3)}~,
						\\
						g_3(q)\ &\coloneqq\ \frac{3(q-3)(q-2)(37q^4-74q^3-523q^2+560q+2100)}{(2q-7)(2q-5)(2q-3)(2q-1)(2q+3)(2 q+5)}~,
						\\
						g_4(q)\ &\coloneqq\ \frac{36(7q^6+21q^5-103q^4-241q^3+586q^2+710q-1400)}{(2 q-5)(2q-3)(2q-1)(2q+3)(2q+5)(2q+7)}~,
						\\
						g_5(q)\ &\coloneqq\ \frac{3(q+3)(q+4)(37q^4+222q^3-79q^2-1236q+1128)}{(2q-3)(2q-1)(2q+3)(2q+5)(2q+7)(2q+9)}~,
						\\
						g_6(q)\ &\coloneqq\ \frac{18(q+3)(q+4)(q+5)(q+6)(q^2+5q-4)}{(2q-1)(2q+3)(2q+5)(2q+7)(2q+9)(2q+11)}~,
						\\
						g_7(q)\ &\coloneqq\ \frac{(q+3)(q+4)(q+5)(q+6)(q+7)(q+8)}{(2q+3)(2q+5)(2q+7)(2q+9)(2q+11)(2q+13)}
					\end{aligned}
				\end{equation}
			\end{subequations}
		\end{itemize}
		for all $n,q\geq2$.

		One can show by substitution that
		\begin{equation}\label{eq:solCnqq<=n}
			\begin{aligned}
				C^{n,q}_1(a)\ &=\ \frac{C^{n,2}_1(a)}{10\sfP_2^{-2}(a)}(2q+1)\big[\sfP_q^{-2}(a)+\sfP_q^{-2}(-a)\big]
				\\
				&\kern1cm+\frac{C^{n,3}_1(a)}{14\sfP_3^{-2}(a)}(2q+1)\big[\sfP_q^{-2}(a)-\sfP_q^{-2}(-a)\big]
			\end{aligned}
		\end{equation}
		for all $n\geq 2$, $2\leq q\leq n$\footnote{$C_1^{n,n}(a)$ is determined by the recursion relation with $q=n-2$} and $a\neq0$ since it solves the recursion relation~\eqref{eq:CRecursionp=1} for $q<n$.

		Finally, by taking two derivatives of $C_1^{n,q}(a)$ in~\eqref{eq:defOfCp} with respect to $a$ and evaluating at $a=\rmi$, one can easily see that
		\begin{equation}\label{eq:C3toC1}
			\frac1{N(q)}\int_{-1}^1\rmd x\,\frac1{(1+x^2)^3}\sfP^2_n(x)\sfP_q^2(x)\ \eqqcolon\ C_3^{n,q}(\rmi)\ =\ -\tfrac18(\partial^2_a+a\partial_a)\big|_{a=\rmi}C_1^{n,q}(a)
		\end{equation}
		all for all $n,q\geq 2$.

		\paragraph{Proof of~\eqref{eq:intF1P}.}
		Using~\eqref{eq:xP}, we can rewrite the integral in~\eqref{eq:intF1P} as
		\begin{equation}\label{eq:PPintegral}
			\begin{aligned}
				\int_{-1}^1\rmd x\,\sfP_n^2(x)F_1(n,x)\ &=\
				\begin{cases}
					\sum_{l=-10}^{10}\chi^{n,l}\int_{-1}^1\rmd x\,\frac1{(1+x^2)^3}\sfP_n^2(x)\sfP_{n+l}^2(x) & \efor n\geq12
					\\
					\sum_{l=-n+2}^{10}\chi^{n,l}\int_{-1}^1\rmd x\,\frac1{(1+x^2)^3}\sfP_n^2(x)\sfP_{n+l}^2(x) & \efor 2\leq n\leq11
				\end{cases}
				\\
				&=\ \sum_{l=-10}^{10}\chi^{n,l}N(n+l)C_3^{n,n+l}(\rmi)~,
			\end{aligned}
		\end{equation}
		where the $\chi^{n,l}$ are independent of $x$, and $N$ was defined in~\eqref{eq:NormalisationOfIntPP}. The reason the first line splits into two cases is that when using~\eqref{eq:xP} to eliminate the explicit $x$-dependence in $x^w\sfP_n^2(x)$, one finds that $x^w\sfP_n^2(x)$ can be written as $\sum_{l\geq-w}^w\caZ^l\sfP_{n+l}^2(x)$ for $n\geq w+2$, or as $\sum_{l\geq n-2}^w\caZ^l\sfP_{n+l}^2(x)$ for $n<w+2$ for some constants $\caZ^l$ since $\sfP_1^2(x)=0$.

		We can reformulate~\eqref{eq:PPintegral} in terms of $\partial^2_a\big|_{a=\rmi}C_1^{n,2}(a)$, $\partial^2_a\big|_{a=\rmi}C_1^{n,3}(a)$, $\partial_a\big|_{a=\rmi}C_1^{n,2}(a)$, $\partial_a\big|_{a=\rmi}C_1^{n,3}(a)$, $ C_1^{n,2}(\rmi)$, and $C_1^{n,3}(\rmi)$ by first using~\eqref{eq:CRecursionp=3} to write $C_3^{n,n+l}(\rmi)$ with $l>0$ in terms of those with $l\leq0$ and then using~\eqref{eq:C3toC1} and~\eqref{eq:solCnqq<=n}. We obtain
		\begin{equation}
			\begin{aligned}
				&\int_{-1}^1\rmd x\,\sfP_n^2(x)F_1(n,x)
				\\
				&\kern.5cm=\ \frac{32}{175}(n-1)(2n^2+n-1)\big[(1+2n)\sfP_n^{-2}(a)+a(3+n)\sfP_{n+1}^{-2}(a)\big]
				\\
				&\kern1.5cm\times\big\{-7(1+(-1)^n)\big[2\big(\partial_a^2-a\partial_a\big)C_1^{n,2}(a)+n(n+1)C_1^{n,2}(a)\big]
				\\
				&\kern1.5cm+5a(1-(-1)^n)\big[2\big(\partial_a^2+a\partial_a\big)C_1^{n,3}(a)+\left(n^2+n-2\right)C_1^{n,3}(a)\big]\big\}\bigg|_{a=\rmi}
			\end{aligned}
		\end{equation}
		for all $n\geq2$. We now claim that
		\begin{subequations}\label{eq:identityC1n2C1n3AndDerivatives}
			\begin{eqnarray}
				\big[2\big(\partial_a^2-a\partial_a\big)C_1^{n,2}(a)+n(n+1)C_1^{n,2}(a)\big]\Big|_{a=\rmi}\! &=&\! 0~,\label{eq:identityC1n2C1n3AndDerivatives:a}
				\\
				\big[2\big(\partial_a^2+a\partial_a\big)C_1^{n,3}(a)+\left(n^2+n-2\right)C_1^{n,3}(a)\big]\Big|_{a=\rmi}\! &=&\! 0\label{eq:identityC1n2C1n3AndDerivatives:b}
			\end{eqnarray}
		\end{subequations}
		for all $n\geq2$. We will only verify the second equation since the first one can be proved in a similar manner.

		To verify~\eqref{eq:identityC1n2C1n3AndDerivatives:b}, we start with an equivalent statement
		\begin{equation}\label{eq:identityC1n2C1n3AndDerivatives:aSwapedIndices}
			\big[2\big(\partial_a^2+a\partial_a\big)C_1^{3,n}(a)+\left(n^2+n-2\right)C_1^{3,n}(a)\big]\Big|_{a=\rmi}\ =\ 0
		\end{equation}
		for all $n\geq 2$. One can check explicitly that~\eqref{eq:identityC1n2C1n3AndDerivatives:aSwapedIndices} holds for $n=2,\ldots,5$. We will then prove by induction that the statement is also true for $n >5$. We first assume that the statement~\eqref{eq:identityC1n2C1n3AndDerivatives:aSwapedIndices} is true for $n<\bar n$ for a particular value of $\bar n>5$ as the induction hypothesis. To show that~\eqref{eq:identityC1n2C1n3AndDerivatives:aSwapedIndices} is true for $n=\bar n$, we use the recursion relation~\eqref{eq:CRecursionp=1} to write the left-hand side of~\eqref{eq:identityC1n2C1n3AndDerivatives:aSwapedIndices} as
		\begin{equation}\label{eq:ProofOfIdentityC1n2C1n3AndDerivatives:bInductionStep1}
			\begin{aligned}
				&\big[2\big(\partial_a^2+a\partial_a\big)C_1^{3,\bar{n}}(a)+\left(\bar{n}^2+\bar{n}-2\right)C_1^{3,\bar{n}}(a)\big]\Big|_{a=\rmi}
				\\
				&=\ -\frac{\bar n^2+\bar n-2}{f^+(\bar n-2)}\big[f^-(\bar n-2)C_1^{3,\bar n-4}(a)+f^0(a,\bar n-2)C_1^{3,\bar n-2}(a)\big]
				\\
				&\kern1cm-\frac{2a}{f^+(\bar n-2)}\big[f^-(\bar n-2)\partial_aC_1^{3,\bar n-4}(a)+f^0(a,\bar n-2)\partial_aC_1^{3,\bar n-2}(a)-2aC_1^{3,\bar n-2}(a)\big]
				\\
				&\kern1cm -\frac{2}{f^+(\bar n-2)}\big[f^-(\bar n-2)\partial^2_aC_1^{3,\bar n-4}(a)+f^0(a,\bar n-2)\partial^2_aC_1^{3,\bar n-2}(a)
				\\
				&\kern2cm-2C_1^{3,\bar n-2}(a)-4a\partial_aC_1^{3,\bar n-2}(a)\big]\bigg|_{a=\rmi}\,.
			\end{aligned}
		\end{equation}
		By the induction hypothesis, we can now use \eqref{eq:identityC1n2C1n3AndDerivatives:aSwapedIndices}, when $n=\bar n-2$ and $\bar n-4$, to write the second derivative of $C^{3,\bar n-2}_1(a)$ and $C^{3,\bar n-4}_1(a)$ as their first derivative and no-derivative at $a=\rmi$. We obtain
		\begin{equation}
			\begin{aligned}
				&\big[2\big(\partial_a^2 +a\partial_a\big)C_1^{3,n}(a)+\left(n^2+n-2\right)C_1^{3,n}(a)\big]\Big|_{a=\rmi}
				\\
				&=\ -\frac{2}{f^+(\bar n-2)}\big[(4\bar n-6)f^-(\bar n-2)C_1^{3,\bar n-4}(a)+(2\bar n-1)f^0(a,\bar n-2)C_1^{3,\bar n-2}(a)
				\\
				&\kern1cm-4a\partial_aC_1^{3,\bar n-2}(a)\big]\bigg|_{a=\rmi}\,.
			\end{aligned}
		\end{equation}
		The terms in the square brackets on the right-hand side can easily be shown to be zero by induction \footnote{$\bar n=4,\ldots,7$ can be shown to be zero explicitly. We first assume that the statement holds for $\bar n <\bar n'$ for a particular value of $\bar n'\geq 7$ as the induction hypothesis. To proof the statement for $\bar n=\bar n'$, one can use the recursion relation~\eqref{eq:CRecursionp=1} (similar to how we obtain \eqref{eq:ProofOfIdentityC1n2C1n3AndDerivatives:bInductionStep1}) to write the statement when $\bar n=\bar n'$ in terms of those with $\bar n<\bar n'$. Then, one can see that by the induction hypothesis the statement holds for $\bar n=\bar n'$.}. Therefore, we can conclude that~\eqref{eq:identityC1n2C1n3AndDerivatives:aSwapedIndices} is true by induction, which in turn implies that~\eqref{eq:identityC1n2C1n3AndDerivatives:b} is true.

		\paragraph{Proof of~\eqref{eq:intF2P}.}
		To prove~\eqref{eq:intF2P}, we first write a product of associated Legendre polynomials as
		\begin{subequations}
			\begin{equation}\label{eq:defOfBCoefficient}
				\frac{\sfP_p^2(x)\sfP_{q}^2(x)}{1-x^2}\ =\ \sum_{w\geq 2}\frac{1}{N(w)}B_{p,q}^w\sfP_w^2(x)~,
			\end{equation}
			where
			\begin{equation}\label{eq:defOfB}
				B_{p,q}^w\ \coloneqq\ \int_{-1}^1\rmd x\,\frac{\sfP_p^2(x)\sfP_q^2(x)\sfP_w^2(x)}{1-x^2}
			\end{equation}
		\end{subequations}
		for all $p,q\geq 2$. This is similar to the definition~\eqref{eq:defOfCp} of $C_p^{n,q}$.

		We can now use~\eqref{eq:defOfBCoefficient} and~\eqref{eq:xP} to write the right-hand side of~\eqref{eq:intF2P} as
		\begin{equation}
			\begin{aligned}
				\int_{-1}^1\rmd x\,\sfP_n^2(x)F_2(n,j,x)\ &=\
				\sum_{l=-w+2}^8\sum_{w\geq2}^{9}\caY^{n,j,w,l}\int_{-1}^1\rmd x\,\frac{\sfP_{w+l}^2(x)\sfP_n^2(x)}{(1+x^2)^3}
				\\
				&\kern1cm+\sum_{l=-8}^8\sum_{w\geq10}\caY^{n,j,w,l}\int_{-1}^1\rmd x\,\frac{\sfP_{w+l}^2(x)\sfP_n^2(x)}{(1+x^2)^3}
				\\
				&=\
				\sum_{l=-w+2}^8\sum_{w\geq2}\caY^{n,j,w,l}N(w+l)C_3^{n,w+l}~,
			\end{aligned}
		\end{equation}
		where $n\geq2$ and $2\leq j\leq n-2$, and $\caY^{n,j,w,l}$ are independent of $x$. The reason for the splitting of the summation over $w$ in the first equality is the same as in~\eqref{eq:PPintegral}. Again we can repeat the calculation as in the case of~\eqref{eq:intF1P} and use~\eqref{eq:identityC1n2C1n3AndDerivatives} to write everything in terms of $\partial^2_a\big|_{a=\rmi}C_1^{n,2}(a)$, $\partial^2_a\big|_{a=\rmi}C_1^{n,3}(a)$, $\partial_a\big|_{a=\rmi}C_1^{n,2}(a)$, and $\partial_a\big|_{a=\rmi}C_1^{n,3}(a)$. After a lengthy but straightforward calculation, one realises that the right-hand side of~\eqref{eq:intF2P} can be simplified to
		\begin{subequations}\label{eq:defOfcaF}
			\begin{equation}
				\begin{aligned}
					\int_{-1}^1\rmd x\,\sfP_n^2(x)F_2(n,j,x)\ &=\ \frac8{7(n-1)(n+2)}\big\{\caF_1(n,j)\partial_aC_1^{n,3}(a)
					\\
					&\kern1cm-[a\caF_1(n,j)-(m+2)(m-1)\caF_2(n,j)]\partial^2_aC_3^{n,3}(a)\big\}
					\\
					&\kern1cm+\frac8{5n(n+1)}\big\{\caF_3(n,j) \partial_aC_1^{n,2}(a)
					\\
					&\kern1cm+[a\caF_3(n,j)+m(m+1)\caF_4(n,j)]\partial^2_aC_1^{n,2}(a)\big\}\bigg|_{a=\rmi}\,,
				\end{aligned}
			\end{equation}
			where
			\begin{equation}
				\begin{aligned}
					\caF_1(n,j)\ &\coloneqq\ \sum_{w\geq1}\big\{\sfP_{2w+1}^{-2}(\rmi)\big[(4 w+3)(j^2-jn-5n^2-n^3-2n^2w+2n+4w^2
					\\
					&\kern1cm
					+14w+13)B_{j,n-j}^{2w+1}+(2w+3)(4w+1)(n-j-1)B_{j,n-j+1}^{2w}
					\\
					&\kern1cm+(j-1)(2 w+3)(4w+1)B_{j+1,n-j}^{2w}
					\\
					&\kern1cm-(4w+3)(j-1)(n-j-1)B_{j+1,n-j+1}^{2w+1}\big]
					\\
					&\kern1cm+i\sfP_{2 w}^{-2}(\rmi)\big[-(n-2)(n+2)(2 w-1)(4 w+3)B_{j,n-j}^{2w+1}
					\\
					&\kern1cm+(4w+1)(n-j-1)(j-n^2-n-2w+2)B_{j,n-j+1}^{2w}
					\\
					&\kern1cm-(4w+1)(j-1)(j+n^2+2w-2)B_{j+1,n-j}^{2w}\big]\big\}
				\end{aligned}
			\end{equation}
			and
			\begin{equation}
				\begin{aligned}
					\caF_2(n,j)\ &\coloneqq\ \sum_{w\geq1}\big\{\sfP_{2w}^{-2}(\rmi)\big[(2w-1)(4w+3)B_{j,n-j}^{2w+1}+(4w+1)(n-j-1)B_{j,n-j+1}^{2w}
					\\
					&\kern1cm+(4w+1)(j-1)B_{j+1,n-j}^{2w}\big]-\rmi\sfP_{2w+1}^{-2}(\rmi)(4 w+3)(n+2w+5)B_{j,n-j}^{2w+1}\big\}
				\end{aligned}
			\end{equation}
			and
			\begin{equation}
				\begin{aligned}
					\caF_3(n,j)\ &\coloneqq\ \sum_{w\geq1}\big\{i\sfP_{2 w}^{-2}(\rmi)\big[-(4w+1)\big(j^2-jn+8n+5n^2+n^3-4nw-2n^2w
					\\
					&\kern1cm+4w^2-6w+3\big)B_{j,n-j}^{2w}+(2w-1)(4w+3)(n-j-1)B_{j,n-j+1}^{2w+1}
					\\
					&\kern1cm+(2w-1)(4w+3)(j-1)B_{j+1,n-j}^{2w+1}
					\\
					&\kern1cm+(4w+1)(j-1)(n-j-1)B_{j+1,n-j+1}^{2w}\big]
					\\
					&\kern1cm+\sfP_{2w+1}^{-2}(\rmi)\big[(- (n^2+2n+4)(2w+3)(4w+1)B_{j,n-j}^{2w}
					\\
					&\kern1cm+(4w+3)(n-j-1)(j+n^2+n+2w+4)B_{j,n-j+1}^{2w+1}
					\\
					&\kern1cm+(4w+3)(j-1)(-j+n^2+2n+2w+4)B_{j+1,n-j}^{2w+1}\big])\big\}
				\end{aligned}
			\end{equation}
			and
			\begin{equation}
				\begin{aligned}
					\caF_4(n,j)\ &\coloneqq\ \sum_{w\geq1}\big\{\rmi\sfP_{2w+1}^{-2}(\rmi)\big[(3+2w)(1+4w)B_{j,n-j}^{2w}-(4w+3)(n-j-1)B_{j,n-j+1}^{2w+1}
					\\
					&\kern1cm-(j-1)(4w+3)B_{j+1,n-j}^{2w+1}\big]-(4w+1)(n-2 w+3)B_{j,n-j}^{2w}\sfP_{2w}^{-2}(\rmi)\big\}~,
				\end{aligned}
			\end{equation}
		\end{subequations}
		where $n\geq2$ and $2\leq j\leq n-2$. Note that we have used
		\begin{equation}\label{eq:xPminus2}
			(q+3)\sfP_{q+1}^{-2}(x)\ =\ (2q+1)x\sfP_q^{-2}(x)-(q-2)\sfP_{q-1}^{-2}(x)
		\end{equation}
		for all $q\in\IZ$ to write $P_{2w+v}^{-2}(i)$ for some integer $v$ in terms of $P_{2w}^{-2}(i)$ and $P_{2w+1}^{-2}(i)$. Next, one can show that
		\begin{subequations}\label{eq:caFAsTotalDerivative}
			\begin{equation}
				\begin{aligned}
					\caF_1(n,j)\ &=\ \sum_{w \geq1}\big\{\big[\tfrac12\rmi(4w+1)(2n^2+n-2w-4)\sfP^{-2}_{2w}(\rmi)+(w-1)(4w+1)\sfP^{-2}_{2w-1}(\rmi)\big]
					\\
					&\kern1cm\times\int_{-1}^1\rmd x\,\partial_x\big[\sfP_j^{2}(x)\sfP_{n-j}^{2}(x)\sfP_{2w}^{2}(x)\big]
					\\
					&\kern1cm-\tfrac12(4w+3)\sfP^{-2}_{2w+1}(\rmi)\int_{-1}^1\rmd x\,\partial_x\big[(1-j)\sfP_{j+1}^2(x)\sfP_{n-j}^2(x)\sfP_{2w+1}^2(x)
					\\
					&\kern1cm+(1-n+j)\sfP_j^2(x)\sfP_{n-j+1}^2(x)\sfP_{2w+1}^2(x)\big]\big\}
				\end{aligned}
			\end{equation}
			and
			\begin{equation}
				\caF_2(n,j)\ =\ \sum_{w\geq1}\big\{-(4w+1)\sfP^{-2}_{2w}(\rmi)\int_{-1}^1\rmd x\,\partial_x\big[\sfP_j^{2}(x)\sfP_{n-j}^2(x)\sfP_{2w}^2(x)\big]\big\}
			\end{equation}
			and
			\begin{equation}
				\begin{aligned}
					\caF_3(n,j)\ &=\ \sum_{w\geq1}\big\{\big[-\tfrac12(4w+3)(2n^2+3n+2w+5)\sfP^{-2}_{2w+1}(\rmi)-\tfrac12\rmi(2w-1)(4w+3)\sfP^{-2}_{2w}(\rmi)\big]
					\\
					&\kern1cm\times\int_{-1}^1\rmd x\,\partial_x\big[\sfP_j^2(x)\sfP_{n-j}^2(x)\sfP_{2w+1}^2(x)\big]
					\\
					&\kern1cm+\tfrac12\rmi(4w+5)\sfP^{-2}_{2w+2}(\rmi)\int_{-1}^1\rmd x\,\partial_x\big[(1-j)\sfP_{j+1}^2(x)\sfP_{n-j}^2(x)\sfP_{2w+2}^2(x)
					\\
					&\kern1cm+(1-n+j)\sfP_j^2(x)\sfP_{n-j+1}^2(x)\sfP_{2w+2}^2(x)\big]\big\}
				\end{aligned}
			\end{equation}
			and
			\begin{equation}
				\caF_4(n,j)\ =\ \sum_{w\geq1}\big\{\rmi(4w+3)\sfP^{-2}_{2w+1}(\rmi)\int_{-1}^1\rmd x\,\partial_x\big[\sfP_j^2(x)\sfP_{n-j}^2(x)\sfP_{2w+1}^2(x)\big]\big\}
			\end{equation}
		\end{subequations}
		for all $n\geq2$ and $2\leq j\leq n-2$. Indeed, as an example, let us verify the expression for $\caF_2(n,j)$ in~\eqref{eq:caFAsTotalDerivative}; the verification of the remaining expressions is follows similar lines. The right-hand side of $\caF_2(n,j)$ in~\eqref{eq:caFAsTotalDerivative} can be written as
		\begin{equation}
			\begin{aligned}
				\text{RHS}_{\caF_2}\ &=\ \sum_{w\geq1}\big\{ - \sfP^{-2}_{2w}(\rmi)\big[2(w+1)(n+2w+3)B_{j,n-j}^{2w-1}+(2w-1)(n-2w+2)B_{j,n-j}^{2w+1}
				\\
				&\kern1cm+(4w+1)(1-n+j)B_{j,n-j+1}^{2w}+(4w+1)(1-j)B_{j+1,n-j}^{2 w}\big]\big\}
			\end{aligned}
		\end{equation}
		for all $n\geq2$ and $2\leq j\leq n-2$, where we have used~\eqref{eq:dP/dx} to compute the derivative and used~\eqref{eq:defOfB} to perform the integration. Now we perform a shift $w\rightarrow w+1$ to the term involving $B_{j,n-j}^{2w-1}$ and so,
		\begin{equation}
			\begin{aligned}
				\text{RHS}_{\caF_2}\ &=\ \sum_{w\geq1}\big\{-2\sfP^{-2}_{2w+2}(\rmi)(w+2)(n+2 w+5)B_{j,n-j}^{2w+1}
				\\
				&\kern1cm-\sfP^{-2}_{2w}(\rmi)\big[(2w-1)(n-2 w+2)B_{j,n-j}^{2w+1}+(4w+1)(1-n+j)B_{j,n-j+1}^{2w}
				\\
				&\kern1cm+(4w+1)(1-j)B_{j+1,n-j}^{2w}\big]\big\}
				\\
				&=\ \sum_{w\geq1}\big\{\sfP_{2w}^{-2}(\rmi)\big[(2w-1)(4w+3)B_{j,n-j}^{2w+1}+(4w+1)(n-j-1)B_{j,n-j+1}^{2w}
				\\
				&\kern1cm+(4w+1)(j-1)B_{j+1,n-j}^{2w}\big]-\rmi\sfP_{2w+1}^{-2}(\rmi)(4 w+3)(n+2w+5)B_{j,n-j}^{2w+1}\big\}\,,
			\end{aligned}
		\end{equation}
		where~\eqref{eq:xPminus2} is used to obtain the second equality. This agrees with the definition of $\caF_2(n,j)$ in~\eqref{eq:defOfcaF}. Note that we do not need to change the range of the summation since $B^1_{j,n-j}=0$.

		Finally, since all of the expressions in~\eqref{eq:caFAsTotalDerivative} are total derivatives, they all vanish as the associated Legendre polynomials vanish at $x=\pm1$.

	\end{body}

\end{document}